%% file: manuscript.tex
\documentclass[a4paper]{elsart}
\usepackage{changebar,dsfont,amsmath,tabularx,color}

\usepackage{pifont}
\usepackage[numbers,sort&compress]{natbib}
\usepackage[final]{graphicx}
\usepackage[colorlinks=true,citecolor=green,linkcolor=blue,pagecolor=black,urlcolor=red]{hyperref}
\usepackage{hypernat}
\DeclareGraphicsExtensions{.eps}
\definecolor{blue}{rgb}{0.2,0.2,0.4}
\definecolor{red}{rgb}{0.793,0.238,0.336}
\definecolor{green}{rgb}{0.0238,0.4,0.0}
\definecolor{black}{rgb}{0,0,0}

\newcommand{\refeq}[1]{(\ref{#1})}

\newcommand{\reffig}[1]{Fig.~\ref{#1}}

\usepackage[utf8]{inputenc}  
\newif\ifpdf
\ifx\undefined\pdfoutput \pdffalse \else \pdftrue \fi
\ifpdf
  \DeclareGraphicsExtensions{.pdf}
\fi
\DeclareGraphicsExtensions{.eps}

\begin{document}
\begin{frontmatter}

\title{Recurrence Plots for the Analysis of Complex Systems}

\author{Norbert Marwan\corauthref{cor}},
\corauth[cor]{Corresponding author.}
\author{M. Carmen Romano},
\author{Marco Thiel},
\author{J\"urgen Kurths}
\address{Nonlinear Dynamics Group, Institute of Physics, 
University of Potsdam, \\Potsdam 14415, Germany\\
marwan@agnld.uni-potsdam.de}

\begin{abstract}
\end{abstract}
\end{frontmatter}

\tableofcontents
\setcounter{tocdepth}{4}
\clearpage

\section*{List of Abbreviations and Symbols}\label{sec:ListofAbbreviationsandSymbols}
\input{abbreviations.tex}

\newpage
\section{Introduction}\label{sec:Introduction}
\input{introduction.tex}

\section{Theory}\label{sec:Theory}
\input{theory.tex}

\section{Methods}\label{sec:Methods}

Now we will use the concept of recurrence for
the analysis of data and to study dynamical systems.
Nonlinear data analysis is based on the study of phase space trajectories. 
At first, we introduce the concept of phase 
space reconstruction (Subsec.~\ref{sec:PhaseSpaceTrajectories}) 
and then give a technical and brief
historical review on recurrence plots (Subsec.~\ref{sec:RecurrencePlots}).
This part is followed by the bivariate extension to cross
recurrence plots (Subsec.~\ref{sec:CrossRecurrencePlots}) and 
the multivariate extension to
joint recurrence plots (Subsec.~\ref{sec:JointRecurrencePlots}). 
Then we describe measures of
complexity based on recurrence/cross recurrence plots 
(Subsec.~\ref{sec:RQA}) and
how dynamical invariants can be derived from RPs
(Subsec.~\ref{sec:Invariants}). 
Moreover, the potential of RPs for the analysis of spatial data, the detection of
UPOs, detection and quantification of different kinds of 
synchronisation and the creation of surrogates to test 
for synchronisation is presented
(Subsecs.~\ref{sec:SpatialData}--\ref{sec:UPOs}). 
Before we describe several applications, we end the
methodological section considering the influence of noise
(Subsec.~\ref{sec:Noise}).

Most of the described methods and procedures are available
in the {\it CRP toolbox} for Matlab$^{\tiny{\textregistered}}$
(provided by TOCSY: 
\href{http://tocsy.agnld.uni-potsdam.de}{http://tocsy.agnld.uni-potsdam.de}).

    \subsection{Trajectories in phase space}\label{sec:PhaseSpaceTrajectories}
    \input{meth_phasespace.tex}

    \subsection{Recurrence plot (RP)}\label{sec:RecurrencePlots}
    \input{meth_recplots.tex}

    %

    \subsection{Cross recurrence plot (CRP)}\label{sec:CrossRecurrencePlots}
    \input{meth_crp.tex}

    \subsection{Joint recurrence plot (JRP)}\label{sec:JointRecurrencePlots}
\input{meth_jrp.tex}

    \subsection{Measures of complexity (recurrence quantification analysis, RQA)}\label{sec:RQA}
    \input{meth_rqa.tex}

    \subsection{Dynamical invariants derived from RPs}\label{sec:Invariants}
    \input{meth_invariants.tex}

    \subsection{Extension to spatial data}\label{sec:SpatialData}
    \input{meth_spatialdata.tex}

    \subsection{Synchronisation analysis by means of recurrences}\label{sec:Synchro}
    \input{meth_synchro.tex}

    \subsection{Information contained in RPs}\label{sec:Reconstruction}
    \input{meth_recons.tex}

    \subsection{Recurrence based surrogates to test for synchronisation }\label{sec:Surrogates}
    \input{meth_surrogates.tex}

    \subsection{Localisation of unstable periodic orbits by RPs}\label{sec:UPOs}
    \input{meth_upos.tex}

    \subsection{Influence of noise on RPs}\label{sec:Noise}
    \input{meth_noise.tex}


\section{Applications}\label{sec:Applications}
\input{appl_general.tex}

    \subsection{RQA analysis in neuroscience}\label{sec_lamex2}
    \input{appl_rqa.tex}

    \subsection{RQA analysis of financial exchange rates}\label{sec:appl_exchangerates}
    \input{appl_exchangerates.tex}

    \subsection{Damage detection using RQA}\label{sec:appl_damagedetection}
    \input{appl_damagedetection.tex}

    \subsection{Time scale alignment of geophysical borehole data}\label{sec:appl_rescaling}
    \input{appl_rescaling.tex}

    \subsection{Finding of nonlinear interrelations in palaeo-climate archives}\label{sec:appl_crp_synchro}
    \input{appl_crp_synchro.tex}

    \subsection{Automatised computation of $K_2$ applied to the stability of extra-solar planetary systems}\label{sec:appl_planets}
    \input{appl_planets.tex}

    \subsection{Synchronisation analysis of experimental data by means of RPs}\label{sec:appl_jrp_synchro}
    \input{appl_synchro.tex}

\section*{Acknowledgments}
The authors would like to thank 
W.~Anishschenko, 
D.~Armbruster, 
S.~Boccaletti,
U.~Feudel, 
P.~Grassberger, 
C.~Grebogi, 
A.~Groth,
J.~L.~Hudson, 
H.~Kantz, 
I.~Kiss, 
G.~Osipov, 
M.~Palu\v s, 
U.~Parlitz, 
A.~Pikovsky,
P.~Read, 
M.~Rosenblum, 
U.~Schwarz, 
Ch.~L.~Webber Jr.,
M.~Zaks, 
J.~Zbilut, 
C.~Zhou
Y.~Zou
for the fruitful discussions that have helped to write this report. 
This work was partly supported by grants from 
project MAP AO-99-030 (contract \#14592) of the Microgravity Application Program/Biotechnology from the Human Spaceflight Program of the European Space Agency (ESA),
Network of Excellence {\it BioSim} of the European Union, contract LSHB-CT-2004-005137\&\#65533Biosym,
DFG priority program 1114 {\it Mathematical methods for time series analysis and digital image processing},
DFG priority program 1097 {\it Geomagnetic Variations -- Spatio-temporal structure, processes, and effects on system Earth},
DFG special research programme SFB 555 {\it Complex Nonlinear Processes}
and the Helmholtz Centre for Mind and Brain Dynamics Potsdam. 

Recurrence plot related software (e.\,g.~CRP toolbox) 
used in this work is available at
\href{http://tocsy.agnld.uni-potsdam.de}{http://tocsy.agnld.uni-potsdam.de}.
A web resource about RPs can be found at 
\href{http://www.recurrence-plot.tk}{http://www.recurrence-plot.tk}.

\clearpage

\begin{appendix}
\section{Mathematical models}\label{apdx:models}
\input{apdx_models.tex}

\section{Algorithms}\label{apdx:algorithms}                                                                                                                                                                                                                                                                                                                                                                                                     
\input{apdx_algorithms.tex}

\end{appendix}

\clearpage
\bibliographystyle{unsrt}
\bibliography{mybibs,misc,rp}
\addcontentsline{toc}{section}{References}
The most relevant references for this report are
marked with \ding{115}.

\end{document}

%% file: abbreviations.tex
%
%
%

\begin{tabbing}
ENSOmax	\= \kill
$\langle x \rangle$	\>mean of series $x$, expectation value of $x$\\
$\bar{x}$	\>series $x$ normalised to zero mean and standard deviation of one\\
$\hat{x}$	\>estimator for $x$\\
$\delta(\cdot)$	\>delta function ($\delta(x)= \{1 \,|\, x = 0;\ 0 \,|\, x\not= 0\}$)\\
$\delta(\cdot,\cdot)$	\>Kronecker delta function ($\delta(x,y)= \{1\, |\, x = y;\ 0 \,|\, x\not= y\}$)\\
$\partial_t(\cdot)$	\>derivative with respect to time ($\partial_t(x) = \frac{d}{dt}$)\\
$\Delta t$	\>sampling time\\
$\varepsilon$	\>radius of neighbourhood (threshold for RP computation)\\
$\lambda$		\>Lyapunov exponent\\
$\mu(\cdot)$	\>probability measure\\
$\Omega$	\>frequency\\
$\pi$		\>order pattern\\
$\Phi$		\>phase\\
$\sigma$	\>standard deviation\\
$\Theta(\cdot)$	\>Heaviside function ($\Theta(x)= \{1 \,|\, x>0;\ 0 \,|\, x \leq 0\}$) \\
$\tau$		\>time delay (index-based units)\\
$\xi$		\>white noise\\
$\mathcal{A}$	\>a measurable set\\
$B_{\vec{x}_i}(\varepsilon)$		\>the $\varepsilon$-neighbourhood around the point $\vec{x}_i$ on the trajectory\\
$\mathrm{cov}(\cdot)$	\>(auto)covariance function\\
$C_d(\varepsilon)$	\>correlation sum for a system of dimension $d$ and by using a threshold $\varepsilon$\\
$CC(\varepsilon)$	\>cross correlation sum for two systems using a threshold $\varepsilon$\\
$\mathbf{CR}(\varepsilon)$	\>cross recurrence matrix between two phase space trajectories\\
			\>by using a neighbourhood size $\varepsilon$\\
$CPR$		\>synchronisation index based on recurrence probabilities\\
CRP			\>cross recurrence plot\\
CS			\>complete synchronisation\\
$\mathbf{D}$	\>distance matrix between phase space vectors\\
$D$			\>distance\\
$D_1$		\>information dimension\\
$D_2$		\>correlation dimension\\
$D_P$		\>point-wise dimension\\
$DET$		\>measure for recurrence quantification: determinism\\
$DET_{\tau}$		\>measure for recurrence quantification: determinism of the $\tau^{\text{th}}$\\
			\>diagonal in the RP\\
$DIV$		\>measure for recurrence quantification: divergence\\
$d$			\>dimension of the system\\
$ENTR$		\>measure for recurrence quantification: entropy\\
FAN			\>fixed amount of nearest neighbours (neighbourhood criterion)\\
FNN			\>false nearest neighbour\\
GS			\>generalised synchronisation\\
$H_q$		\>Renyi entropy of order $q$\\
$h_D$		\>topological dimension\\
$I_q$		\>generalised mutual information (redundancy) of order $q$\\
$i,j,k$		\>indices\\
$JC(\varepsilon_1,\varepsilon_2)$	\>joint correlation sum for two systems using thresholds $\varepsilon_1$ and $\varepsilon_2$\\
$JK_2$		\>joint Renyi entropy of 2$^{\text{nd}}$ order\\
$JPR$		\>joint probability of recurrence\\
JRP			\>joint recurrence plot\\
$K_2$		\>Renyi entropy of 2$^{\text{nd}}$ order (correlation entropy)\\
$L$			\>measure for recurrence quantification: average line length of \\
			\>diagonal lines\\
$L_{\tau}$		\>measure for recurrence quantification: average line length of \\
			\>diagonal lines of the $\tau^{\text{th}}$ diagonal in the RP\\
$L_{\text{max}}$	\>measure for recurrence quantification: length of the longest \\
			\>diagonal line\\
$L_p$-norm	\>vector norm, e.g.~the Euclidean norm ($L_2$-norm), Maximum \\
			\>norm ($L_{\infty}$-norm)\\
$l_{\text{min}}$	\>predefined minimal length of a diagonal line\\
$LAM$		\>measure for recurrence quantification: laminarity\\

LOI			\>line of identity (the main diagonal line in a RP, $\mathbf{R}_{i,\,i}=1$)\\
LOS			\>line of synchronisation (the distorted main diagonal line in a CRP)\\
$MFNN$		\>Mutual false nearest neighbours\\
$m$			\>embedding dimension\\
$N$			\>length of a data series\\
$N_{\text{l}}$, $N_{\text{v}}$	\>number of diagonal/vertical lines\\
$N_{\text{n}}$		\>number of (nearest) neighbours\\
$\mathds{N}$		\>set of natural numbers\\
OPRP				\>order patterns recurrence plot\\
$\mathbf{P}_{i,j}$	\>probability to find a recurrence point at $(i,j)$\\
$P(\cdot)$			\>histogram\\
$P(\varepsilon,l)$	\>histogram or frequency distribution of line lengths\\
$p(\cdot)$			\>probability\\
$p(\varepsilon,\tau)$	\>probability that the trajectory recurs after $\tau$ time steps\\
$p(\varepsilon,l)$		\>probability to find a line of exactly length $l$\\
$p_c(\varepsilon,l)$	\>probability to find a line of at least length $l$\\
PS			\>phase synchronisation\\
$Q(\tau)$			\>CRP symmetry measure\\
$q(\tau)$			\>CRP asymmetry measure\\
$q$			\>order\\
$\mathds{R}$	\>set of real numbers\\
$\mathcal{R}$	\>set of recurrence points\\
$\mathbf{R}(\varepsilon)$	\>recurrence matrix of a phase space trajectory by using\\
			\>a neighbourhood size $\varepsilon$\\
$R$			\>mean resultant length of phase vectors (synchronisation measure)\\
RATIO		\>measure for recurrence quantification: ratio between $DET$ and $RR$\\
RP			\>recurrence plot\\
RQA			\>recurrence quantification analysis\\
$RR$, $REC$	\>measure for recurrence quantification: recurrence rate (percent \\
			\>recurrence)\\
$RR_{\tau}$		\>measure for recurrence quantification: recurrence rate of the $\tau^{\text{th}}$ \\
			\>diagonal in the RP\\
$S$			\>synchronisation index based on joint recurrence\\
$T^{(1)}$	\>recurrence time of $1^{st}$ type\\
$T^{(2)}$	\>recurrence time of $2^{nd}$ type\\
$T_{\text{ph}}$		\>phase period\\
$T_{\text{rec}}$	\>recurrence period\\
$TREND$		\>measure for recurrence quantification: trend\\
$TT$		\>measure for recurrence quantification: trapping time\\
$V_{\text{max}}$	\>measure for recurrence quantification: length of the longest vertical line\\
$v_{\text{min}}$	\>predefined minimal length of a vertical line\\
$\mathcal{X}$	\>a measurable set\\
\end{tabbing}

%% file: introduction.tex
%
%
%
%

\begin{flushright}
\begin{minipage}[b]{.4\columnwidth}
{\it El poeta tiene dos obligaciones sagradas: 
partir y regresar.\\
(The poet has two holy duties: to set out and to return.)}
\begin{flushright}
{\footnotesize \it Pablo Neruda}
\end{flushright}

\end{minipage}
\end{flushright}

If we observe the sky on a hot and humid day in summer,
we often ``feel'' that a thunderstorm is brewing. When
children play, mothers often know instinctively when a
situation is going to turn out dangerous. Each time we
throw a stone, we can approximately predict where it is
going to hit the ground. Elephants are able to find food 
and water during times of drought.
These predictions are not based
on the evaluation of long and complicated sets of
mathematical equations, but rather on two facts which are
crucial for our daily life: 
\begin{enumerate}
\item similar situations often evolve in a similar way; 
\item some situations occur over and over again.
\end{enumerate}
The first fact is linked to a certain
determinism in many real world systems. Systems of very
different kinds, from very large to very small time-space scales can
be modelled by (deterministic) differential equations.  On
very large scales we might think of the motions of
planets or even galaxies, on intermediate scales of a
falling stone and on small scales of the firing of
neurons. All of these systems can be described by the same
mathematical tool of differential equations. They behave
deterministically in the sense that in principle we can
predict the state of such a system to any precision and
forever once the initial conditions are exactly known. Chaos
theory has taught us that some systems -- even though
deterministic --  are very sensitive to fluctuations and
even the smallest perturbations of the initial conditions
can make a precise prediction on long time scales
impossible. Nevertheless, even for these chaotic systems
short-term prediction is practicable.

The second fact is
fundamental to many systems and is probably one of
the reasons why life has developed memory. Experience
allows remembering similar situations, making predictions
and, hence, helps to survive. But remembering similar
situations, e.\,g., the hot and humid air in summer which
might eventually lead to a thunderstorm, is only helpful
if a system (such as the atmospheric system) returns or
recurs to former states. Such a recurrence is a
fundamental characteristic of many dynamical systems.

They can indeed be used to study the properties of many systems, from
astrophysics (where recurrences have actually been introduced) over
engineering, electronics, financial markets, population dynamics, epidemics
and medicine to brain dynamics. The methods described in this review are
therefore of interest to scientists working in very different areas of
research.

The formal concept of recurrences was introduced by Henri
Poincar\'e in his seminal work from 1890
\cite{poincare1890}, for which he won a prize sponsored by
King Oscar~II of Sweden and Norway on the occasion of his
majesty's 60$^{\text{th}}$ birthday. Therein, Poincar\'e
did not only discover the ``homoclinic tangle'' which lies
at the root of the chaotic behaviour of orbits, but he
also introduced (as a by-product) the concept of
recurrences in conservative systems. When speaking about
the restricted three body problem he mentioned: ``In this
case, neglecting some exceptional trajectories, the
occurrence of which is infinitely improbable, it can be
shown, that the system recurs infinitely many times as
close as one wishes to its initial state.'' (translated 
from \cite{poincare1890}) Even though
much mathematical work was carried out in the following years,
Poincar\'e's pioneering work and his discovery of
recurrence had to wait for more than 70 years for the
development of fast and efficient computers to be
exploited numerically. The use of powerful computers
boosted chaos theory and allowed to study new and exciting
systems. Some of the tedious computations needed to use
the concept of recurrence for more practical purposes
could only be made with this digital tool.

In 1987, Eckmann et al.~introduced the method of {\it recurrence plots
(RPs)} to visualise the recurrences of dynamical systems.
Suppose we have a trajectory $\{\vec x_i\}_{i=1}^N$ of a
system in its phase space \cite{eckmann87}. The
components of these vectors could be, e.\,g., the position and
velocity of a pendulum or quantities such as temperature,
air pressure, humidity and many others for the atmosphere.
The development of the systems is then described by a
series of these vectors, representing a trajectory in an
abstract mathematical space. Then, the corresponding RP
is based on the following recurrence matrix
\begin{equation}\label{eq_rp0}
\mathbf{R}_{i,j} = 
\begin{cases}
1: \vec{x}_i \approx \vec{x}_j\\
0: \vec{x}_i \not\approx \vec{x}_j
\end{cases}
\quad i,j=1, \ldots, N.
\end{equation} 
where $N$ is the number of considered states and
$\vec{x}_i \approx \vec{x}_j$ means equality up to
an error (or distance) $\varepsilon$.
Note that this $\varepsilon$ is essential as systems
often do not recur exactly to a formerly visited state but
just approximately. Roughly
speaking, the matrix compares the states of a system at
times $i$ and $j$. If the states are similar, this is
indicated by a one in the matrix, i.\,e.~$\mathbf{R}_{i,j}=1$. 
If on the other hand the states are
rather different, the corresponding entry in the matrix is
$\mathbf{R}_{i,j}=0$. So the matrix tells us when similar
states of the underlying system occur. This report shows
that much more can be concluded from the recurrence
matrix, Eq.~(\ref{eq_rp0}). But before going into details, we use
Eckmann's representation of the recurrence matrix, to give
the reader a first impression of the patterns of recurrences
which will allow studying dynamical systems and their
trajectories. 

%
%
%
%

\begin{figure}[htbp] 
\centering \includegraphics[width=\columnwidth]{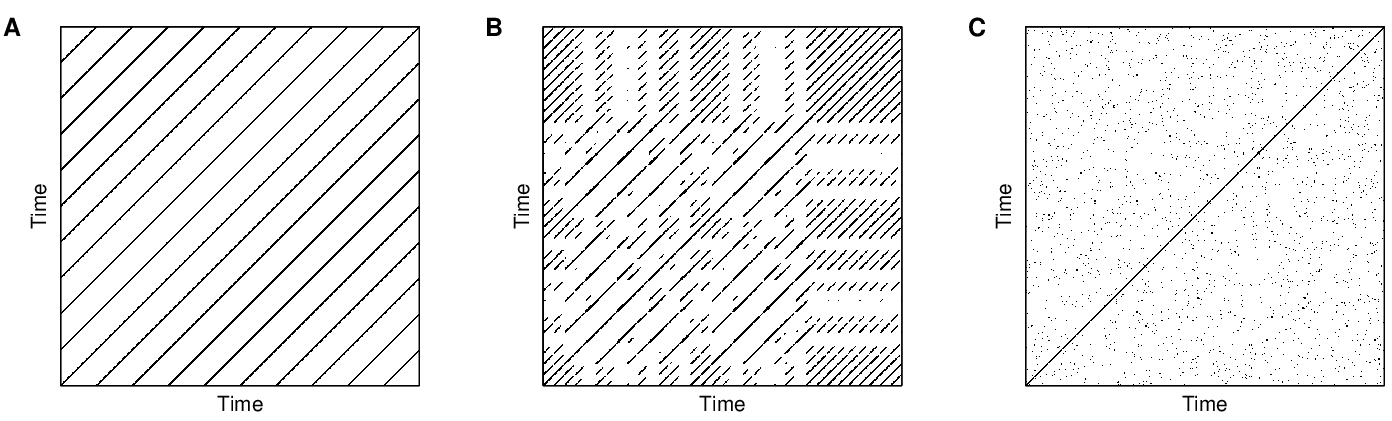} 
\caption{Recurrence plots of (A) a periodic motion with one frequency,
(B) the chaotic R\"ossler system (Eq.~(\ref{eq_roessler})
with parameters $a=b=0.2$ and $c=5.7$) and
(C) of uniformly distributed noise.}\label{fig_rps_intro} 
\end{figure}

Let us consider the RPs of three
prototypical systems, namely of a periodic motion on a
circle (Fig.~\ref{fig_rps_intro}A), of the chaotic
R\"ossler system (Fig.~\ref{fig_rps_intro}B), and of
uniformly distributed, independent noise
(Fig.~\ref{fig_rps_intro}C). In all systems 
recurrences can be observed, but the patterns of the plots are rather
different. The periodic motion is reflected by long and
non-interrupted diagonals. The vertical distance between
these lines corresponds to the period of the oscillation.
The chaotic R\"ossler system also leads to diagonals which
are seemingly shorter. There are also certain vertical
distances, which are not as regular as in the case of the
periodic motion. However, on the upper right, there is a
small rectangular patch which rather looks like the RP of
the periodic motion. We will see later (Sec.~\ref{sec:UPOs}) that this structure
really is a (nearly) periodic motion on the attractor of
the R\"ossler system, which is called an unstable periodic
orbit (UPO). The RP of the uncorrelated stochastic signal consists of
many single black points. The distribution of the points
in this RP looks rather erratic. Reconsidering all three cases,
we might conjecture that the shorter the diagonals in the
RP, the less predictable the system. This conjecture was
already made by Eckmann et al., who suggested that the inverse of
the longest diagonal (except the main diagonal for which
$i=j$) is proportional to the largest Lyapunov exponent of
the system \cite{eckmann87}. Later it will be shown
how the diagonal lines in the RP are related to the predictability
of the system more precisely (Subsec.~\ref{sec:Invariants}).
This very first visual
inspection indicates that the structures found in RPs are
closely linked to the dynamics of the underlying system.

Scientists working in various fields have made use of RPs. 
Applications of RPs can be found in numerous
fields of research such as 
astrophysics \cite{kurths94,asghari2004,zolotova2006}, 
earth sciences \cite{marwan2002npg,marwan2003climdyn,march2005}, 
engineering \cite{elwakil99,nichols2006}, 
biology \cite{giuliani96,manetti2001}, 
cardiology or neuroscience \cite{zbilut91,marwan2002herz,naschitz2004b,thomasson2001,marwan2004,acharya2005}.

This report will summarise recent developments of how to
exploit recurrences to gain understanding of dynamical
systems and measured data. We believe that much more can
be learned from recurrences and that the full potential of
this approach is not yet tapped. This overview can by no
means be complete, but we hope to introduce this powerful
tool to a broad readership and to enthuse scientists to
apply it to their data and systems.

Most of the described methods and procedures are available
in the {\it CRP toolbox} for Matlab$^{\tiny{\textregistered}}$
(provided by TOCSY: 
\href{http://tocsy.agnld.uni-potsdam.de}{http://tocsy.agnld.uni-potsdam.de}).

%% file: theory.tex
%
%
%
%

Recurrence is a fundamental characteristic of many dynamical systems and was
introduced by Poincar\'e in 1890 \cite{poincare1890}, as mentioned in
Sec.~\ref{sec:Introduction}. 
In the following century, much progress has been made in the theory of dynamical
systems. Especially, in the last decades of the 20$^{\text{th}}$ century,
triggered by the development of fast and efficient computers, new and
deep-rooted mathematical structures have been discovered in this field. It has
been recognised that in a larger context recurrences are part of one of three
broad classes of asymptotic invariants \cite{katok95}: 
\begin{enumerate}
\item growth of the number of orbits of various kinds and of the
complexity of orbit families (an important invariant of the orbit growth is
the topological entropy);
\item types of recurrences; and 
\item asymptotic distribution and statistical behaviour of orbits.
\end{enumerate}
The first two classes are of purely topological nature;
the last one is naturally related to ergodic
theory.

Of the different types of recurrences which form part of the second class
of invariants, the Poincar\'e recurrence is of particular interest to this work. It is based on the
Poincar\'e Recurrence Theorem \cite{katok95} (Theorem 4.1.19):

\begin{quote}
Let $T$ be a measure-preserving transformation of a probability 
space $(\mathcal{X},\mu)$ and let $\mathcal{A} \subset \mathcal{X}$ be 
a measurable set.\footnote{Here $\mu$ is a Borel measure on a separable metrisable space $\mathcal{X}$. Note that these assumptions are rather weak 
from a practical point of view. Such a measure preserving function is 
obviously given in Hamiltonian systems and also for all points on (the $\omega$-limit set of) a chaotic attractor.} 
Then for any natural number $N \in \mathds{N}$
\begin{equation}\label{eq_poincareTheorem} 
\mu \Bigl(\bigl\{x\in \mathcal{A} \, |\  \{T^n(x)\}_{n\ge N}\subset \mathcal{X} \backslash \mathcal{A}\bigr\} \Bigr)=0. 
\end{equation}
\end{quote}

Here we give the rather short proof of this theorem:

\begin{quote}

Replacing $T$ by $T^N$ in Eq.~(\ref{eq_poincareTheorem}), 
we find that it is enough to prove the statement for $N=1$. 
The set 
\begin{equation}
\tilde{\mathcal{A}} := \left\{ x \in \mathcal{A} \, |\ \{T^n(x)\}_{n\in\mathds{N}}
        \subset \mathcal{X} \backslash \mathcal{A} = \mathcal{A}
        \cap \left(\bigcap\limits_{n=1}^{\infty}T^{-n}(\mathcal{X} \backslash \mathcal{A})\right)
\right\}  \nonumber
\end{equation}
is measurable. $T^{-n}(\tilde{\mathcal{A}})\cap\tilde{\mathcal{A}}=\emptyset$ for every 
$n$ and hence 
\begin{equation}
T^{-n}(\tilde{\mathcal{A}})\cap T^{-m}(\tilde{\mathcal{A}})=\emptyset \quad \forall\, m,n\in\mathds{N}. \nonumber
\end{equation}
$\mu(T^{-n}(\tilde{\mathcal{A}}))=\mu(\tilde{\mathcal{A}})=0$ 
since $T$ preserves $\mu$. Thus $\mu(\tilde{\mathcal{A}})=0$ since 
\begin{equation}
1 = \mu(\mathcal{X})\ge\mu\left(\bigcup^{\infty}_{n=0}T^{-n}(\tilde{\mathcal{A}})\right) =
   \sum^{\infty}_{n=0}\mu\left(T^{-n}(\tilde{\mathcal{A}})\right) =
   \sum^{\infty}_{n=0}\mu(\tilde{\mathcal{A}}). \nonumber
\end{equation}
That means, that if we have a measure preserving transformation, the trajectory will
eventually come back to the neighbourhood of any former point with probability one.
\end{quote}

However, the theorem only guarantees the existence of recurrence but does not 
tell how long it takes the system to recur. Especially
for high dimensional complex systems the recurrence time might be extremely long. For
the Earth's atmosphere the recurrence time has been estimated to be about $10^{30}$~years, 
which is many orders of magnitude longer than the time the universe exists so
far \cite{vandenDool94}. 

Moreover, the {\it first return} of a {\it set} is defined as follows: if
$\mathcal{A}\subset \mathcal{X}$ is a measurable set of a measurable (probability) dynamical system
$\left\{\mathcal{X},\mu,T\right\}$, the first return of the set $\mathcal{A}$ is given by
\begin{equation} 
\tau (\mathcal{A})=\min\left\{n>0:T^n \mathcal{A} \cap \mathcal{A} \not= \emptyset\right\}.
\end{equation} 
Generically, for hyperbolic systems the recurrence or first return time appears to
exhibit certain universal properties \cite{balakrishnan2001}: 
\begin{enumerate}
\item the recurrence time has an exponential limit distribution; \label{consequ_1}
\item successive recurrence times are independently distributed; \label{consequ_2}
\item as a consequence of (\ref{consequ_1}) and (\ref{consequ_2}), the sequence of successive recurrence times has a
limit law that is a Poisson distribution.
\end{enumerate}
These properties, which are also well-known characteristics of certain stochastic systems, such as finite aperiodic Markov chains \cite{feller49,cox94,pitskel91}, have been
rigourously established for deterministic dynamical systems exhibiting sufficiently
strong mixing \cite{sinai70,hirata93,collet92}.
They have also been shown valid for a wider class of systems that remains, however,
hyperbolic \cite{hirata95}.

Recently, recurrences and return times have been studied
with respect to their statistics \cite{hirata99,penne99} and linked to various
other basic characteristics of dynamical systems, such as the Pesin's dimension \cite{afraimovich97}, the
point-wise and local dimensions \cite{afraimovich2000,afraimovich2003,gao99} 
or the Hausdorff dimension \cite{barreira2001}.
Also the multi-fractal properties of return time statistics have been studied
\cite{hadyn2002,saussol2003a}. Furthermore, it has been shown that recurrences are
related to Lyapunov exponents and to various entropies \cite{saussol2002,saussol2003b}. 
They have been linked to rates of mixing \cite{young99}, and the
relationship between the return time statistics of continuous and discrete systems
has been investigated \cite{balakrishnan2000}.
It is important to emphasise that RPs, Eq.~(\ref{eq_rp0}),  
can help to understand and also provide a visual impression of these fundamental characteristics.

\begin{figure}[htbp]
\centering{\includegraphics[width=0.5\columnwidth]{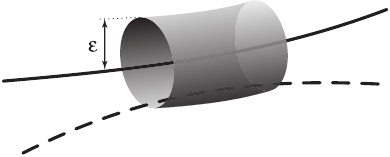}}
\caption{A diagonal line in a RP corresponds with a section
of a trajectory (dashed) which stays within an 
$\varepsilon$-tube around another section (solid).}\label{fig_eps_tube}
\end{figure}

However, for the study of RPs also the first class of the 
asymptotic invariants is important, namely
invariants which are linked to the growth of the number of orbits of various kinds
and of the complexity of orbit families. In this
report we consider the return times and especially focus on the times at which these
recurrences occur, and for how long the trajectories evolve close to each other (the length 
of diagonal structures in recurrence plots will be linked to these times): a central 
question will concern the interval of time, that a trajectory stays within an 
$\varepsilon$-tube around another section of
the trajectory after having recurred to it (Fig.~\ref{fig_eps_tube}). This time interval depends on the
divergence of trajectories or orbit growth of the respective system.

The most
important numerical invariant related to the orbit growth is the topological entropy $h_D$.
It represents the exponential growth rate for the number of orbit segments
distinguishable with arbitrarily fine but finite precision. The topological entropy 
$h_D$ describes, roughly speaking, the total exponential complexity of
the orbit structure with a single number. We just present the discrete case here (see
also \cite{katok95}).

Let $F:\mathcal{X}\rightarrow \mathcal{X}$ be a continuous map of a compact metric
space $X$ with distance function $D$. We define an increasing sequence of metrics
$D^F_n,\;n=1,2,3,\ldots$, starting from $D^F_1=D$ by
\begin{equation}\label{eq_distance}
D^F_n(x,y)=\max_{0\le i\le n-1} D\left(F^i(x),F^i(y)\right).
\end{equation}
In other words, $D^F_n$ measures the distance between the orbit segments
$I^n_x=\left\{x,\ldots,F^{n-1}x\right\}$ and
$I^n_y=\left\{y,\ldots,F^{n-1}y\right\}$. We denote the open 
ball around $x$ by $B_F(x,\varepsilon,n)=\left\{y\in \mathcal{X}\mid D^F_n(x,y)<\varepsilon\right\}$.

A set $\mathcal{E}\subset \mathcal{X}$ is
said to be $(n,\varepsilon)$-spanning if 
$\mathcal{X}\subset\bigcup_{x \in \mathcal{E}}B_F(x,\varepsilon,n)$. 
Let $S_D(F,\varepsilon,n)$ be the minimal cardinality of an
$(n,\varepsilon)$-spanning set, or equivalently the cardinality of a minimal
$(n,\varepsilon)$-spanning set. This quantity gives the minimal number of initial
conditions whose behaviour approximates up to time $n$ the behaviour of {\it any}
initial condition up to $\varepsilon$. Consider the exponential growth rate for that
quantity 
\begin{equation}
h_D(F,\varepsilon)=\overline{\lim\limits_{n\to\infty}}
          \frac{1}{n}\log S_D(F,\varepsilon,n) ,
\end{equation}
where $\overline{\lim\limits_{n\to\infty}}$ denotes the supremum limit. Note that 
$h_D(F,\varepsilon)$ does not decrease with $\varepsilon$. Hence, the
topological entropy $h_D(F)$ is defined as 
\begin{equation} 
h_D(F)=\lim\limits_{\varepsilon\to 0}h_D(F,\varepsilon).
\end{equation}
It has been shown that if $D'$ is another
metric on $X$, which defines the same topology as $D$, then $h_{D'}(F)=h_D(F)$ and
the topological entropy will be an invariant of topological conjugacy \cite{katok95}. Roughly
speaking, this shows that a change of the coordinate system does not change the
entropy. This is highly relevant, as it suggests that some of the structures in RPs do not depend on the special choice of the metric. The entropy $h_D$ allows 
characterising of dynamical systems with respect to their ``predictability'', e\,g., periodic systems are characterised by $h_D=0$. If the system becomes more 
irregular, $h_D$ increases. Chaotic systems typically yield $0<h_D<\infty$, whereas time 
series of stochastic systems have infinite $h_D$.

Recurrences are furthermore related to UPOs
and the topology of the attractor \cite{gilmore2002}. In Subsec.~\ref{sec:UPOs} 
we will describe this relationship in detail.

These considerations show that recurrences are deeply rooted in the theory of 
dynamical systems. Much of the efforts have been dedicated to the study of 
recurrence times. Additionally, in the late 1980's Eckmann et al.~have 
introduced RPs and the recurrence matrix. This matrix contains much 
information about the underlying dynamical system and can be exploited 
for the analysis of measured time series. Much of this report is devoted 
to the analysis of time series based on this matrix. We show how these 
methods are linked to theoretical concepts and show their respective applications.

%% file: meth_phasespace.tex
%
%
%

The states of natural or technical systems typically change in
time, sometimes in a rather intricate manner. The study of such 
complex dynamics is an important task in numerous scientific disciplines and their applications. 
Understanding, describing and forecasting such changes is of utmost 
importance for our daily life. The prediction of the weather, earthquakes or 
epileptic seizures are only three out of many examples.

Formally, a dynamical system is given by (1) a (phase) space (2) a continuous 
or discrete time and (3) a time-evolution law. 
The elements or ``points'' of the phase space represent possible states 
of the system. Let us assume that the state of such a system at a fixed 
time $t$ can be specified by $d$ components (e.\,g., in the case of a harmonic 
oscillator, these components could be its position and velocity). 
These parameters can be considered to form a vector 
\begin{equation}
\vec x (t)=\left(x_1(t),\,x_2(t),\ldots,\,x_d(t)\right)^{\text{T}}
\end{equation}
in the $d$-dimensional phase space of the system. In the most general 
setting, the time-evolution law is a rule that allows determining 
the state of the system at each moment of time $t$ from its states 
at all previous times. Thus the most general time-evolution law is 
time dependent and has infinite memory. However, we will restrict to 
time-evolution laws which enable calculating all future states given 
a state at any particular moment. For time-continuous systems the 
time evolution is given by a set of differential equations
\begin{equation}\label{eq_dynamics}
\dot{\vec x}(t) = \frac{d\vec x(t)}{dt} = \vec F(\vec x(t)),
\quad F:\mathds{R}^d\rightarrow\mathds{R}^d.
\end{equation}
The vectors $\vec x(t)$ define a trajectory in phase space.

In experimental settings, typically not all relevant components to construct 
the state vector are known or cannot be measured. Often we are 
confronted with a time-discrete measurement of only one observable. 
This yields a scalar and discrete time series $u_i=u(i\Delta t)$, 
where $i=1,\ldots,N$ and $\Delta t$ is the sampling rate of the 
measurement. In such a case, the phase space has to be reconstructed 
\citep{packard80,takens81}. A frequently used method for the 
reconstruction is the time delay method:
\begin{equation}\label{eq_embedding}
\hat{\vec x}_i = \sum_{j=1}^{m} u_{i+(j-1)\tau} \, \vec e_j,
\end{equation}
where $m$ is the embedding dimension and $\tau$ is the time delay. 
The vectors $\vec e_i$ are unit vectors and span an orthogonal coordinate
system ($\vec{e}_i\cdot\vec{e}_j=\delta_{i,j}$).
If $m \geq 2D_2+1$, where $D_2$ is the correlation dimension of the attractor,
Takens' theorem and several extensions of it, guarantee the existence of 
a diffeomorphism between the original and the reconstructed attractor 
\citep{takens81,sauer91}. This means that both attractors can be 
considered to represent the same dynamical system in different coordinate systems.

For the analysis of time series, both embedding parameters, the dimension 
$m$ and the delay $\tau$, have to be chosen appropriately. Different 
approaches for the estimation of the smallest sufficient embedding 
dimension (e.\,g.~the false nearest neighbours algorithm~\cite{kantz97}), as well as for an 
appropriate time delay $\tau$ (e.\,g. the auto-correlation
function, the mutual information function; cf.~\cite{cao97,kantz97}) have been proposed.

Recurrences take place in a systems phase space. In order to analyse 
(univariate) time series by RPs, Eq.~(\ref{eq_rp0}), we will reconstruct in the following the phase 
space by delay embedding, if not stated otherwise.

%% file: meth_recplots.tex
%
%
%
\subsubsection{Definition}\label{sec:DefRecurrencePlots}

\begin{figure}[bt]
\centering \includegraphics[width=\columnwidth]{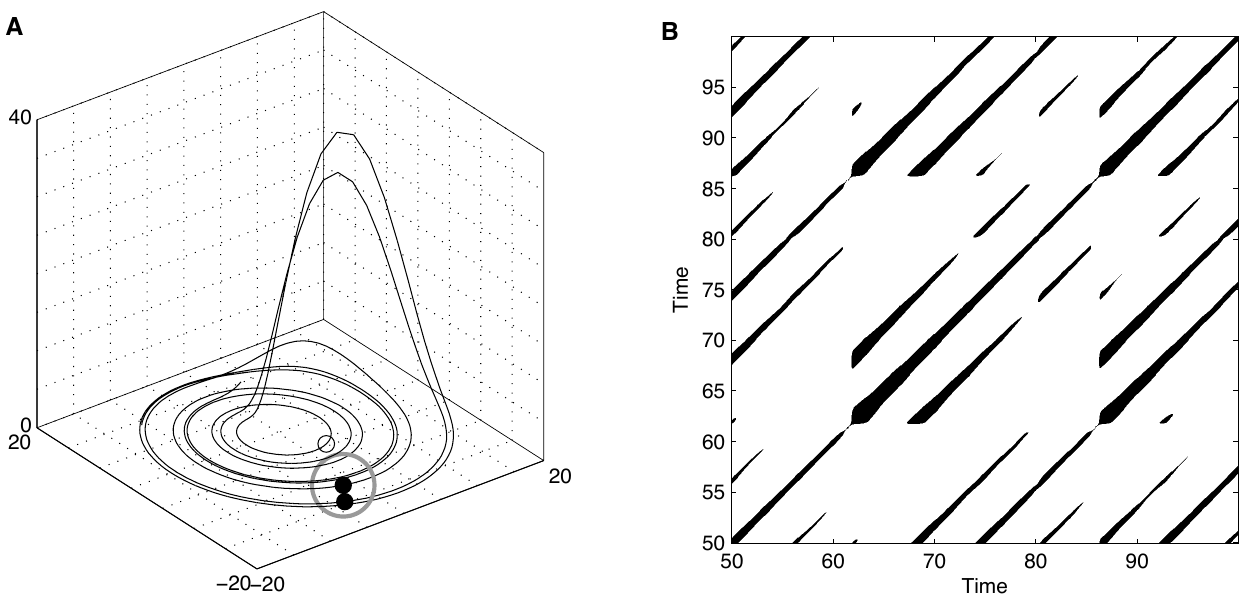} 
\caption{(A) Segment of the phase space trajectory of the R\"ossler system,
Eqs.~(\ref{eq_roessler}), with $a=0.15$, 
$b=0.20$, $c=10$, by using
its three components and (B) its corresponding recurrence plot.
A phase space vector at $j$ which falls into the neighbourhood 
(grey circle in (A)) of a given phase space vector at $i$ is 
considered as a recurrence point (black point on the trajectory in (A)). 
This is marked with a black point in the RP at the position $(i,j)$. A 
phase space vector outside the neighbourhood (empty circle in (A)) leads 
to a white point in the RP. The radius of the neighbourhood for the RP is 
$\varepsilon=5$; $L_2$-norm is used.}\label{fig_constrRP}
\end{figure}

As our focus is on recurrences of states of a dynamical system, we define
now the tool which measures recurrences of a trajectory  
$\vec x_i \in \mathds{R}^d$ in phase space: the {\it recurrence plot},
Eq.~(\ref{eq_rp0}) \cite{eckmann87}. The RP efficiently 
visualises recurrences 
(\reffig{fig_constrRP}A) and can be formally expressed by
the matrix
\begin{equation}\label{eq_rp}
\mathbf{R}_{i,j}(\varepsilon) = 
\Theta\left(\varepsilon-\left\|\vec x_{i} - \vec x_{j}\right\|\right),
\quad i,j=1, \ldots, N,
\end{equation}
where $N$ is the number of measured points $\vec x_i$, 
$\varepsilon$ is a threshold distance, $\Theta (\cdot)$ 
the Heaviside function 
(i.\,e.~$\Theta(x)=0$, if $x<0$, and $\Theta(x)=1$ otherwise)
and $\left\| \cdot \right\|$ is a norm.
For $\varepsilon$-recurrent states, i.\,e.~for states 
which are in an $\varepsilon$-neighbourhood, we introduce 
the following notion 
\begin{equation}\label{eq_vectorequal}
\vec x_{i} \approx \vec x_{j} \quad \Longleftrightarrow \quad \mathbf{R}_{i,j} \equiv 1.
\end{equation}
The recurrence plot (RP) is obtained by plotting the recurrence matrix,
Eq.~(\ref{eq_rp}), and using different colours for its binary
entries, e.\,g., plotting a black dot at the coordinates $(i,j)$, if 
$\mathbf{R}_{i,j} \equiv 1$, and a white dot, if  
$\mathbf{R}_{i,j} \equiv 0$.
Both axes of the RP are time axes and show rightwards and upwards
(convention). Since $\mathbf{R}_{i,i} \equiv 1 \, |_{i=1}^N$ 
by definition, the RP has always a black main diagonal line, the 
{\it line of identity (LOI)}. Furthermore, the RP is 
symmetric by definition with respect to the main diagonal, 
i.\,e.~$\mathbf R_{i,j} \equiv \mathbf R_{j,i}$ 
(see Fig.~\ref{fig_constrRP}).

\begin{figure}[htbp]
\centering \includegraphics[width=9cm]{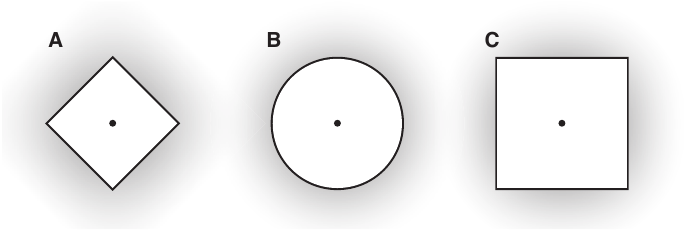} 
\caption{Three commonly used norms for the neighbourhood  with the same radius
around a point (black dot) exemplarily shown for the two-dimensional phase space: 
(A) $L_1$-norm, (B) $L_2$-norm and (C) $L_{\infty}$-norm.}\label{fig_norm}
\end{figure}

In order to compute an RP, an appropriate {\it norm} has to be chosen. The most
frequently used norms are the $L_1$-norm, the $L_2$-norm (Euclidean
norm) and the $L_{\infty}$-norm (Maximum or Supremum norm). Note
that the neighbourhoods
of these norms have different shapes (\reffig{fig_norm}).
Considering a fixed $\varepsilon$,
the $L_{\infty}$-norm finds the most, the $L_1$-norm
the least and the $L_2$-norm an intermediate amount of neighbours.
To compute  RPs, the $L_{\infty}$-norm is often applied, because 
it is computationally faster and allows to study some features in RPs
analytically.

\subsubsection{Selection of the threshold $\varepsilon$}\label{sec:Selectionofthethreshold}

A crucial parameter of an RP is the {\it threshold} $\varepsilon$.
Therefore, special attention has to be required for its choice.
If $\varepsilon$ is chosen too small, there may be 
almost no recurrence points and we cannot learn anything about 
the recurrence structure of the underlying system. On the other 
hand, if $\varepsilon$ is chosen too large, almost every point is a
neighbour of every other point, which leads to a lot of artefacts.
A too large $\varepsilon$ includes also points into the neighbourhood
which are simple consecutive points on the trajectory. This effect is called 
{\it tangential motion} and causes thicker and longer 
diagonal structures in the RP as they actually are.
Hence, we have to find a compromise for the value of $\varepsilon$. Moreover, the influence 
of noise can entail choosing a larger threshold, because noise would distort 
any existing structure in the RP. At a higher threshold, this structure may 
be preserved (see Subsec.~\ref{sec:Noise}).

Several ``rules of thumb'' for the choice of the threshold $\varepsilon$ have been advocated in the 
literature, e.\,g., a few per cent of the maximum phase space diameter has been 
suggested \citep{mindlin92}. Furthermore, it should
not exceed 10\% of the mean or the maximum phase space diameter
\citep{koebbe92,zbilut92}.

A further possibility is to choose $\varepsilon$ according to the recurrence 
point density of the RP by seeking a scaling region in the recurrence 
point density \citep{zbilut2002b}. However, this may not be suitable for 
non-stationary data. For this case it was proposed to choose $\varepsilon$ 
such that the recurrence point density is approximately 1\% \cite{zbilut2002b}.

Another criterion for the choice of $\varepsilon$ takes into account that a measurement
of a process is a composition of the real signal and some observational noise with 
standard deviation $\sigma$ \cite{thiel2002}. In order to get similar 
results as for the noise-free situation, $\varepsilon$
has to be chosen such that it is five times larger than the standard deviation 
of the observational noise, i.\,e.~$\varepsilon > 5\,\sigma$ (cf.~Subsec.~\ref{sec:Noise}).
This criterion holds for a wide class of processes.

For (quasi-)periodic processes, the diagonal structures 
within the RP can be used in order to determine an 
optimal threshold \cite{matassini2002a}. For this purpose, 
the density distribution of recurrence points along the 
diagonals parallel to the LOI is considered (which corresponds
to the diagonal-wise defined $\tau$-recurrence rate $RR_{\tau}$, 
Eq.~(\ref{eq_rr_star})). From such a density plot, the number
of significant peaks $N_{\text{p}}$ is counted. Next, the average
number of neighbours $N_{\text{n}}$, Eq.~(\ref{eq_number_of_neighbours}), 
that each point has, is computed. The threshold $\varepsilon$ should
be chosen in such a way that $N_p$ is maximal and $N_{\text{n}}$ 
approaches $N_{\text{p}}$. Therefore, a good choice of $\varepsilon$
would be to minimise the quantity 
\begin{equation}
\beta(\varepsilon) = \frac{|N_{\text{n}}(\varepsilon) - N_{\text{p}}(\varepsilon)|}{N_{\text{n}}(\varepsilon)}.
\end{equation}
This criterion minimises the fragmentation and thickness of 
the diagonal lines with respect to the threshold, which can be
useful for de-noising, e.\,g., of acoustic signals.
However, this choice of $\varepsilon$
may not preserve the important distribution of the diagonal lines in
the RP if observational noise is present (the estimated threshold
can be underestimated).

Other approaches use a fixed recurrence point density. 
In order to find an $\varepsilon$ which corresponds to a fixed recurrence 
point density $RR$ (or recurrence rate, Eq.~(\ref{eq_rr})), the cumulative distribution 
of the $N^2$ distances between each pair of vectors $P_c(D)$ can be used. The $RR^{\text{th}}$ percentile
is then the requested $\varepsilon$ (e.\,g.~for $RR=0.1$ the threshold $\varepsilon$ is 
given by $\varepsilon = D$ with $P_c(D) = 0.1$).
An alternative is to fix the number of 
neighbours for every point of the trajectory. In this case, the threshold is actually
different for each point of the trajectory, i.\,e.~$\varepsilon=\varepsilon(\vec x_i)=\varepsilon_i$ 
(cf.~Subsec.~\ref{sec:VariationsRecurrencePlots}). The advantage of the latter two methods 
is that both of them preserve the recurrence point density and allow to compare RPs 
of different systems without the necessity of normalising the time series beforehand.

The choice of $\varepsilon$ depends strongly on the considered 
system under study.
However, all kinds of dynamical invariants derived from RPs 
(cf.~Subsec.~\ref{sec:Invariants}) can only be obtained 
in the limit $\varepsilon \rightarrow 0$.

\subsubsection{Structures in RPs}\label{sec:StructuresinRecurrencePlots}

\begin{figure}[bthp]
\centering \includegraphics[width=\columnwidth]{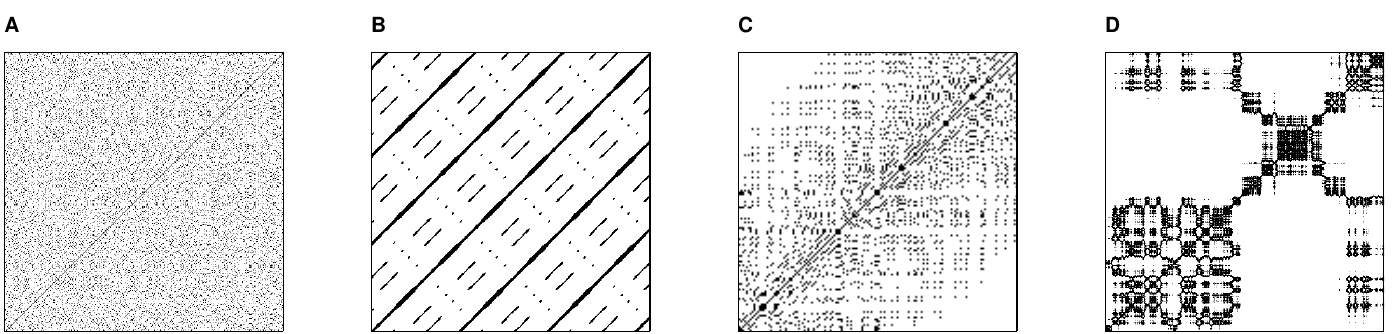} 
\caption{Characteristic typology of recurrence plots: 
(A) homogeneous (uniformly distributed white noise), (B) periodic (super-positioned
harmonic oscillations), (C) drift (logistic map corrupted
with a linearly increasing term  $x_{i+1}=4 x_i (1-x_i) + 0.01\,i$, cf.~\reffig{fig_RP_log}D) 
and (D) disrupted (Brownian motion). 
These examples illustrate how different RPs can be. 
The used data have the length 400 (A, B, D) and 150 (C), respectively;
RP parameters are $m=1$, $\varepsilon=0.2$ (A, C, D) and
$m=4$, $\tau=9$, $\varepsilon=0.4$ (B); $L_2$-norm.}\label{fig_RPexamples}
\end{figure}

As already mentioned, the initial purpose of 
RPs was to provide a visual inspection of trajectories especially in a
higher dimensional phase space. RPs yield important insights into the time 
evolution of these trajectories, because typical patterns
in RPs are linked to a specific behaviour of the system.
Large scale patterns in RPs, designated in \cite{eckmann87} as 
{\it typology}, can be classified in 
{\it homogeneous}, {\it periodic}, {\it drift} and {\it disrupted} ones
\cite{eckmann87,marwan2003diss}:

\begin{itemize}
\item {\it Homogeneous} RPs are typical of stationary systems in which 
the relaxation times are short in comparison with the time spanned by
the RP. An example of such an RP is that of a stationary random time series (\reffig{fig_RPexamples}A).
\item Periodic and quasi-periodic systems have RPs with diagonal oriented, 
{\it periodic} or {\it quasi-periodic} recurrent structures
(diagonal lines, checkerboard structures). \reffig{fig_RPexamples}B shows the 
RP of a periodic system with two harmonic frequencies and with a frequency ratio of four 
(two and four short lines lie between the continuous diagonal lines). Irrational 
frequency ratios cause more complex quasi-periodic recurrent structures
(the distances between the diagonal lines are different). 
However, even for oscillating systems whose oscillations are not easily
recognisable, RPs can be useful
(cf.~unstable periodic orbits, Subsec.~\ref{sec:UPOs}). 
\item A {\it drift} is caused by systems with slowly varying parameters, 
i.\,e.~non-stationary systems. The RP pales away from the LOI (\reffig{fig_RPexamples}C). 
\item Abrupt changes in the dynamics as well as extreme events
cause {\it white areas or bands} in the RP (\reffig{fig_RPexamples}D). 
RPs allow finding and assessing extreme and rare
events easily by using the frequency of their recurrences.
\end{itemize}

A closer inspection of the RPs reveals also small-scale structures, the {\it texture}
\cite{eckmann87}, which 
can be typically classified in {\it single dots},
{\it diagonal lines} as well as {\it vertical} and {\it horizontal 
lines} (the combination of vertical and horizontal lines obviously forms 
rectangular clusters of recurrence points); in addition, even {\it bowed lines} may occur
\cite{eckmann87,marwan2003diss}:

\begin{itemize}
\item {\it Single, isolated recurrence points} can occur if states are rare, 
if they persist only for a very short time, or fluctuate strongly. However, 
they are not a unique sign of randomness or noise (e.\,g.~in maps). 

\item A {\it diagonal line} 
$\mathbf{R}_{i+k,j+k} \equiv 1 \, |_{k=0}^{l-1}$ 
(where $l$ is the length of the diagonal line)
occurs when a segment of the trajectory runs almost in parallel to another segment 
(i.\,e.~through an $\varepsilon$-tube around the other segment, 
Fig.~\ref{fig_eps_tube}) for $l$ time units:
\begin{equation}\label{eq_diagonalline0}
\vec x_{i} \approx \vec x_{j},\ \vec x_{i+1} \approx \vec x_{j+1},\ \ldots \,,\ \vec x_{i+l-1} \approx \vec x_{j+l-1}.
\end{equation}
A diagonal line of length $l$ is then defined by
\begin{equation}\label{eq_diagonalline}
\left( 1 - \mathbf{R}_{i-1,j-1} \right) \left( 1 - \mathbf{R}_{i+l,j+l} \right) \prod_{k=0}^{l-1} \mathbf{R}_{i+k,j+k}
     \equiv 1.
\end{equation}
The length of this diagonal line is determined by the duration of such 
similar local evolution of the trajectory segments. The direction of these 
diagonal structures is parallel to the LOI (slope one, angle $\pi/4$). 
They represent trajectories which evolve through the same $\varepsilon$-tube for a
certain time. Since the definition of the R\'enyi 
entropy of second order $K_2$ uses the time how long trajectories evolve 
in an $\varepsilon$-tube, the existence of a relationship between the length of the 
diagonal lines and $K_2$ (and even the sum of the positive Lyapunov exponents, Eq.~(\ref{eq_ruelle})) 
is plausible (cf.~invariants, Subsec.~\ref{sec:Invariants}).
Note that there might be also diagonal structures
perpendicular to the LOI, representing parallel segments of the trajectory running with opposite
time directions, 
i.\,e.~$\vec x_{i} \approx \vec x_{j}, \vec x_{i+1} \approx \vec x_{j-1}, \ldots$ 
(mirrored segments). This is often a hint for an inappropriate 
embedding. 

\item A {\it vertical (horizontal) line} 
$\mathbf{R}_{i,j+k} \equiv 1 \, |_{k=0}^{v-1}$
(with $v$ the length of the vertical line)
marks a time interval in which a state does not change or changes very slowly:
\begin{equation}
\vec x_{i} \approx \vec x_{j},\ 
\vec x_{i} \approx \vec x_{j+1},\ \ldots\,,\ 
\vec x_{i} \approx \vec x_{j+v-1}.
\end{equation}
The formal definition of a vertical line is
\begin{equation}\label{eq_verticalline}
 \left( 1 - \mathbf{R}_{i,j-1} \right) \left( 1 - \mathbf{R}_{i,j+v} \right) \prod_{k=0}^{v-1} \mathbf{R}_{i,j+k}
     \equiv 1.
\end{equation}
Hence, the state is trapped for some time. This is a typical
behaviour of laminar states (intermittency) \cite{marwan2002herz}. 

\item {\it Bowed lines} are lines with a non-constant slope. The shape of a bowed line depends on the local time
relationship between the corresponding close trajectory segments (cf.~Eq.~(\ref{eq:los_line})).

\end{itemize}

Diagonal and vertical lines are the base for a quantitative analysis of 
RPs (cf.~Subsec.~\ref{sec:RQA}).

\begin{figure}[tbhp]
\centering{\includegraphics[width=0.47\columnwidth]{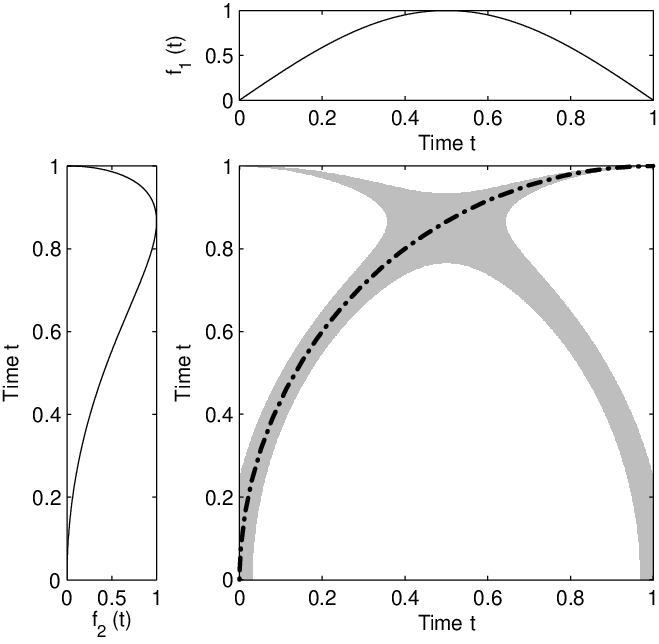}}
\caption{Detail of a recurrence plot for a trajectory $f(t) = \sin(\pi\,t)$ 
whose sub-sections $f_1(t)$ and $f_2(t)$ undergo different 
transformations in the time scale: $T_1(t) = t$ and $T_2(t) = 1-\sqrt{1-t^2}$.
The resulting bowed line (dash-dotted line) has the slope $b(t) = \frac{1-t}{\sqrt{1-(1-t)^2}}$, which
corresponds with a segment of a circle.}\label{fig:segment_circle}
\end{figure}

More generally, the line structures in RPs exhibit locally the
time relationship between the current close
trajectory segments \citep{marwan2005}.
A line structure in an RP of length $l$ corresponds to the closeness of 
the segment $\vec x(T_1(t))$ to another segment $\vec x(T_2(t))$, where
$T_1(t)$ and $T_2(t)$ are two local and in general different time scales 
which preserves
$\vec x(T_1(t)) \approx \vec x(T_2(t))$ for some (absolute) time $t=1, \dots, l$. 
Under some assumptions (e.\,g., piece-wise existence of an inverse of the 
transformation $T(t)$, the two segments visit the same area in the phase 
space), a line in the RP can simply be expressed by the time-transfer
function (Fig.~\ref{fig:segment_circle})
\begin{equation}\label{eq:los_line}
\vartheta(t) = T_2^{-1}\left(T_1(t)\right).
\end{equation} 
Particularly, we find that the local slope $b(t)$ of a line in an RP represents 
the local time derivative of the inverse second time scale $T_2^{-1}(t)$ 
applied to the first time scale $T_1(t)$
\begin{equation}\label{eq:los}
b(t) = \partial_t T_2^{-1}\left(T_1(t)\right) = \partial_t \vartheta(t). 
\end{equation} 
This is a fundamental relation between the local slope $b(t)$ of
line structures in an RP and the time scaling of the corresponding
trajectory segments. As special cases, we find that the slope  $b=1$ 
(diagonal lines) corresponds to 
$T_1 = T_2$, whereas $b=\infty$ (vertical lines) corresponds to 
$T_2 \rightarrow 0$, i.\,e.~the second trajectory segment evolves infinitely slow
through the $\varepsilon$-tube around first trajectory segment.
From the slope $b(t)$ of a line in an RP we can infer the 
relation $\vartheta(t)$ between two segments of  $\vec x(t)$
($\vartheta(t) = \int b(t) dt$).
Note that the slope $b(t)$ depends only on
the transformation of the time scale and is independent of the 
considered trajectory $\vec x(t)$ \citep{marwan2005}. 

This feature is, e.\,g., used in the application of CRPs as a tool for
the adjustment of time scales of two data series \citep{marwan2004nova,marwan2005} 
and will be discussed in Subsec.~\ref{sec:CrossRecurrencePlots}.

Summarising the explanations about typology and texture, we establish the
following list of features and give the corresponding
qualitative interpretation (Tab.~\ref{tab:typology}).

\begin{table}[htbp]
\caption{Typical patterns in RPs and their meanings.}\label{tab:typology}
\begin{tabularx}{\columnwidth}{p{9pt}p{4.4cm}X}
\hline
&Pattern & Meaning\\
\hline
\hline
(1) & Homogeneity & the process is stationary\\
(2) & Fading to the upper left and lower right corners & non-stationary data; the process 
      contains a trend or a drift\\
(3) & Disruptions (white bands)  & non-stationary data; some states are rare or far from the normal;
      transitions may have occurred\\
(4) & Periodic/quasi-periodic patterns & cyclicities in the process; the time distance between periodic
      patterns (e.\,g.~lines) corresponds to the period; different distances between
      long diagonal lines reveal quasi-periodic processes\\
(5) & Single isolated points & strong fluctuation in the process;
      if only single isolated points occur, the process may be an uncorrelated random 
      or even anti-correlated process\\
(6) & Diagonal lines (parallel to the LOI) & the evolution of states is similar 
      at different epochs; the process could be deterministic;
      if these diagonal lines occur beside single isolated points, the process could be chaotic (if 
      these diagonal lines are periodic, unstable periodic orbits can be observed)\\
(7) & Diagonal lines (orthogonal to the LOI) & the evolution of states is similar 
      at different times but with reverse time; 
      sometimes this is an indication for an insufficient embedding\\
(8) & vertical and horizontal lines/clusters & some states do not change or change 
      slowly for some time; indication for laminar states\\
(9) & Long bowed line structures & the evolution of states is similar at different epochs
      but with different velocity; the dynamics of the system could be changing\\
\hline
\end{tabularx}
\end{table}

Another problem is the LOI; 
some authors exclude it from the RP. This may be useful for the quantification
of RPs, which we will discuss later. It can also be motivated by the 
definition of the Grassberger-Procaccia correlation sum \citep{grassberger83} 
(or generalised 2$^{\text{nd}}$ order correlation integral)
which was introduced for the determination of the correlation dimension 
$D_2$ and is closely related to RPs:
\begin{equation}\label{eq_corrsum}
C_2(\varepsilon) = \frac{1}{N^2}
             \sum_{ \substack{i,j=1\\ i\not=j}
	     }^{N}
             \Theta\left(\varepsilon-\left\|\vec x_{i} - \vec x_{j}\right\|\right),
             \qquad \vec x \in \mathds{R}^d.
\end{equation}
Eq.~(\ref{eq_corrsum}) excludes the comparisons of $\vec x_i$ with itself.
Nevertheless, since the threshold value $\varepsilon$ is finite (and normally
about $10$\% of the mean phase space radius), further long diagonal lines
can occur directly below and above the LOI for smooth or high resolution data.
Therefore, the diagonal lines in a small corridor around the LOI correspond to 
the {\it tangential motion} of the phase space trajectory, but
not to different orbits. Thus, for the estimation of invariants of the
dynamical system, it is
better to exclude this entire corridor and not only the LOI. 
This step corresponds to suggestions to exclude the tangential motion
as it is done for the computation of the correlation dimension
(known as Theiler correction or Theiler window \cite{theiler86})
or for the alternative estimators of Lyapunov exponents \cite{kurths87,gao94} 
in which only those phase space points are considered that fulfil 
the constraint $|j-i| \ge w$. Theiler has suggested using 
the auto-correlation 
time as an appropriate value for $w$ \cite{theiler86}, and Gao and Zheng 
state that $w=(m-1)\tau$ would be sufficient \cite{gao94}. However, 
in a representation of an RP it is better to picture the LOI. This LOI has also 
importance when applications of cross recurrence plots will be discussed (Sec.~\ref{sec:appl_rescaling}).

The visual interpretation of RPs requires some experience. 
RPs for paradigmatic systems provide an instructive introduction into characteristic 
typology and texture (e.\,g.~Fig.~\ref{fig_RPexamples}). However,
a quantification of the obtained structures is necessary for a more objective investigation
of the considered system (see Subsec.~\ref{sec:RQA}).

\begin{figure}[bbbbht]
\centering \includegraphics[width=\columnwidth]{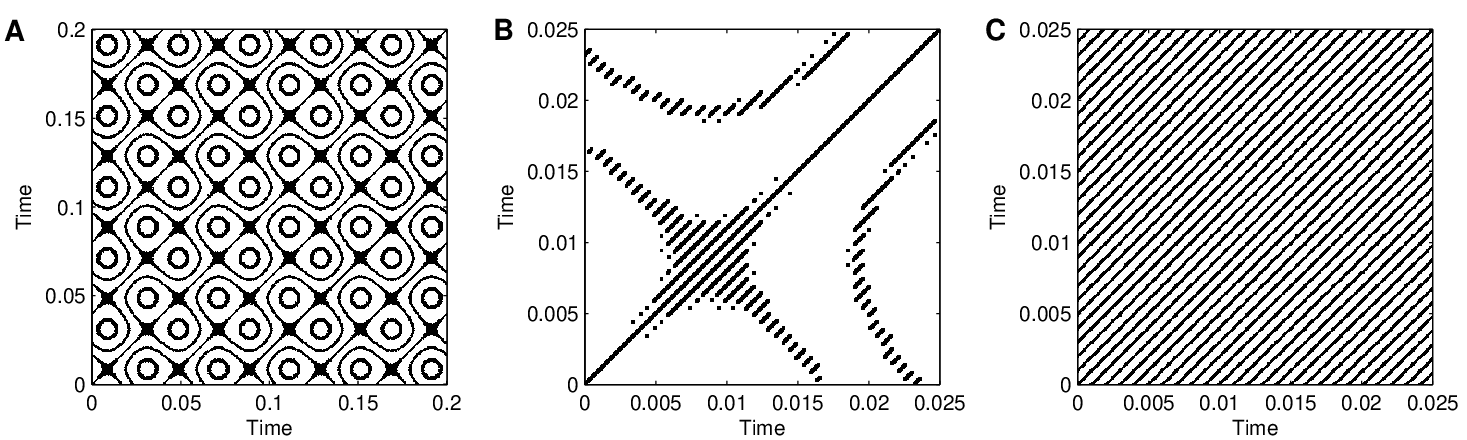} 
\caption{(A) Characteristic patterns of gaps (all white areas) 
in a recurrence plot of a modulated
harmonic oscillation $\cos(2 \pi 1000t + 0.5 \sin(2\pi 25 t))$,
which is sampled with a sampling frequency of 1~kHz.
(B) Magnified detail of the RP presented in (A), which obtains the
actual periodic line structure due to the oscillation, but disturbed by
extended white areas. 
(C) Corresponding RP as shown in (B), but for a higher sampling rate
of 10~kHz. The periodic line structure due to the oscillation
covers now the entire RP. Used RP parameters: $m=3$, $\tau=1$, $\varepsilon=0.05\sigma$, $L_{\infty}$-norm.
}\label{fig_samplingeffect}
\end{figure}

The previous statements hold for systems, whose characteristic 
frequencies are much lower
than the sampling frequency of their observation. If the sampling frequency
is only one magnitude higher than the system's frequencies, and their ratio
is not a multiple of an integer, and consequently some recurrences
will not be found although they should be there \cite{facchini2005}.
This discretisation effect yields in extended characteristic gaps in the recurrence plot,
those appearances depend on the modulations of the systems frequencies (Fig.~\ref{fig_samplingeffect}). 

%
%
%

\subsubsection{Influence of embedding on the structures in RPs}\label{sec:Influenceofembedding}

\begin{figure}[bth]
\centering \includegraphics[width=\columnwidth]{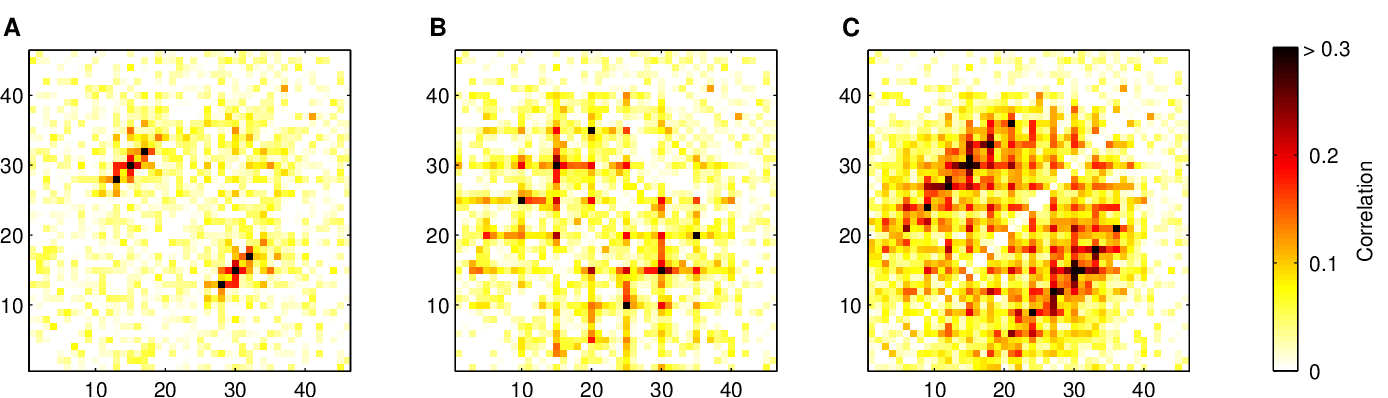} 
\caption{Correlation between a single recurrence point at $(15,30)$ (marked
with grey circle) and other recurrence points in an RP for an uncorrelated 
Gaussian noise (estimated from 2,000 realisations). The embedding parameters are
(A) $m=1$, $\tau=1$, $\varepsilon=0.38$, (B) $m=3$, $\tau=5$, 
$\varepsilon=1.22$ and (C) $m=5$, $\tau=2$, $\varepsilon=1.62$, 
which preserve an approximately constant recurrence point density (0.2).
}\label{fig_rp_redundancy}
\end{figure}

In the case that only a scalar time series has been measured, the phase space has to be 
reconstructed, e.\,g., by means of the delay embedding technique. However, the 
embedding can cause a considerable amount of spurious correlations
in the regarded system, which are then reflected in the RP (\reffig{fig_rp_redundancy}).
This effect can even yield distinct diagonally oriented structures
in an RP of a time series of uncorrelated values if the embedding is high, although
diagonal structures should be extremely rare for such uncorrelated data.

In order to understand this, we consider uncorrelated Gaussian noise $\xi_i$ with standard deviation $\sigma$ and
compute analytically correlations that are induced by a non-appropriate embedding.
Because the considered process is uncorrelated, the correlations detected 
afterwards must be due to the method of embedding.

Using time delay embedding, Eq.~(\ref{eq_embedding}), 
with embedding dimension $m$ and delay $\tau$, 
a vector in the reconstructed phase space
is given by
\begin{equation} 
\vec{x}_i=\sum\limits_{k=0}^{m-1}\xi_{i+k\tau}\vec{e}_{k}.
\end{equation} 
The distance between each pair of these vectors is
$\mathbf{D}_{i,j} = \left\|\vec x_{i} - \vec x_{j}\right\|$. 
Moving $h$ steps ahead in time (i.\,e.~along a diagonal line
in the RP) the respective distance is $\mathbf{D}_{i+h,j+h} = \left\|\vec x_{i+h} - \vec x_{j+h}\right\|$.
For convenience, the auto-covariance function of $\mathbf{D}_{i,j}^2$ will be computed 
instead of computing the auto-covariance function of $\mathbf{D}_{i,j}$. Using the $L_2$-norm, 
the auto-covariance function is
\begin{eqnarray}\label{eq_ACF_RP}
\mathrm{cov}_{\mathbf{D}^2}(h,j-i) & = &
            \left< \left( \sum\limits_{k=0}^{m-1} \left(\xi_{i+k\tau}-\xi_{j+k\tau}\right)^2 - E \right)  \right. \nonumber \\
           & & \left. \left( \sum\limits_{k=0}^{m-1} \left(\xi_{i+h+k\tau}-\xi_{j+h+k\tau}\right)^2 - E \right) \right>,
\end{eqnarray}
where
\begin{equation}
E=\left<\sum\limits_{k=0}^{m-1}\left(\xi_{i+k\tau}-\xi_{j+k\tau}\right)^2\right> = 2 \sigma^2 m \left(1 - \delta_{0,j-i}\right)
\end{equation}
is the expectation value and $\delta_{i,j}$ is the Kronecker delta ($\delta_{i,j} = 1$ if $i=j$, and 
$\delta_{i,j} = 0$ if $i \ne j$). Setting $p=j-i$ and assuming $p>0$ and $h> 0$ 
to avoid trivial cases, we find \cite{thiel2006}
\begin{equation}
\mathrm{cov}_{\mathbf{D}^2}(h,p) = \sum\limits_{k=0}^{m-1}(m-k) \left( 8 \delta_{k\tau,h} +
      2 \delta_{k\tau, p+h} + \delta_{k\tau, p-h} \right).
\end{equation} 
This equation shows that there will be peaks in the auto-covariance function 
if $h$, $p+h$ or $p-h$ are equal to one of the first $m-1$ multiples of $\tau$. 
These peaks are not present when no embedding is used ($m=1$). 
Such spurious correlations induced by embedding lead to modified small-scale structures in
the RP: an increase of the embedding dimension cleans up the RP from single 
recurrence points (representatives for the uncorrelated states) and emphasises 
the diagonal structures as diagonal lines (representatives for the correlated states). 
This, of course, influences any quantification of RPs, which is based on 
diagonal lines. Hence, we should be careful in interpreting and 
quantifying structures in RPs of measured systems. If the embedding 
dimension is, e.\,g., inappropriately high, spurious long diagonal lines 
will appear in the RP. In order to avoid this problem, the embedding parameters
have to be chosen carefully, or alternatively quantification measures which are 
independent of the embedding dimension have to be
used (cf.~Subsec.~\ref{sec:Invariants}).

The spurious correlations in RPs due to embedding can also be understood
from the fact that an RP computed with  
any embedding dimension can be derived from an RP computed without embedding ($m=1$).
Consider, e.\,g., $m=2$ with certain $\tau$ and the maximum norm. 
A recurrence point at $(i,j)$ will occur if
\begin{equation}
\vec{x}_i \approx \vec{x}_j \quad \Longleftrightarrow \quad \max\left( | x_i - x_j | , | x_{i+\tau} - x_{j+\tau} | \right) < \varepsilon.
\end{equation}
This is the same as
$x_i \approx x_j$ and $x_{i+\tau} \approx x_{j+\tau}$
and corresponds to two recurrence points at $(i,j)$ and $(i+\tau, j+\tau)$ 
in an RP without embedding. Thus, a recurrence point for a reconstructed
trajectory with an embedding dimension $m$ is 
\begin{equation}\label{eq_redundant_RPs}
\mathbf{R}_{i,j}^{(m)} = \mathbf{R}_{i,j}^{(1)} \cdot
                                   \mathbf{R}_{i + \tau,j + \tau}^{(1)} \cdot
                                    \ldots \cdot 
                                    \mathbf{R}_{i + (m-1)\tau,j + (m-1)\tau}^{(1)},
\end{equation}
where $\mathbf{R}^{(1)}$ is the RP without embedding (or parent RP)
and $\mathbf{R}^{(m)}$ is the RP for embedding dimension $m$ \citep{march2005}.
The entry at $(i,j)$ in the recurrence matrix $\mathbf{R}^{(m)}$ consists of
information at times $(i + \tau,j + \tau), \ldots,  (i + (m-1)\tau,j + (m-1)\tau)$.
If the threshold $\varepsilon$ is large enough, spurious recurrence points 
along the line $(i+k,j+k)$ for $k = 0\, \ldots, (m-1)\tau$ can appear.
It is clear that, e.\,g., in the case of a stochastic signal which is 
embedded in a high-dimensional space, such diagonal lines in an 
RP may feign a non-existing determinism.

\subsubsection{Modifications and extensions}\label{sec:VariationsRecurrencePlots}

In the original definition of RPs, the neighbourhood is
a ball (i.\,e.~$L_2$-norm is used) and its radius
is chosen in such a way that it contains a fixed amount $N_{\text{n}}$ of states $\vec x_j$ 
\citep{eckmann87}. With such a neighbourhood,
the radius $\varepsilon=\varepsilon_{i}$ changes for each $\vec x_i$
and $\mathbf{R}_{i,j}\not=\mathbf{R}_{j,i}$ because the
neighbourhood of $\vec x_i$ is in general not the same as that of $\vec x_j$.
This leads to an asymmetric RP, but all columns of
the RP have the same recurrence density (\reffig{fig_rp_lorenz}D).
Using this neighbourhood criterion, $\varepsilon_i$ can be adjusted in such a way 
that the recurrence point density has a fixed predetermined value
(i.\,e.~$RR=N_{\text{n}}/N$). This neighbourhood 
criterion is denoted as 
{\it fixed amount of nearest neighbours (FAN)}.
However, the most commonly used neighbourhood is that with a
fixed radius $\varepsilon_i=\varepsilon, \forall i$. For RPs
this neighbourhood was firstly used in \cite{zbilut91}. A fixed 
radius means that $\mathbf{R}_{i,j}=\mathbf{R}_{j,i}$ resulting in
a symmetric RP.
The type of neighbourhood that should preferably be used depends on the 
purpose of the analysis. Especially for the later 
introduced cross recurrence plots (Subsec.~\ref{sec:CrossRecurrencePlots})
and the detection of generalised synchronisation (Subsec.~\ref{sec:DetectionOfGS}), 
the neighbourhood with a FAN will 
play an important role.

In the literature further variations of RPs
have been proposed (henceforth we assume $\vec x \in \mathds{R}^d$):
\begin{itemize}
\item Instead of plotting the recurrence matrix (Eq.~\ref{eq_rp}), the distances
    \begin{equation}\label{eq_DP}
    \mathbf{D}_{i,j}=\|\vec x_{i} - \vec x_{j}\|
    \end{equation}
    can be plotted (\reffig{fig_rp_lorenz}H). Although this is not an RP,
    it is sometimes called {\it global recurrence plot}
    \citep{webberweb} or {\it unthresholded recurrence plot}
    \citep{iwanski98}. The name {\it distance plot} would be perhaps more appropriate.

    A practical modification is the {\it unthresholded recurrence plot} defined
    in terms of the correlation sum $C_2$, Eq.~(\ref{eq_corrsum}), 
    \begin{equation}\label{eq_UTRP}
    \mathbf{U}_{i,j}=C_2\left(\|\vec x_{i} - \vec x_{j}\| \right),
    \end{equation}
    where the values of the correlation sum with respect to the distance
    $\|\vec x_{i} - \vec x_{j}\|$ are used \citep{march2005}.
    Applying a threshold $\varepsilon$ to such an unthresholded RP reveals
    an RP with a recurrence point density which is exactly $\varepsilon$.

    These representations can also help in studying phase space trajectories. 
    Moreover, they may help to find an appropriate threshold value $\varepsilon$.
\item Iwanski and Bradley defined a variation of an RP with a corridor
    threshold $[\varepsilon_{\text{in}},\varepsilon_{\text{out}}]$ (\reffig{fig_rp_lorenz}E) \cite{iwanski98},
    \begin{equation}
    \mathbf{R}_{i,j}([\varepsilon_{\text{in}},\,\varepsilon_{\text{out}}]) = 
        \Theta\left(\left\|\vec x_{i} - \vec x_{j}\right\|-\varepsilon_{\text{in}}\right)\cdot \Theta\left(\varepsilon_{\text{out}}-\left\|\vec x_{i} - \vec x_{j}\right\|\right).
    \end{equation}
    Those points $\vec x_j$ are considered to be recurrent that fall into the
    shell with the inner radius $\varepsilon_{\text{in}}$ and the outer radius
    $\varepsilon_{\text{out}}$. 
   An advantage of such a {\it corridor thresholded recurrence plot}
    is its increased robustness against recurrence points coming from
    the tangential motion. However,
    the threshold corridor removes the inner points in broad diagonal lines, which
    results in two lines instead of one. These RPs are, therefore, not directly suitable
    for a quantification analysis.
    The shell as a neighbourhood was used in an attempt to
    compute Lyapunov exponents from experimental time series \citep{eckmann86}.
\item Choi et al.~introduced the {\it perpendicular recurrence plot}
     (\reffig{fig_rp_lorenz}F) \cite{choi99}
    \begin{equation}
    \mathbf{R}_{i,j}(\varepsilon) =
    \Theta\left(\varepsilon-\left\|\vec x_{i} - \vec x_{j}\right\|\right) \cdot \delta \left( \dot{\vec{x}}_i \cdot (\vec x_{i} - \vec x_{j}) \right),
    \end{equation}
    with $\delta$ denoting the Delta function ($\delta(x) = 1$ if $x=0$, and 
    $\delta(x) = 0$ otherwise). This RP contains
    only those points $\vec x_j$ that fall into the neighbourhood of 
    $\vec x_i$ and lie in the $(d-1)$-dimensional
    subspace of $\mathds{R}^d$ that is perpendicular to the phase space trajectory 
    at $\vec x_i$. These points correspond locally to those lying on a Poincar\'e section.  
    This criterion cleans up the RP more effectively from recurrence points based on the tangential 
    motion than the previous corridor thresholded RPs. This kind of RP is more efficient 
    for estimating invariants and is more robust for the detection of UPOs (if they exist).

\item The {\it iso-directional recurrence plot}, 
    introduced by Horai and Aihara \cite{horai2002},
    \begin{equation}
      \mathbf{R}_{i,j}(\varepsilon,T) = 
            \Theta\left(\varepsilon-\left\|\left(\vec x_{i+T} - \vec x_{i}\right)
            - \left(\vec x_{j+T} - \vec x_{j}\right) \right\|\right),
    \end{equation}
    is another variant which takes the direction of the trajectory evolution
    into account.
    Here a recurrence is related to neighboured trajectories which 
    run parallel and in the same direction. The authors introduced 
    an additional {\it iso-directional neighbours plot}, which is simply 
    the product between the common RP and the iso-directional 
    RP \cite{horai2002}
    \begin{equation}
      \mathbf{R}_{i,j}(\varepsilon,T) = 
            \Theta\left(\varepsilon-\left\|\vec x_{i} - \vec x_{j}\right\|\right) \cdot 
            \Theta\left(\varepsilon-\left\|\left(\vec x_{i+T} - \vec x_{i}\right)
            - \left(\vec x_{j+T} - \vec x_{j}\right) \right\|\right).
    \end{equation}
    The computation of this special recurrence plot is simpler than 
    that of the perpendicular RP. Although the cleaning of the 
    RP from false recurrences is better than in the common RP, 
    it does not reach the quality of a perpendicular RP. 
    A disadvantage is the additional parameter $T$ which has to be 
    determined carefully in advance (however, it seems that this 
    parameter can be related to the embedding delay $\tau$).

\item It is also
    possible to test each state with a pre-defined amount $k$ of subsequent states
    \citep{atay99,zbilut91,koebbe92}
    \begin{equation}
    \mathbf{R}_{i,j}(\varepsilon) = 
        \Theta\left(\varepsilon-\left\|\vec x_{i} - \vec x_{i+i_0+j-1}\right\|\right), 
        \quad i=1, \ldots, N-k, \ j=1, \ldots, k.
    \end{equation}
    This reveals an $(N-k) \times k$-matrix which does not have to be square (\reffig{fig_rp_lorenz}J). 
    The $y$-axis represents the time distances to the following recurrence points but
    not their absolute time. All diagonally oriented structures in the common RP are now 
    projected to the horizontal direction. For $i_0=0$, the LOI,
    which was the main diagonal line in the original RP, is now the horizontal line which
    coincides with the $x$-axis.
    With non-zero $i_0$, the RP contains recurrences of a
    certain state only in the pre-defined time interval after time $i_0$ \citep{koebbe92}.
    
    This representation of recurrences may be more intuitive than that of 
    the original RP because the consecutive
    states are not oriented diagonally. However, such an RP represents only the first
    $(N-k)$ states. Mindlin and Gilmore proposed the {\it close returns 
    plot} \cite{mindlin92} which is, in fact, such an RP exactly for dimension one. 
    Using this definition of RP, a first quantification approach of RPs 
    (or ``close returns plots'') was introduced (``close returns histogram'', 
    recurrence times; cf.~Subsec.~\ref{sec:RQA_diagonal_lines}). 
    It has been used for the investigation of periodic 
    orbits and topological properties of strange attractors 
    \cite{lathrop89,tufillaro90,mindlin92}.
\item The {\it windowed} and {\it meta recurrence plots} have been suggested 
    as tools to investigate an external force or non-stationarity in a system
    \cite{manuca96,casdagli97}. The first ones are obtained by covering 
    an RP with $w \times w$-sized squares (windows) and by
    averaging the recurrence points that are contained in 
    these windows \cite{casdagli97}. Consequently, a windowed recurrence plot
    is an $N_w \times N_w$-matrix, where $N_w$ is the floor-rounded $N/w$,
    and consists of values which are not limited to zero and one (this suggests
    a colour-encoded representation). These values
    correspond to the {\it cross correlation sum}, Eq.~(\ref{eq_xcorrsum}),
    \begin{equation}\label{eq_metaRP}
    \mathbf{C}_{I,J}(\varepsilon) = \frac{1}{w^2}
             \sum_{i,j=1}^{w} 
             \mathbf{R}_{i+(I-1)w,j+(J-1)w}(\varepsilon)\,, 
             \quad I,J=1, \ldots, \frac{N}{w}
    \end{equation}
    between sections in $\vec x$ with
    length $w$ and starting at $(I-1)w+1$ and $(J-1)w+1$ (for cross correlation
    integral cf.~\cite{kantz94}). Windowed RPs can be useful for the
    detection of transitions or large-scale patterns in RPs of very long data series.

    The {\it meta recurrence plot},
    as defined in \cite{casdagli97}, is a distance matrix derived from
    the cross correlation sum, Eq.~(\ref{eq_metaRP}),
    \begin{equation}
    \mathbf{D}_{I,J}(\varepsilon) = \frac{1}{\varepsilon^d}\left(
             \mathbf{C}_{I,I}(\varepsilon) +
             \mathbf{C}_{J,J}(\varepsilon) -
             2\, \mathbf{C}_{I,J}(\varepsilon)\right).
    \end{equation}
    By applying a further threshold value to $\mathbf{D}_{I,J}(\varepsilon)$
    (analogous to Eq.~\refeq{eq_rp}), a black-white dotted representation
    is also possible.  
    
    Manuca and Savit have gone one step further by using quotients from the
    cross correlation sum to form a {\it meta phase space} \cite{manuca96}.
    From this meta phase space a recurrence or non-recurrence
    plot is created, which can be used to characterise non-stationarity
    in time series. 

\item Instead of using the spatial closeness between phase space trajectories, 
    {\it order patterns recurrence plots (OPRP)} are based on order patterns $\pi$ for the definition
    of a recurrence. An order pattern $\pi$ of dimension $d$ is defined by the
    discrete order sequence of the data series $x_i$ and has the length $d$. For $d=3$
    we get, e.\,g., six different order patterns (Fig.~\ref{fig_orderpatterns}). Using these order
    patterns, the data series $x_i$ is symbolised by order patterns:

    $$x_i, x_{i-\tau_1}, \ldots, x_{i-\tau_d-1} \ \rightarrow \ \pi_i.$$

\end{itemize}
\begin{figure}[htbp]
\centering{\includegraphics[width=0.75\columnwidth]{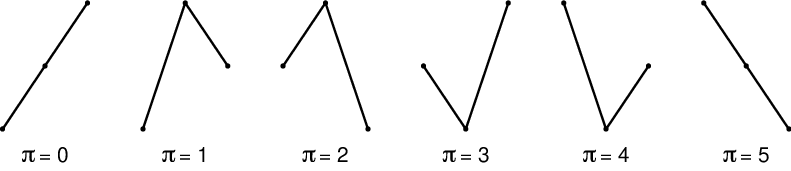}}
\caption{Order patterns for dimension $d=3$. They describe a specific 
rank order of three data points and can be used to exhibit a recurrence
by means of local rank order sequence.}\label{fig_orderpatterns}
\end{figure}
    The order patterns recurrence plot (\reffig{fig_rp_lorenz}G) is then defined by the pair-wise test
    of order patterns \citep{groth2004}:
    \begin{equation}
    \mathbf{R}_{i,j}(d) =\delta\left(\pi_{i},\ \pi_{j}\right).
    \end{equation}
    Such an RP represents those times, when specific rank order sequences 
    in the system recur. Its main advantage is its robustness with respect to
    non-stationary data. Moreover, it increases the applicability of
    cross recurrence plots (cf.~Subsec.~\ref{sec:CrossRecurrencePlots}).

    A hybrid between a common RP and an OPRP is the {\it ordinal recurrence plot}
    (\reffig{fig_rp_lorenz}I) \cite{groth2006diss}:
    \begin{eqnarray}
    \mathbf{R}_{i,j}(d) &=& \Theta\left( s (x_{i+\tau} - x_{i})\right)
                         \Theta\left( s (x_{j+\tau} - x_{j})\right) \cdot  \nonumber \\
                       & & {} \Theta\left( s (x_{i+\tau} - x_{j})\right)
                         \Theta\left( s (x_{j+\tau} - x_{i})\right),
    \ \text{for} \  s \in \{-1,+1\}.
    \end{eqnarray}
    It looks whether two states are close and, additionally, whether such both
    states grow or shrink simultaneously (Fig.~\ref{fig_ordinals}).

\begin{figure}[htbp]
\centering{\includegraphics[width=0.75\columnwidth]{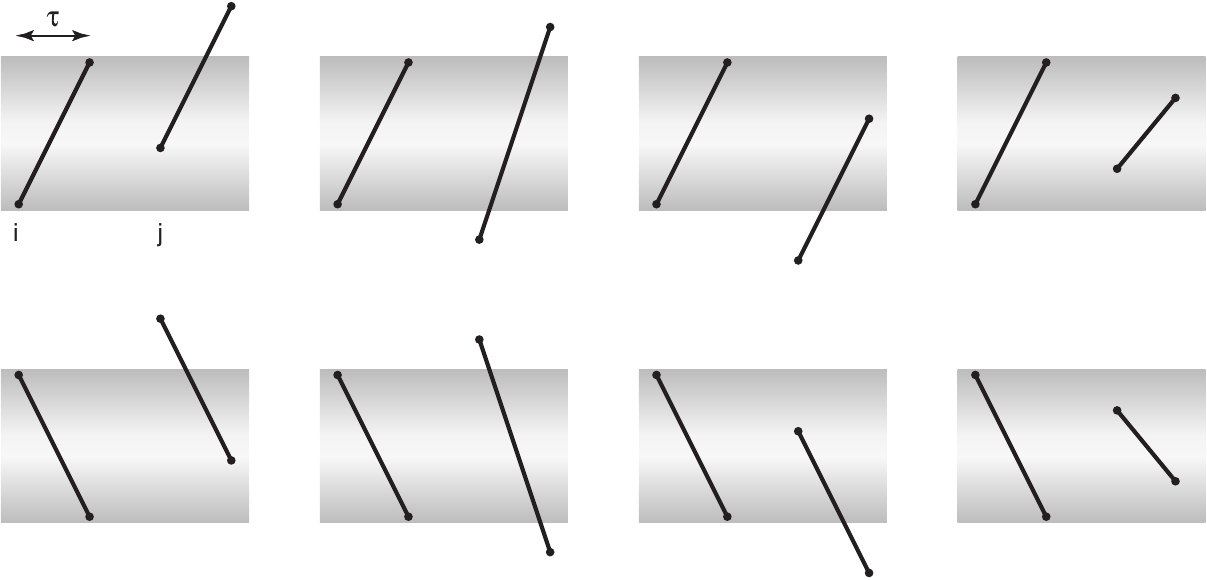}}
\caption{Ordinal cases which are considered to be a recurrence
in an ordinal recurrence plot.}\label{fig_ordinals}
\end{figure}

\begin{figure}[hbtp]
\centering \includegraphics[width=\columnwidth]{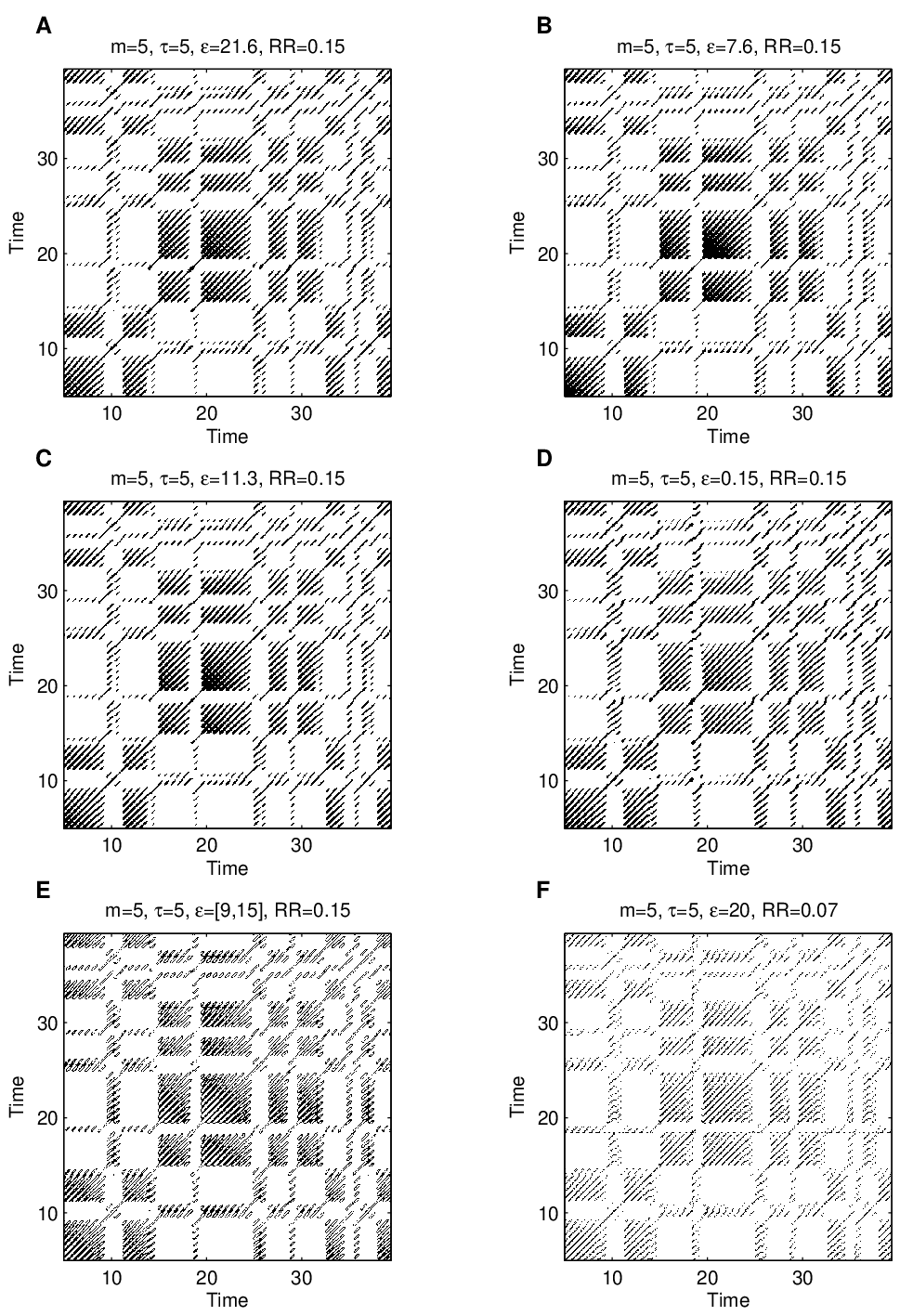} 
\caption{Continued on p.~\pageref{fig_rp_lorenz}.}
\end{figure}

Furthermore, the term {\it recurrent plots} can be found for RPs in the 
literature (e.\,g.~\cite{balasubramaniam2000}). However, this term should not
be used for RPs (it is sometimes used for return time plots).
Finally, it should be mentioned that the term {\it recurrence plots}
is sometimes used for another representation not related to RPs (e.\,g.~\cite{huang2002}).

\addtocounter{figure}{-1}
\begin{figure}[tbp]
\includegraphics[width=\columnwidth]{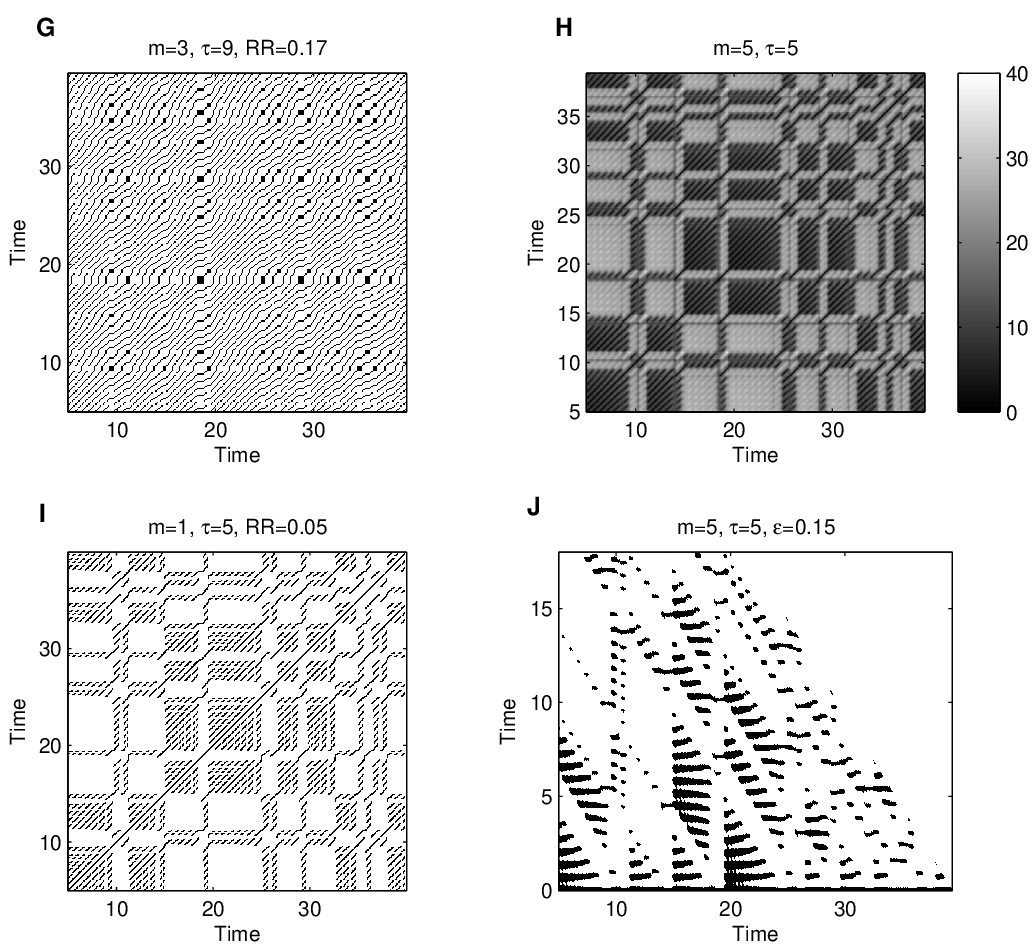} 
\caption{Examples of various definitions of RPs for a section of 
the $x$-component of the Lorenz system, Eqs.~(\ref{eq_lorenz}), $r=28$, 
$\sigma=10$, $b=\frac{8}{3}$, sampling time $\Delta t = 0.03$ \cite{lorenz63}: 
(A) RP by using the $L_{\infty}$-norm,
(B) RP by using the $L_{1}$-norm, 
(C) RP by using the $L_{2}$-norm,
(D) RP by using a fixed amount of nearest neighbours (FAN), 
(E) RP by using a threshold corridor $[\varepsilon_{\text{in}},\varepsilon_{\text{out}}]$,
(F) perpendicular RP ($L_{2}$-norm), 
(G) order patterns RP ($m=3$, $\tau_{1,\,2,\,3}=9$),
(H) distance plot (unthresholded RP, $L_{2}$-norm),
(I) ordinal RP ($m=1$, $\tau=5$), 
(J) RP where the $y$-axis represents the relative time distances to the next recurrence points
but not their absolute time (``close returns plot'', $L_{2}$-norm).
Except for (F), (G), (H) and (I), the 
parameter $\varepsilon$ is chosen in such a way that the recurrence point 
density (RR) is approximately the same. The embedding parameters ($m=5$ and $\tau=5$) correspond
to an appropriate time delay embedding.}\label{fig_rp_lorenz}
\end{figure}

The selection of a specific definition of the RP depends on the problem
and on the kind of system or data. Perpendicular
RPs are highly recommended for the quantification analysis based on diagonal
structures, whereas corridor thresholded RPs are not suitable for this task.
Windowed RPs are appropriate for the visualisation of the long-range behaviour 
of rather long data sets. 
If the recurrence behaviour for the states $\vec x_i$ within a pre-defined section
$\{\vec{x}_{i+i_0},\,\ldots,\, \vec{x}_{i+i_0+k}\}$ of the phase
space trajectory is of special interest,
an RP with a horizontal LOI will be suitable.
We will use the standard definition of 
RP, Eq.~(\ref{eq_rp}), according to \cite{eckmann87} in this report. 

It should be emphasised again that the recurrence of states is a 
fundamental concept in the analysis of dynamical systems.
Besides RPs, there are some other methods 
that use recurrences: e.\,g.~recurrence time statistics
\citep{balakrishnan2000,gao99,kac47}, 
first return map \citep{lathrop89}, space time separation plot \citep{provenzale92} or
recurrence based measures for the detection of non-stationarity (closely related to the 
recurrence time statistics, \cite{kennel97,rieke2002}).

The pair-wise test between all elements of a series or of two different series, can
also be found in other methods much earlier than 1987. There are some methods 
rather similar to RPs developed in several 
fields. To our knowledge, first ideas go back to the 70ies, where the {\it dynamic time
warping} was developed for speech recognition. The purpose was to
match or align two sequences \citep{rabiner93, sakoe78}.
For a similar purpose, but for genome sequence alignment, the {\it dot
matrix}, {\it similarity plot} or {\it sequence matrix} (several terms
for the same thing) were introduced \citep[e.\,g.][]{kruskal1983,maizel1981,vihinen88}. 
Remarkable variations thereof are {\it dot plots} and {\it link plots}, which
were developed to detect structures in text and computer codes \citep{bernstein1991},
and {\it similarity plots}, developed for the analysis of changing images \citep{cutler2000}.
The {\it self-similarity matrix} and the {\it ixegram} represent similarities
or distances of features in long data series, like frequencies, histograms or
specific descriptors. They are used to recognise specific structures in
music and the ixegram is part of the MPEG-7 standard \citep{casey2000,foote2001}. 
{\it Contact maps} are used to visualise the complex structures of folded proteins
and to reconstruct them (e.\,g.~by measuring the distances between C atoms) and
go back to the early 70ies (e.\,g.~\cite{chan1989,domany2000,holm1993}).
We have already mentioned the {\it close returns plots} as a special case of the RPs for
$m=1$, but they should be
listed here too. They were introduced to recognise periodic orbits and topological
properties of attractors and differ from RPs in the use of
consecutive states as the $y$-axis instead of all states \citep{mindlin92}.

%% file: meth_crp.tex
%
%
%
%

\begin{figure}[b]
\centering \includegraphics[width=\columnwidth]{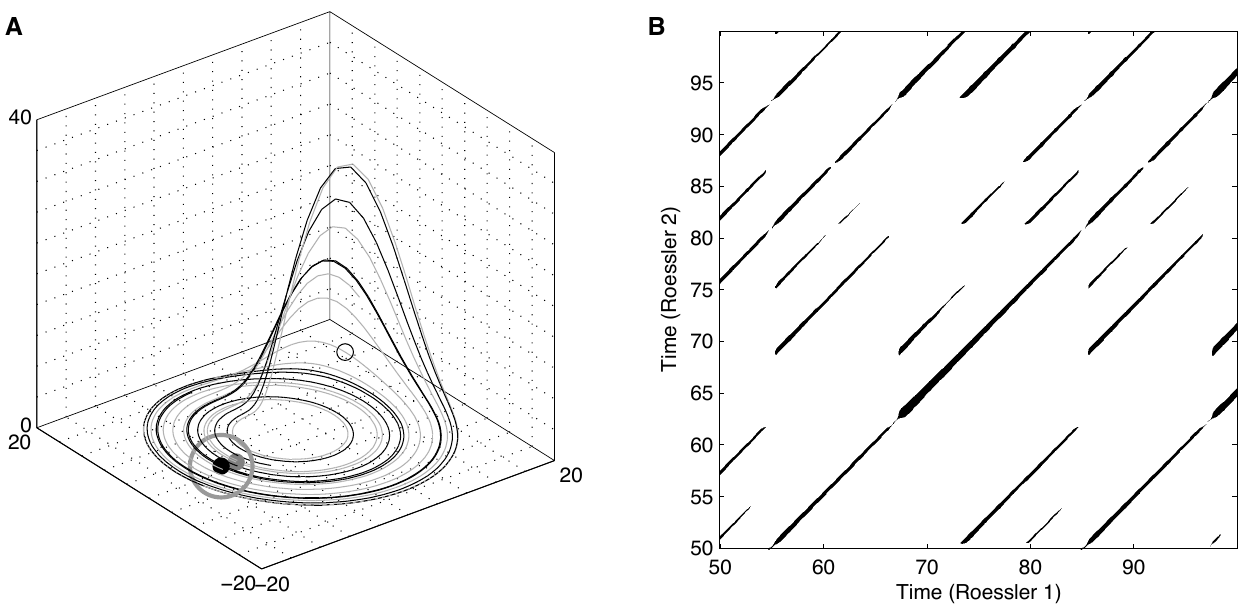} 
\caption{(A) Phase space trajectories of two coupled R\"ossler systems, 
Eqs.~(\ref{eq_2roessler1}) and (\ref{eq_2roessler2}), with $a=0.15$, 
$b=0.20$, $c=10$, $\nu=0.015$ and $\mu=0.01$ by using their three components
(black and grey line correspond to the first and second oscillator).
In (B) the corresponding CRP is shown ($L_2$ norm and $\varepsilon=3$ is used).
(A) If a phase space vector of the second R\"ossler system at $j$ (grey point on the grey line) 
falls into the neighbourhood (grey circle) of a phase space vector of the first R\"ossler 
system at $i$, in the CRP (B) at the location $(i,j)$ a black point will occur.}\label{fig_constrCRP}
\end{figure}

The {\it cross recurrence plot (CRP)} is a bivariate extension of the RP and was introduced to analyse 
the dependencies between two different systems by comparing their states \citep{zbilut98,marwan2002pla}. 
They can be considered as a generalisation of the linear cross-correlation function. 
Suppose we have two dynamical systems, each one represented by the trajectories 
$\vec{x}_i$ and $\vec{y}_i$ in a $d$-dimensional phase space (Fig.~\ref{fig_constrCRP}A). 
Analogously to the RP, Eq.~(\ref{eq_rp}), the corresponding cross recurrence 
matrix (Fig.~\ref{fig_constrCRP}B) is defined by 
\begin{equation}\label{eq_crp}
\mathbf{CR}^{\vec{x},\vec{y}}_{i,j}(\varepsilon) = 
\Theta\left(\varepsilon-\|\vec{x}_i-\vec{y}_j\|\right), \qquad i=1,\ldots,N, \ j=1,\ldots,M, 
\end{equation}
where the length of the trajectories of $\vec{x}$ and $\vec{y}$ is not required 
to be identical, and hence the matrix $\mathbf{CR}$ is not necessarily square. 
Note that both systems are represented in the same phase space, because a CRP
looks for those times when a state of the first system recurs to one of the other system. 
Using experimental data, it is sometimes difficult to reconstruct the phase space.
If the embedding parameters are estimated from both time series, but are not equal, 
the higher embedding should be chosen. However, the data under consideration should be from the
same (or a very comparable) process and, actually, should represent the same 
observable. Therefore, the reconstructed phase space should be the same.
An exception is the {\it order patterns cross recurrence plot}, where 
the values of the states are not directly compared but the local rank order sequence of the two
systems \cite{groth2005}. Then both systems can be represented by different
observables (or time series of very different amplitudes).

This bivariate extension
of the RP was introduced for the {\it cross recurrence quantification} \cite{zbilut98}. 
Independently, the concept of CRPs also surfaces in an approach to
study interrelations between time series \cite{marwan99}.
The components of $\vec{x}_i$ and $\vec{y}_i$ are usually normalised before computing the cross 
recurrence matrix. Other possibilities are to use a fixed amount of neighbours 
(FAN) for each $\vec{x}_i$ or to use order patterns \citep{groth2004}. 
This way the components of $\vec{x}_i$ and $\vec{y}_i$ do not 
need to be normalised. The latter 
choices of the neighbourhood have the additional advantage of working 
well for slowly changing trajectories (e.\,g.~drift).

Since the values of the main diagonal $\mathbf{CR}_{i,i}\,|_{i=1}^N$ are not
necessarily one, there is usually not a black main diagonal (Fig.~\ref{fig_constrCRP}B). 
Apart from that,
the statements given in the subsection about all the structures in RPs
(Subsec.~\ref{sec:StructuresinRecurrencePlots}, Tab.~\ref{tab:typology}) 
hold also for CRPs. The lines which are
diagonally oriented are here of major interest too.
They represent segments on both trajectories, which run 
parallel for some time. The frequency and length of these lines 
are obviously related to a certain similarity between the 
dynamics of both systems. A measure based on the lengths of
such lines can be used to find nonlinear interrelations between two
systems (cf.~Subsecs.~\ref{sec:RQA} and ~\ref{sec:appl_crp_synchro}), 
which cannot be detected by
the common cross-correlation function \cite{marwan2002pla}.

\begin{figure}[b]
\centering \includegraphics[width=\columnwidth]{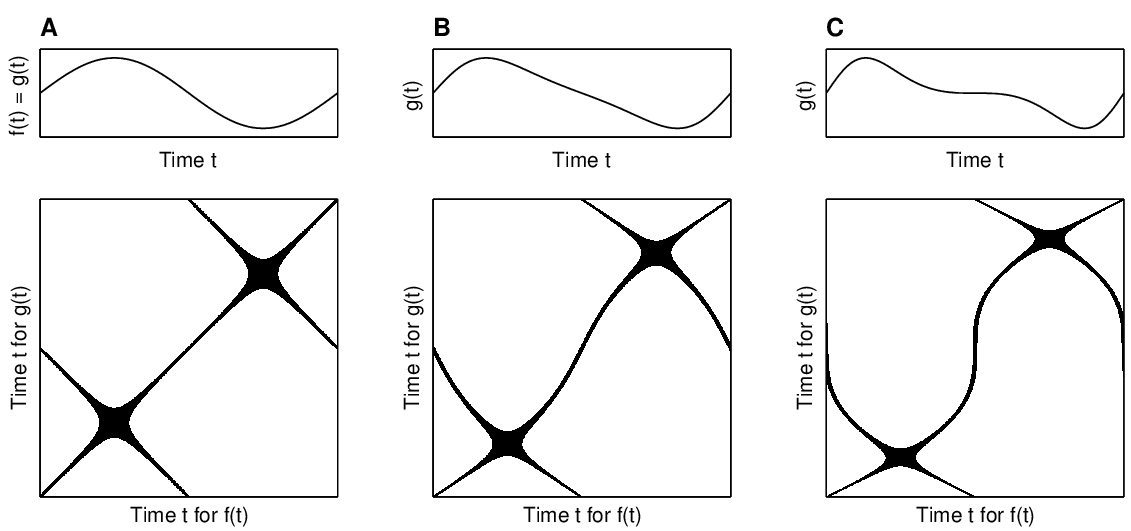} 
\caption{CRPs of sine functions $f(t)=\sin(\varphi t)$
and $g(t)=\sin(\varphi t + a  \sin(\psi t))$ with fixed frequencies 
$\varphi$ and $\psi$, whereas (A) $a=0$, (B) $a=0.5$  and (C)
$a=1$. The variation in the time domain
leads to a deformation of the synchronisation line.
The CRPs are computed without embedding.}\label{fig_crp_los}
\end{figure}

An important advantage of CRPs is that they reveal the local
difference of the dynamical evolution of close trajectory 
segments, represented by bowed lines.
A time dilatation or time
compression of one of the trajectories causes 
a distortion of the diagonal lines (cf.~remarks about the
relationship between the slope of RP lines and local time
transformations in Subsec.~\ref{sec:StructuresinRecurrencePlots}). 
Assuming two identical trajectories, the CRP coincides with the 
RP of one trajectory and contains the main black diagonal or line of identity (LOI). 
If the values of the second trajectory are slightly modified, the LOI
will become somewhat disrupted and is called line of synchronisation (LOS). 
However, if we do not modify the amplitudes but
stretch or compress the second trajectory slightly, the LOS will
still be continuous but not a straight line with slope one (angle of $\pi/4$).
This line can rather become bowed (\reffig{fig_crp_los}). As we have already seen
in the Subsec.~\ref{sec:StructuresinRecurrencePlots},
the local slope of lines in an RP as well as in a CRP corresponds to the
transformation of the time axes of the two considered trajectories, 
Eq.~(\ref{eq:los}) \cite{marwan2005}.
A time shift between the trajectories causes a dislocation of the LOS. Hence,
the LOS may lie rather far from the main diagonal of the CRP. As we will 
see in the following example, 
the LOS allows finding the non-parametric rescaling function 
between different time series.

\subsubsection*{Example: Time scale alignment of two harmonic functions with changing frequencies}

\begin{figure}[b]
\centering \includegraphics[width=\columnwidth]{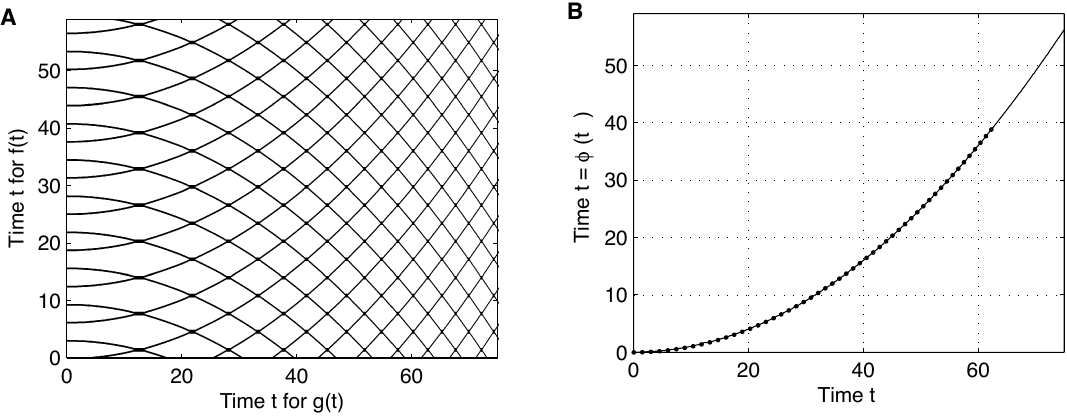}
\caption{\label{fig_crp_sin}
(A) Cross recurrence plot of two sine functions $f(t)=\sin(\varphi t)$
and $g(t)=\sin(\psi t^2)$, which is the base for the determination
of the rescaling function between both data series. The CRP was created without
embedding and using FAN $\varepsilon=0.1$ (constant recurrence density 
of 10\,\%). (B) Time-transfer function (black) determined from the CRP. 
It has the expected parabolic shape of the
squared coherence in the time domain (dotted line: square 
function).}
\end{figure}

\begin{figure}[htbp]
\centering \includegraphics[width=.8\columnwidth]{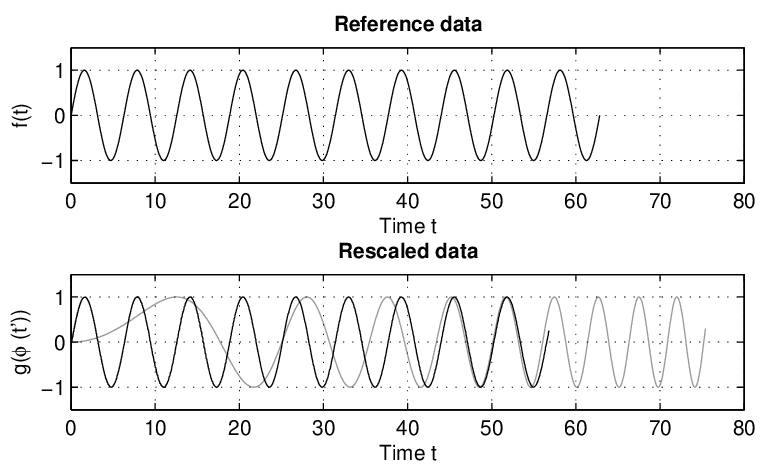}
\caption{\label{fig_crp_sin_data}
Reference data series (upper panel) and rescaled data 
series before (grey) and after (black) rescaling by using 
the time-transfer function presented in Fig.~\ref{fig_crp_sin} (lower panel).}
\end{figure}

We consider two sine functions $f(t)=\sin(\varphi t)$
and $g(t')=\sin(\psi t')=\sin(\psi t^2)$, where the time scale
of the second sine differs from the first by a quadratic transformation
($t' = t^2$) and has a frequency of $\psi=0.01 \, \varphi$.
Such a nonlinear changing of time scales can be found in nature,
e.\,g., with increasing depth, sediments in a lake undergo an 
increasing amount of pressure resulting in compression (cf.~Subsec.~\ref{sec:appl_rescaling}).
It can be assumed that both data series come from the same process and were
subjected to different deposital compressions (e.\,g.~a squared
or exponential
increasing of the compression).  Hence, their CRP contains a bowed
LOS (Fig.\,\ref{fig_crp_sin}A). We have used no embedding 
and a varying threshold $\varepsilon$,
such that the CRP contains a constant recurrence density of 10\,\% (FAN). 
In order to find the non-parametrical 
time-transfer function $t=\phi(t')$, the LOS has to be resolved
from the CRP. The resulting rescaling 
function has the expected squared shape $t=\phi(t')=0.01 \, t'^2$ 
(grey dashed curve in 
Fig.\,\ref{fig_crp_sin}B). Substituting the time scale $t'$
in the second data series $g(t')$ by this rescaling function 
$t=\phi(t')$, we get a set of aligned data $f(t)$ and $g(t)$
with the non-parametric rescaling function $t=\phi(t')$
(Fig.\,\ref{fig_crp_sin_data}). The aligned data series are
now approximately the same.

%% file: meth_jrp.tex
\begin{figure}[t]
\centering \includegraphics[width=\columnwidth]{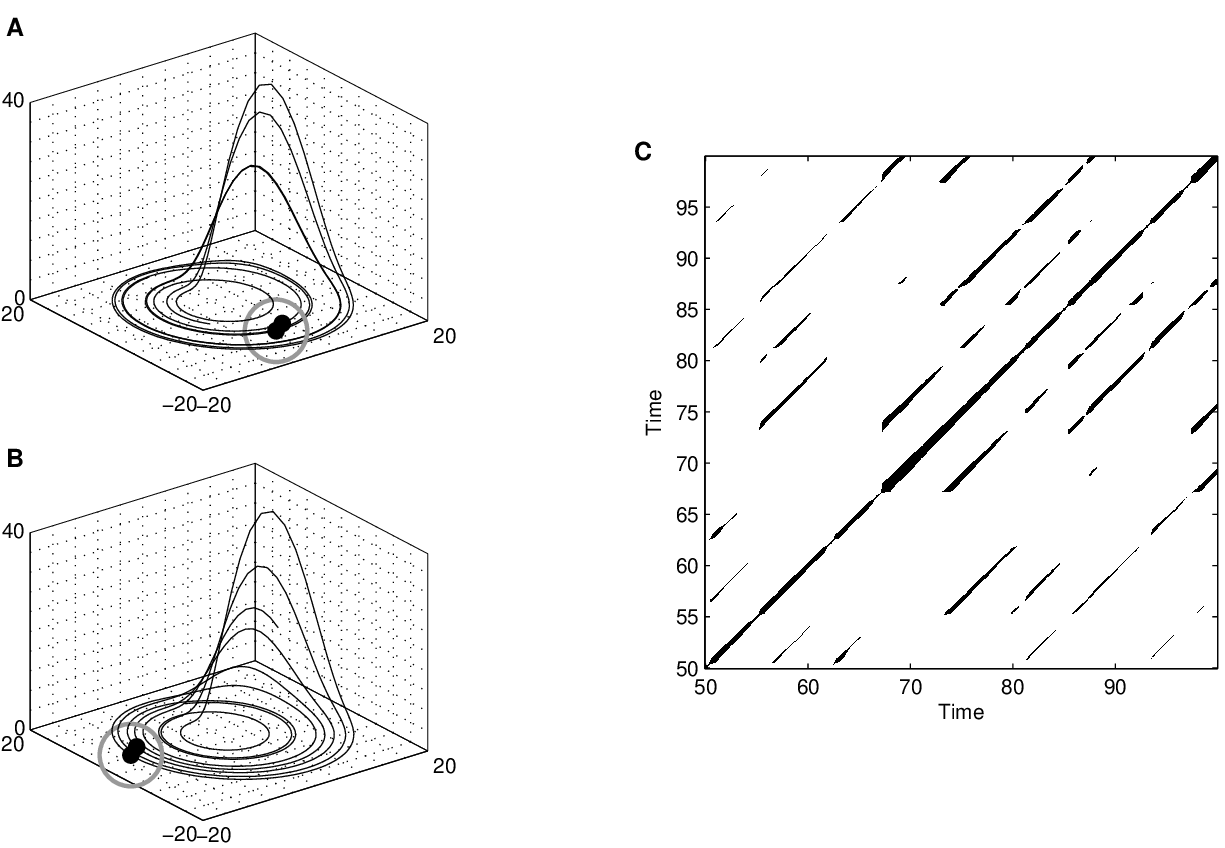} 
\caption{(A, B) Phase space trajectories of two coupled R\"ossler systems, 
Eqs.~(\ref{eq_2roessler1}) and (\ref{eq_2roessler2}), with $a=0.15$, 
$b=0.20$, $c=10$, $\nu=0.015$ and $\mu=0.01$. In (C) the corresponding JRP is shown
($L_2$ norm and $\varepsilon=5$ is used for both systems).
If two phase space vectors of the second R\"ossler system at $i$ and $j$ 
are neighbours (black points in (B)) {\it and} if two phase space 
vectors of the first R\"ossler system at same $i$ and $j$ are also 
neighbours (black points in (A)), a black point in the JRP at the location $(i,j)$ 
will occur.}\label{fig_constrJRP}
\end{figure}

As we have seen in the previous section, the bivariate
extension of RPs to CRPs allows studying the
relationship between two different systems by examining the
occurrence of similar {\it states}. However, CRPs cannot be
used for the analysis of two physically different time
series, because the difference between two vectors with
different physical units or even different phase space
dimension does not make sense.

Another possibility to compare different systems is to
consider the  {\it recurrences} of their trajectories in
their respective phase spaces  separately and look for the
times when both of them recur simultaneously,  i.\,e.~when
a {\it joint recurrence} occurs. By means of this
approach,  the individual phase spaces of both systems are
preserved. Formally,  this corresponds to an extension of
the phase space to $\mathds{R}^{d_x+d_y}$,  where $d_x$ and
$d_y$ are the phase space dimensions of the corresponding 
systems, which are in general different (i.\,e.~it
corresponds to the direct product of the individual phase
spaces).  Furthermore, two different thresholds for each
system,  $\varepsilon^{\vec x}$ and $\varepsilon^{\vec y}$,
are considered, so that  the criteria for choosing the
threshold (Subsec.~\ref{sec:Selectionofthethreshold})  can
be applied separately,  respecting the natural measure of
both systems. Hence, it is intuitive to introduce  the {\it
joint recurrence matrix} (Fig.~\ref{fig_constrJRP}) for two
systems $\vec x$ and $\vec y$
\begin{equation}\label{eq_jrp}
\mathbf{JR}^{\vec{x},\vec{y}}_{i,j}(\varepsilon^{\vec x},\varepsilon^{\vec y}) = 
\Theta\left(\varepsilon^{\vec x}-\|\vec{x}_i-\vec{x}_j\|\right) 
\Theta\left(\varepsilon^{\vec y}-\|\vec{y}_{i}-\vec{y}_{j}\|\right), \quad i,j=1,\ldots,N,
\end{equation}
or, more generally, for $n$ systems $\vec{x}_{(1)}$, $\vec{x}_{(2)}$, \dots, $\vec{x}_{(n)}$ 
and using Eq.~(\ref{eq_rp}), the {\it multivariate joint recurrence matrix} can be introduced
\begin{equation}\label{eq_jrp_general}
\mathbf{JR}^{\vec{x}_{(1,\ldots,n)}}_{i,j}(\varepsilon^{\vec{x}_{(1)}}, \ldots, \varepsilon^{\vec{x}_{(n)}}) = 
\prod_{k=1}^n \mathbf{R}^{\vec{x}_{(k)}}_{i,j}(\varepsilon^{\vec{x}_{(k)}}), \qquad i,j=1,\ldots,N.
\end{equation}
In this approach, a recurrence will take place if a point
$\vec x_j$ on the first trajectory returns to the
neighbourhood of a former point $\vec x_i$,  and {\it
simultaneously} a point $\vec y_j$ on the second trajectory
returns to the neighbourhood of a former point $\vec y_i$.
That means, that  the joint probability that both
recurrences (or $n$ recurrences, in the multidimensional
case) happen simultaneously in their respective phase
spaces are studied. 
In such a definition of a recurrence it is not necessary that the
recurrence occurs at same states of the considered systems.

The graphical representation of the matrix
$\mathbf{JR}_{i,j}$ is called {\it joint recurrence plot 
(JRP)}. The definition of the RP,
Eq.~(\ref{eq_rp}), is a special case of the  definition of
the JRP for only one system.

This way, if the systems are physically different
(e.\,g.~they may have different phase space dimensions
$d_1$, \dots, $d_n$ or may be reconstructed from  different
physical observables), the joint recurrences are still
well-defined, in contrast to
the cross recurrences, Eq.~(\ref{eq_crp}). Additionally, the JRP is invariant
under permutation of the coordinates in one or both  of the
considered systems.

Moreover, a delayed version of the joint recurrence matrix can be introduced
\begin{equation}\label{eq_delayed_JRP}
\mathbf{JR}^{\vec{x},\vec{y}}_{i,j}(\varepsilon^{\vec x},\varepsilon^{\vec y}, \tau) = 
\mathbf{R}^{\vec{x}}_{i,j}(\varepsilon^{\vec x}) \mathbf{R}^{\vec{y}}_{i+\tau,j+\tau}(\varepsilon^{\vec y}), \qquad i,j=1,\ldots,N-\tau,
\end{equation}
which is very useful for the analysis of interacting
delayed systems (e.\,g.~for  lag synchronisation)
\cite{rosenblum1997,sosnovtseva1999}, or even for systems
with feedback (cf.~Subsec.~\ref{sec:Synchro}). 

The JRP can also be computed by using a fixed amount of
nearest neighbours. Then, each single RP which contributes
to the final JRP is computed by using the same number of
nearest neighbours.

\subsection*{Example: Comparison between CRPs and JRPs}

\begin{figure}[htbp]
\centering \includegraphics[width=\columnwidth]{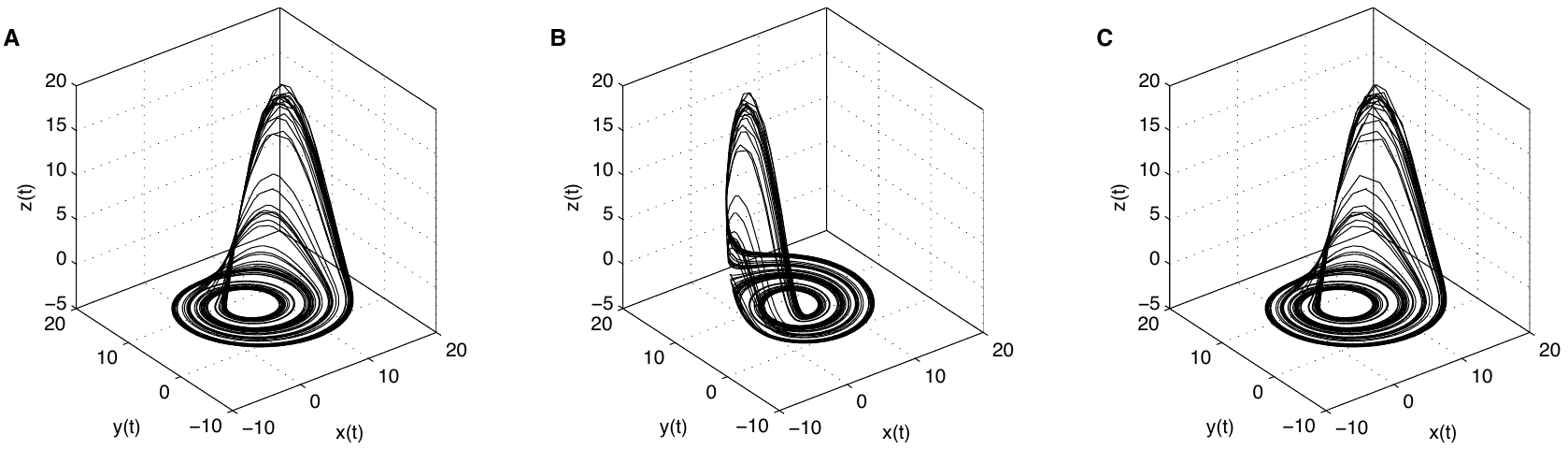} 
\caption{(A) Phase space trajectory of the R\"ossler system (Eqs.~\ref{eq_roessler}, 
with $a=0.15$, $b=0.2$ and $c=10$). 
(B) Same trajectory as in (A) but rotated on the $z$-axis by $\frac{3}{5}\pi$.
(C) Same trajectory as in (A) but time scale transformed by $\tilde t = t^2$.
}\label{fig_rotated_roessler}
\end{figure}

In order to illustrate the difference between CRPs and
JRPs, we consider the phase space trajectory of  the
R\"ossler system in three different situations: the
original trajectory (Fig.~\ref{fig_rotated_roessler}A), the
trajectory rotated on the $z$-axis
(Fig.~\ref{fig_rotated_roessler}B) and the trajectory under
a parabolic stretching/compression of the time scale
(Fig.~\ref{fig_rotated_roessler}C).  These three
trajectories look very similar; one of them is rotated and
the other one contains  another time parametrisation (but
looks identical to the original trajectory in phase
space).

\begin{figure}[htbp]
\centering \includegraphics[width=\columnwidth]{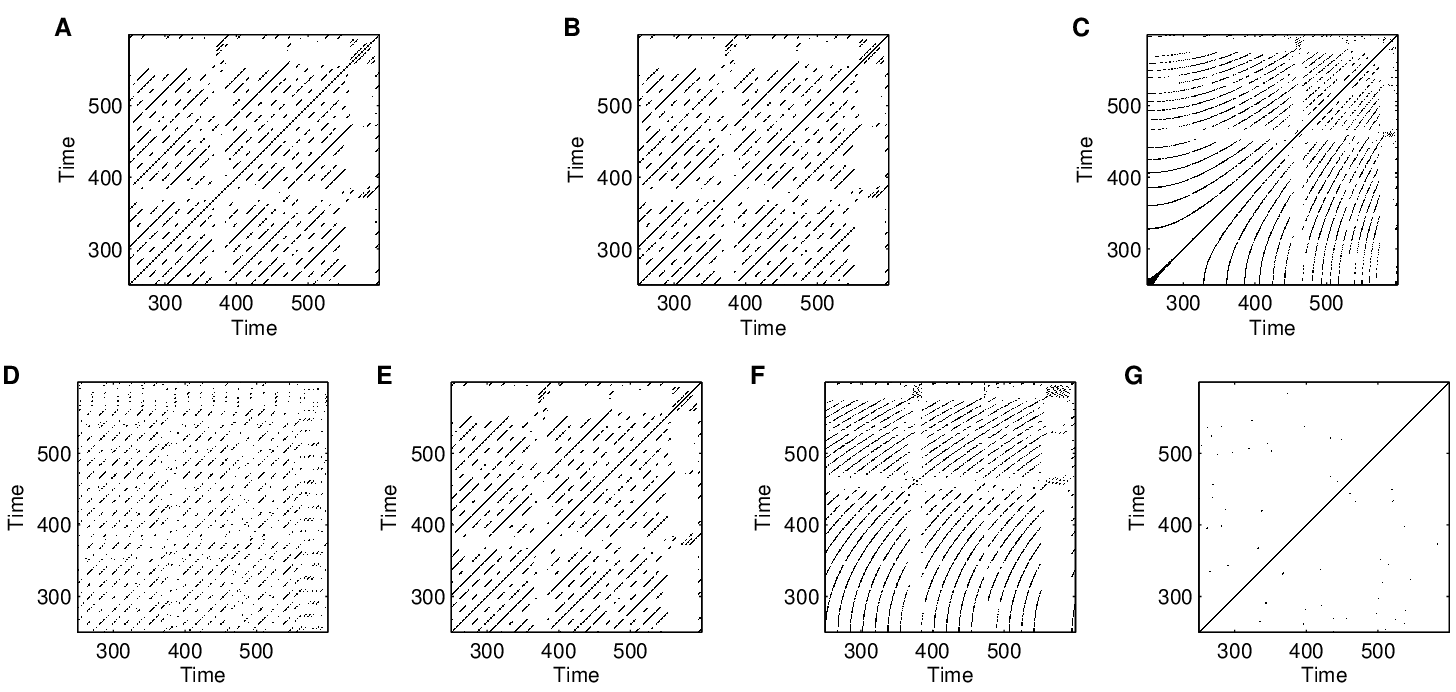} 
\caption{RPs of the (A) original trajectory of the R\"ossler system, (B) of the rotated 
trajectory and (C) of the stretched/compressed trajectory.
(D) CRP and (E) JRP of the original and rotated trajectories and
(F) CRP and (G) JRP of the original and stretched/compressed trajectories.
The threshold for recurrence is $\varepsilon=1$.
}\label{fig_crp_jrp_example}
\end{figure}

At first, let us look at the RPs of these three
trajectories. The RP of the original trajectory is
identical to the RP of the rotated one, as expected
(Fig.~\ref{fig_crp_jrp_example}A and B). The RP of the
stretched/compressed trajectory looks different than the RP
of the  original trajectory
(Fig.~\ref{fig_crp_jrp_example}C): it contains bowed lines,
as the recurrent structures are shifted and stretched in
time with respect to the original RP.

Now we calculate the CRP between the original trajectory and
the rotated one (Fig.~\ref{fig_crp_jrp_example}D) and
observe, that it is rather different from the  RP of the
original trajectory (Fig.~\ref{fig_crp_jrp_example}A). This
is because in the CRP the difference between each pair of
vectors is computed, and this difference is not invariant
under rotation of one of the systems. Hence, a rotation of
the reference system of one trajectory changes the CRP. 
Therefore, the CRP cannot detect that both trajectories are
identical up to  a rotation. In contrast, the JRP of the
original trajectory and the rotated one
(Fig.~\ref{fig_crp_jrp_example}E) is identical to the RP of
the original trajectory (Fig.~\ref{fig_crp_jrp_example}A).
This is  because the JRP considers joint recurrences,
i.\,e.~recurrences  which occur simultaneously in both
systems, and they are invariant under affine
transformations. 

The CRP between the original trajectory and the 
stretched/compressed one contains the bowed LOS, which 
reveals the functional shape of the parabolic transformation 
of the time scale (Fig.~\ref{fig_crp_jrp_example}F). Note that the CRP
represents the times at which both trajectories visit the
same region of the phase space. On the other hand, the JRP
of these trajectories is almost empty 
(Fig.~\ref{fig_crp_jrp_example}G) because
the recurrence structure of both systems is now different.
Both trajectories have different time scales, and hence,
there are almost no joint recurrences. Therefore, the JRP
is not able to detect  the time transformation applied to
the trajectory, even though the shape of the phase space
trajectories is very similar.

To conclude we can state that CRPs are more
appropriate to investigate relationships between the parts
of the same system which have been subjected to different physical
or mechanical processes, e.\,g., two borehole cores in a lake
subjected to different compression rates (see
Subsec.~\ref{sec:appl_rescaling}). On the other hand, JRPs are
more appropriate for the investigation of two interacting
systems which influence each other, and hence, adapt to
each other, e.\,g., in the framework of phase and generalised
synchronisation (see
Subsecs.~\ref{sec:Synchro},~\ref{sec:appl_jrp_synchro}).

%
%
%

%% file: meth_rqa.tex
%
%
%
%

In order to go beyond the visual impression yielded by RPs, 
several measures of complexity which quantify the 
small-scale structures in RPs (Subsec.~\ref{sec:Selectionofthethreshold}), 
have been proposed in \cite{zbilut92,webber94,marwan2002herz} and are known as 
{\it recurrence quantification analysis (RQA)}.
These measures are based on the
recurrence point density and the diagonal and vertical line structures of the RP. 
A computation of these measures in small windows
(sub-matrices) of the RP moving along the LOI yields the time 
dependent behaviour of these variables. Some studies based on RQA
measures show that they are able to identify bifurcation
points, especially chaos-order transitions \citep{trulla96}.
The vertical structures in the RP are related to intermittency
and laminar states. Those measures quantifying the vertical structures
enable also to detect chaos-chaos transitions \citep{marwan2002herz}.

In the following, we focus on the application of the RQA to RPs, 
but it is important to emphasise that the RQA can also be analogously applied to CRPs and JRPs. 
However, some of the interpretations of the RQA given below are not valid for CRPs.
Henceforth, we assume that the RP is calculated by using a 
fixed threshold $\varepsilon$ (hence the RP is symmetric).

First, we introduce several such measures and then their potentials
and limits for the identification of changes are discussed.

\subsubsection{Measures based on the recurrence density}

The simplest measure of the RQA is the {\it recurrence rate} ($RR$) or {\it per cent 
recurrences}
\begin{equation}\label{eq_rr}
RR(\varepsilon) = \frac{1}{N^2} \sum_{i,j=1}^N \mathbf{R}_{i,j}(\varepsilon),
\end{equation}
which is a measure of the
density of recurrence points in the RP. Note that it corresponds to the definition of the
correlation sum, 
Eq.~(\ref{eq_corrsum}), 
except that the LOI is usually not included.
Furthermore, in the limit $N \to \infty$, $RR$ is the probability that 
a state recurs to its $\varepsilon$-neighbourhood in phase space. On the other hand, the $RR$
of CRPs corresponds to the cross correlation sum \citep{kantz94}
\begin{equation}\label{eq_xcorrsum}
CC_2(\varepsilon) = \frac{1}{N^2}
             \sum_{i,j=1}^{N} \mathbf{CR}_{i,j}(\varepsilon),
\end{equation}
and the $RR$ of JRPs of $n$ systems to the joint correlation sum
\begin{equation}\label{eq_jcorrsum}
JC_2(\varepsilon^{(1)},\dots,\varepsilon^{(n)}) = \frac{1}{N^2}
             \sum_{i,j=1}^{N}
             \prod_{k=1}^n \mathbf{R}^{\vec{x}_{(k)}}_{i,j}(\varepsilon^{(k)}).
\end{equation}

The value 
\begin{equation}\label{eq_number_of_neighbours}
N_{\text{n}}(\varepsilon) = \frac{1}{N} \sum_{i,j=1}^N \mathbf{R}_{i,j}(\varepsilon)
\end{equation}
is simply the
{\it average number of neighbours} that each point on the trajectory has in its 
$\varepsilon$-neighbourhood.

\subsubsection{Measures based on diagonal lines}\label{sec:RQA_diagonal_lines}

The next measures are based on the histogram $P(\varepsilon,l)$ of diagonal lines of length $l$, i.\,e.
\begin{equation}\label{eq_P_of_l}
P(\varepsilon,l) = \sum_{i,j=1}^N\bigl(1-\mathbf R_{i-1,j-1}(\varepsilon)\bigr)\bigl(1-\mathbf R_{i+l,j+l}(\varepsilon)\bigr)\prod_{k=0}^{l-1}\mathbf R_{i+k,j+k}(\varepsilon).
\end{equation}
 In the following subsections we omit the symbol $\varepsilon$ from the RQA measures for the sake of simplicity (i.\,e.~$P(l)=P(\varepsilon,l)$).

Processes with uncorrelated or weakly correlated, stochastic or chaotic 
behaviour cause none or very short diagonals,
whereas deterministic processes cause longer diagonals and less single,
isolated recurrence points. Therefore, the ratio of recurrence points that
form diagonal structures (of at least length $l_{\min}$) to all recurrence points 
\begin{equation}\label{eq_det}
DET = \frac{\sum_{l=l_{\min}}^N l\, P(l)}{\sum_{l=1}^N l\, P(l)}
\end{equation}
is introduced as a measure for {\it determinism}
(or predictability) of the
system. The threshold $l_{\min}$ excludes the diagonal lines which are
formed by the tangential motion of the phase space
trajectory. For $l_{\min}=1$ the determinism is one.
The choice of $l_{\min}$ could be made in a similar way as the choice of the
size for the Theiler window \cite{theiler86}, but we
have to take into account that the histogram $P(l)$ can become
sparse if $l_{\min}$ is too large, and, thus, the reliability of $DET$
decreases.

A diagonal line of length $l$ means that a segment of the trajectory 
is rather close during $l$ time steps to another segment of the trajectory at a different time;
thus these lines are related to the divergence of the trajectory segments. 
The {\it average diagonal line length}
\begin{equation}\label{eq_l}
L = \frac{\sum_{l=l_{\min}}^N l\, P(l)}{\sum_{l=l_{\min}}^N P(l)}
\end{equation}
is the average time that two segments of the trajectory are close to each other,
and can be interpreted as the mean 
prediction time.

Another RQA measure considers the length $L_{\text{max}}$ of the longest diagonal line found in the RP, or its inverse, the {\it divergence},
\begin{equation}\label{eq_lmax}
L_{\max} = \max \left(\left\{l_i\right\}_{i=1}^{N_l}\right) 
\quad\text{respectively}\quad
DIV = \frac{1}{L_{\max}}.
\end{equation}
where $N_l=\sum_{l \geq l_{\min}} P(l)$ is the the total number of diagonal lines.
These measures are related to the exponential divergence of
the phase space trajectory. The faster the trajectory segments
diverge, the shorter are the diagonal lines and the higher
is the measure $DIV$.

Eckmann et al.~have stated that ``the length of the diagonal lines 
is related to the 
largest positive Lyapunov exponent'' if there 
is one in the considered system \citep{eckmann87}. 
Different approaches have been
suggested in order to use these lengths, Eqs.~(\ref{eq_l}) and (\ref{eq_lmax}), to estimate the
largest positive Lyapunov exponent, such as computing the $DIV$ \citep{trulla96}
or the average of the inverse of the half lengths of the diagonals
(using perpendicular RPs) \cite{choi99}.
However, the relationship between these measures and
the positive Lyapunov exponent is not as simple as it was mostly stated
in the literature. As already mentioned, the $K_2$ entropy is related
with the (cumulative) frequency distribution of the lengths of the diagonal 
lines and, therefore, with the lower limit of the sum of the positive 
Lyapunov exponents. In Subsec.~\ref{sec:Invariants} we explain this relationship in detail and
show how $L_{\max}$ can be used as an estimator for $K_2$ (and, hence,
for the lower limit of the sum of the positive Lyapunov exponents).

The measure {\it entropy} refers to the Shannon entropy 
of the probability $p(l) = \frac{P(l)}{N_l}$ to find a diagonal line 
of exactly length $l$ in the RP,
\begin{equation}
ENTR = -\sum_{l=l_{\min}}^{N} p(l) \ln p(l).
\end{equation}
$ENTR$ reflects the complexity of the RP in respect of the 
diagonal lines, e.\,g.~for uncorrelated 
noise the value of $ENTR$ is rather small, indicating its low complexity.

The measures introduced up to now, $RR$, $DET$, $L$ etc.~can also be 
computed separately for each diagonal parallel to the LOI.
Henceforth, RQA measures for a certain line parallel to the LOI and
with distance $\tau$ from the LOI are called $\tau$-recurrence rate, 
$\tau$-determinism etc., and the measures are marked with a 
subscribed index, like  $RR_{\tau}$, $DET_{\tau}$, $L_{\tau}$ etc.
Following this procedure, we need to define the number of diagonal lines $P_{\tau}(l)$ of length $l$ on
each diagonal $\mathbf{RR}_{i,i+\tau}$ parallel to the LOI.
$\tau=0$ corresponds to the main diagonal, $\tau>0$ to diagonals above 
and $\tau<0$ diagonals below the LOI (i.\,e.~$\mathbf{RR}_{i+|\tau|,i}$), which represent
positive and negative time delays, respectively.

The $\tau$-recurrence rate for those diagonal lines with distance
$\tau$ from the LOI is then
\begin{equation}\label{eq_rr_star}
RR_{\tau} = \frac{1}{N-\tau} \sum_{i=1}^{N-\tau} \mathbf{R}_{i,i+\tau} = \frac{1}{N-\tau} \sum_{l=1}^{N-\tau} l\, P_{\tau}(l).
\end{equation}
This measure corresponds to the {\it close returns histogram} introduced 
for quantifying close returns plots \cite{lathrop89}.
It can be considered as a {\it generalised auto-correlation function}, as it also
describes higher order correlations between the points of the trajectory
in dependence on $\tau$. A further advantage with respect to the linear
auto-correlation function is that  $RR_{\tau}$ can be
determined for a trajectory in phase space and not only for a single
observable of the system's trajectory. It can be interpreted as
the probability that a state recurs to its $\varepsilon$-neighbourhood
after $\tau$ time steps.

Analogous to the RQA, the $\tau$-determinism
\begin{equation}
DET_{\tau} = \frac{\sum_{l=l_{min}}^{N-\tau} l\, P_{\tau}(l)}
{\sum_{l=1}^{N-\tau} l\, P_{\tau}(l)}
\end{equation}
is the proportion of recurrence points forming
diagonal lines longer than $l_{min}$ to all recurrence 
points, and the $\tau$-average diagonal line length
\begin{equation}
L_{\tau} = \frac{\sum_{l=l_{min}}^{N-\tau} l\, P_{\tau}(l)}
{\sum_{l=l_{min}}^{N-\tau} P_{\tau}(l)}
\end{equation}
is the mean length of the diagonal structures on the
considered diagonal parallel to the LOI. The 
$\tau$-entropy can be applied to the diagonal-wise 
consideration as well.

These diagonal-wise computed measures, $RR_{\tau}$, $DET_{\tau}$ and $L_{\tau}$,
over time distance $\tau$ from the LOI can be used, e.\,g., to determine
the Theiler window. 
This diagonal-wise determination of the RQA measures plays
an important role in the analysis of CRPs as well.
Long diagonal structures in the CRP reveal a similar 
time evolution of the trajectories of both processes. It is obvious that 
a progressively increased similarity between both processes causes an 
increase of the recurrence point density along the main diagonal 
$\mathbf{CR}_{i,i} \,|_{i=1}^N$
until finally the LOI appears and the CRP becomes an RP. Thus,
the occurrence of diagonal lines in a CRP can be used in
order to benchmark the similarity between the considered processes.
Using this approach
it is possible to assess the similarity in the dynamics of two different 
systems in dependence on a certain time delay. 

The $\tau$-recurrence $RR_{\tau}$ of a CRP reveals the probability of the occurrence of 
similar states in both systems with a certain delay $\tau$.
$RR_{\tau}$ has a high value for systems whose trajectories often visit 
the same phase space regions.

As already mentioned, stochastic as well as strongly fluctuating 
processes cause none or only
short diagonals, whereas deterministic processes cause longer 
diagonals. If two deterministic processes have the same or
similar time evolution, i.\,e.~parts of the phase space trajectories 
visit the same phase space regions for certain times, the
amount of longer diagonals increases and the amount of shorter 
diagonals decreases. The $\tau$-determinism $DET_{\tau}$
of a CRP is related to the similar time evolution of the
systems' states. The measure $L_{\tau}$ 
quantifies the duration of the similarity in the dynamics of both systems. 
A high coincidence of both trajectories increases the length of these diagonals. 

Considering CRPs, smooth trajectories with long auto-correlation times will
result in a CRP with long diagonal structures, even if
the trajectories are not linked to each other (this effect
corresponds to the tangential motion of one trajectory).
In order to avoid counting such ``false'' diagonals, the lower
limit for the diagonal line length $l_{min}$ should be of the
order of the auto-correlation time.

\begin{figure}[htbp]
\centering \includegraphics[width=.7\textwidth]{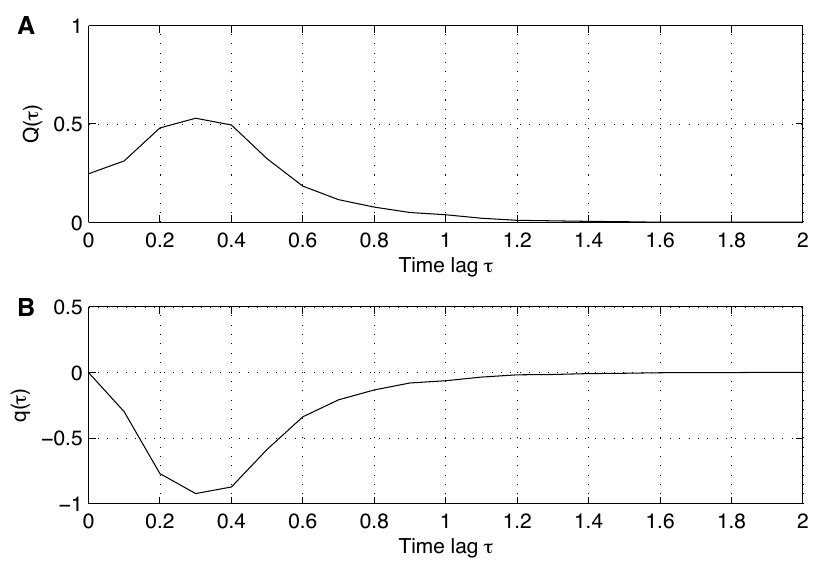}
\caption{Measure of symmetry (A) and asymmetry (B) applied on the diagonal-wise
computed recurrence rate $RR_{\tau}$ of the CRP of two mutually coupled R\"ossler systems in the 
funnel regime, Eqs.~(\ref{eq_2roessler2_1}) and 
(\ref{eq_2roessler2_2}), for $a=0.2925$, $b=0.1$, $c=8.5$, $\nu=0.02$, with a slight time lag.
(A) The maxima for the symmetry measure $Q(\tau)$ at a time lag of $k=0.3$ reveals the
synchronisation between both systems after the detected lag. 
(B) The negative value of the asymmetry measure $q(\tau)$ at this lag reveals
that the second R\"ossler oscillator leads the first one.}\label{fig_diagonal_RQA}
\end{figure}

By applying a measure of symmetry and asymmetry on the $\tau$-RQA 
measures (for a small range $0 \leq \tau \ll N$), e.\,g.~on $RR_{\tau}$,
\begin{equation}
Q(\tau) = \frac{RR_{\tau} + RR_{-\tau}}{2}, \quad \text{and} \quad q(\tau) = \frac{RR_{\tau} - RR_{-\tau}}{2},
\end{equation}
we can simply quantify interrelations between two systems and are able to 
determine which system leads the other (Fig.~\ref{fig_diagonal_RQA};
this is similar to an approach for the detection of event synchronisation proposed
in \cite{quiroga2002}).

%
%
%
%
%
%
%
%
%

Summarising, we can state that
high values of $RR_{\tau}$ indicate a high probability of occurrence
of the same state in both systems, and high values of $DET_{\tau}$ and $L_{\tau}$ 
indicate a long time span, in which both systems visit the same region of phase space.
Therefore, $DET_{\tau}$ and $L_{\tau}$ are sensitive to fast and strongly fluctuating data.
It is important to emphasise
that these parameters are statistical measures and that their validity 
increases with the size of the CRP, i.\,e.~with the length of the regarded trajectory.

The consideration of an additional CRP
\begin{equation}
\mathbf{CR}^{-}_{i,j} = \Theta \left(\varepsilon-\left\|\vec x_{i} + \vec y_{j}\right\|\right)
\end{equation}
with a negative signed second trajectory $-\vec{y}_j$
allows distinguishing correlation and anti-correlation 
between the considered trajectories \citep{marwan2002pla}.
In order to recognise the measures for both possible CRPs, the superscript index $+$ is added to the measures 
for the positive linkage and the superscript index $-$ for the negative linkage, e.\,g.~$RR^+_{\tau}$ and $RR^-_{\tau}$.

Another approach used to study positive and negative relations between 
the considered trajectories involves the composite measures for the
$\tau$-recurrence rate
\begin{equation}
RR^c_{\tau} = \frac{1}{N-k} \sum_{j-i=k} \left( \mathbf{CR}^{+}_{i,j} - \mathbf{CR}^{-}_{i,j} \right),
\end{equation}
the $\tau$-determinism
\begin{equation}
DET^c_{\tau} = DET^+_{\tau} - DET^-_{\tau},
\end{equation}
and the $\tau$-average diagonal length
\begin{equation}
L^c_{\tau} = L^+_{\tau} - L^-_{\tau},
\end{equation}
as it was used in \cite{marwan2003climdyn}.
This presentation is similar to the time-dependent presentation 
of the cross correlation function (but with the important difference that
the $\tau$-RQA measures consider also higher order moments) and 
is more intuitive than the separate representation of $RR_{\tau}^+$, $RR_{\tau}^-$ etc.
However, for the investigation of interrelations based on even functions, 
these composite measures are not suitable.

A further substantial advantage of applying the $\tau$-RQA on CRPs is the capability 
to find also nonlinear similarities in short and non-stationary 
time series with high noise levels as they typically occur, e.\,g., in
life or earth sciences. In these cases, using a fixed amount 
of nearest neighbours is more appropriate than a fixed threshold 
$\varepsilon$. Also the use of OPRPs or
JRPs is appropriate for the analysis of this kind of data.

Note that the $\tau$-RQA measures as functions of the distance $\tau$
to the main diagonal are also 
important for the quantification of RPs.
For example, the measure $RR_{\tau}$ can be used
to find UPOs in low-dimensional chaotic 
systems \citep{gilmore98,lathrop89,mindlin92}. Since periodic orbits 
are more closely related to the occurrence of longer diagonal structures, 
the measures  $DET_{\tau}$ and $L_{\tau}$ are more suitable candidates
for this kind of study. The measure $RR_{\tau}$ has been already used
in \cite{eckmann87} for the study of non-stationarity in the data. 
Beyond this, $RR_{\tau}$ can be applied to analyse synchronisation between 
oscillators (Subsec.~\ref{sec:Synchro}).

Another RQA measure is the {\it trend},
which is a linear regression coefficient over the recurrence point density $RR_{\tau}$ 
of the diagonals parallel to the LOI as a function of the
time distance between these diagonals and the LOI
\begin{equation}
TREND = \frac{\sum_{\tau=1}^{\tilde N} (\tau-\tilde N/2)(RR_{\tau}-\langle RR_{\tau}\rangle)}
{\sum_{\tau=1}^{\tilde N} (\tau-\tilde N/2)^2}.
\end{equation}
It provides information about non-stationarity in the process, 
especially if a drift is present in the analysed trajectory. 
The computation excludes the edges of the RP ($\tilde N < N$) 
because of the lack of a sufficient number of recurrence points. 
The choice of $\tilde N$ depends on the studied system. Whereas
$N - \tilde N>10$ should be sufficient for noise, this difference should be much
larger for a system with some auto-correlation length (ten times the order of 
magnitude of the auto-correlation time should be enough). 
It should be noted that if the time dependent RQA 
(measures computed in sliding windows) is used, $TREND$ will depend
strongly on the size of the window and may yield ambiguous results 
for different window sizes.

A further measure, the {\it ratio}, has been defined as the ratio 
between $DET$ and $RR$ \citep{webber94}. It can be computed based on the 
number $P(l)$ of diagonal lines of length $l$ as follows
\begin{equation}
RATIO = N^2\,\frac{\sum_{l=l_{min}}^N l\,P(l)}{\left(\sum_{l=1}^N l\,P(l)\right)^2}.
\end{equation}
A heuristic study of physiological time series has revealed that this ratio
can be used to uncover transitions in the dynamics; during certain types of 
qualitative transitions $RR$ decreased, 
whereas $DET$ remained constant \citep{webber94}.

The RQA measures discussed so far, are based on diagonal structures in the RP.
In Subsec.~\ref{sec:vert_rqa}, this quantitative view is extended to vertical (horizontal) structures and further
measures of complexity based on the distribution of the vertical lines are proposed.

\subsubsection*{Example: Finding nonlinear interrelations by applying RQA on CRPs}\label{sec_crpex1}

This example shows the ability of CRPs to find nonlinear interrelations
between two processes,
which cannot be detected by means of linear tests \cite{marwan2002pla}. 
We consider linear correlated noise (auto-regressive process)
which is nonlinearly coupled with the $x$-component
of the Lorenz system (for standard 
parameters $\sigma=10$, $r=28$, $b=8/3$ and a time resolution of 
$\Delta t = 0.01$ \citep{lorenz63,argyris94}):
\begin{equation}\label{eq_ar}
y_i=0.86 \, y_{i-1} + 0.500 \, \xi_i + \kappa \, x_i^2,
\end{equation}
where $\xi$ is Gaussian white noise and $x_i$ ($x(t) \rightarrow x_i$, $t=i\,\Delta t$)
is normalised with respect to the standard deviation.
The data length is 8,000 points and the coupling $\kappa$
is realised without any lag. 

\begin{figure}[htb]
\centering \includegraphics[width=12cm]{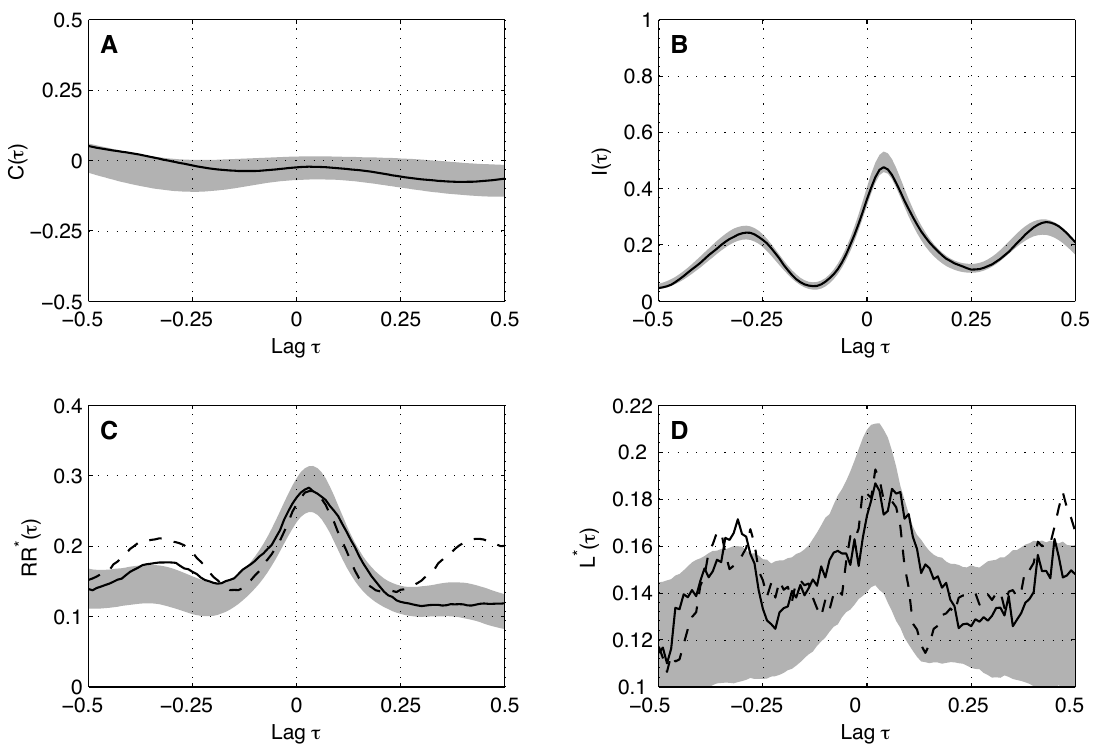} 
\caption{(A) Cross correlation $C(\tau)$, (B) mutual information $I(\tau)$, (C)
$\tau$-recurrence rate $RR_{\tau}$ and (D) $\tau$-average line length $L_{\tau}$ for the 
forced auto-regressive process and the forcing function; 
the curves represent the measures for one realisation as functions of the 
delay $\tau$ for a coupling $\kappa=0.2$.
In (C) and (D) the solid lines show positive relation;
the dashed lines show negative relation. 
The grey bands mark the $2\sigma$ margin of the distributions
of the measures gained from 500 realisations. The lag $\tau$ and
the average line length $L_{\tau}$ have units of time \cite{marwan2002pla}.
}\label{fig_crqa_lorenz}
\end{figure}

As expected, due to the nonlinear linkage the cross 
correlation analysis between $x$ and $y$ does not reveal 
a significant linear correlation between these data series
(\reffig{fig_crqa_lorenz}A). However, the mutual information as a well-established
measure to detect nonlinear dependencies \cite{kantz97}
shows a strong dependence between $x$ and $y$ at a delay of 
$0.05$ (\reffig{fig_crqa_lorenz}B). The CRP based 
$\tau$-recurrence rate $RR_{\tau}$ and $\tau$-average diagonal length $L_{\tau}$ exhibit
maxima at a lag of about $0.05$ for $RR^+$/$L^+$ and $RR^-_{\tau}$/$L^-_{\tau}$ 
and additionally at $0.45$ and $-0.32$ for $RR^-_{\tau}$/$L^-_{\tau}$
(\reffig{fig_crqa_lorenz}C, D). The maxima around $0.05$ 
for the $+$ and $-$ measures are a strong indication of a nonlinear
relationship between the data.
The delay of approximately $0.05$ stems from the auto-correlation of $y$ and 
approximately corresponds to its correlation time $\Delta t/\ln{0.86}=0.066$.
The maxima at $0.45$ and $-0.32$ correspond to the half mean period
of the Lorenz system.
Since the result is rather independent of the sign of the second data, 
the found interrelation is of the kind of an even function.
500 realisations of the AR model have been used in order to estimate the distributions
of the measures. The $2\sigma$ margins of these distributions can be used
to assess the significance of the results.

%
%
%

Due to the rapid fluctuation of $y$, the number of long 
diagonal lines in the CRP decreases. Therefore, measures based on these
diagonal structures, especially $DET_{\tau}$, 
do not work well with
such heavily fluctuating data. However, we can infer that the 
measures $RR_{\tau}$ as well as $L_{\tau}$ (though less significant for
rapidly fluctuating data) are suitable for finding a nonlinear 
relation between the considered data series $x$ and $y$, where the 
linear analysis is not able to detect such a relation.
Furthermore, this technique is applicable to rather short 
and non-stationary data compared to the mutual information.

\subsubsection{Measures based on vertical lines}\label{sec:vert_rqa}

\begin{figure}[htb]
\centering \includegraphics[width=0.3\columnwidth]{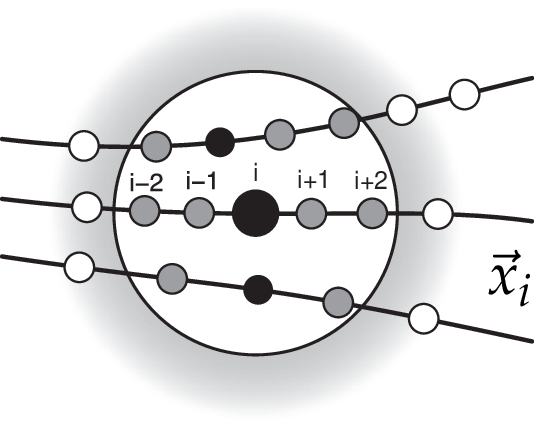} 
\caption{States of subsequent time may fall into the neighbourhood of the state 
at time $i$, pretending artificial recurrences (grey dots). This is
called tangential motion and such points are referred to as
sojourn points.}\label{fig_tangential_motion}
\end{figure}

A vertical line of length $v$ starting at the coordinates $(i,j)$ 
of a RP is given by the condition (\ref{eq_verticalline}).
In continuous time systems discretised with sufficiently high time resolution
and with an appropriate large threshold $\varepsilon$, a large part of
these vertical lines usually correspond to the tangential motion of 
the phase space trajectory (Fig.~\ref{fig_tangential_motion}). 
However, not all 
elements of these sets belong to the tangential motion. 
For example, even though there is no tangential motion
in maps, we find vertical lines in their RPs, e.~g. in the presence of laminar states
in intermittent regimes. 
Furthermore, in systems with two different time scales, 
we might find vertical lines because of the finite size of the 
threshold $\varepsilon$, and not because of tangential motion. 
Sometimes, the points belonging to the tangential motion 
are called {\it sojourn points} \cite{gao99}. But we will
not use this term, because it is a bit misleading, and actually
those points forming vertical lines but which do not belong 
to the tangential motion would be the sojourn points.

The total number of vertical lines of the length $v$ in the RP 
is then given by the histogram
\begin{equation}
P(v)=\sum_{i,j=1}^N\bigl(1-\mathbf R_{i,j}\bigr)\bigl(1-\mathbf R_{i,j+v}\bigr)\prod_{k=0}^{v-1}\mathbf R_{i,j+k}.
\end{equation}

Analogous to the definition of the determinism, Eq.~\refeq{eq_det}, the
ratio between the recurrence points forming the vertical structures and the entire
set of recurrence points can be computed,
\begin{equation}\label{eq_lam}
LAM = \frac{\sum_{v=v_{min}}^{N}vP(v)}{\sum_{v=1}^{N}vP(v)},
\end{equation}
and is called \textit{laminarity}. The computation of $LAM$ is
realised for those $v$ that exceed a minimal length $v_{min}$ in order to
decrease the influence of the tangential motion. For
maps, $v_{min}=2$ is an appropriate value. $LAM$ represents the
occurrence of laminar states in the system without describing
the length of these laminar phases. $LAM$ will decrease if the RP
consists of more single recurrence points than vertical
structures.

The average length of vertical structures is given by (cf.~Eq.~\refeq{eq_l})
\begin{equation}
TT = \frac{\sum_{v=v_{min}}^{N} v P(v)} {\sum_{v=v_{min}}^{N} P(v)},
\end{equation}
and is called \textit{trapping time}. Its computation requires also
the consideration of a minimal length $v_{min}$, as in the case of $LAM$. 
$TT$ estimates the mean time that the system will abide at a specific state or how long the state will be trapped.

Finally, the {\it maximal length of the vertical lines} in the RP 
\begin{equation}
V_{\max}=\max\left(\left\{v_l\right\}_{l=1}^{N_v}\right)
\end{equation}
can be regarded, analogously to the standard measure
$L_{\max}$ ($N_v$ is the absolute number of vertical lines).

In contrast to the RQA measures based on diagonal lines, these measures 
are able to find chaos-chaos transitions \citep{marwan2002herz}.
Hence, they allow for the investigation of 
intermittency, even for rather short and non-stationary data series.
Furthermore, since for periodic dynamics the measures quantifying vertical structures are zero,
chaos-order transitions can also be identified.


\subsubsection*{Example: Comparison of measures based on diagonal and vertical lines}

\begin{figure}[tb]
\centering \includegraphics[width=9.5cm]{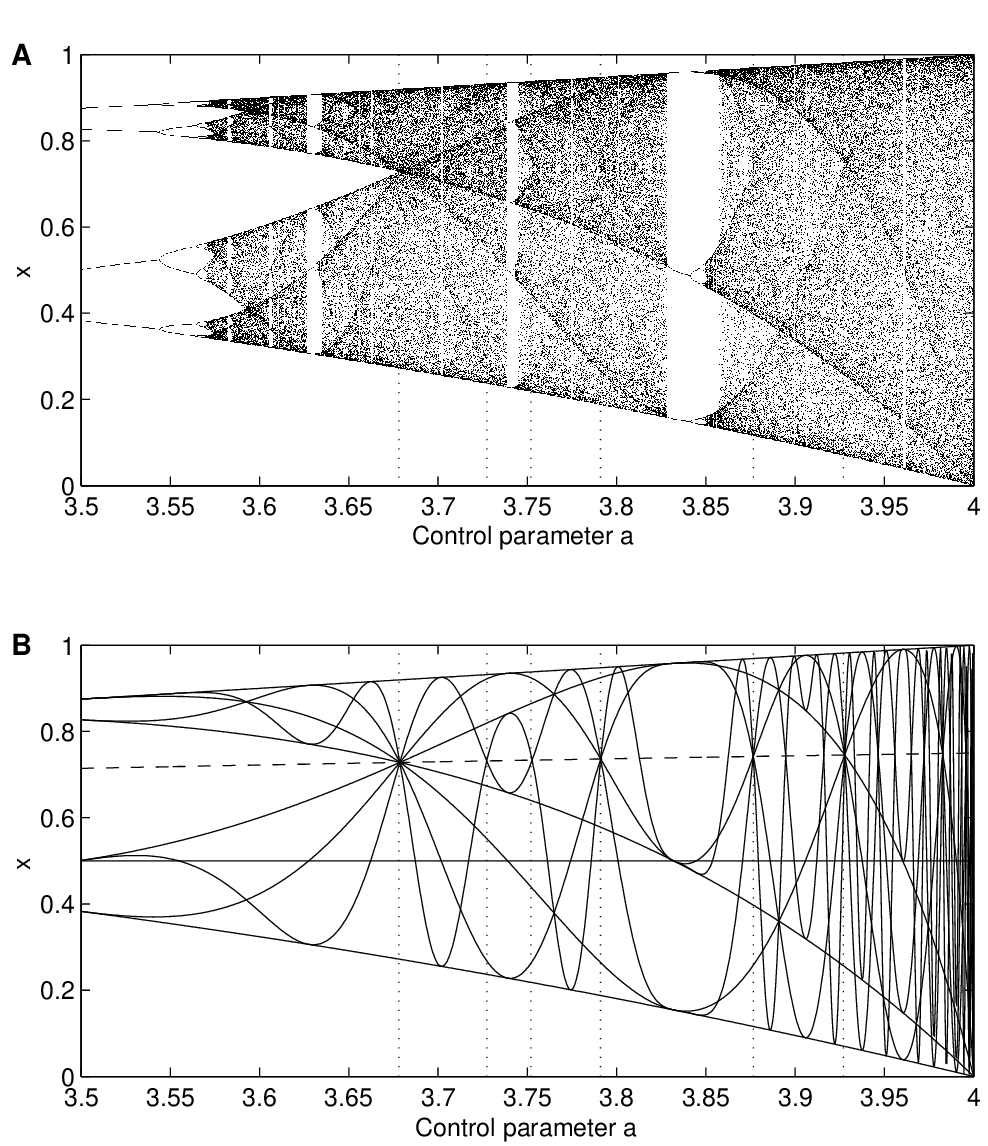} 
\caption{(A) Bifurcation diagram of the logistic map.
(B) Low ordered supertrack functions $s_i(a)\,|_{i=1}^{10}$ and
the fixed point of the logistic map $1-1/a$ (dashed). 
Their intersections represent periodic windows, band merging and
laminar states. The dotted lines show a choice
of points, which represent band merging and laminar phases 
($a=3.678$, $3.727$, $3.752$, $3.791$, $3.877$, $3.927$)
\cite{marwan2002herz}.}\label{fig_log}
\end{figure}

Next, we illustrate the application of the RQA for the logistic map,
Eq.~(\ref{eq_logistic_map}), 
and compare the measures based on diagonal with the ones based one vertical structures. 
We generate for each value of the control parameter $a\in [3.5,4]$, with $\Delta a=0.0005$ a separate 
time series (\reffig{fig_log}) of the rather short length $1,000$. In the
analysed range of $a$, various dynamical regimes and transitions between them occur,
e.\,g., accumulation points, periodic and chaotic states, band
merging points, period doublings, inner and outer crises,
i.\,e.~system \refeq{eq_logistic_map} generates various order-chaos, chaos-order as
well as chaos-chaos transitions.
\citep{collet80}.


Useful tools for studying the chaotic behaviour of the logistic map 
are the \textit{supertrack functions}, which are recursively generated
from
\begin{equation}\label{eq_supertrack}
s_{i+1}(a)=a \, s_{i}(a)\bigl(1-s_i(a)\bigr), \quad s_0(a)=\frac{1}{2}, \quad i=1,\,2,\,\ldots
\end{equation}
$s_i(a)$ represent the functional dependence of 
stable states at a given iteration number $i$ on the control parameter $a$
\citep{oblow88}.
The intersection of $s_{i}(a)$ with $s_{i+j}(a)$ indicates the
occurrence of a $j$-period cycle, and the intersection of $s_{i}(a)$ 
with the fixed-point $(1-1/a)$ of the logistic map indicates 
the point of an unstable singularity, i.\,e.~laminar behaviour
(\reffig{fig_log}, intersection points are marked with dotted lines).

\begin{figure}[htbp]
\includegraphics[width=6cm]{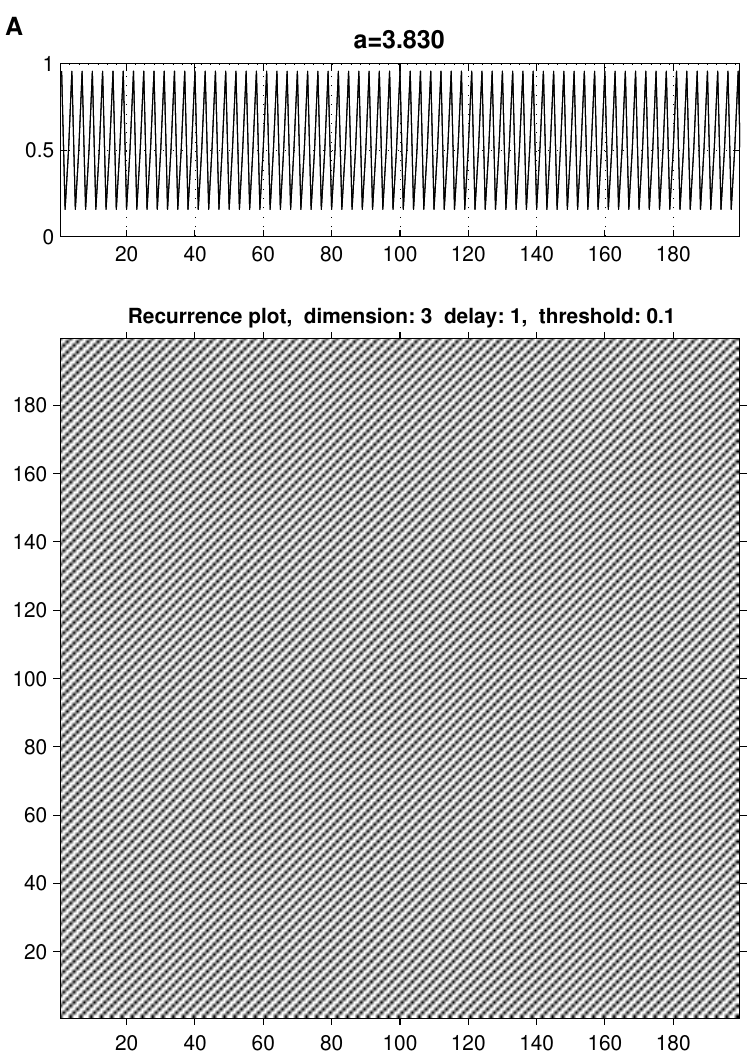} 
\hspace*{0.75cm}
\includegraphics[width=6cm]{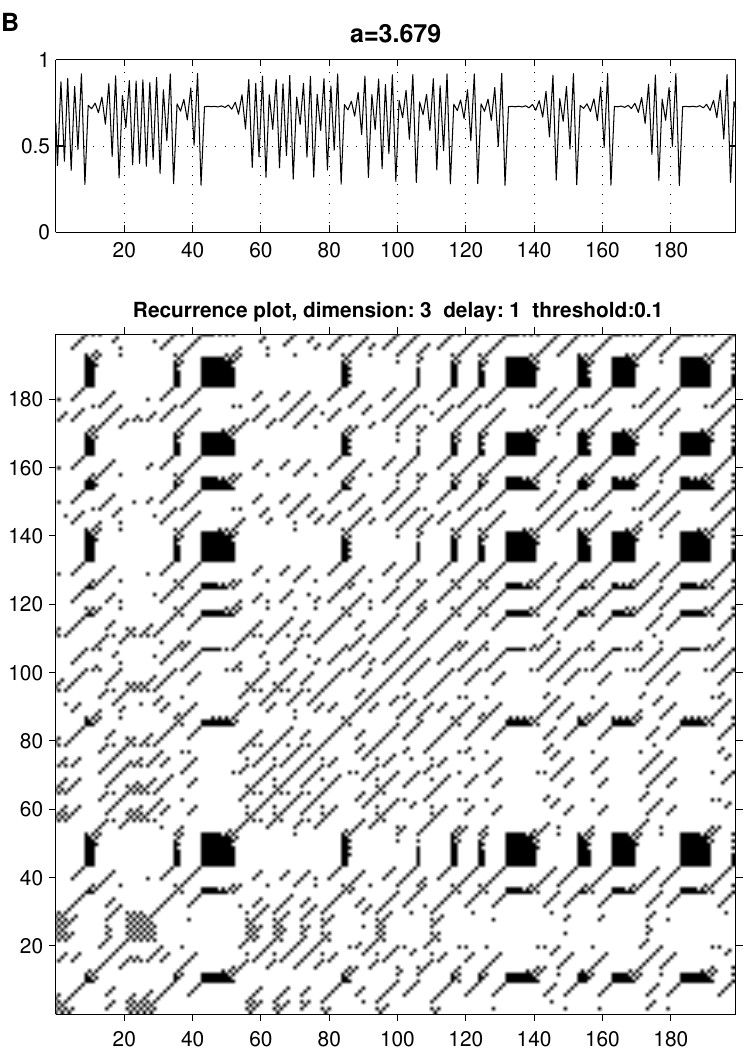} 

\vspace{22pt}
\centering \includegraphics[width=6cm]{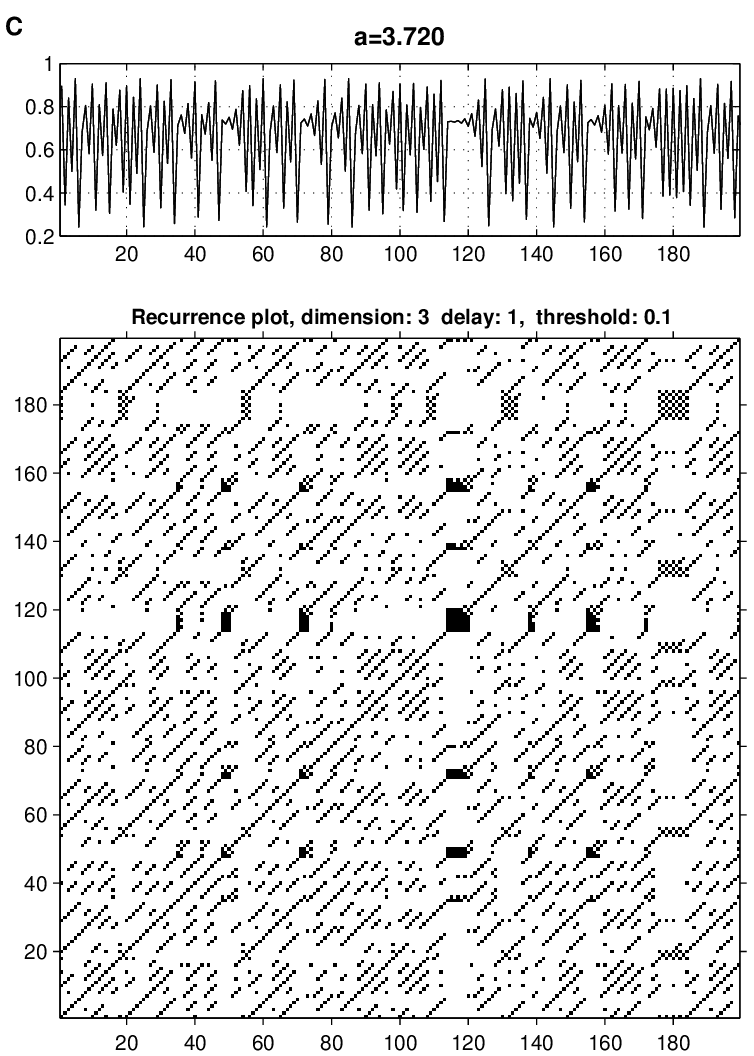} 
\hspace*{0.75cm}
\centering \includegraphics[width=6cm]{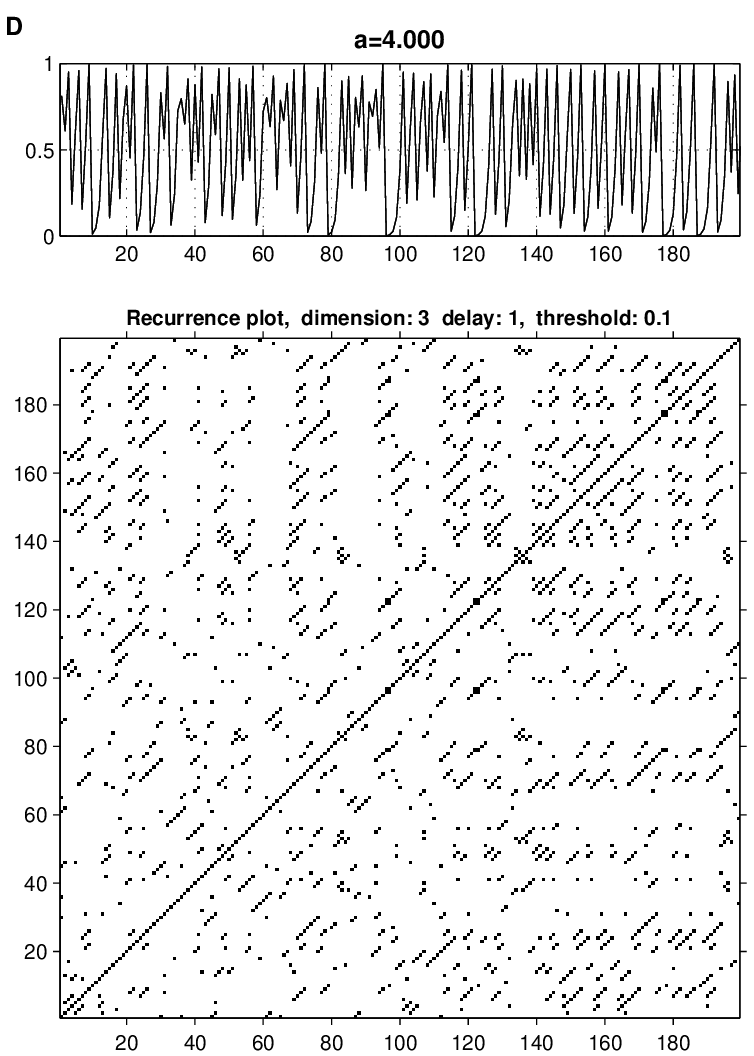} 

\caption{RPs of the logistic map for
various values of the control parameters $a$: (A) periodic-3-window $a=3.830$, (B) band merging $a=3.679$,
(C) supertrack intersection $a=3.720$ and (D) chaos (exterior crisis) $a=4$;
with embedding dimension $m=3$, 
time delay $\tau=1$ and threshold 
$\varepsilon=0.1\sigma$ \cite{marwan2002herz}.}\label{fig_RP_log}
\end{figure}

\begin{figure}[htbp]
\centering \includegraphics[width=\columnwidth]{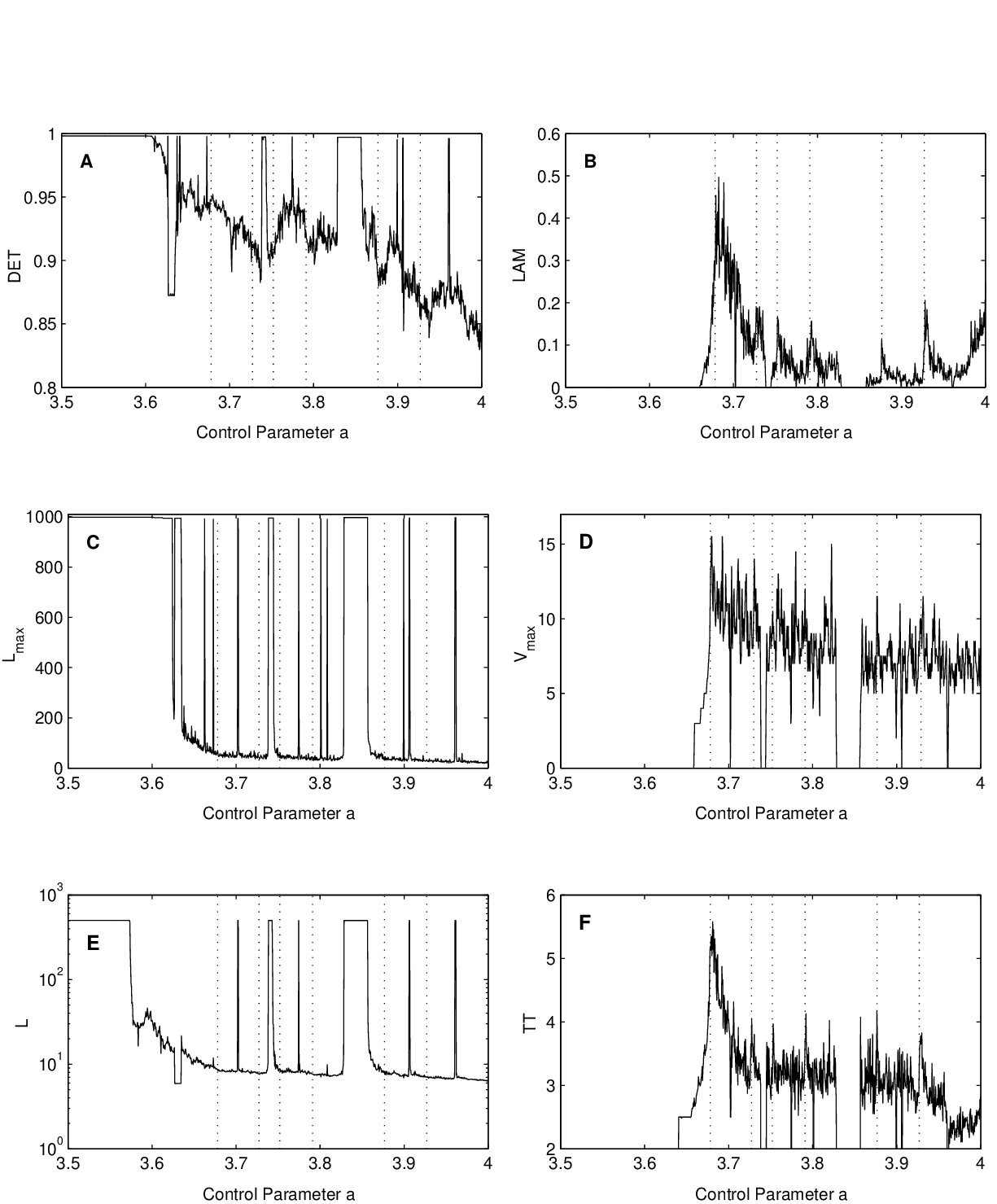} 
\caption{Selected RQA measures $DET$, $L_{\max}$ and $L$
and the measures $LAM$, $V_{\max}$ and $TT$. 
The vertical dotted lines indicate some of the points, at which band merging 
and laminar behaviour occur (cf.~\reffig{fig_log}), whereby not all of them have been
marked. Whereas (A) $DET$, (C) $L_{\max}$ and (E) $L$ show 
periodic-chaos/chaos-periodic transitions (maxima), (B)
$LAM$, (D) $V_{\max}$ and (F) $TT$ exhibit in addition to those 
transitions (minima) chaos-chaos transitions (maxima). The differences
between $LAM$ and $V_{\max}$ are caused by the fact that $LAM$ measures
only the amount of laminar states, whereas $V_{\max}$ measures the
maximal duration of the laminar states. Although some peaks of $V_{\max}$
and $TT$ are not at the dotted lines, they correspond to laminar states
(not all can be marked) \cite{marwan2002herz}.}\label{fig_rqa_log}
\end{figure}
%

Although an embedding is not necessary for maps (i.\,e.~$m=1$), we use here
an embedding of $m=3$ and $\tau=1$ in order to compare the
results with \cite{trulla96}.
The cut-off distance $\varepsilon$ is selected
to be 10\% of the standard deviation of the 
data series. Smaller values would lead to a better distinction of
small variations (e.\,g.~the range before the accumulation point
consists of small variations), but the recurrence point
density would decrease and hence, the statistics of
continuous structures in the RP would become poor.

RPs for various values of the control parameter $a$ exhibit 
already visually specific characteristics (\reffig{fig_RP_log}). Periodic states 
cause continuous and periodic diagonal lines in the RP, but no vertical
or horizontal lines (\reffig{fig_RP_log}A). On the other hand, chaos-chaos transitions, as
band merging points,
inner crises or regions of intermittency represent states with 
short laminar behaviour and cause vertically
and horizontally spread black areas in the RP (\reffig{fig_RP_log}B, C). 
Moreover, diagonal lines occur. 
The fully developed chaotic state ($a=4$) causes a
rather homogeneous RP with numerous single points and some short (in comparison
to the length of the LOI), diagonal or vertical lines (\reffig{fig_RP_log}D).
Vertical (and horizontal) lines occur much more frequently at supertrack crossing
points (chaos-chaos transitions) than in other chaotic
regimes (\reffig{fig_log}).

The measures $DET$, $L$ and $L_{\max}$, which are basing on the diagonal lines,
show clear maxima at the periodic-chaos/chaos-periodic transitions. The measure $L_{\max}$
finds all of such transitions, but $DET$ and $L$ do not detect all of them
(\reffig{fig_rqa_log}A, C, E and Tab.~\ref{tab:comparison_RQA}).
However, they all are not able to detect chaos-chaos transitions.
But, the chaos-chaos transitions to the laminar states are identified
by the measures $LAM$, $TT$ and $V_{\max}$, which are based on
the vertical structures (\reffig{fig_rqa_log}B, D, F and Tab.~\ref{tab:comparison_RQA}). 
These measures show distinct maxima or peaks at
the chaos-chaos transitions. Furthermore, the measures fall to zero within the period windows,
hence, the chaos-order transitions can also be identified.
Since vertical lines occur much more frequently at inner crisis,
band merging points and in regions of intermittency (i.\,e.~laminar states) than in other chaotic
regimes, $TT$ and $V_{\max}$ grow up significantly at those points. This can also be seen
by looking at the supertrack functions (\reffig{fig_log}B). 
Although $LAM$ also reveals laminar states, it
is quite different from the other two measures because it does not increase 
at inner crises (Tab.~\ref{tab:comparison_RQA}). 
Noise, of course, influences these results. For small noise
levels, the transitions can still be identified. Moreover, $LAM$ is more robust against 
noise than $TT$ and $V_{\max}$ \cite{marwan2002herz}.


The behaviour of these measures regarding the control parameter $a$
is similar to some of formerly proposed measures of complexity
\citep{saparin94,wackerbauer94}. 
The R\'enyi dimension $D_q$ of order $q<0$,
the fluctuation complexity as well as the re-normalised entropy exhibit
local maxima in regions of intermittency, a rapid increase at inner crises
and a rapid decrease and increase at
the transitions between chaos and periodic windows. The difference
between the formerly proposed measures and $LAM$, $TT$ and $V_{\max}$ 
is the amount of data points needed. It is important to emphasise 
that for the methods proposed in \cite{saparin94} and \cite{wackerbauer94}, 
more than 100,000 data points are needed, whereas 1,000 data points 
are enough for the measures based on RPs.

\begin{table}
\caption{Comparison of RQA measures based on diagonal ($DET$, $L$ and $L_{\max}$)
and vertical structures ($LAM$, $TT$ and $V_{\max}$) regarding 
periodic-chaos/chaos-periodic transitions (PC/CP), 
chaos-chaos transitions (band merging -- BM and
inner crisis -- IC) and laminar states.}\label{tab:comparison_RQA}
\centering \begin{tabular}{lllll}
\hline
measure         &PC/CP transitions   &BM and IC      &laminar states\\
\hline
\hline
$DET$           &increases              &--             &--\\
$L$             &increases              &--             &--\\
$L_{\max}$      &increases              &--             &--\\
$LAM$           &drops to zero           &--             &increases\\
$TT$            &drops to zero           &increases      &increases\\
$V_{\max}$      &drops to zero           &increases      &increases\\
\hline
\end{tabular}
\end{table}

%% file: meth_invariants.tex
%
%
%
%

The RQA measures introduced in Subsec.~\ref{sec:RQA}
are rather heuristic but describe RPs quantitatively
and are especially helpful to find various transitions in dynamical systems. However, their
major drawback is that they are typically not invariant with respect to the 
embedding used to reconstruct the phase space trajectory, i.\,e.~the values 
of the RQA measures depend rather strongly on the embedding parameters. 
Therefore, it is important to know, whether  typical invariants in nonlinear dynamics, like entropies
or dimensions, can be inferred from the recurrence matrix, too.

Let us start with the generalised correlation sum (the generalisation of 
Eq.~(\ref{eq_corrsum})),
\begin{equation}\label{eq_gen_cor_int}
C_q(\varepsilon) = \left\{  
\frac{1}{N} \sum_{i=1}^N 
\left[
\frac{1}{N} \sum_{j=1}^N \Theta\left(\varepsilon - \| \vec{x}_i - \vec{x}_j \|\right)
\right]^{q-1}
\right\}^{\frac{1}{q-1}},
\end{equation} 
which can be used to estimate generalised entropies, dimensions, mutual 
information and redundancies
(e.\,g.~\cite{grassberger83a,grassberger83b,pawelzik1987,kurths87,prichard1995}). 
From Eq.~(\ref{eq_gen_cor_int}) it is obvious that the density of
recurrence points in an RP can also be used to estimate these
invariants. However, some of these invariants, such as the mutual 
information and the generalised entropies, are related to further 
features of RPs and not just the density of recurrence points.

As we have mentioned in Subsec.~\ref{sec:RQA}, an important  ingredient for
the computation of the RQA measures is the  length-distribution of diagonal
lines $P(\varepsilon,l)$ in the RP, Eq.~(\ref{eq_P_of_l}),  because it
encodes main properties of the system, such as predictability and measures
of complexity. Diagonal lines in the RP represent co-moving segments of
different parts of the trajectory  $\vec x_{i+k}$ and $\vec x_{j+k}$ for
some $k = 1,\ldots, l$, Eqs.~(\ref{eq_diagonalline0}) and (\ref{eq_diagonalline}).
The longer the trajectories move within an $\varepsilon$-tube (cf.~Fig.~\ref{fig_eps_tube}), the longer
the diagonal lines in the RP will be. As the time  for which trajectories
starting at close  initial conditions move within an $\varepsilon$-tube is
related to the  inverse of the largest Lyapunov exponent, it can be supposed
that also  the length of the diagonal lines in an RP can be linked to  the
predictability of the underlying system. Since the introduction of RPs a
relationship between the length of the diagonal lines and the maximal
Lyapunov exponent has been stated \citep{eckmann87}. As already mentioned, several attempts have
been tried to heuristically fix this relationship as a direct inverse
relation between the maximal or averaged line length and the Lyapunov
exponent \citep{trulla96,atay99}. Choi et al.~introduced a measure based on
the width and the absolute number of diagonal lines and related this measure
to the largest Lyapunov exponent \cite{choi99}. However, we show next that
the distribution of diagonal lines is not directly related to the maximal
Lyapunov exponent but to the correlation entropy
\citep{faure98,thiel2003,thiel2004a,march2005}. The formal relationship between the correlation
entropy and the Lyapunov exponents \cite{ruelle1978,beck1992} is 
\begin{equation}\label{eq_ruelle}
K_2 \le \sum_{\lambda_i>0}\lambda_i,
\end{equation}
where $\lambda_i$ denote the Lyapunov exponents.
Moreover, the algorithm for
the estimation of these  invariants gives some justification for the ad hoc
measures  of the RQA.


\subsubsection{Correlation entropy and correlation dimension}\label{sec:k2_d2}

At first, the definition of the {\it R\'enyi entropy} of second order is
recalled in order to deduce how it is linked to the distribution of diagonal
lines in the RP. Let us consider a trajectory $\vec x(t)$ in the basin of
an  attractor in a $d$-dimensional phase space. We divide the
phase space into $d$-dimensional hyper-cubes of size $\varepsilon$. Then 
$p_{i_1,\ldots,i_l}(\varepsilon)$ denotes the joint probability that $\vec
x(t=1\,\Delta t)$  is in the $\varepsilon$-box $i_1$, $\vec x(t=2\,\Delta
t)$ is in the box $i_2$, \ldots, and  $\vec x(t=l\,\Delta t)$ is in the box
$i_l$. The second order R\'enyi entropy  ({\it correlation entropy})
\citep{grassberger83b,renyi1970} is then defined by
\begin{equation}\label{eq_k2}
K_2=-\lim_{\Delta t\to 0}\lim_{\varepsilon \to 0}\lim_{l \to \infty}
     \frac{1}{l\Delta t}\ln \sum_{i_1,\ldots,i_l} p^2_{i_1,\dots,i_l}(\varepsilon).
\end{equation}

Assuming that the system is ergodic, which is always the case for chaotic
systems  as they are mixing, we obtain
\begin{equation}\label{eq_ergodic}
\sum_{i_1,\ldots,i_l}p^2_{i_1,\ldots,i_l}(\varepsilon)=
    \frac{1}{N}\sum_{t=1}^N p_{i_1(t),\ldots,i_l(t+ (l-1) \Delta t)}(\varepsilon),
\end{equation}
where $i_1(t)$ is the box that the $t^{th}$ data point is in, $i_2(t+\Delta
t)$ is the box that the $(t+\Delta t)^{th}$ data point is in, 
etc.~\cite{prichard1995}. Furthermore, approximating
$p_{i_1(t),\ldots,i_l(t+(l-1)\Delta )}(\varepsilon)$ by the probability 
$p_t(\varepsilon,l)$ of finding a sequence of $l$ points in boxes  of size
$\varepsilon$ centred at the points $\vec x(t), \dots, \vec x(t+(l-1)
\Delta t)$, we can state  \begin{equation} \frac{1}{N}\sum_{t=1}^N
p_{i_1(t),\ldots,i_l(t+ (l-1) \Delta t)}(\varepsilon) \approx
\frac{1}{N}\sum_{t=1}^N p_t(\varepsilon,l). \end{equation}

Moreover, $p_t(\varepsilon,l)$ can be expressed by means of the 
recurrence matrix 
\begin{equation}\label{eq_p_of_l}
p_t(\varepsilon,l) = 
    \lim_{N \to \infty}\frac{1}{N}\sum_{s=1}^N \prod_{k=0}^{l-1}\mathbf{R}_{t+k,s+k}(\varepsilon).
\end{equation}

Based on Eqs.~(\ref{eq_k2})--(\ref{eq_p_of_l}), an estimator for the second 
order R\'enyi entropy can be found by means of the RP
\begin{equation}\label{eq_estimator_k2}
\hat{K_2}(\varepsilon,l)=
     -\frac{1}{l\,\Delta t} \ln \left( p_c(\varepsilon,l) \right)=-\frac{1}{l\,\Delta t} \ln 
     \left(\frac{1}{N^2}\sum_{t,s=1}^N \prod_{k=0}^{l-1} \mathbf{R}_{t+k,s+k}(\varepsilon)\right) ,
\end{equation}
where $p_c(\varepsilon,l)$ is the probability to find a diagonal of at 
least length $l$ in the RP. Therefore, if
we plot $p_c(\varepsilon,l)$ in a logarithmic scale versus $l$, we should obtain a 
straight line with slope $-\hat{K}_2(\varepsilon)\Delta t$ for large $l$.

On the other hand, the $l$-dimensional 2$^{\text{nd}}$-order correlation 
sum 
\begin{equation}\label{eq_C2}
C_2(\varepsilon,l)=\lim_{N \to \infty}\frac{1}{N^2}\sum_{i,j=1}^N 
\Theta\left(\varepsilon - 
\sqrt{\sum_{k=0}^{l-1}\left\|\vec x_{i+k}- \vec x_{j+k}\right\|^2}
\right).
\end{equation}
can be used in the definition for $K_2$ \citep{grassberger83a}
\begin{equation}\label{eq_estimator_k2_from_C2}
K_2(\varepsilon,l) =
-\lim_{\Delta t\to 0}\lim_{\varepsilon \to 0}\lim_{l \to \infty}
     \frac{1}{l\Delta t}\ln C_2(\varepsilon,l).
\end{equation}
Due to the exponential divergence of the trajectories, the condition in Eq.~(\ref{eq_C2})
\begin{equation*}
\sum_{k=0}^{l-1}\left|\left|\vec x_{i+k}-\vec x_{j+k}\right|\right|^2 \le \varepsilon^2
\end{equation*}
is essentially equivalent to
\begin{equation*}
\left|\left|\vec x_{i+k}-\vec x_{j+k}\right|\right| < \varepsilon \quad \text{for} \quad k=1,\ldots,l,
\end{equation*}
and, therefore,  equivalent to the product in Eq.~(\ref{eq_p_of_l}).

If only a scalar time series is available, 
embedding is used, Eq.~(\ref{eq_embedding}). Taking the embedding 
dimension $l$ (here $l$ is used for the 
embedding dimension instead of $m$ for didactical reasons) and time 
delay $\tau=1$, the $l$-dimensional correlation sum can be estimated by
\begin{equation}
\tilde C_2(\varepsilon,l)=\lim_{N \to \infty}\frac{1}{N^2}\sum_{i,j=1}^N 
\Theta\left(\varepsilon - 
\sqrt{\sum_{k=0}^{l-1}\left|x_{i+k}- x_{j+k}\right|^2}
\right).
\end{equation}
Then, an estimator of $K_2$ (Grassberger-Procaccia (G-P) algorithm) can be obtained by
\begin{equation}
\tilde{K}_{2}(\varepsilon,l)=\frac{1}{\Delta t}\ln\frac{\tilde C_2(\varepsilon,l)}{\tilde C_2(\varepsilon,l+1)}.
\end{equation}

Moreover, $C_2(\varepsilon,l)$  scales like
$C_2(\varepsilon,l) \sim \varepsilon^{D_2}$ \citep{grassberger83c}, what leads
with Eq.~(\ref{eq_estimator_k2_from_C2}) to
\begin{equation}\label{eq_C2_scales_e_K2}
C_2(\varepsilon,l)\sim \varepsilon^{D_2} \e^{-l\,\Delta t\, K_2}.
\end{equation}

Due to the similarity between the approaches using RPs and the
correlation sum, we state the basic relation
\begin{equation}\label{eq_relationship_K2}
p_c(\varepsilon,l) \approx \sum_{i_1,\ldots,i_l} p^2(i_1,\ldots,i_l)\approx \tilde C_2(\varepsilon,l) \sim \varepsilon^{D_2} 
\e^{-l\,\Delta t\, K_2} .
\end{equation} 

The difference between both approaches is that in $p_c(\varepsilon,l)$ we include information 
about $l$ vectors, whereas in $\tilde C_2(\varepsilon,l)$ we have just information about $l$ 
coordinates. Besides this, in the RP approach $l$ is a length in the plot, whereas in the 
G-P algorithm $l$ means the embedding dimension.

The relationship~(\ref{eq_relationship_K2}) also allows to  estimate $D_2$ from 
$p_c(\varepsilon,l)$. Considering Eq.~(\ref{eq_relationship_K2}) for two different thresholds 
$\varepsilon$ and $\varepsilon+\Delta \varepsilon$ and dividing both of them, we get
\begin{equation}\label{eq_corrdim_RP}
\hat D_2(\varepsilon) = \ln\left(\frac{p_c(\varepsilon,l)}{p_c(\varepsilon+\Delta\varepsilon,l)}\right)/
                        \ln\left(\frac{\varepsilon}{\varepsilon+\Delta\varepsilon}\right),
\end{equation}
which is an estimator of the {\it correlation dimension} $D_2$ \citep{grassberger83b}. 
Analogously to Eq.~(\ref{eq_estimator_k2}), the joint R\'enyi entropy of second order 
can be estimated using the probability to find a diagonal of at least length $l$ 
in the JRP instead of the RP of a single system. This extension of the estimator 
of Eq.~(\ref{eq_estimator_k2}) is useful for the analysis of two or more interacting 
systems, as will be shown in Subsec.~\ref{sec:Synchro}.
The joint R\'enyi entropy of second order is defined as
\begin{equation}\label{eq_JK2}
JK_2=-\lim_{\Delta t\to 0}\lim_{\varepsilon \to 0}\lim_{l \to \infty} 
    \frac{1}{l\Delta t}\ln\sum_{\substack{i_1,\ldots,i_l\\j_1,\ldots,j_l}}p^2_{i_1,\ldots,i_l,j_1,\ldots,j_l}(\varepsilon),
\end{equation}
where $p_{i_1,i_2,\ldots,i_l,j_1,j_2,\ldots,j_l}(\varepsilon)$ is the joint probability 
that $\vec x(t=\Delta t)$ is in box $i_1$, $\vec x(t=2\Delta t)$ is in box 
$i_2$, \ldots, $\vec x(t=l\Delta t)$ is in box $i_l$ and simultaneously 
$\vec y(t=\Delta t)$ is in box $j_1$, $\vec y(t=2\Delta t)$ is in box 
$j_2$, \ldots, and $\vec y(t=l\Delta t)$ is in box $j_l$. Using again the 
ergodicity of the system, 
\begin{equation}
\sum_{\substack{i_1,\ldots,i_l\\j_1,\ldots,j_l}}p^2_{i_1(t),\ldots,i_l(t+(l-1)\Delta t),j_1(t),\ldots,j_l(t+(l-1)\Delta t)} = 
    \frac{1}{N}\sum_{t=1}^N p_{i_1,\ldots,i_l,j_1,\ldots,j_l}
\end{equation} 
can be stated. On the other hand, the following approximation
\begin{eqnarray}
\lefteqn{p_{i_1(t),\ldots,i_l(t+(l-1)\Delta t),j_1(t),\ldots,j_l(t+(l-1)\Delta t)} \approx {} }  \\ \nonumber
& &{} \frac{1}{N}\sum_{s=1}^N \prod_{m=0}^{l-1}\Theta(\varepsilon^{\vec x}-\|\vec x_{t+m}-\vec x_{s+m}\|)\Theta(\varepsilon^{\vec y}-\|\vec y_{t+m}-\vec y_{s+m}\|)
\end{eqnarray}
can be made (cf.~Eqs.~(\ref{eq_ergodic}) and (\ref{eq_p_of_l})). Then, substituting this 
expression in Eq.~(\ref{eq_JK2}), the estimator
for the joint R\'enyi entropy of second order is
\begin{equation}\label{JK2_est}
\widehat{JK_2}(\varepsilon^{\vec x},\varepsilon^{\vec y},l) = 
    -\frac{1}{l\Delta t}\ln \underbrace{\left(\frac{1}{N^2}\sum_{t,s=1}^N \prod_{k=0}^{l-1}\mathbf{JR}_{t+k,s+k}(\varepsilon^{\vec x},\varepsilon^{\vec y})\right)}_{\ast}.
\end{equation}
Note, that $\ast$ is the probability $p_c(\varepsilon^{\vec x},\varepsilon^{\vec y},l)$ 
to find a diagonal of at least length $l$ in the JRP. Hence, the logarithmic representation
of $p_c(\varepsilon^{\vec x},\varepsilon^{\vec y},l)$ versus $l$ reveals a straight line 
for small thresholds $\varepsilon^{\vec x},\varepsilon^{\vec y}$ and long lines, whose 
slope is equal the joint R\'{e}nyi entropy multiplied with the sampling 
time interval $\Delta t$.

\subsubsection*{Example: Estimation of invariants by recurrences in chaotic systems}\label{sec:Invariants_examples}

We illustrate the algorithm based on RPs to estimate 
$K_2$ and $D_2$ by applying it to two prototypical nonlinear systems: 
the Bernoulli map, Eq.~(\ref{eq_bernoulli}), and the 
R\"ossler system, Eqs.~(\ref{eq_roessler}),
with parameters $a=b=0.2$, $c=5.7$ and sampling rate 0.2.
The RPs are computed without using embedding, i.\,e.~taking the original components.

\begin{figure}[bthp]
\begin{center}
\includegraphics[width=1.0\textwidth]{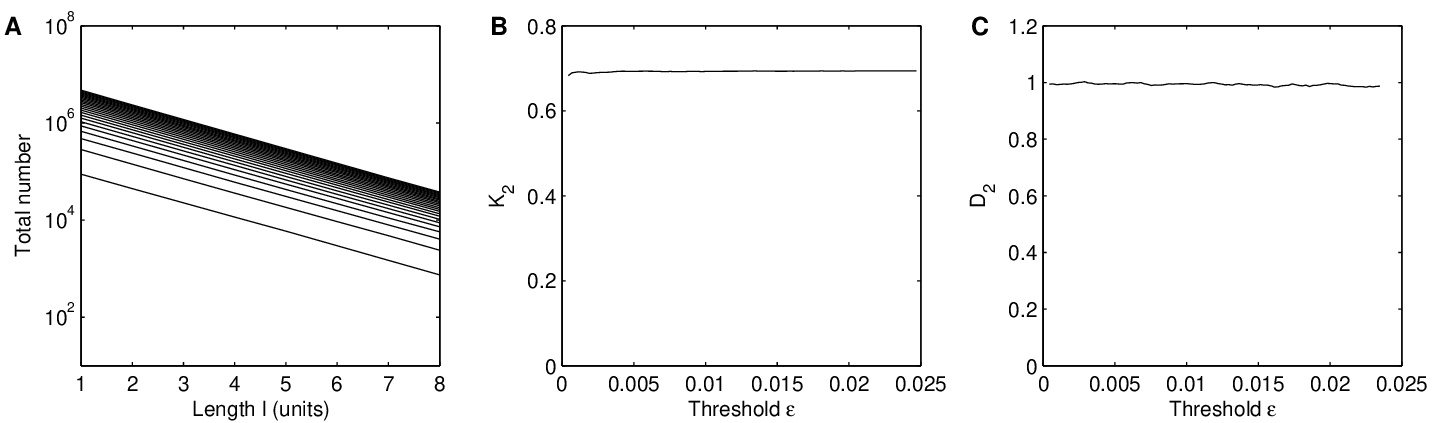}
\caption{(A) Total number of diagonal lines
of at least length $l$ in the RP of the Bernoulli map. 
Each histogram is computed for a different threshold $\varepsilon$, 
from 0.000436 (bottom) to 0.0247 (top). 10,000 data points have been used for the computation.
(B) Estimate of $K_2$ in dependence on $\varepsilon$.
(C) Estimate of $D_2$ in dependence on $\varepsilon$.
}\label{fig_bernoulli_k2_d2_eps2}
\end{center}
\end{figure}

The distribution of the diagonal lines with at
least length $l$ in the RP of the Bernoulli map
is computed for 100 different values of the 
threshold $\varepsilon = [0.000436, 0.0247]$.
The plot of the logarithm of this length distribution
reveals straight parallel lines 
for the different values of $\varepsilon$ 
(Fig.~\ref{fig_bernoulli_k2_d2_eps2}A).
The slope of these lines is an estimate for $K_2$
(Fig.~\ref{fig_bernoulli_k2_d2_eps2}B).
The obtained estimate is $\hat K_2=0.6929 \pm 0.0016$, 
which is very close to the theoretical value $K_2=\ln 2 \approx 0.6931$. 
For $\hat D_2$ we obtain a value of $0.9930 \pm 0.0098$, which 
is also close to the theoretical value of $D_2=1$.  These 
results confirm numerically the relationships presented 
in Subsec.~\ref{sec:k2_d2}.

%
%
%
%
%

\begin{figure}[bthp]
\begin{center}
\includegraphics[width=1.0\textwidth]{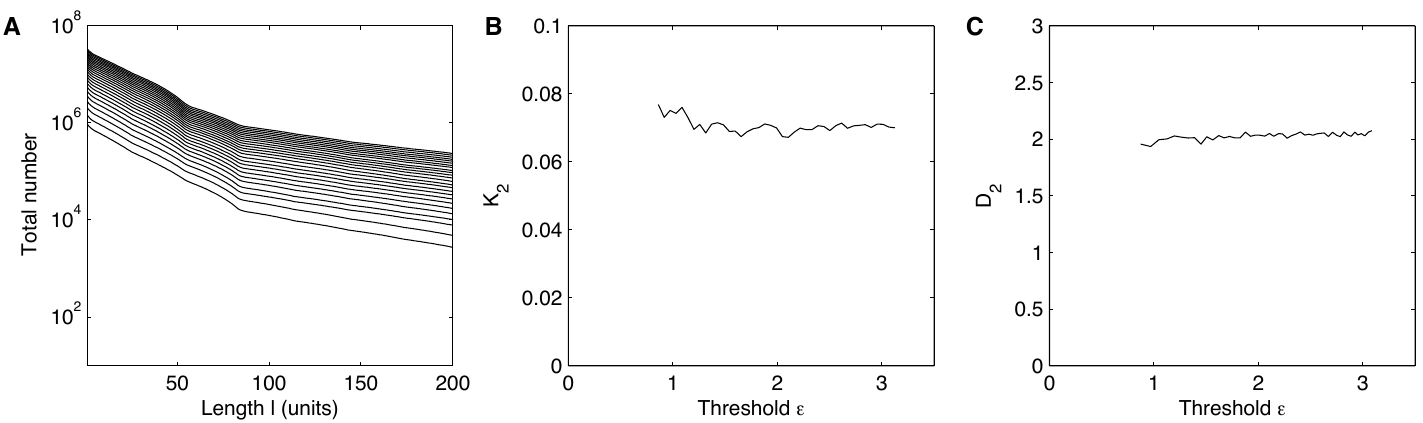}
\caption{(A) Total number of diagonal lines
of at least length $l$ in the RP of the R\"ossler system. Each 
histogram is computed for a different threshold $\varepsilon$,
from 0.862 (bottom) to 6.465 (top). 
(B) Estimate of $K_2$ in dependence on $\varepsilon$ ($K_2 = 0.069 \pm 0.003$). 
10,000 data points have been used for the computation.
(C) Estimate of $D_2$ in dependence on $\varepsilon$ ($D_2 = 2.03 \pm 0.03$). 
200,000 data points have been used for the computation.
}\label{fig_roessler_k2_d2_eps2}
\end{center}
\end{figure}

In the case of the R\"ossler system (Fig.~\ref{fig_rps_intro}B), the most remarkable finding
is the existence of {\it two} well differentiated scaling regions
for the distribution of diagonal lines of at least length $l$ 
(Fig.~\ref{fig_roessler_k2_d2_eps2}A): for $1 \le l \le 84$ 
the slope is about 3--4 times larger than the slope for $l>84$ 
(the time interval between two points of the integrated 
trajectory is 0.2, hence $l=84$ corresponds to 16.8). 
As $K_2$ is defined for $l \to \infty$ the second slope 
yields the estimation of the entropy, which is 
$\hat K_2=0.069 \pm 0.003$ (Fig.~\ref{fig_roessler_k2_d2_eps2}B). 
Note, that $K_2$ is a lower bound for the sum of the positive Lyapunov 
exponents and the estimated value of $K_2$ 
is close to the largest Lyapunov exponent of the R\"ossler 
system, which is approximately 0.072 \cite{alligood96}.

%
%
%
%
%
%

%
%
%
%

However, the slope of the first part of the curve is interesting too, 
as it is also independent of $\varepsilon$. The region $1\le l \le 84$ 
characterises the short term dynamics of the system up to three cycles 
around the attractor and corresponds in absolute units to a time of 
$t=16.8$. These three cycles reflect a characteristic period of 
the R\"ossler system that is called {\it recurrence period} 
$T_{\text{rec}}$. It is different from the dominant 
{\it phase period} $T_{\text{ph}}$, which is given by the 
dominant frequency of the power 
spectrum. $T_{\text{rec}}$, however, is given by the 
recurrences to the same state in phase space.

For predictions on time scales below the recurrence period,
the second slope of $\ln( \hat p_c(\varepsilon,l))$ versus $l$ 
gives a better predictability than the first slope 
(more than three times better), which means that
there exist two time scales that characterise the attractor. 
The first slope is greater than the second one because it is 
more difficult to predict the next step if we have only 
information about a piece of the trajectory 
for less than one recurrence period. 
Once we have scanned the trajectory for more than 
$T_{\text{rec}}$, the predictability increases and hence, 
the slope of $p_c(\varepsilon,l)$ in the 
logarithmic plot decreases. Hence the first slope, as well 
as the time at which the second 
slope begins, reveal important characteristics of the attractor. Note that
even though the first slope is not a dynamical invariant (it changes if one uses
embedding coordinates), most of the initial conditions ``diverge'' at a rate given by the first
slope and not by the second one.

%
%

\begin{figure}[bthp]
\begin{center}
\includegraphics[width=0.4\textwidth]{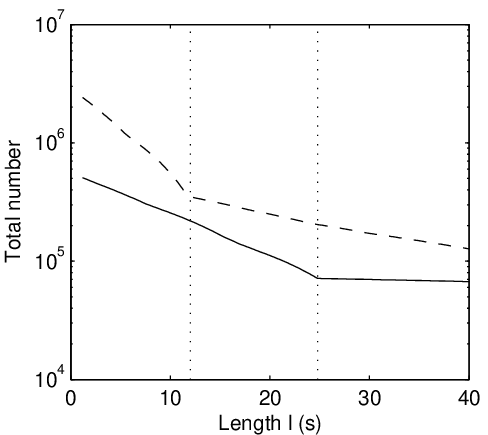}
\caption{Total number of diagonal lines
of at least length $l$ in the RP of the R\"ossler system
with $c=9$ (solid) and $c=30$ (dashed). The two scaling regions are separated
at approximately 12 and 24.8, which corresponds 
to $T_{\text{rec}} = 2T_{\text{ph}}$ and $T_{\text{rec}} = 4T_{\text{ph}}$, 
respectively.
}\label{fig_roessler_hist_L}
\end{center}
\end{figure}

The relationship between the transition of the 
scaling regions and the phase period can also be seen 
in attractors of different shape. For a R\"ossler system with fixed 
$a=b=0.1$, but varied parameter $c$, the form of the
attractor as well as $T_{\text{rec}}$ and $T_{\text{ph}}$
change. For example, $c = 9$ yields $T_{\text{rec}}=2T_{\text{ph}}$
(with $T_{\text{ph}} \approx 6$), and $c=30$, gives 
$T_{\text{rec}}=4T_{\text{ph}}$ (with $T_{\text{ph}} \approx 6.2$). 
In both cases
the length of the first scaling region ($T_{\text{rec}} = 12$ and 
$24.8$, resp.) corresponds as expected to
$2T_{\text{ph}}$ and $4T_{\text{ph}}$, respectively 
(Fig.~\ref{fig_roessler_hist_L}). The existence of {\it two} scaling regions in the
R\"ossler system is a new and striking point of the recurrence
analysis that cannot be observed with the method proposed by
Grassberger and Procaccia to estimate $K_2$ \cite{grassberger83a}. 
The existence of the two different scaling regions
can also be found in other non-hyperbolic systems,
such as the Lorenz oscillator, Eqs.~(\ref{eq_lorenz}). 
This is an indication of the non-hyperbolic nature of 
such systems and fits well with results obtained 
by other approaches \cite{anishchenko2004}.

Next, we estimate $D_2$ for the R\"ossler system for various choices
of the threshold $\varepsilon$ (Fig.~\ref{fig_roessler_k2_d2_eps2}D)
using Eq.~(\ref{eq_corrdim_RP}) and the average
over lines of length $l$ which correspond to the {\it first} scaling region.
The estimated value using 200,000 data points is $\hat
D_2=1.86\pm 0.04$. This result is in accordance with the estimation of
$D_2$ by the G-P algorithm given in \cite{raab2001},
where the value $1.81$ was obtained. 
Restricting the average in $l$ to the {\it second} scaling region, we
obtain the slightly higher value $\hat D_2=2.03\pm 0.03$, which is in
accordance with the value $D_2=2.06 \pm 0.02$ obtained in 
\cite{grassberger83a,huebner93}. Note that the extent and the onset of
a scaling region in $D_2(\varepsilon)$ may lead to problems in the
$D_2$ estimation \cite{ding93}. Furthermore, the number of points needed
to estimate $D_2$ accurately is much larger than the one needed for
the estimation of $K_2$ (in the examples shown here, we used 10,000
data points for the estimation of $K_2$ and 200,000 data points for
the estimation of $D_2$).

\begin{figure}[bthp]
\centering
\includegraphics[width=1.0\textwidth]{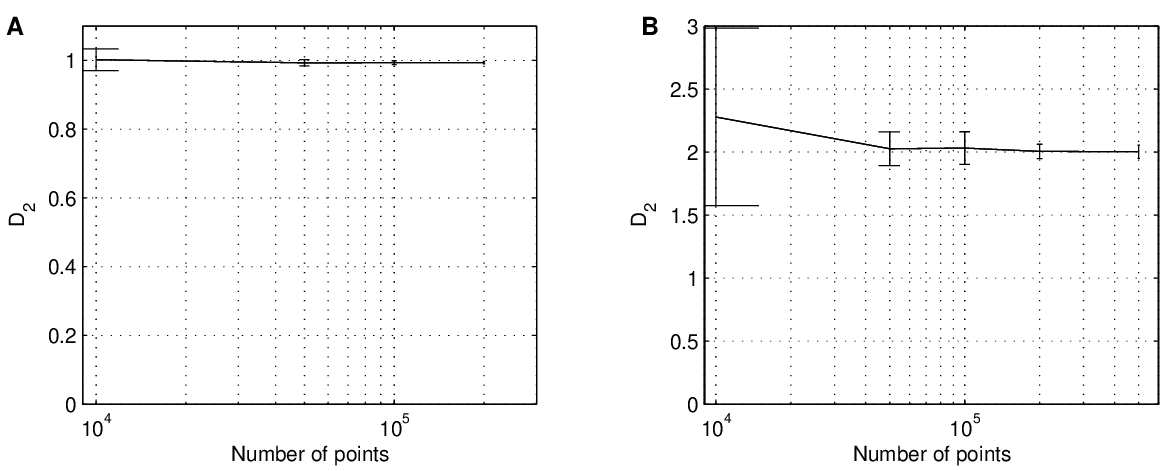}
\caption{Estimation of $D_2$ by means of RPs in dependence on 
the number of points of the trajectory used for the computation 
for (A) the Bernoulli map and (B) the R\"ossler system (second scaling
region used).
The error bars indicate the standard deviation of $\hat D_2(\varepsilon)$.
}\label{fig_d2_number}
\end{figure}

Therefore, it is interesting to investigate the dependence of 
the estimated value for $D_2$ on the number of points of the 
trajectory used for the computation. Let us consider the Bernoulli 
map and the R\"ossler system (Fig.~\ref{fig_d2_number}).
We estimate $D_2$
by the mean value over 50 different values of the threshold
$\varepsilon$ in the same range as
in Figs.~\ref{fig_bernoulli_k2_d2_eps2} and ~\ref{fig_roessler_k2_d2_eps2}, 
respectively. For the Bernoulli map, the estimation of $D_2$
is rather accurate for already 10,000 data points. On the other 
hand, for the R\"ossler system, at least 50,000 data points are 
necessary for a more precise estimation. 
This result confirms the expectation that the higher the dimension
or the more complex a system is, the more 
data points are necessary for its characterisation \cite{kantz97}.

%
%
%

\begin{figure}[bthp]
\centering
\includegraphics[width=0.6\textwidth]{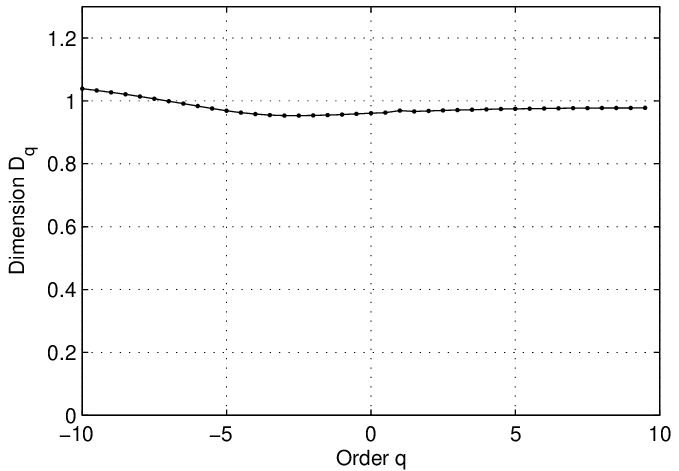}
\caption{Estimation of $D_q$ by means of RPs for the Bernoulli 
map (20,000 data points have been used for the computation).
}\label{fig_gen_dim_bernoulli}
\end{figure}

As mentioned at the beginning of this section, it is also possible 
to estimate the generalised dimensions \cite{grassberger83b} 
from the recurrence matrix by
\begin{equation}\label{eq_gen_dim}
D_q=\lim_{\varepsilon \to 0}\frac{\log(C_q(\varepsilon))}{\log \varepsilon},
\end{equation}
with $C_q$ given by Eq.~(\ref{eq_gen_cor_int}). 
The estimation of $D_q$ for the simple case of the Bernoulli map 
reveals $D_q \approx 1 \ \,\forall \, q$ (Fig.~\ref{fig_gen_dim_bernoulli}),
which corresponds to the theoretical value $D_q = 1 \ \,\forall \, q$.
However, it must be noted that the problems that arise for the estimation 
of $D_q$ for $q \le 1$ are also present in the recurrence approach. 
In order to improve the accuracy of the estimation more 
sophisticated methods, such as the Enlarged Box Algorithm 
\cite{pastorsatorras1996} should be used.  


\subsubsection{Recurrence times and point-wise dimension}\label{sec:information_dim}

Another invariant characteristic, the {\it point-wise dimension}, can be
also estimated by  using recurrence times \cite{gao99}. Recurrence times
can be directly  picked up from the RP. First, we denote the set of points
of the trajectory $\vec x$ which fall into the
$\varepsilon$-neighbourhood  $B_{\vec{x}_i}(\varepsilon)$ of an arbitrary
chosen point at $i$ with
\begin{equation}\label{eq_set_recpoints}
\mathcal{R}_{i} = \{\vec x_{j_1}, \vec x_{j_2},\, \dots\, |\ \mathbf{R}_{i,j_k} = 1 \}.
\end{equation}
The elements of this set correspond to the {\it recurrence points} of the
$i^{\text{th}}$  column $\{{\mathbf R}_{i,j}\}_{j=1}^N$ of an RP. The corresponding
recurrence times between these recurrence points  ({\it recurrence times
of first type}, according to the notation  given in \cite{gao99}) are
$\{T_k^{(1)}=j_{k+1}-j_k\}_{k \in \mathds{N}}$. Due to possible tangential
motion, some of the recurrence points in $\mathcal{R}_i$ correspond to
recurrence times $T_k^{(1)} = 1$ (Fig.~\ref{fig_tangential_motion}). 
However, in order to obtain the real 
recurrence times (Poincar{\'e} recurrence times), such points must be discarded.  One
approach is to remove all consecutive recurrence points  with $T_k^{(1)} =
1$ from the set $\mathcal{R}_i$. This results in a new set 
$\mathcal{R}'_i=\{\vec x_{j'_1}, \vec x_{j'_2}, \dots\}$. Then, the
recurrence times ({\it recurrence times of second type} according to
\cite{gao99}) $\{T_k^{(2)}=j'_{k+1}-j'_k\}_{k \in \mathds{N}}$ are
calculated from the remaining recurrence points (i.\,e.~from
$\mathcal{R}'_i$). Hence, $T^{(2)}$ measures vertically the time distance between the
beginning of (vertically) subsequent recurrence structures in the RP. An alternative
estimator for the recurrence times $T^{(2)}$ is the  average of the lengths of
the {\it white} vertical lines at  a specific column of the RP. For
systems with less laminar structures in the RP (i.\,e.~$LAM$ and $TT$ tend
to zero),  the distribution of such an average almost coincides with the
distribution of $T^{(2)}$ as defined by \cite{gao99}. However, for systems
with laminar states (e.\,g.~logistic map for certain values of the control
parameter), $T^{(2)}$ as defined by \cite{gao99} over-estimates the recurrence
times and $T^{(2)}$ computed from the white vertical lines under-estimates the
recurrence times.

Based on the recurrence times $T^{(1)}$ and $T^{(2)}$, it is possible to estimate 
the {\it point-wise dimension} $D_P(i)$, which is defined by 
\begin{equation}\label{eq_pointw_dim}
\mu\left(B_{\vec{x}_i}(\varepsilon)\right) \sim \varepsilon^{D_P(i)},
\end{equation}
where $\mu\left(\cdot\right)$ is the probability measure (cf.~Eq.~(\ref{eq_poincareTheorem})).
The measure
$\mu\left(B_{\vec{x}_i}(\varepsilon)\right)$ can be estimated by the
frequency at which the neighbourhood of the point at $i$ is visited by the
trajectory, which is  the reciprocal of the {\it mean recurrence time}
$\langle T^{(1)} \rangle$,  i.\,e.~$\mu\left(B_{\vec{x}_i}(\varepsilon)\right)
= \mu\left(\mathcal{R}_{i}\right) \approx  1/\langle T^{(1)} \rangle$ and thus
\cite{gao99}
\begin{equation}\label{eq_pointw_dim_T1}
\langle T^{(1)}(\varepsilon) \rangle \sim \varepsilon^{-D_P(i)}.
\end{equation}
This measure depends on the chosen point $\vec{x}_i$, because some parts
of the attractor are visited more frequently than others.

Analogously, we can state that the measure of the set of neighbours of
$\vec x_i$ without points belonging to the tangential motion $\mathcal{R}'_{i}$ is
$\mu(\mathcal{R}'_{i}) \approx 1/\langle T^{(2)} \rangle$.  When the points
belonging to the tangential motion represent a zero-dimensional set,  $\mu(\mathcal{R}'_{i}) =
\mu(\mathcal{R}_{i}) \sim \varepsilon^{D_P(i)}$.  When these points
represent a one-dimensional set,  $\mu(\mathcal{R}'_{i}) \sim
\varepsilon^{D_P(i)-1}$.  Hence, for discrete maps and continuous systems
with small $\varepsilon$,
\begin{equation} 
\langle T^{(2)}(\varepsilon) \rangle \sim \varepsilon^{-D_P(i)},
\end{equation} 
and for continuous systems with large $\varepsilon$,
\begin{equation}\label{eq_pointw_dim_T2}
\langle T^{(2)}(\varepsilon) \rangle \sim \varepsilon^{-(D_P(i)-1)}. 
\end{equation} 

Furthermore, the mean recurrence times can be used to analyse transient
and non-stationary dynamics, computing them in sliding  windows and
monitoring how $\langle T^{(1)}(\varepsilon) \rangle$ and 
$\langle T^{(2)}(\varepsilon) \rangle$ change in different blocks of sub-data
sets. When analysing experimental data, often only scalar time series are
available. Hence, the time delay embedding technique must be used. In this
case, the disadvantage of using $\langle T^{(1)}(\varepsilon) \rangle$ to
analyse the non-stationarity, is that it sensitively depends on the used
embedding parameters. Hence, in this case
$\langle T^{(2)}(\varepsilon) \rangle$ yields more robust results.


\subsubsection{Generalised mutual information (generalised redundancies)}\label{sec:mi}

The {\it mutual information} quantifies the amount of information that we 
obtain from the measurement of one variable on another. It has become 
a widely applied measure to quantify dependencies within or between 
time series (auto and cross mutual information).
The time delayed generalised mutual information (redundancy) $I_q(\tau)$ of a system
$\vec x_i$ is defined by \citep{renyi1970}
\begin{equation}
I_q^{\vec x}(\tau) = 2 H_q - H_q(\tau).
\end{equation}
$H_q$ is the $q^{\text{th}}$-order R\'enyi entropy of $\vec x_i$ and
$H_q(\tau)$ is the $q^{\text{th}}$-order joint R\'enyi entropy of $\vec x_i$ 
and $\vec x_{i+\tau}$
\begin{equation}\label{eq_renyi_entropy_q}
H_q=-\ln\sum\limits_{k}p_k^q,\qquad
H_q(\tau)=-\ln\sum\limits_{k,l}p_{k,l}^q(\tau),
\end{equation} 
where $p_k$ is the probability that $\vec x_i$ is in the
$k^{\text{th}}$ box and $p_{k,l}(\tau)$ is the joint probability that $\vec x_i$ is
in box $k$ and $\vec x_{i+\tau}$ is in box $l$ 
(note that we use here the same neighbourhood definition as for 
Eq.~(\ref{eq_k2})). 
Assuming that the system under consideration is ergodic and approximating the probability
that $\vec x_i$ is in the $k^{\text{th}}$ box of the partition by the probability to find a point of the trajectory
in a box of size $\varepsilon$ centred at $\vec x_i$,
the recurrence matrix can 
be used to estimate Eq.~(\ref{eq_renyi_entropy_q}) in the case $q=2$,
\begin{equation}
\hat H_2=-\ln \left(\frac{1}{N^2}\sum_{i,j=1}^N\mathbf R_{i,j}\right)
\end{equation}
and
\begin{equation}
\hat H_2(\tau) = 
-\ln \left(\frac{1}{N^2}\sum_{i,j=1}^N\mathbf R_{i,j}\mathbf R_{i+\tau,j+\tau}\right) =
-\ln \left(\frac{1}{N^2}\sum_{i,j=1}^N\mathbf{JR}^{\vec x, \vec x}_{i,j}(\tau)\right),
\end{equation}
where $\mathbf{JR}^{\vec x, \vec x}_{i,j}(\tau)$ denotes the delayed joint 
recurrence matrix, Eq.~(\ref{eq_delayed_JRP}).
The second order generalised mutual information can be estimated then
by means of RPs \citep{thiel2003}
\begin{equation}\label{eq_mi_rp}
\hat I_2^{\vec x}(\tau)= 
              \ln \left(\frac{1}{N^2}\sum\limits_{i,j=1}^N \mathbf{JR}_{i,j}^{\vec x, \vec x}(\tau)\right) - 
              2 \ln \left(\frac{1}{N^2} \sum\limits_{i,j=1}^N \mathbf{R}_{i,j}\right).
\end{equation}

\begin{figure}[tbhp]
\begin{center}
\includegraphics[width=\textwidth]{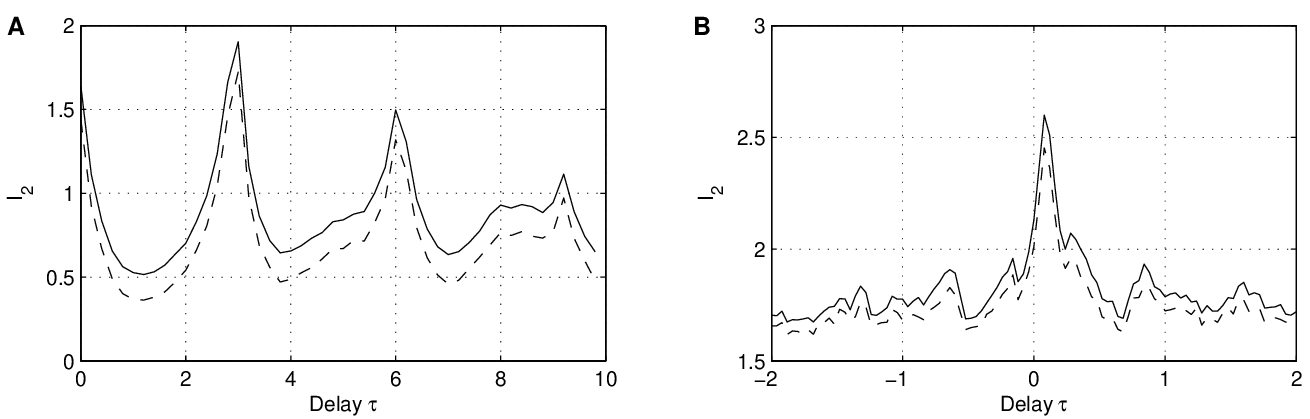}
\caption{(A) Mutual information for the $x$-component of the R\"ossler 
system, Eqs.~(\ref{eq_roessler}), with $a=0.15$, $b=0.2$ and $c=10$,
and time step $\Delta t=0.2$,
computed by the RP method (solid line) and 
the correlation sum approach (dashed line).
(B) Cross mutual information between the $x$ and $y$-components of the Lorenz system, 
Eqs.~(\ref{eq_lorenz}), for $\sigma=10$, $b=\frac{5}{3}$ and $r=28$ 
and time step $\Delta t=0.04$,
computed by the RP method (solid line) and 
the correlation sum approach (dashed line).
The difference between the two estimations is due to the used Euclidean norm;
using the maximum norm, both curves coincide.
}\label{fig_mi_roessler}
\end{center}
\end{figure}

As mentioned in the introduction of this section, the mutual information
can be estimated by means of the correlation sum \cite{liebert1989}. 
The joint distribution can be considered as the distribution in a 
two-dimensional embedding space \cite{kantz94}. 
Hence, 
\begin{equation}\label{mi_corr_sum}
\tilde I_2^{\vec x}(\tau)=\ln C_2^{\tau}(\varepsilon) - 2\ln C_2(\varepsilon),
\end{equation} 
where $C_2^{\tau}(\varepsilon)$ is the correlation dimension 
of the embedded two dimensional vectors 
$\vec{x}_i(\tau) = (u_i, u_{i+\tau})^T$, and $C_2(\varepsilon)$ 
is the correlation sum of the scalar time series $u_i$.
This corresponds to Eq.~(\ref{eq_mi_rp}) in the special case 
that the dimension of $\vec x$ is equal to one (i.\,e.~that 
it is a scalar time series) and for the maximum norm 
(Fig.~\ref{fig_mi_roessler}A). The advantage using 
the RP approach is that we can compute the mutual information 
for an entire phase space
vector and not only for a single component.

Moreover, the generalised cross mutual information of second order 
of two different systems $\vec x$ and $\vec y$ 
\begin{equation}\label{I2}
I_2^{\vec{x},\vec{y}} = H_2^{\vec{x}}+H_2^{\vec{y}}-H_2^{\vec{x},\vec{y}}.
\end{equation}  
can be estimated by means of RPs as well:
\begin{equation}
\hat I_2^{\vec{x},\vec{y}} = -\ln \frac{1}{N^2}\sum_{i,j=1}^N \mathbf R_{i,j}^{\vec x} -\ln \frac{1}{N^2}\sum_{i,j=1}^N \mathbf R_{i,j}^{\vec y} + \ln \frac{1}{N^2}\sum_{i,j=1}^N \mathbf{JR}^{\vec x,\vec y}_{i,j}.
\end{equation} 
However, $\hat I_2^{\vec x, \vec y}$ depends on the special choices 
of the thresholds $\varepsilon^{\vec x}$ and $\varepsilon^{\vec y}$. 
If,  e.\,g., the phase space diameter of $\vec x$ is very large in 
comparison with the one of $\vec y$, then the value for $\varepsilon^{\vec x}$ 
should be chosen larger than the value for $\varepsilon^{\vec y}$. 
One possibility to equilibrate the difference between the sizes of both 
phase spaces is to choose
the thresholds in such a way that the recurrence 
rate of $\vec x$ is equal to the recurrence rate of $\vec y$ 
(as described in Subsec.~\ref{sec:Selectionofthethreshold}), i.\,e.
\begin{equation}
\frac{1}{N^2}\sum_{i,j=1}^N \mathbf R_{i,j}^{\vec x}=\frac{1}{N^2}\sum_{i,j=1}^N \mathbf R_{i,j}^{\vec y} =RR.
\end{equation}
Then, the estimate of $I_2^{\vec x, \vec y}$ can be written as
\begin{equation}\label{eq_2mi_RP}
\tilde I_2^{\vec x, \vec y}=\ln \frac{1}{N^2}\sum_{i,j=1}^N \mathbf{JR}_{i,j} -
                             2 \ln RR.
\end{equation}
Using the delayed joint recurrence matrix, Eq.~(\ref{eq_delayed_JRP}),
the cross mutual information can be computed as a function of the time delay
(Fig.~\ref{fig_mi_roessler}B).
The cross mutual information is also called cross redundancy \cite{prichard1995}. 


\subsubsection{Influence of embedding on the invariants estimated by RPs}\label{embedding_inv}

We have seen in the last subsections, that it is possible to 
estimate basic dynamical invariants by means of RPs. As usually 
in nonlinear time series analysis, if the time 
evolution of only one component of the state vector is 
observed, the trajectory in the phase 
space has to be reconstructed by means of an appropriate embedding before computing 
the dynamical invariants (cf.~Subsec.~\ref{sec:PhaseSpaceTrajectories}).
However, we show in this section, that this additional step is not necessary 
for the estimation of  
$K_2$ and $D_2$ by means of RPs.

As we have shown in Subsec.~\ref{sec:k2_d2}, the estimation 
of $K_2$ and $D_2$ is based on the probability 
$p_c(\varepsilon,l)$ to find a diagonal of at 
least length $l$ in an RP.
Using embedded vectors, Eq.~(\ref{eq_embedding}), with embedding 
dimension $m$ and delay $\tau$, and considering the maximum norm 
for the estimation of $p_c(\varepsilon,l)$, we get
\begin{equation}\label{eq_p_of_l_embedding}
p_c^{(m,\tau)}(\varepsilon,l)=
\frac{1}{N^2}\sum\limits_{i,j=1}^{N}
\prod\limits_{k=0}^{l-1}
    \Theta\left(\varepsilon-\max\limits_{s=0,\ldots,m-1}
    \left|x_{i+k+s\tau}-x_{j+k+s\tau}\right|\right).
\end{equation}
The product in Eq.~(\ref{eq_p_of_l_embedding}) can be written as 
\begin{equation}
\begin{split}
\prod\limits_{k=0}^{l-1}
    \Theta\bigl(\varepsilon-\max\limits_{s=0,\ldots,m-1}
        \left|x_{i+k+s\tau}-x_{j+k+s\tau}\right|\bigr) 
= \\
\Theta\bigl(\varepsilon-\max\limits_{ m=0,\cdots,l-1 \atop s=0,\ldots,m-1}
        \left|x_{i+k+s\tau}-x_{j+k+s\tau}\right|\bigr).
\end{split}
\end{equation}
Hence, substituting this into Eq.~(\ref{eq_p_of_l_embedding}), we obtain
\begin{equation}\label{eq_p_of_l_embedding2}
p_c^{(m,\tau)}(\varepsilon,l)=
\frac{1}{N^2}\sum\limits_{i,j=1}^{N}
\Theta\left(\varepsilon-\max\limits_{k=0,\ldots, \atop l-1 + (m-1)\tau}\left|x_{i+k}-x_{j+k}\right|\right).
\end{equation}
From this equation it follows that for different embedding $(m,\tau)$ and 
$(m', \tau')$ the following relationship holds
\begin{equation}\label{eq_p_of_l_indep}
p_c^{(m,\tau)}(\varepsilon,l)=p_c^{(m',\tau')}(\varepsilon,l') \quad \text{if} \quad  
l+(m-1)\tau = l'+(m'-1)\tau',
\end{equation}
provided that $l,l'>\tau, \tau'$ and $m,m'\ge 1$ \cite{thiel2004a}.
Hence, the decay of $p_c^{(m,\tau)}(\varepsilon,l)$ is the same for different
embedding. The curve is only shifted to larger $l$'s if the dimension is decreased. 
Since the slope of $\ln p_c^{(m,\tau)}(\varepsilon,l)$ for large $l$'s is
used in order to the estimate $K_2$ (see Subsec.~\ref{sec:k2_d2}), the estimated value is
independent of the choice of the embedding parameters.

Analogously,  substituting
\begin{equation}
\ln\left(\frac{p_c^{(d,\tau)}(\varepsilon_1,l)}{p_c^{(d,\tau)}(\varepsilon_2,l)}\right)=
\ln\left(\frac{p_c^{(d',\tau)}(\varepsilon_1,l')}{p_c^{(d',\tau)}(\varepsilon_2,l')}\right)
\end{equation} 
in Eq.~(\ref{eq_corrdim_RP}), we see that the estimation of $D_2$ by means of 
RPs is also independent of the choice of the embedding parameters.
Hence, this is one important 
advantage of the RPs method compared to others.

%% file: meth_spatialdata.tex
%
%
%
%

The concept of recurrence is not only restricted to univariate time series.
It is clear that recurrence is also a basic phenomenon in spatio-temporal dynamical systems.
But even for snapshots of such high-dimensional dynamics, such as images,
we can expect recurrent structures. However, RPs cannot be directly applied to 
spatial data. One possibility to study the
recurrences of spatial data is to separate these
objects into many one-dimensional data series, and to apply 
the recurrence analysis separately to each of these series \cite{vasconcelos2006}. 
Another possibility, suggested in this section, is the extension of
the temporal approach of RPs to a spatial one \cite{marwan2006pla}. With this step we 
focus on the RPs' potential to determine similar (recurrent) 
epochs in data.

For a $d$-dimensional (Cartesian) space, we define
a spatial recurrence plot by
\begin{equation}
\mathbf{R}_{\vec \imath,\vec \jmath} = \Theta\left(\varepsilon-\left\|\vec x_{\vec \imath} - \vec x_{\vec \jmath}\right\|\right),
\quad \vec x_{\vec \imath} \in \mathds{R}^m,\ \vec \imath,\,\vec \jmath \in \mathds{N}^d,
\end{equation}
where $\vec \imath$ is the $d$-dimensional coordinate and $\vec x_{\vec \imath}$
is the phase-space vector at the location given by coordinate $\vec \imath$.
This means, we consider each direction in space as a single embedding vector 
of dimension $m$, but compare each of them with all others.
The resulting RP has now the dimension $2 \times d$ and cannot 
be visualised anymore. However, its quantification is still possible.

Analogously to the one-dimensional case, where the LOI is a one-dimensional 
line (Subsec.~\ref{sec:DefRecurrencePlots}), a similar 
diagonal-oriented, $d$-dimensional 
structure in the $n$-dimensional recurrence plot ($n=2\,d$), the
{\it hyper-surface of identity} (HSOI), can be defined:
\begin{equation}
\mathbf{R}_{\vec \imath,\vec \jmath} 
     \equiv  1 \quad \forall\  \vec \imath = \vec \jmath.
\end{equation}

In the special case of a two-dimensional image with scalar
values instead of phase-space vectors, i.\,e.~$m=1$, we have 
\begin{equation}
\mathbf{R}_{i_1,i_2,j_1,j_2} =
     \Theta\left(\varepsilon-\left|x_{i_1,i_2} - x_{j_1,j_2}\right|\right),
\end{equation}
which is in fact a four-dimensional recurrence plot, and its HSOI 
is a two-dimensional plane.

In two-dimensional RPs, the recurrence quantification is 
based on line structures. Thus, the definition of equivalent 
structures in higher-dimensional RPs is crucial for their 
quantification analysis. 

Analogously to the definition of diagonal lines, Eq.~(\ref{eq_diagonalline}), 
a diagonal hyper-surface of size $l$ ($\vec l = (l,\dots,l),
\vec l \in \mathds{N}^d$) is then defined by
\begin{equation}\label{eq_diagonalline_x}
 \left( 1 - \mathbf{R}_{\vec \imath - \vec 1,\vec \jmath - \vec 1} \right) 
 \left( 1 - \mathbf{R}_{\vec \imath + \vec l,\vec \jmath + \vec l} \right) 
     \prod_{\substack{k_1,\dots,k_d=0}}^{l-1} \mathbf{R}_{\vec \imath+\vec k,\vec \jmath+\vec k}
     \equiv 1,
\end{equation}
and a vertical hyper-surface of size $v$ ($\vec v = (v,\dots,v),
\vec v \in \mathds{N}^d$) is given by
\begin{equation}\label{eq_verticalline_x}
 \left( 1 - \mathbf{R}_{\vec \imath,\vec \jmath - \vec 1} \right) 
 \left( 1 - \mathbf{R}_{\vec \imath,\vec \jmath + \vec v} \right) 
     \prod_{\substack{k_1,\dots,k_d=0}}^{v-1} \mathbf{R}_{\vec \imath,\vec \jmath+\vec k}
     \equiv 1.
\end{equation}

Using these definitions, we can compute the frequency distributions 
of the sizes of diagonal and vertical hyper-surfaces in the spatial RP.
The recurrence quantification measures, as defined in Subsec.~\ref{sec:RQA},
can be applied to these distributions, and hence, these quantification measures
are now suitable for characterising spatial data as well.

\subsubsection*{Example: RQA of spatial data}

\begin{figure}[bp] 
\centering \includegraphics[width=\columnwidth]{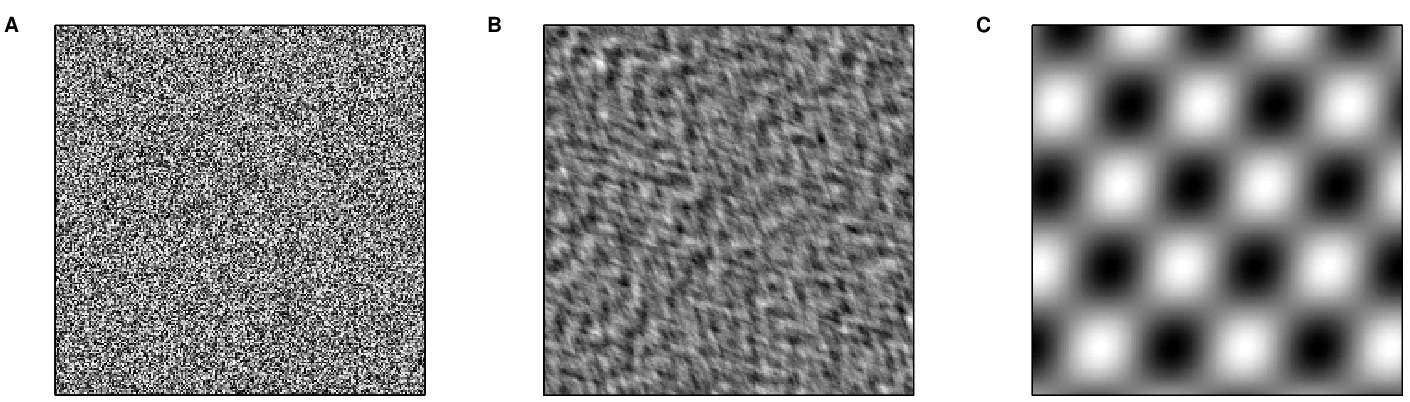} 
\caption{Two-dimensional examples representing (A) uniformly distributed 
white noise, (B) a two-dimensional auto-regressive process and (C) 
periodically recurring structures.}\label{fig_data_sample}
\end{figure}

To illustrate the extension of temporal RPs to higher-dimensional
spatial RPs, we consider
three prototypical examples in 2D (i.\,e.~images). The first image (A)
is uniformly distributed white noise, the second image (B) is 
the result of a static two-dimensional auto-regressive process (2$^{\text{nd}}$ order, AR2),
Eq.~(\ref{eq_2D-ar2}), and the
third image (C) represents periodically recurrent structures (Fig.~\ref{fig_data_sample}).
All these examples have a geometric size of $200 \times 200$ pixels, 
and the values of the spatial series
are normalised to a mean of zero and a standard
deviation of one.

The resulting RPs are four-dimensional matrices of size 
$200 \times 200 \times 200 \times 200$, and can hardly be visualised.
However, we can reduce their dimension by one in order to visualise these RPs:
we consider only that part of the RP, where 
$i_2 = j_2$ (the resulting $200 \times 200 \times 200$ cube is a 
hyper-surface of the four-dimensional RP along the LOI). 

\begin{figure}[tthp] 
\centering 
\includegraphics[width=\columnwidth]{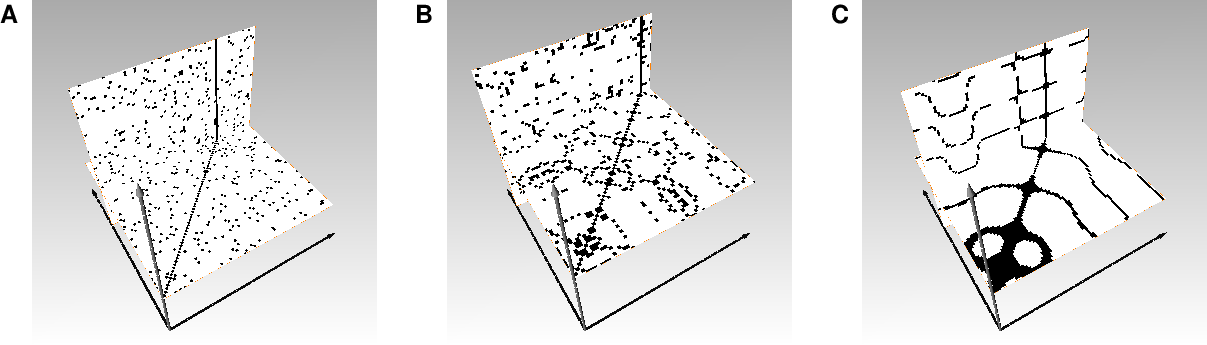} 
\caption{Slices through three-dimensional subsections of four-dimensional 
RPs of the two-dimensional
examples shown in Fig.~\ref{fig_data_sample}. As known from
one-dimensional data, (A) random data causes homogeneous RPs consisting
of single points, (B) correlations in data cause extended structures and
(C) periodic data causes periodically occurring structures in the 
RPs ($\varepsilon = 0.2$)}\label{fig_rp2_sample}
\end{figure}

The features obtained in higher-dimensional RPs can be interpreted analogously
to the case of two-dimensional RPs. Single points correspond
to strongly fluctuating, uncorrelated data, as it is typical for noise
(Fig.~\ref{fig_rp2_sample}A).
Correlations in the data cause extended structures, which can be
lines, planes or even cuboids (Fig.~\ref{fig_rp2_sample}B).
Periodically recurrent patterns in data induce periodic line and
plane structures in the spatial RP (Fig.~\ref{fig_rp2_sample}C).

The recurrence quantification of these 4D RPs confirms these observations.
Except from the recurrence rate $RR$, the other recurrence quantification 
measures discriminate clearly between the three features 
(Tab.~\ref{tab_res_examples}). Because all images were normalised 
to the same standard deviation, the recurrence rate $RR$ is roughly 
the same for all examples. 
For the random image (A) the determinism $DET_{HS}$ and
laminarity $LAM_{HS}$ tend to zero, what is expected, because the values
in the image heavily fluctuate even between adjacent pixels. 
For the 2D-AR2 image (B), $DET_{HS}$ and $LAM_{HS}$ are slightly above zero, revealing
the correlation between adjacent pixels. The last example (C)
has, as expected, the highest values in $DET_{HS}$ and $LAM_{HS}$, because
same structures occur several times in this image and the image is 
rather smooth. Although the variation
in $DET_{HS}$ and $LAM_{HS}$ seems to be similar, there is a significant difference
between both measures. Whereas $LAM_{HS}$ represents the probability that a
specific value will not change over spatial variation (what results
in extended same-coloured areas in the image), $DET_{HS}$ measures 
the probability that similar changes in the image recur. The value obtained for $LAM_{HS}$ is twice the one obtained
for $DET_{HS}$ in the 2D-AR2 image. Hence, there are more areas
without changes in the image than such with typical, recurrent changes.
In contrast, $DET_{HS}$ is higher than $LAM_{HS}$ for the periodic image, because
it contains characteristic changing structures which recur several
times but do not have a constant value.

\begin{table}
\caption{Recurrence quantification measures for prototypical examples in 2D.
For the minimal size of the diagonal and 
vertical planes $l_{min}=3$ and $v_{min}=4$
is used.}\label{tab_res_examples}
\centering \begin{tabular}{lrrr}
\hline
Example&        $RR$&   $DET$&  $LAM$\\
\hline
\hline
(A) noise       &0.218& 0.007&  0.006\\
(B)     2D-AR2  &0.221& 0.032&  0.065\\
(C)     periodic&0.219& 0.322&  0.312\\
\hline
\end{tabular}
\end{table}

%% file: meth_synchro.tex
%
%
%
%

In this section, the relationship between 
recurrences and synchronisation of complex systems will be discussed.
This relationship can be used in order to detect
different transitions to synchronisation and also to detect
synchronisation in cases where known methods fail.

There are three basic types of synchronisation in coupled complex
systems: 
if the trajectories of the systems evolve due to coupling on the same
trajectory, they are {\it completely synchronised}.
If there exists a functional relationship between the
systems, they are {\it generalised synchronised}, and if their
phases adapt to each other so that they evolve in the same
manner, the systems are {\it phase synchronised}. 

If two systems synchronise in some way, their recurrences in
phase space are not independent from each other. Hence,
comparing the recurrences of each single system with the
joint recurrences of the entire system, we can expect to get indications
about their synchronisation type and degree.

On the other hand, RPs reflect the different time scales
which play a crucial role in the dynamics of a system. For example,
the RP of a periodic trajectory (Fig.~\ref{fig_rps_intro}A)
consists of uninterrupted diagonal lines equally spaced. The
distance between the diagonal lines is the period of the
trajectory. In the case of white noise
(Fig.~\ref{fig_rps_intro}C), the RP consists of mostly
isolated points uniformly distributed, thus, the distances
between the recurrence points are uniformly distributed,
indicating that there is no predominant time scale in the
series. Hence, there is an equivalence
between the distances which separate the recurrent points in
the RP and the characteristic time scales of the considered
system. This correspondence is crucial in order to analyse phase
synchronisation, where the phases or time scales of the
interacting systems are locked.

In this section, we give first a brief overview about
synchronisation of chaotic systems. Then, the
detection of different kinds of transitions to synchronisation by
means of joint recurrences is presented. At last, some
indices for the detection of phase and generalised
synchronisation based on recurrences are introduced.

\subsubsection{Synchronisation of chaotic systems}\label{sec:sync_theory}

Chaotic systems defy the concept of synchronisation due to the
high sensitivity to slightly different initial conditions.
However, it has been demonstrated that this kind of systems
can synchronise. 

The first studies about synchronisation of chaotic systems consider {\it
complete synchronisation (CS)}. In this case, coupled identical  chaotic
systems which start at different initial conditions evolve onto the same
trajectory \cite{fujisaka1983,afraimovich1986,pecora1990}.  However, under
experimental conditions it is difficult and mostly impossible to have two
fully identical systems. Usually, there is some mismatch between the
parameters of the systems under consideration. Hence, it is important to
study synchronisation between non-identical systems. Starting with two
uncoupled non-identical oscillators and increasing the coupling strength, a
rather weak degree of synchronisation may occur, where the phases and
frequencies of the chaotic oscillators become locked, whereas their
amplitudes often remain almost uncorrelated. This behaviour is denoted by {\it phase
synchronisation (PS)}. 

The phase of a chaotic autonomous oscillator
is closely related to its zero Lyapunov exponent because it
corresponds to the translation along the chaotic trajectory.
Hence, a perturbation in this direction neither decays nor
grows. This property makes the adjustment of the phases of two
chaotic oscillators (or of one oscillator and a force)
possible. If two chaotic oscillators are not coupled, the two
zero Lyapunov exponents will be linked to the individual phases.
Increasing the coupling strength, PS can be obtained by the
transition of one of the zero Lyapunov exponents to negative
values, indicating the establishment of a relationship between
the phases \cite{pikovskyBook2001}.

If the coupling strength
between non-identical chaotic oscillators is further increased,
a strong dependence between the amplitudes will then be
established, so that the states of both oscillators become almost
identical but shifted in time, i.\,e.~$\vec x(t) \approx \vec
y(t+\tau)$ \cite{rosenblum1997}. This regime is called {\it
lag synchronisation (LS)}. The transition to LS has also been
related to the transition of a positive Lyapunov exponent to
negative values. Actually, LS sets in after the zero crossing
of the Lyapunov exponent. After the onset of LS, a further
increase of the coupling strength leads to a decrease of the
time lag $\tau$ between the trajectories of the oscillators.
Hence, the oscillators tend to be {\it almost synchronised},
i.\,e.~$\vec x(t) \approx \vec y(t)$. 

Note that the above
descriptions of synchronisation transitions and their
connection with the changes in the Lyapunov spectrum is valid
for phase coherent oscillators, for which a phase can be
defined as a monotonously increasing function of time.
However, for non-phase coherent chaotic oscillators, this
definition may not be possible and the crossing of the zero
Lyapunov exponent to negative values may not be an indicator
for the onset of PS \cite{boccaletti2002}.

The question about synchronisation of coupled systems which are
essentially different has been addressed first in
\cite{afraimovich1986,rulkov1995}. In this case, there is in general no
trivial manifold in  phase space which attracts the systems' trajectories.
It has been shown, that these systems can synchronise in a more general
way, namely $\vec y=\vec f(\vec x)$, where $\vec f$ is a transformation
which maps asymptotically the trajectories of $\vec x$ into the ones of
the attractor $\vec y$, leading to {\it generalised synchronisation (GS)}.
The properties of the function $\vec f$ depend on the features of the
systems $\vec x$ and $\vec y$, as well as on the attraction properties of
the synchronisation manifold $\vec y=\vec f(\vec x)$. In most cases,
evidence of GS has been provided for unidirectional coupling schemes.
However, examples of bidirectionally coupled systems that undergo GS also
exist, as e.\,g.~shown in \cite{kocarev_nolta95,boccaletti2000}.

All these different types of synchronisation between complex systems have
been demonstrated in numerous laboratory experiments
\cite{pecora1990,kittel1998,wiggeren1998,kocarev1992,ticos2000,allaria2001,kiss2001}
and in natural systems \cite{schaefer1998,blasius1999,tass1998}.

\subsubsection{Detection of synchronisation transitions} 

\begin{figure}[htbp]
\centering\includegraphics[width=0.5\textwidth]{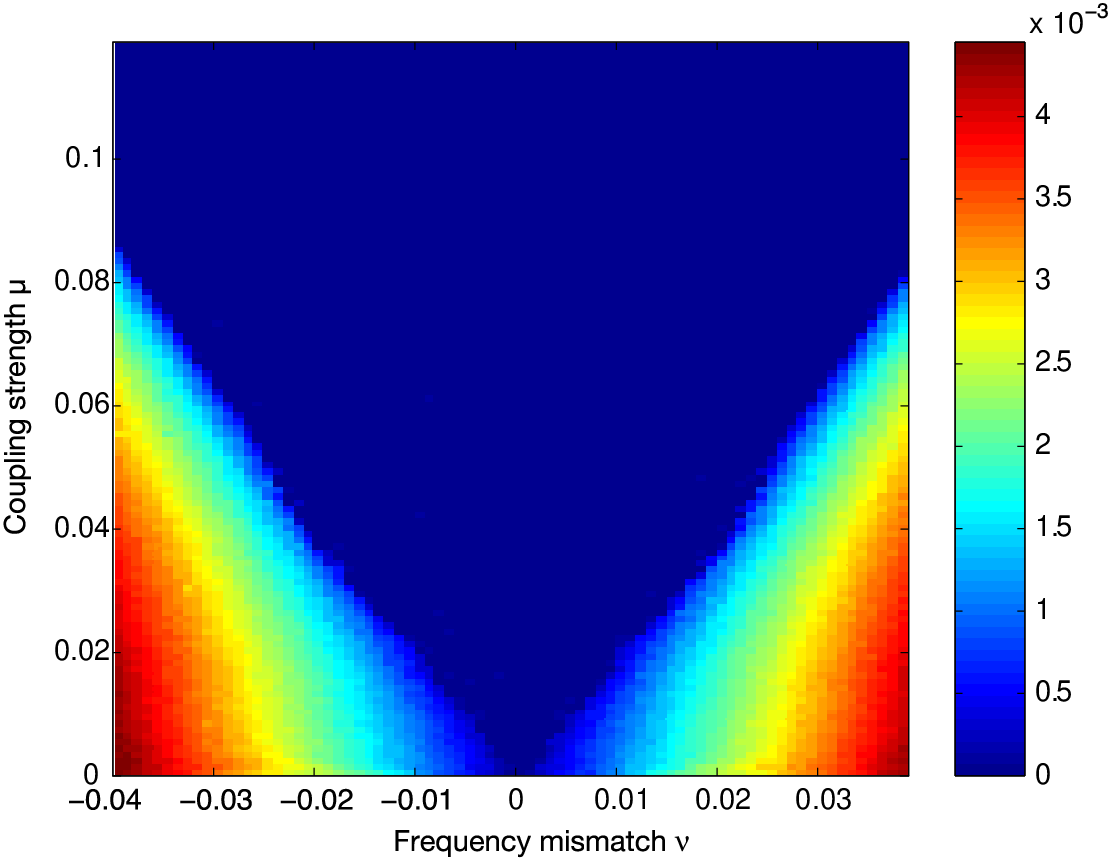}
\caption{The difference $|\Delta \Omega|$ between the mean 
frequencies of two R\"ossler oscillators in dependence on 
the frequency mismatch $\nu$ and coupling strength $\mu$ shows the
well-known Arnold tongue \cite{romano2004}.
}\label{fig_freq_diff}
\end{figure}

As we have seen in Subsec.~\ref{sec:Invariants}, it is
possible to estimate the joint R\'enyi entropy by means of JRPs 
(cf.~Eq.~(\ref{JK2_est})). In this section, the transitions to PS
and LS will be characterised by means of the joint R\'enyi entropy.  To
exemplify this, we consider the prototypical chaotic system of  two
mutually coupled R\"ossler oscillators, Eqs.~(\ref{eq_2roessler1}) and
(\ref{eq_2roessler2}) with  $a=0.15$, $b=0.20$, $c=10$, for the coupling
parameters $\nu \in [-0.04,0.04]$  and $\mu \in [0.0,0.12]$, for which the
two oscillators undergo transitions to PS. The difference of the mean
frequencies $\Delta \Omega=\Omega_1-\Omega_2$ of both oscillators reveals
the well-known Arnold tongue (Fig.~\ref{fig_freq_diff};  the mean
frequencies $\Omega_1$ and $\Omega_2$ are calculated as proposed in
\cite{pikovskyBook2001}). 

%
%
%

\begin{figure}[htbp]
\includegraphics[width=\textwidth]{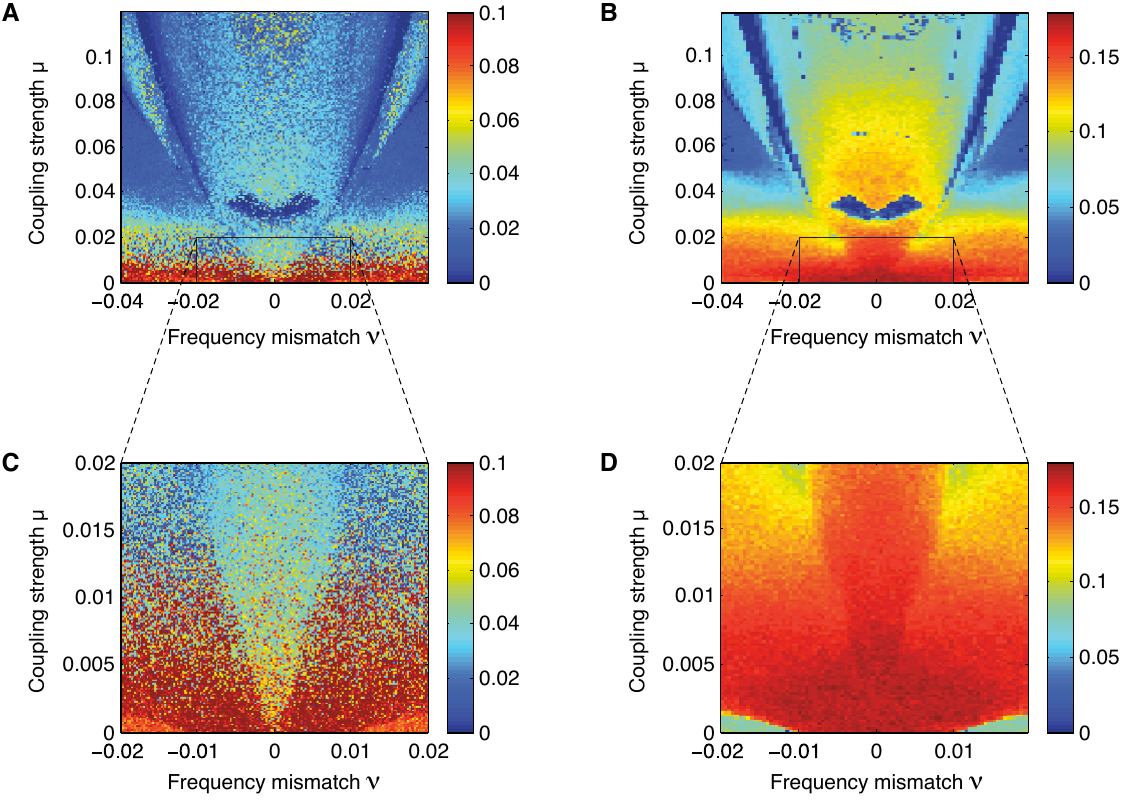}
\caption{(A) Estimate of the joint R\'enyi entropy $\widehat {JK}_2$ of 
two R\"ossler oscillators in dependence on the frequency mismatch $\nu$
and coupling strength $\mu$.
(B) Sum of the positive Lyapunov exponents of the two 
R\"ossler oscillators in dependence on the frequency mismatch $\nu$ and 
coupling strength $\mu$. 
(C) and (D) magnifications of the joint R\'enyi entropy (C) and
the sum of the positive Lyapunov exponents (B) 
for low values of the coupling strength: The tip of the Arnold tongue 
cannot be clearly distinguished by the sum of the positive Lyapunov 
exponents, but by the joint R\'enyi entropy
\cite{romano2004}.
}\label{fig_k2_lyap_2roess}
\end{figure}

Next, we estimate $JK_2$ based on JRPs, Eq.~(\ref{JK2_est}), in the same
parameter range for both oscillators. The results also reflect the Arnold 
tongue (Fig.~\ref{fig_k2_lyap_2roess}A), but exhibit more details  than
$\Delta \Omega$ (Fig.~\ref{fig_freq_diff}).

\begin{itemize} 
\item First, two ``borders'' in the upper part of the 
plot can be observed ($\mu>0.04$): the outer one
corresponds to the border of the Arnold tongue, i.\,e.~inside
this border the oscillators are in PS, whereas outside they
are not. Both borders have very low values of $\widehat
{JK_2}$, i.\,e.~the behaviour of the system is rather regular
there, even periodic in small regions on both borders. This is
a remarkable fact, because it means that for relatively high
coupling strengths the transition to PS is 
chaos-period-chaos, since inside the tongue is $\widehat
{JK}_2 > 0$, indicating a chaotic regime. 

\item Inside the
Arnold tongue, for coupling strengths $\mu$ between
approximately $0.025$ and $0.04$, a region can be found (which
looks like two eyes), where the value of $\widehat {JK}_2$ is
(almost) zero, i.\,e.~the region is periodic or quasi-periodic.

\item For $\mu \ge 0.03$ the region inside the Arnold tongue
is ``more chaotic'' (larger $\widehat {JK}_2$) than outside
the tongue. This is surprising, as we would expect that if
both oscillators are synchronised, the behaviour of the whole
system becomes more and more regular for increasing coupling.
\end{itemize} 

Hence, by means of $JK_2$ estimated by recurrences, new 
characteristics of the transition to PS can be found.

In order to validate these results, the
formal relationship between $K_2$ and the Lyapunov
exponents, Eq.~(\ref{eq_ruelle}), can be used.
As two coupled systems can be regarded as a single 
system, the joint R\'enyi entropy corresponds to the R\'enyi
entropy of the whole system. Therefore, Eq.~(\ref{eq_ruelle})
is also valid for $JK_2$, considering the sum over the
positive Lyapunov exponents of both sub-systems. 

We calculate the Lyapunov spectrum of the whole system defined by
Eqs.~(\ref{eq_2roessler1}) and (\ref{eq_2roessler2}), by using the
equations (i.\,e.~not estimated from the time series). As $JK_2$ is
bounded from above by the sum of the positive Lyapunov exponents,
Eq.~(\ref{eq_ruelle}), the plot of $\sum_{\lambda_i>0}\lambda_i$ shows
qualitatively the same structures as the plot of $\widehat {JK}_2$
(Fig.~\ref{fig_k2_lyap_2roess}A, B). It is noteworthy, that $JK_2$ was
estimated from the time series of both oscillators, consisting of $10,000$
data points and with a sampling rate corresponding to $30$ data points per
oscillation, whereas for the computation of the sum of the positive
Lyapunov exponents, Eqs.~(\ref{eq_2roessler1}) and (\ref{eq_2roessler2}) 
were used \cite{romano2004}.  Hence, the technique to estimate the
predictability of the system in parameter space based on JRPs is quite
appropriate and yields robust and reliable results even in cases, where
the equations governing the system are not known.

However, there is one qualitative difference between both approaches: for
$\mu \in [0,0.006]$ the tip of the Arnold  tongue can be clearly
identified by $\widehat {JK}_2$, but it cannot be distinguished by
considering the sum of the positive Lyapunov exponents
(Fig.~\ref{fig_k2_lyap_2roess}C, D). This is due to the fact, that the
equality $K_2 = \sum_{\lambda_i>0}\lambda_i$ holds only for hyperbolic systems, but
the 6-dimensional system, Eqs.~(\ref{eq_2roessler1}) and
(\ref{eq_2roessler2}), is not hyperbolic, and hence only the relation
(\ref{eq_ruelle}) holds. Hence, $JK_2$ provides important
complementary information  to the sum of the positive Lyapunov exponents.

\subsubsection{Detection of PS by means of recurrences}\label{sec:DetectionOfPS} 

In order to detect phase synchronisation (PS), now we use the relationship between the (vertical)
distances which separate diagonal lines in an RP, and the time scales
which characterise the dynamical system.

\begin{figure}[htbp]
\centering \includegraphics[width=0.7\textwidth]{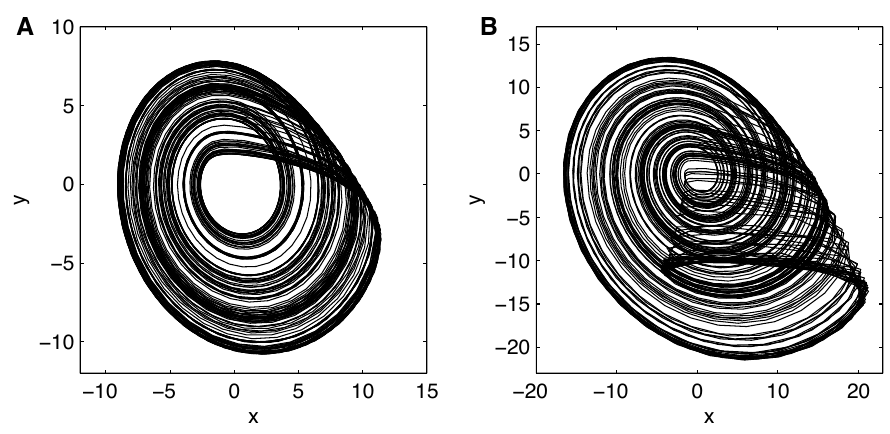}
\caption{(A) Projection of the R\"ossler attractor, Eqs.~(\ref{eq_roessler}),
with $a=0.15$, $b=0.2$ and $c=10$, on the $(x,y)$-plane. It looks like 
a smeared limit cycle. Hence, a phase can be assigned 
to the chaotic oscillator just by means of the angle 
of rotation around the fixed point, Eq.~(\ref{phase}). 
(B) Projection of the R\"ossler attractor in the funnel 
regime, Eqs.~(\ref{eq_roessler}), 
with $a=0.3$, $b=0.4$ and $c=7.5$, on the $(x,y)$-plane. There is no point around 
which the trajectory oscillates. Hence, the definition 
of the phase by Eq.~(\ref{phase}) is not appropriate in 
this case.}\label{fig_roessler_projections}
\end{figure}

The straightforward procedure to detect PS in complex oscillators is to
estimate explicitly their phases. If the system has a dominant peak in
the power spectrum, there are several methods to define the phase. One
possibility is to project the trajectory on an appropriate plane, so that
the projection looks like a smeared limit cycle with well-defined
rotations around a centre (Fig.~\ref{fig_roessler_projections}A).  Then
the phase can be identified with the angle of rotation
\cite{rosenblum1996,pikovskyBook2001}  \begin{equation}\label{phase}
\Phi(t)=\arctan \frac{y(t)}{x(t)}. \end{equation}

%
%

But a pre-requisite for the appropriate application of these
techniques is that the trajectory in a certain projection plane moves
around an origin. This condition may not be met  if the signal has a
broad-band spectrum, which is typical for non-coherent signals (Fig.~\ref{fig_roessler_projections}B)
\cite{rosenblum1996,pikovskyBook2001,rosenblum2002}. 

One approach to overcome this problem is to use an ensemble of
well-defined oscillators which act as a filter to find out the frequency
of the complex signal \cite{rosenblum2002}. However, depending on which
coordinate used to couple the complex system to the ensemble of
oscillators, the value obtained for the frequency can vary. Another
definition of the phase, based on the general idea of the curvature, has
recently been proposed to treat such systems, e.\,g., the R\"ossler system
in the non-coherent funnel regime \cite{osipov2003}. However, this
approach is in general limited to lower dimensional systems and is rather
sensitive to high levels of noise. 

\begin{figure}
\centering \includegraphics[width=0.55\textwidth]{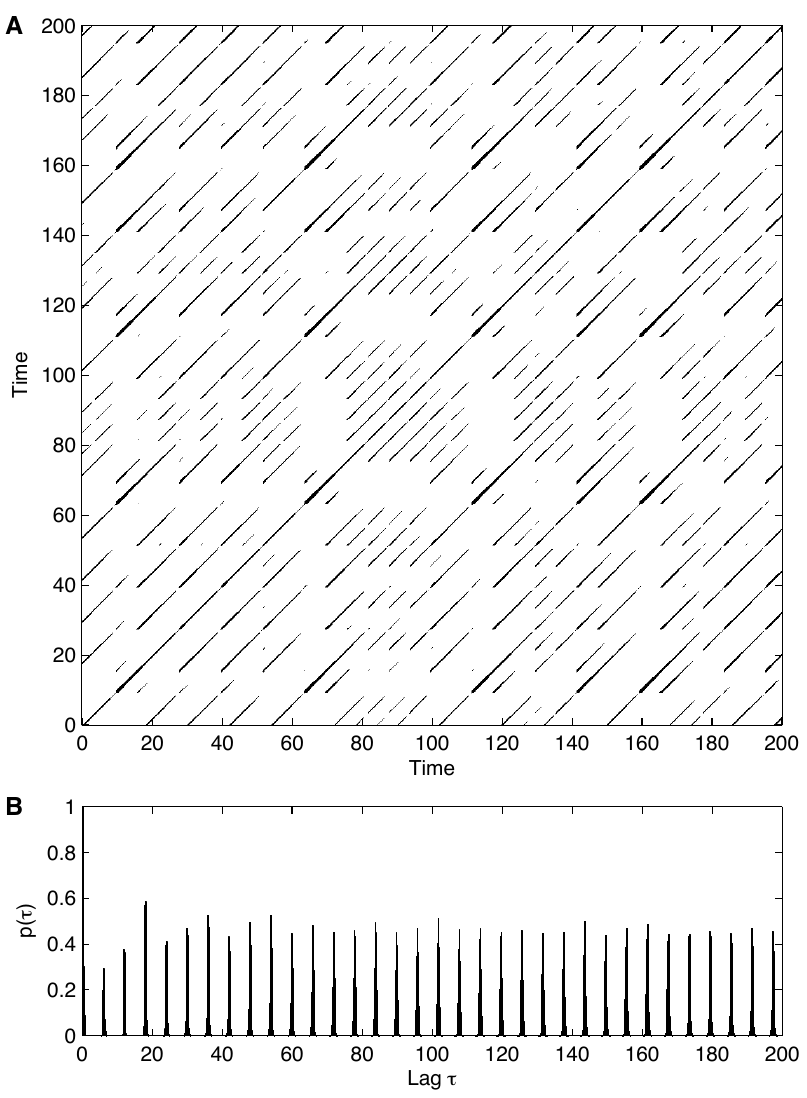}
\caption{(A) RP of the R\"ossler system, Eqs.~(\ref{eq_roessler}), 
for $a=0.15$, $b=0.2$ and $c=10$. 
(B) Corresponding probability of recurrence $p(\tau)$, i.\,e.~$RR_{\tau}$. 
}\label{fig_rp_rr_roessler}
\end{figure}

Alternatively, the concept of recurrence can be used to detect indirectly
PS  in a wide class of chaotic systems and even for time series  corrupted
by noise, where other methods fail \cite{romano2005}. As demonstrated in
Subsec.~\ref{sec:RQA_diagonal_lines}, diagonal lines in the RP indicate the existence of
some determinism in the system under consideration. The vertical distances
between these diagonal lines reflect the characteristic time scales of the
system. In contrast to periodic dynamics, for a chaotic oscillator the
diagonal lines are interrupted due to the divergence of nearby
trajectories. Furthermore, the distances between the diagonal lines are
not constant, i.\,e.~we find a distribution of distances,  reflecting the
different time scales present in the chaotic system
(Fig.~\ref{fig_rp_rr_roessler}). 

If two oscillators are in PS, the distances between diagonal lines in
their respective RPs coincide, because their phases, and hence their time
scales adapt to each other. However, the amplitudes of oscillators, which
are only PS but  not in GS or CS, are in general uncorrelated. Therefore,
their RPs are not identical. However, if the probability that the first
oscillator recurs after $\tau$ time steps is high, then the probability
that the second oscillator recurs after the same time interval will be
also high, and vice versa. Therefore, looking at the probability
$p(\varepsilon, \tau)$ that the system recurs to the
$\varepsilon$-neighbourhood of a former point $\vec x_i$ of the trajectory
after $\tau$ time steps and comparing $p(\varepsilon, \tau)$ for both
systems allows detecting and quantifying PS properly. $p(\varepsilon, \tau)$
can be estimated directly from the RP by using the  diagonal-wise
calculated $\tau$-recurrence rate, Eq.~(\ref{eq_rr_star}),
\begin{equation}\label{eq_p_of_tau} 
\hat{p}(\varepsilon, \tau) =
RR_{\tau}(\varepsilon) = \frac{1}{N-\tau}\sum_{i=1}^{N-\tau}\Theta(\varepsilon-\|\vec
x_{i}-\vec x_{i+\tau}\|). 
\end{equation} 
As already stated in Subsec.~\ref{sec:RQA_diagonal_lines}, $\hat{p}(\varepsilon, \tau)$ can
be considered as a {\it generalised auto-correlation function}, revealing
also higher order correlations between the points of the trajectory
in dependence on $\tau$. In order to simplify the notation,
henceforth we will use $p(\tau)$ instead of $\hat{p}(\varepsilon, \tau)$.

Studying the coincidence of the positions of the maxima of $p_{\tau}$ for
two coupled systems $\vec x$ and $\vec y$, PS can be identified. More
precisely, the correlation coefficient between  $p_{\tau}^{\vec x}$
and $p^{\vec y}(\tau)$ 
\begin{equation}\label{eq_CPR} 
CPR = \langle \bar p^{\vec x}(\tau) \bar p^{\vec y}(\tau) \rangle, 
\end{equation} 
can be used in order to quantify
PS, where $\bar p^{\vec x}(\tau)$ and $\bar p^{\vec y}(\tau)$ are the probabilities
normalised to zero mean and standard deviation of one. If both systems are
in PS, the probability of recurrence will be maximal at the same time and
$CPR \approx 1$. On the other hand, if the systems are not in PS, the
maxima of the probability of recurrence will not occur simultaneously.
Then we observe a drift and hence expect low values of $CPR$.

\subsubsection*{Example: Detection of PS in mutually coupled R\"ossler oscillators}

\begin{figure}[htbp]
\centering \includegraphics[width=.7\textwidth]{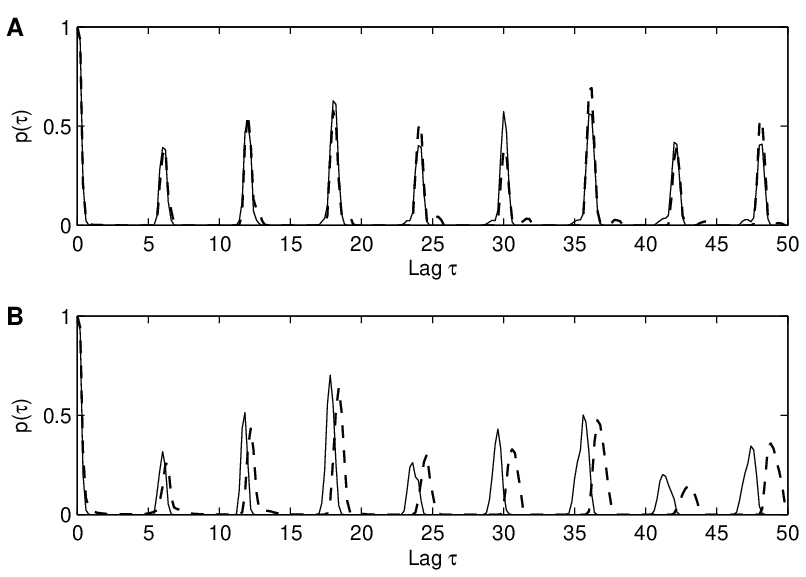}
\caption{Recurrence probability $p(\tau)$ for two mutually coupled 
R\"ossler systems, Eqs.~(\ref{eq_2roessler2_1}) and (\ref{eq_2roessler2_2}),
for $a=0.16$, $b=0.1$, $c=8.5$, in (A) phase synchronised and 
(B) non-phase synchronised regime. Solid line: oscillator 
$\vec x$, dashed line: oscillator $\vec y$.
}\label{fig_p_2roess}
\end{figure}

In order to exemplify how this method works, we consider two mutually
coupled R\"ossler systems, Eqs.~(\ref{eq_2roessler2_1}) and
(\ref{eq_2roessler2_2}), in the phase coherent regime ($a=0.16$, $b=0.1$,
$c=8.5$). According to \cite{osipov2003}, for $\nu=0.02$ and $\mu=0.05$
both systems are in PS. We observe that the local maxima of $p_{\vec
x}(\tau)$ and $p_{\vec y}(\tau)$ occur at $\tau=n \, T$, where $T$ is
the mean period of both R\"ossler systems and $n$ is an integer
(Fig.~\ref{fig_p_2roess}A).  

It is important to emphasise that the heights of  the local maxima are in
general different for both systems if they are only in PS and not in GS,
as we will see later. But the positions of the local maxima of $p(\tau)$
coincide, and the correlation coefficient is $CPR=0.998$. For $\mu=0.02$
the systems are not in PS and the positions of the maxima of $p(\tau)$ do
not coincide anymore (Fig.~\ref{fig_p_2roess}B), clearly indicating that
the frequencies are not locked. In this case, the correlation coefficient
is $CPR=0.115$.

\begin{figure}[htpb] \centering
\includegraphics[width=.7\textwidth]{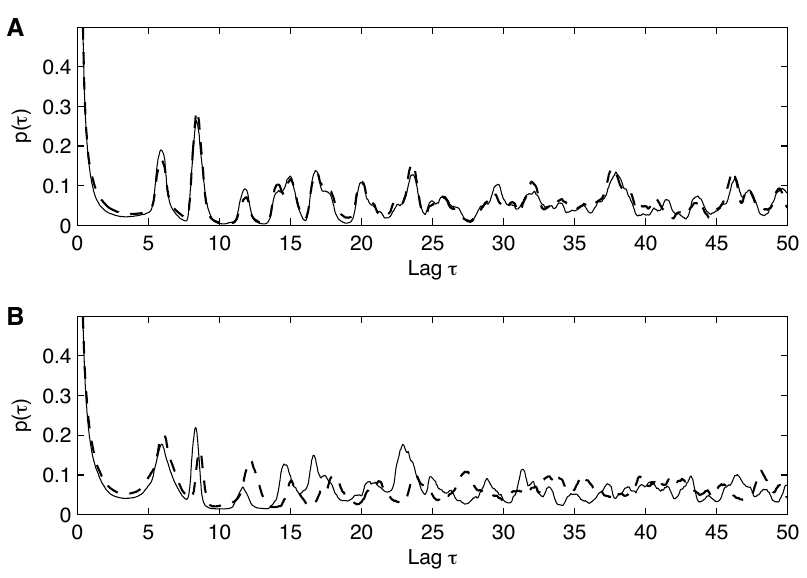} 
\caption{Recurrence
probability $t(\tau)$ for two mutually  coupled R\"ossler systems in
funnel regime, Eqs.~(\ref{eq_2roessler2_1}) and (\ref{eq_2roessler2_2}),
for $a=0.2925$, $b=0.1$, $c=8.5$. (A) $\mu=0.2$ (PS) and (B) $\mu=0.05$
(non-PS).  Solid line: oscillator $\vec x$, dashed line: oscillator  $\vec
y$.}\label{fig_p_2funnel} 
\end{figure}

It is important to emphasise that this method is highly efficient even for
non-phase coherent oscillators, such as two mutually coupled R\"ossler
systems in the funnel regime, Eqs.~(\ref{eq_2roessler2_1}) and 
(\ref{eq_2roessler2_2}), for $a=0.2925$, $b=0.1$, $c=8.5$, $\nu=0.02$.  We
analyse again two different coupling strengths: $\mu=0.2$ and $\mu=0.05$. 
The peaks in $p(\tau)$ (Fig.~\ref{fig_p_2funnel}) are not as
well-pronounced and regular as in the coherent regime, reflecting the different time
scales that play a relevant role and the broad band power spectrum of
these systems. However, for $\mu=0.2$ the positions of the local maxima
coincide for both oscillators (Fig.~\ref{fig_p_2funnel}A), indicating PS,
whereas for $\mu=0.05$ the positions of the local maxima do not coincide
anymore (Fig.~\ref{fig_p_2funnel}B), indicating non-PS. These results are
in accordance with \cite{osipov2003}. 

In the PS case of this latter
example, the correlation coefficient is  $CPR=0.988$,  and in the non-PS
case, $CPR=0.145$. Note that the positions of the  first peaks in
$p(\tau)$ coincide (Fig.~\ref{fig_p_2funnel}B),  although the oscillators
are not in PS. This is due to the small frequency mismatch ($2\nu=0.04$).
However, by means of the index $CPR$ we can distinguish rather well
between both regimes.

Furthermore, the index $CPR$ is able to detect PS even in time
series which are strongly corrupted by noise \cite{romano2005}. 
Additionally, $CPR$ indicates
clearly the onset of PS. In \cite{romano2005}, the results
obtained for $CPR$ in dependence on the coupling strength were
compared with the Lyapunov exponents, as they theoretically
indicate the onset of PS (in the phase-coherent
case). The results obtained with $CPR$ coincide with the ones obtained
by means of the Lyapunov exponents. 

The results obtained with $CPR$ are very
robust with respect to the choice of the threshold $\varepsilon$.
Simulations show that the outcomes are almost independent of the choice of
$\varepsilon$ corresponding to a percentage of black points in the RP
between $1\%$ and $90\%$, even for non-coherent oscillators. The patterns
obtained in the RP of course depend on the choice of $\varepsilon$. But
choosing $\varepsilon$ for both interacting oscillators in such a way that
the percentage of black points in both RPs is the same, the relationship
between their respective recurrence structures does not change for a broad
range of values of $\varepsilon$.

\subsubsection{Detection of GS by means of recurrences}\label{sec:DetectionOfGS} 

Now we demonstrate that it is also possible to detect generalised
synchronisation (GS)  by means of RPs \cite{romano2005}. When the
equations of the system are known, GS can be characterised by the
conditional stability of the driven chaotic oscillator \cite{kocarev1996}.
However, when analysing measured time series, the model equations are
usually not known. Hence, several methods based on the technique of delay
embedding and on  conditional neighbours have been developed to
characterise GS, such as the method of {\it mutual false nearest neighbours} (MFNN)
\cite{rulkov1995,boccaletti2000,arnhold1999}. But systems
exhibiting non-invertibility or wrinkling hamper the detection of GS by
such techniques \cite{so2002}. Further techniques, such as the $\delta^p$
and $\delta^{p,q}$ method, have been developed to overcome these problems
\cite{he2003}. 

We develop an index for GS based on RPs which can also deal with
such problems. It is based on the results presented in
Subsec.~\ref{sec:Reconstruction}: the recurrence matrix of a system
contains the whole information necessary to reconstruct it topologically,
i.\,e.~the recurrence matrix contains all ``essential'' dynamical
information about the system. On the other hand, two systems connected by
a homeomorphism are said to be topologically equivalent. Hence, two such
systems should have a very similar recurrence matrix. 

Let us consider the average probability of recurrence over
time for systems $\vec x$ and $\vec y$, i.\,e.~the recurrence 
rate, Eq.~(\ref{eq_rr}), $RR^{\vec x}$ and $RR^{\vec y}$.
The average probability of joint recurrence over time 
is then given by $RR^{\vec x, \vec y}$, which is the recurrence 
rate of the JRP of systems $\vec x$ and $\vec y$ \cite{romano2004}.
If both systems are independent, the average
probability of the joint recurrence will be
$RR^{\vec x, \vec y} = RR^{\vec x}RR^{\vec y}$. 
On the other hand, for both systems in GS, we expect
approximately the same recurrences, and hence 
$RR^{\vec x, \vec y} \approx RR^{\vec x} = RR^{\vec y}.$
For the computation of the recurrence matrices in the case of
essentially different systems that undergo GS, it is more
appropriate to use a fixed amount of nearest neighbours $N_{\text{n}}$
for each column in the matrix than using a fixed threshold 
(FAN, cf.~Subsec.~\ref{sec:VariationsRecurrencePlots}) 
\cite{arnhold1999}, which corresponds, as already stated, to the original definition
of RPs by Eckmann et al.~\cite{eckmann87}. 
$RR^{\vec x}$ and $RR^{\vec y}$ are then equal and fixed by 
$N_{\text{n}}$, because of Eq.~(\ref{eq_number_of_neighbours}) $RR^{\vec x}=RR^{\vec y}=N_{\text{n}}/N$. Now we call $RR=N_{\text{n}}/N$
and define the coefficient 
\begin{equation*}
S=\frac{RR^{\vec x, \vec y}}{RR}
\end{equation*} 
as an index for GS that varies from $RR$ (independent) to $1$ (GS). 
Furthermore, in order to be able to detect also lag synchronisation (LS) 
\cite{rosenblum1997} with this index, a time lag is included
by using the time delayed JRP, Eq.~(\ref{eq_delayed_JRP}),
\begin{equation}
S(\tau)=\frac{\frac{1}{N^2} \sum_{i,j}^N \mathbf{JR}^{\vec{x},\vec{y}}_{i,j}(\tau)}{RR},
\end{equation}
where $JRP$ is computed by using FAN. 
Then, we introduce an index for GS based
on the average {\it joint probability of recurrence} $JPR$ by choosing
the maximum value of $S(\tau)$ and normalising it,
\begin{equation}\label{sigma} 
JPR=\max\limits_{\tau}\frac{S(\tau)-RR}{1-RR}.
\end{equation}
The index $JPR$ ranges from $0$ to $1$. $RR$
is a free parameter. However, simulations show that the $JPR$
index is rather independent of the choice of $RR$.

\subsubsection*{Example: Detection of GS in a driven Lorenz system}

\begin{figure} 
\centering \includegraphics[width=1.0\textwidth]{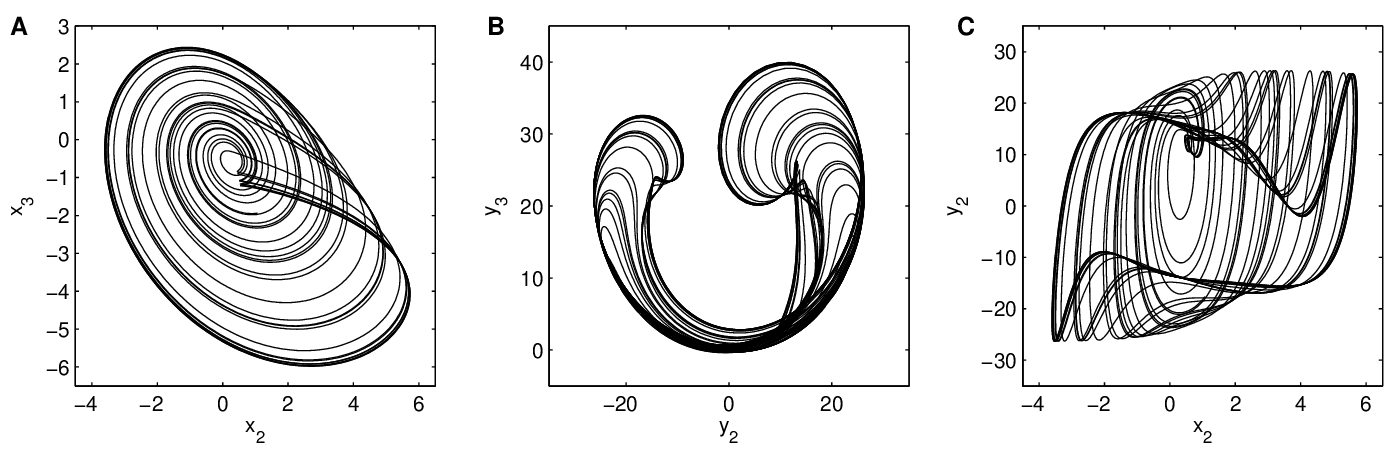} 
\caption{Projection of the
(A) R\"ossler driving system, (B) the driven Lorenz system and 
(C) the plot of the second components of these systems $x_2$ vs.~$y_2$.
} \label{fig_roess_lor}
\end{figure}

To illustrate the application of $JPR$, we analyse an interaction of two
rather different chaotic oscillators, the Lorenz system
(Eqs.~(\ref{lorenz_driven}) with $\sigma=10$,  $r=28$ and $b=\frac{8}{3}$)
driven by a R\"ossler system  (Eqs.~(\ref{roess_driver}) with
$a=0.45$, $b=2$ and $c=4$). Since the driven Lorenz system
is asymptotically stable \cite{kocarev1996},  the systems
(\ref{roess_driver}) and (\ref{lorenz_driven}) are in GS. However, the
shapes of both attractors in phase space are very different
(Fig.~\ref{fig_roess_lor}A, B), and both are neither in LS nor in CS
(Fig.~\ref{fig_roess_lor}C).

To mimic typical experimental situations, we perform
the analysis with just one component of each system. $10,000$ data 
points are used with a sampling time interval
of $0.02$. Using $x_3$ and $y_3$ as observables, we reconstruct
the phase space vectors using delay coordinates, Eq.~(\ref{eq_embedding}),
for $\vec x$: $m=3$, $\tau=85$ and for $\vec y$: $m=3$, $\tau=10$.

\begin{figure}
\centering \includegraphics[width=1.0\textwidth]{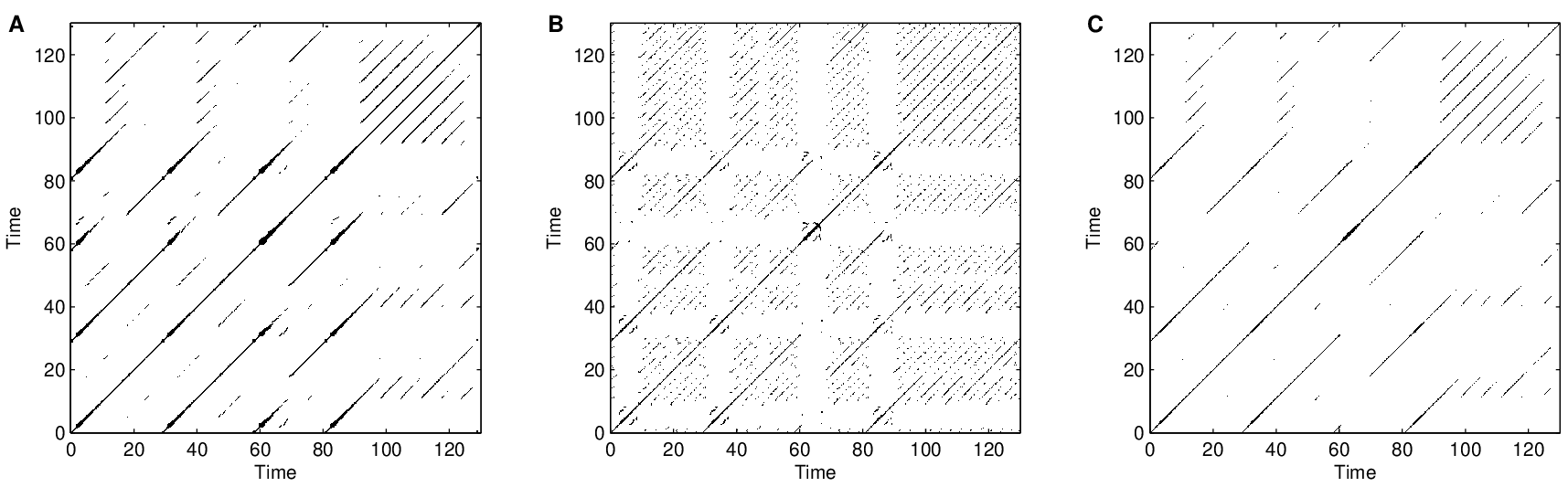}
\caption{RPs of (A) the R\"ossler sub-system, (B) the driven Lorenz sub-system, 
and (C) the JRP of the whole system.}\label{fig_roess_lor_rps_gs}
\end{figure}

The RPs of both sub-systems $\vec x$ and $\vec y$ are rather similar
despite of the  essential difference between the shapes of the attractors
(Fig.~\ref{fig_roess_lor_rps_gs}A, B). Therefore, the structures of
each single RP are also reflected in the JRP
(Fig.~\ref{fig_roess_lor_rps_gs}C) and consequently, its recurrence rate
is rather high.  With the choice $RR=0.02$ the JPR index is $JPR=0.64$
(the value of $JPR$ is similar for other choices of $RR$).

\begin{figure}
\centering \includegraphics[width=1.0\textwidth]{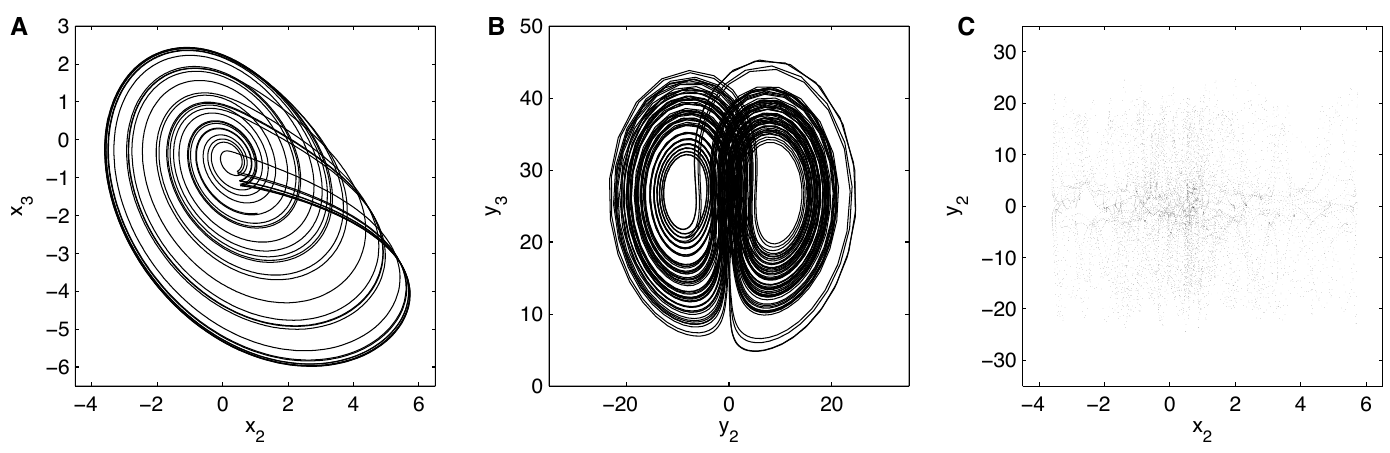}
\caption{Projection of the (A) R\"ossler system, (B) the independent Lorenz system and 
(C) the plot of the second components of these systems $x_2$ vs.~$y_2$.} \label{fig_roess_lor_indep}
\end{figure}

\begin{figure}  
\centering \includegraphics[width=1.0\textwidth]{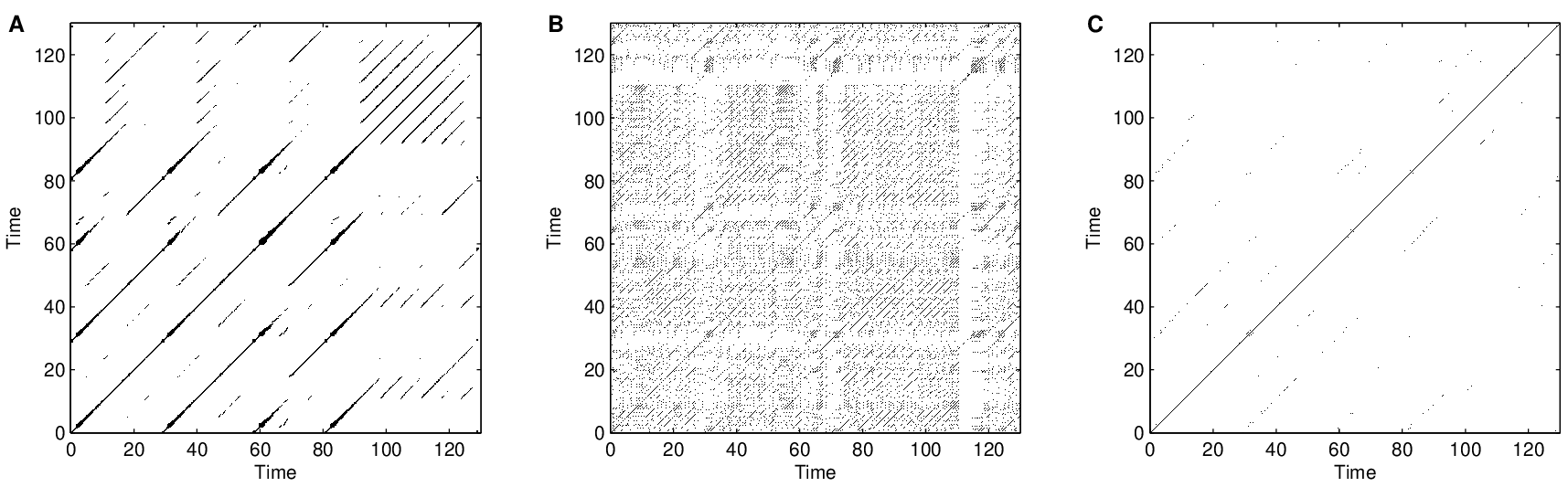}
\caption{RPs of (A) the R\"ossler sub-system, (B) the independent Lorenz sub-system, 
and (C) the JRP of the whole system.}\label{fig_roess_lor_rps_indep}
\end{figure}

In contrast, the RPs of the independent systems ($u=y_1$ in 
Eqs.~(\ref{lorenz_driven}), Fig.~\ref{fig_roess_lor_indep}) 
look rather different (Fig.~\ref{fig_roess_lor_rps_indep}A and B;
using embedding parameters $m=3$ and $\tau=5$ for both systems, and
$RR=0.02$). Therefore, the JRP is almost empty 
(Fig.~\ref{fig_roess_lor_rps_indep}C),
i.\,e.~the mean probability over time for a joint 
recurrence is very small. In this case, the JRP index 
is $JPR=0.03$.

\subsubsection{Comparison with other methods}

As mentioned at the beginning of this section, there are several methods to
detect GS. They might be divided in three main kinds: the {\it conditional stability
approach} \cite{kocarev1996}, the {\it auxiliary system approach} \cite{abarbanel1996}
and the {\it mutual false nearest neighbours (MFNN) based approach} \cite{rulkov1995,boccaletti2000,arnhold1999}.

In the case of the analysis of
experimental time series, the conditional stability approach has the
disadvantage that it requires the estimation of the Lyapunov exponents, which
might be rather difficult by dealing with observed data.
The auxiliary system approach requires an identical copy of the response system,
which might be problematic to obtain in the case of measured data due to noise
influences and inaccuracy in the parameters of the system. In contrast, the
false nearest neighbours approach is more appropriate for the analysis of
experimental data. However, if the systems to be analysed have more than one
predominant time scale, some errors might occur. For example, if the systems
have a slow and a fast time scale, even though the driver and the response
systems are not in GS, the parameter MFNN might be of the order of one. This is
because the distance between the nearest neighbours of points of the trajectory
belonging to the fast time scale is of the order of the size of the attractor.
For example, in the case of experimental data from electrochemical oscillators
which are characterised by a non-phase coherent dynamics, the method of the
false nearest neighbours does not work to detect GS. In contrast, the recurrence
based method detects the onset of GS correctly (cf.~Sec.~\ref{sec:appl_jrp_synchro}).  
Note that the recurrence based method does
not consider the distances between nearest neighbours to detect GS, but marginal
and joint probabilities of recurrence for the driver and response systems. The
recurrence based approach is close to the concept of transfer entropy
\cite{schreiber2000} used for the detection of the direction of the coupling
between systems.

Furthermore, there might be complications in the dynamics of
the systems, such as non-invertibility and wrinkling, which hamper the detection
of GS \cite{so2002}. Some approaches, such as the time series based
implementation of the auxiliary systems approach \cite{parlitz2005} and the
$\delta^{p,q}$ method \cite{he2003} have been introduced to overcome such
problems. The $\delta^{p,q}$ method considers two points of the trajectory as
neighbours, if they are neighbouring not only at one time instant, but during
some longer interval of time. By means of this extension, it is possible to
detect GS also in the complicated cases of non-invertibility and wrinkling.
Introducing an analogous extension in the recurrence based approach, which can
be accomplished increasing the embedding dimension, it is also possible to
detect GS in such more intricated cases. 

With respect to the computational
demand, the recurrence based approach is comparable to the false nearest
neighbour or $\delta^{p,q}$ approaches, as in all cases the
distances between all pairs of points of the trajectories have to be computed, 
which is a $N^2$ problem.

\subsubsection{Onset of different kinds of synchronisation}\label{sec:onset} 

We have seen that the indices $CPR$ and $JPR$ clearly distinguish between
oscillators in PS and oscillators which are not in PS, respectively GS. On the
other hand, the synchronisation indices should not only distinguish between
synchronized and not synchronised regimes, but also indicate clearly the onset
of PS, respectively of GS. 

In order to demonstrate that the recurrence based
indices fulfill this condition, we exemplify their application with two mutually
coupled R\"ossler systems in a phase coherent regime, Eqs.~(\ref{eq_2roessler2_1}),
with $a=0.16$, $c=8.5$ and $\nu=0.02$. We increase the coupling strength $\mu$
continuously and compute for each value of $\mu$ the indices $CPR$, and $JPR$.

\begin{figure} 
\centering
\includegraphics[width=.6\textwidth]{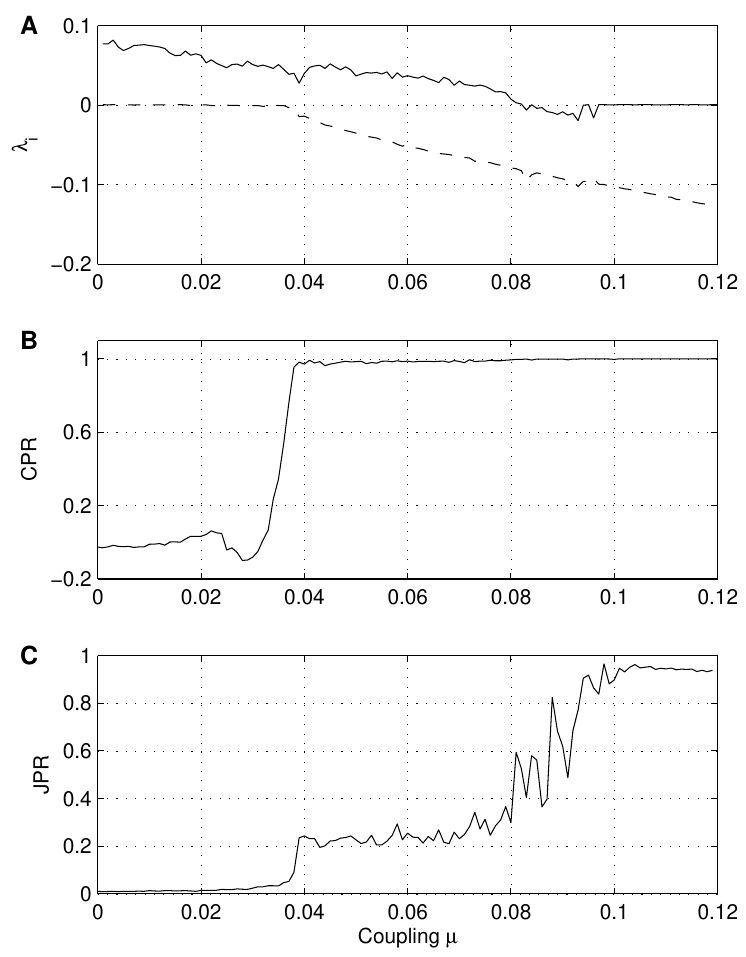}
\caption{
(A) $\lambda_2$ (solid line) and $\lambda_4$ (dashed line) as well as
(B) $CPR$ and
(C) $JPR$ indices
in dependence of the
coupling strength for two mutually coupled R\"ossler systems in the phase
coherent regime.} \label{fig_onset_syn}
\end{figure}

For a not too large but fixed frequency mismatch between both oscillators and
increasing coupling strength, the transitions to PS and LS are reflected in the
Lyapunov spectrum \cite{pikovskyBook2001,boccaletti2002}. If both oscillators
are not in PS, there are two zero Lyapunov exponents, that correspond to the
(almost) independent phases. Increasing the coupling strength, the fourth
Lyapunov exponent $\lambda_4$ becomes negative (Fig.~\ref{fig_onset_syn}A), 
indicating the onset of PS. For higher coupling
strengths, the second Lyapunov exponent $\lambda_2$ crosses zero, which
indicates the establishment of a strong correlation between the amplitudes 
(Fig.~\ref{fig_onset_syn}A). This last transition occurs almost
simultaneously with the onset of LS \cite{rosenblum1997}. Therefore, 
$\lambda_2$ and $\lambda_4$ are considered in order to validate 
the results obtained with $CPR$ and $JPR$.

%

Using the $CPR$ index, the transition to PS is detected when this index becomes of the
order of one (Fig.~\ref{fig_onset_syn}B). The transition to PS occurs at 
approximately $\mu=0.037$, in accordance with the transition of the fourth 
Lyapunov exponent $\lambda_4$ to negative values. The $JPR$ index also exhibits the
transition to PS, although it is an index for GS and LS. This index shows
three plateaus in dependence on the coupling strength
(Fig.~\ref{fig_onset_syn}C), indicating the onset of PS at the beginning of
the second one. On the other hand, $JPR$ clearly indicates the onset of LS,
because it becomes nearly one (third plateau) at approximately $\mu=0.1$
(Fig.~\ref{fig_onset_syn}C), after the transition from hyper-chaos to
chaos, which takes place at approximately $\mu=0.08$
(Fig.~\ref{fig_onset_syn}A). Between $\mu=0.08$ and $\mu=0.1$, the values
of $JPR$ have large fluctuations. This reflects the intermittent lag
synchronisation \cite{rosenblum1997,boccaletti2000b}, where LS is interrupted by
intermittent bursts of no synchronisation.

%% file: meth_recons.tex
%
%
%
%

In the preceding sections it has been shown that the recurrence matrix
contains relevant information about the dynamics of the system under
consideration. Quantifying the structures in the RP,
chaos-period and even chaos-chaos transitions can be detected.
Furthermore, it has been demonstrated that dynamical invariants, such as
the correlation entropy, correlation dimension and mutual
information can be estimated by means of RPs.  On the one hand, all 
relevant dynamical information is
fully preserved in the distance matrix $\mathbf{D}_{i,j}=\|\vec x_i-\vec x_j\|$ 
\cite{mcguire97}. But on the other hand, the possibility to estimate dynamical
invariants suggests that the
RP contains much information about the
underlying system, even though the trajectory is reduced to a matrix
of zeros and ones. We next show that under some general conditions, it 
is possible to reconstruct the underlying attractor from 
the binary RP, at least topologically \cite{thiel2004b}.
This is based on the fundamental property that from the RP of a scalar
time series $\{x_i\}_{i=1}^N$, a new time
series $\{x'_i\}_{i=1}^N$ can be obtained, which has the same rank order as the
original one. Then, from this $\{x'_i\}_{i=1}^N$ we can
reconstruct the attractor by, e.\,g., time delay embedding, Eq.~(\ref{eq_embedding}).
The so reconstructed manifold is topologically equivalent to the original
one, i.\,e.~there exists a homeomorphism between both attractors. However,
$\varepsilon$ must be chosen appropriately to make this reconstruction 
possible.

\begin{figure}[hbtp]
\centering \includegraphics[width=.7\columnwidth]{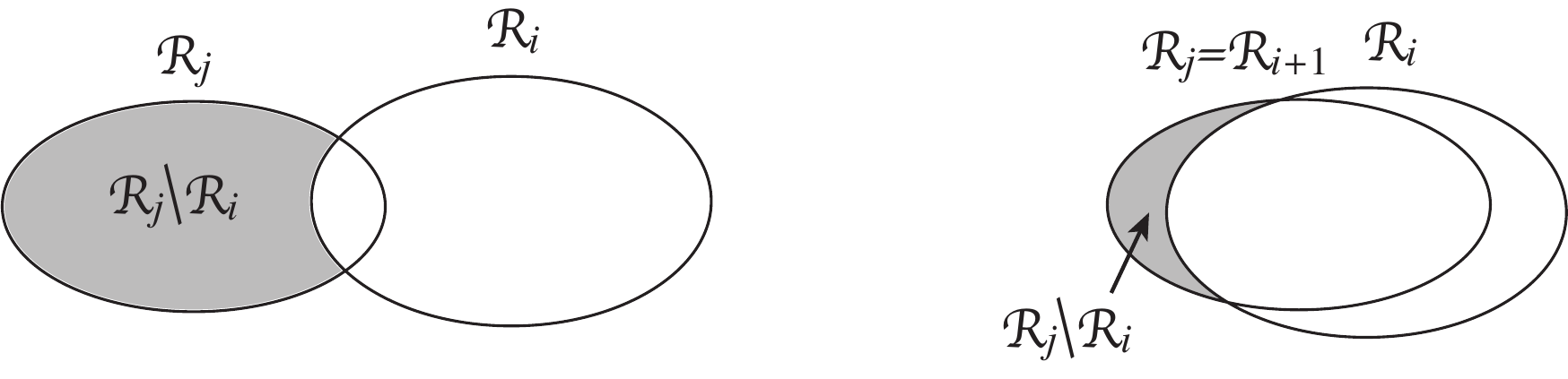} 
\caption{The rank order of a time series can be reconstructed from an RP by an 
algorithm which searches for the minimum $\mathcal{R}_j \setminus \mathcal{R}_i$, 
i.\,e.~the set of neighbours of $x_j$ which are not neighbours of $x_i$. 
If $j=\min_{k}\{\mathcal{R}_k \setminus \mathcal{R}_i\}$, then $i=r_k$ and 
$j=r_{k+1}$ or vice versa, i.\,e.~$i$ and $j$ are consecutive points in the rank ordered time series.
}\label{fig_set_diff}
\end{figure}

The reconstruction algorithm uses the RP as input and yields 
the rank order $\{r_i\}_{i=1}^N$ of the underlying time series 
$\{x_i\}_{i=1}^N$, i.\,e.~$x_{r_1} \le x_{r_2} \le \ldots \le x_{r_N}$ as its output. 
The algorithm is based on the idea that two consecutive 
points $x_{r_i}$ and $x_{r_{i+1}}$ of this rank ordered time series can be supposed to have rather similar 
neighbourhoods $\mathcal{R}_{r_i}$ and $\mathcal{R}_{r_{i+1}}$
(cf.~Eq.~(\ref{eq_set_recpoints}) and Fig.~\ref{fig_set_diff}; $\mathcal{R}_i$ corresponds 
to the recurrence points in the column $i$ of the RP). Hence, the 
algorithm searches for columns in the RP which are ``as similar as possible'', 
reconstructing iteratively the rank order of the time series.

It is important to note  that the reconstruction algorithm
assumes that the maximum distance between two consecutive points
in the rank ordered time series is smaller than 
$\varepsilon$. The reconstruction
algorithm consists of three main phases and a total of nine
steps which are described in detail in Appendix~\ref{apdx:alg_recons}.

\begin{figure}[htbp]
\centering \includegraphics[width=1.0\textwidth]{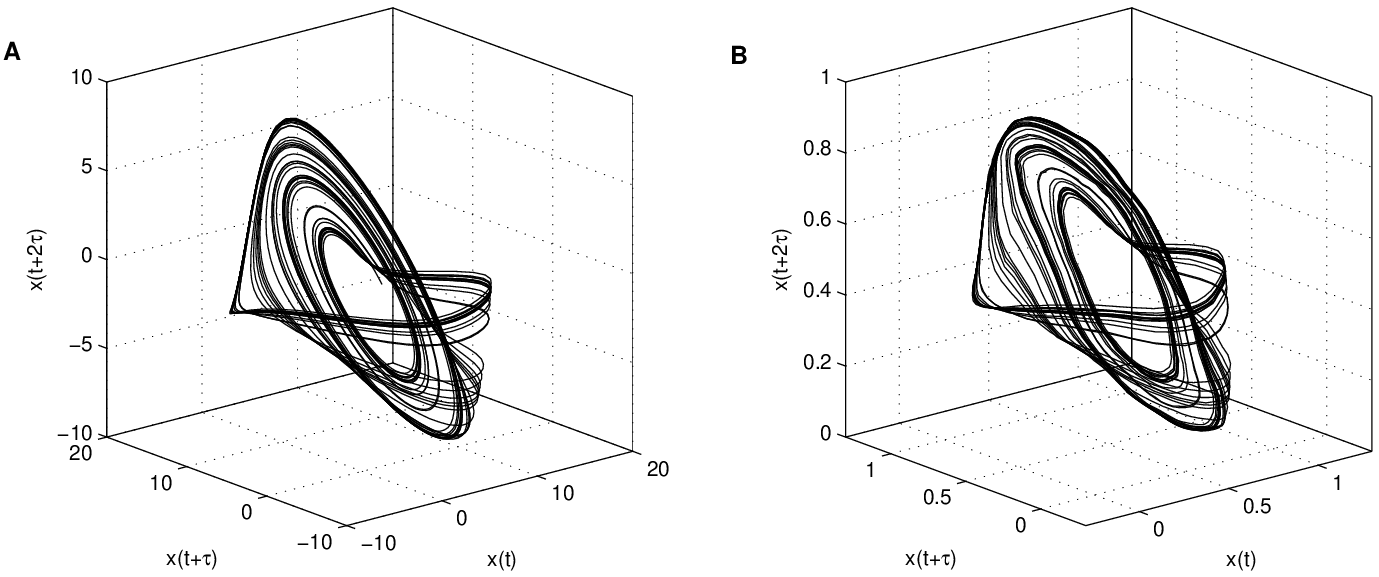}
\caption{Embedded R\"ossler attractor based on the
$x$-component of the original data (A) and 
from the reconstructed $x$-component (B) 
\cite{thiel2004b}.}\label{fig_reconstructed_roessler}
\end{figure}

To illustrate this reconstruction, we consider the 
R\"ossler attractor, Eqs.~(\ref{eq_roessler}), for $a=0.15$,
$b=0.2$ and $c=10$. We compute the RP of the $x$-component and use
it as input for the reconstruction algorithm. Then, we obtain the time series $x'$, which has the
same rank order as $x$. The both attractors derived from
the delay embedded $x$ and $x'$ are rather similar (Fig.~\ref{fig_reconstructed_roessler}).
It is obvious that both attractors are homeomorphic.

The needed number of points for the reconstruction for a given threshold 
$\varepsilon$ can be estimated, assuming that the values are uniformly 
distributed. The distance $D$ between two neighbouring points
in the interval $[x_{\min}, x_{\max}]$ is then exponentially
distributed
\begin{equation}
P(D) = N \, e^{-N\,D},
\end{equation}
where $N$ is the length of the time series. Assuming, without
loss of generality, that $[x_{\min}, x_{\max}]=[0,1]$,
there are $N+1$ intervals, which have to be all smaller
than $\varepsilon$. The probability that the maximal distance between two 
consecutive points in the interval is smaller than $\varepsilon$ is given by 
\begin{equation}\label{cons_points}
p(\varepsilon) = \left(1 - e^{-N\,\varepsilon}\right)^{N+1}.
\end{equation}
Using relation~(\ref{cons_points}), the minimal length of the time series
necessary to reconstruct
(with a probability of about 0.999) the time series 
(and the attractor) from an RP can be estimated.
Using $\varepsilon = 0.1$, only about 90 points are needed
to reconstruct the time series.
In general, the larger $\varepsilon$, the less points are needed to reconstruct 
the time series. However, if $\varepsilon$ is too large, the reconstruction 
algorithm works but cannot distinguish properly different points and possibly 
important fine structures of the system. This problem can be exemplified by 
choosing $\varepsilon=1/2+\delta$ in the case of the time series, which is 
uniformly distributed in the unit interval. Then, all points in the the 
band $[1/2-\delta,1/2+\delta]$ have the same neighbourhoods and hence 
cannot be distinguished.

Note that the reconstruction algorithm is only valid if we have
the RP of a scalar time series, i.\,e.~the reconstruction algorithm
does not work for RPs of an $m$-dimensional
trajectory with $m>1$. It is still an open question, whether it is
possible to reconstruct the attractor from the RP of a higher
dimensional trajectory.  First results indicate that (under some conditions)
it is possible.

%% file: meth_surrogates.tex
%
%
%
%

In Subsec.~\ref{sec:Synchro} we have considered the problems concerning the
phase synchronisation analysis of complex systems when there is more than one 
narrowband time scale, i.\,e.~when their power spectra are rather broad. As
it was shown, RPs help to overcome this problem, allowing to extend the
concept of phase synchronisation to rather complex systems.

Another important problem in the framework of synchronisation analysis is that even
though the synchronisation measures may be normalised, experimental time
series often yield values of these measures which are neither close to 0 nor to 1 but intermediate,
e.\,g., 0.4 or 0.6 and hence are difficult to interpret. This problem can
be overcome if the coupling strength between the interacting systems can be
varied systematically and a rather large change in the measure can be
observed, i.\,e.~if it is an {\it active experiment}
\cite{pikovskyBook2001}. However, much more typical are {\it passive
experiments}, in which it is not possible to change the coupling strength
systematically, e.\,g., the possible synchronisation of the heart beats of a mother
with those of her foetus \cite{vanleeuwen2003}. In some cases, this problem has
been tackled by interchanging the pairs of oscillators, i.\,e.~testing
the same foetus with a different (surrogate) mother, and using them as {\it natural
surrogates}. These surrogates are independent and hence cannot be in
phase synchronisation (PS) with the original system. Thus, if the
synchronisation index, as Eq.~(\ref{eq_CPR}), obtained from the original
data is not significantly higher than the index obtained from the natural
surrogates, there is not sufficient evidence to claim synchronisation. But
even this natural surrogates approach via time series has some drawbacks.
The natural variability and also the frequency of the natural surrogates
are in general different from the original ones. Furthermore, the data
acquisition can be expensive and at least in many cases problematic or
even impossible (e.\,g.~in the climatological interaction between the
El Ni\~no/Southern Oscillation and the North Atlantic Oscillation). In such cases it would
be convenient to perform a hypothesis test based on surrogates generated
by a mathematical algorithm. The null hypothesis, which the surrogates
must be consistent with, is that they are independent realisations of the
same underlying system, i.\,e.~trajectories of the same underlying
system starting at different initial conditions. In this section, we
present a technique for the generation of surrogates, which is based on
RPs. These surrogates mimic the dynamical behaviour of the system and are
consistent with the null hypothesis mentioned above. Then, computing the
synchronisation index between one sub-system of the original system and
another sub-system of the surrogate, and comparing it with the
synchronisation index obtained for the original system, we can test 
if an independent process can give the same index of PS.


As suggested in Subsec.~\ref{sec:Reconstruction}, it is possible to reconstruct topologically the
attractor of the system from its RP. Therefore, the RP contains all topological information about the
underlying attractor. 
Hence, a first idea for the generation of  surrogates is to change the structures in an RP 
consistently with the ones produced by the underlying dynamical system
 (the structures in the RP are linked to dynamical invariants of the underlying system, such as 
the correlation entropy and the correlation dimension, Subsec.~\ref{sec:Invariants}) and then 
reconstruct the trajectory from the modified RP. Furthermore, we use the fact that 
there are identical columns in an RP, i.\,e.~$R_{i,k}=R_{j,k}\ \forall k$. Thus, there are points which are 
not only neighbours (i.\,e.~$\|\vec{x}_i-\vec{x}_j\|<\varepsilon$), but which also share the same 
neighbourhood. These points are called {\it twins}. Twins are special points of the time series as 
they are indistinguishable considering their neighbourhoods, but in general different and, hence, 
have different pasts and -- more important -- different futures. The key idea of how to introduce the 
randomness needed for the generation of surrogates of a deterministic system is that we can 
jump randomly to one of the possible futures of the existing twins. 
The generation of a twin surrogate trajectory $\{\vec {x}^s_i\}_{i=1}^N$ of $\{\vec {x}_i\}_{i=1}^N$
is described in detail in the Appendix~\ref{apdx:alg_twins}.

\begin{figure}[htbp]
\centering
\includegraphics[width=1.0\textwidth]{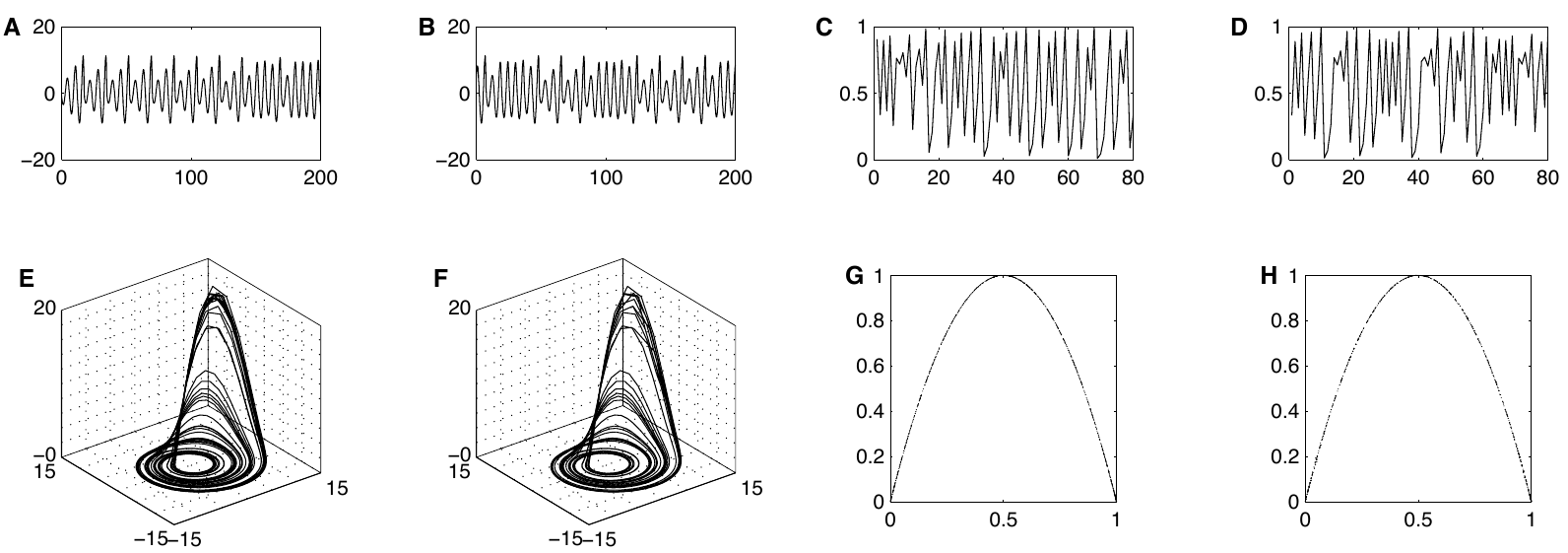}
\caption{Segment of the $x$-component of the 
(A) original R\"ossler system, Eqs.~(\ref{eq_roessler}), 
for $a=0.15$, $b=0.2$ and $c=10$ and 
(B) of one twin surrogate (TS) of the R\"ossler system. 
Trajectory of the (C) R\"ossler system in phase space and
(D) of one TS of the R\"ossler system. 
Segment of the trajectory of the 
(E) logistic map, Eq.~(\ref{eq_logistic_map}), for $a=4.0$ and
(F) of one TS of the logistic map. Phase portrait 
(G) of the logistic map and
(H) of one TS of the logistic map 
(modified after \cite{thiel2005}).}\label{fig_examp_surr}
\end{figure}

This RP based algorithm creates {\it twin surrogates (TS)} which are shadows of a (typical)
trajectory of the system \cite{thiel2005}. In the limit of
an infinitely long trajectory, its TS are
characterised by the same dynamical invariants and the same
attractor. However, if the measure of the attractor can be
estimated from the observed finite trajectory reasonably well, its
surrogates share the same statistics. 
The trajectories of the TS are visually
indistinguishable from the original ones (Fig.~\ref{fig_examp_surr}).
Also their power spectra and
correlation functions are consistent with the ones of the original
system. 

The idea behind TS to test for synchronisation
consists in exchanging one original sub-system with one twin
surrogate. Then, if the synchronisation index obtained for the
original system is not significantly different from the one
computed for the exchanged sub-systems, we have not sufficient
evidence to claim synchronisation. 

Suppose that we have two coupled self-sustained
oscillators $\vec x_i$ and $\vec y_i$.  Then, we generate
$M$ pairs of TS of the joint system, i.\,e.~$\{\vec{x}_i^{s_j}\}_{j=1}^M$ and
$\{\vec{y}_i^{s_j}\}_{j=1}^M$ (for $i=1, \ldots, N$). These surrogates
correspond to independent copies of the joint system, 
i.\,e.~trajectories of the whole system beginning at different initial
conditions. Note, that the coupling between $\vec{x}_i$ and
$\vec{y}_i$ is also mimicked by the surrogates. Next, we
compute the differences between the phases of the original system
$\Delta \Phi_i=|\Phi_{x_i}-\Phi_{y_i}|$ applying, e.\,g., the analytical
signal approach \cite{pikovskyBook2001} and compare them with
$\Delta\Phi^{s_j}_i =|\Phi_{x_i}-\Phi_{y_i}^{s_j}|$.
Then, if $\Delta \Phi_i$
does not differ significantly from $\Delta \Phi^{s_j}_i$ with
respect to some index for PS, the null hypothesis cannot be
rejected and, thus, there is not enough evidence for PS.

\subsection*{Example: Statistical test of the synchronisation analysis of R\"ossler oscillators}

We illustrate the TS technique to test for PS by
considering two non-identical, mutually coupled R\"ossler
oscillators, Eqs.~(\ref{eq_2roessler1}) and (\ref{eq_2roessler2}),
with $a=0.15$, $b=0.20$, $c=10$ and $\nu=0.015$. In this active experiment, the
coupling strength $\mu$ is varied from 0 to 0.08, and a PS index for
the original trajectory is calculated for each value of $\mu$. Next, we
generate 200 TS and compute the PS index between the measured
first oscillator and the surrogates of the second one. As PS index
we use the mean resultant length $R$ of complex phase vectors
\cite{rodriguez99}, which is motivated by Kuramoto's order
parameter \cite{kuramoto84}
\begin{equation}\label{R}
R=\left|\frac{1}{N}\sum\limits_{i=1}^{N}e^{j\Delta\Phi_i}\right|.
\end{equation}
$R$ takes values in the interval from 0 (non PS) to 1 (perfect
PS)~\cite{rodriguez99}. Let $R^{s_i}$ denote the PS index between
the first oscillator and the surrogate $i$ of the second one. To
reject the null hypothesis at a significance value $\alpha$, $R$
must be larger than $(1-\alpha)\cdot 100$\% of all
$R^{s_j}$. This corresponds to computing the
significance level from the cumultative histogram at the 
$(1-\alpha)$ level.

\begin{figure}[htbp]
\begin{center}
\includegraphics[width=\textwidth]{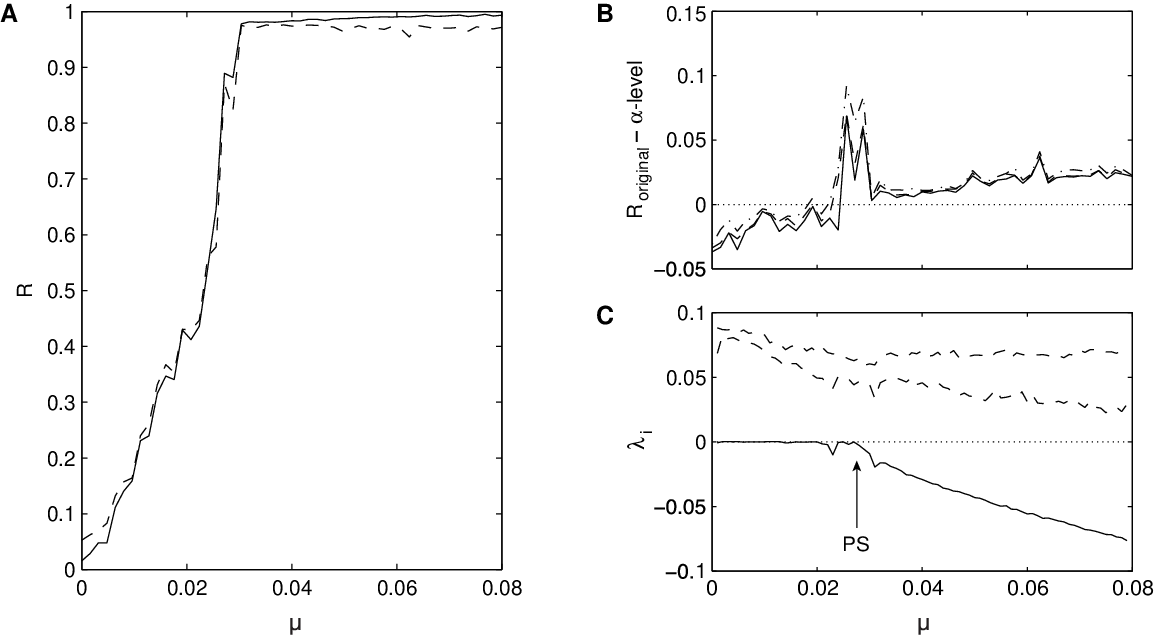}
\caption{ 
(A) Mean resultant length $R$ of the original two mutually coupled 
R\"ossler systems, Eqs.~(\ref{eq_2roessler1}) and (\ref{eq_2roessler2}),
with $a=0.15$, $b=0.20$, $c=10$, a frequency mismatch of $\nu=0.015$ 
(solid), and significance level of 1\% (dashed).
(B) Difference between $R$ of the original data and significance level 
of 1\% (solid), 2\% (dashed) and 5\% (dashed-dotted). 
(C) The four largest Lyapunov exponents for the considered 
6-dimensional system. $\lambda_4$ is highlighted (solid) and the 
arrow indicates the transition to PS. 
}\label{fig_ts_example}
\end{center}
\end{figure}

For $\mu < 0.025$, $R$ of the original system is, as
expected, below the significance level (Fig.~\ref{fig_ts_example}A)
and, hence, the difference is negative (Fig.~\ref{fig_ts_example}B).
For higher values of $\mu$, $R$ exceeds the significance level
(the difference becomes positive). This is in agreement with
the criterion for PS via Lyapunov exponents $\lambda_i$
\cite{pikovskyBook2001}: $\lambda_4$ becomes negative at
$\mu \approx 0.028$ (Fig.~\ref{fig_ts_example}C), which
approximately coincides with the intersection of the curve of $R$
for the original system and the significance level (zero-crossing
of the curves in Fig.~\ref{fig_ts_example}B). Therefore, the PS region 
can be successfully recognised by means of the TS.

Note that also the significance limit increases when the
transition to PS occurs (Fig.~\ref{fig_ts_example}A). As the TS mimic
both the linear and nonlinear characteristics of the system, the
surrogates of the second oscillator have the same
mean frequency in the PS region as the first original oscillator. 
Therefore, $R^{s_j}$ is rather high (it can take on values of up to 
$0.97$).
However, $\Phi_{x_i}$ and $\Phi_{y_i}^{s_j}$ do not
adapt to each other as they are independent, thus, the value of
$R$ for the original system is {\it significantly} higher than the
$R^{s_j}$ and, hence, it indicates PS.
We notice that even though the obtained value for a normalised PS index
is higher than 0.97 (right side of Fig.~\ref{fig_ts_example}A),
this still does not provide conclusive evidence for
PS. The knowledge of
the PS index {\it alone} is not sufficient to infer PS. Especially 
in passive
experiments, the synchronisation analysis should always be
accompanied by a hypothesis test.

%% file: meth_upos.tex
%
%
%
%
%
%
%
%
%
%
%
%
%
%
%
%
%
%
%
%
%

The localisation and quantification of unstable periodic orbits (UPOs) 
in chaotic attractors is very important as an orbit on a chaotic 
attractor is the closure of the set of UPOs which build the skeleton 
of the attractor. Roughly speaking, a trajectory, therefore, can be 
regarded as jumping from one UPO to the next one. Furthermore, the 
set of UPOs in an attractor is a dynamical invariant of the system 
\cite{auerbach1987}, as mentioned in Sec.~\ref{sec:Theory}. 

RPs can be used to easily localise UPOs in chaotic time series
\cite{lathrop89,mindlin92,bradley2002}. The main 
idea is the following: when the trajectory of the system comes close to an 
UPO, it stays in its vicinity for a certain time interval, whose 
length depends on how unstable the UPO is. This is reflected in 
the RP, as the pattern corresponding to periodic movement consists 
of uninterrupted equally spaced diagonal lines (if we consider
the RP of the original trajectory in phase space and not a projection).
Hence, UPOs present in the underlying system can be localised 
by identifying such ``windows'' inside the RP, where the
patterns correspond to a periodic movement.

\begin{figure}[thp]
\centering \includegraphics[width=.6\textwidth]{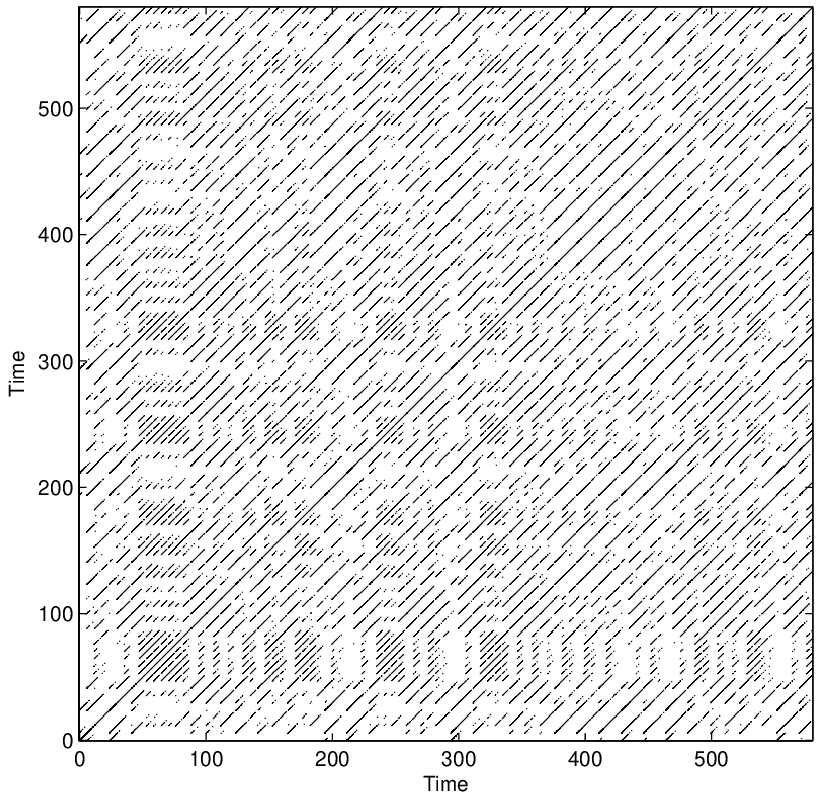}
\caption{RP of the R\"ossler system, Eqs.~(\ref{eq_roessler}), with 
$a=0.15$, $b=0.2$ and $c=10$. 
The three components $x$, $y$, and $z$ were used for the computation, 
as well as the $L_{\infty}$-norm.}\label{fig_rp_roessler}
\end{figure}

\begin{figure}[bhpt]
\centering \includegraphics[width=1.0\textwidth]{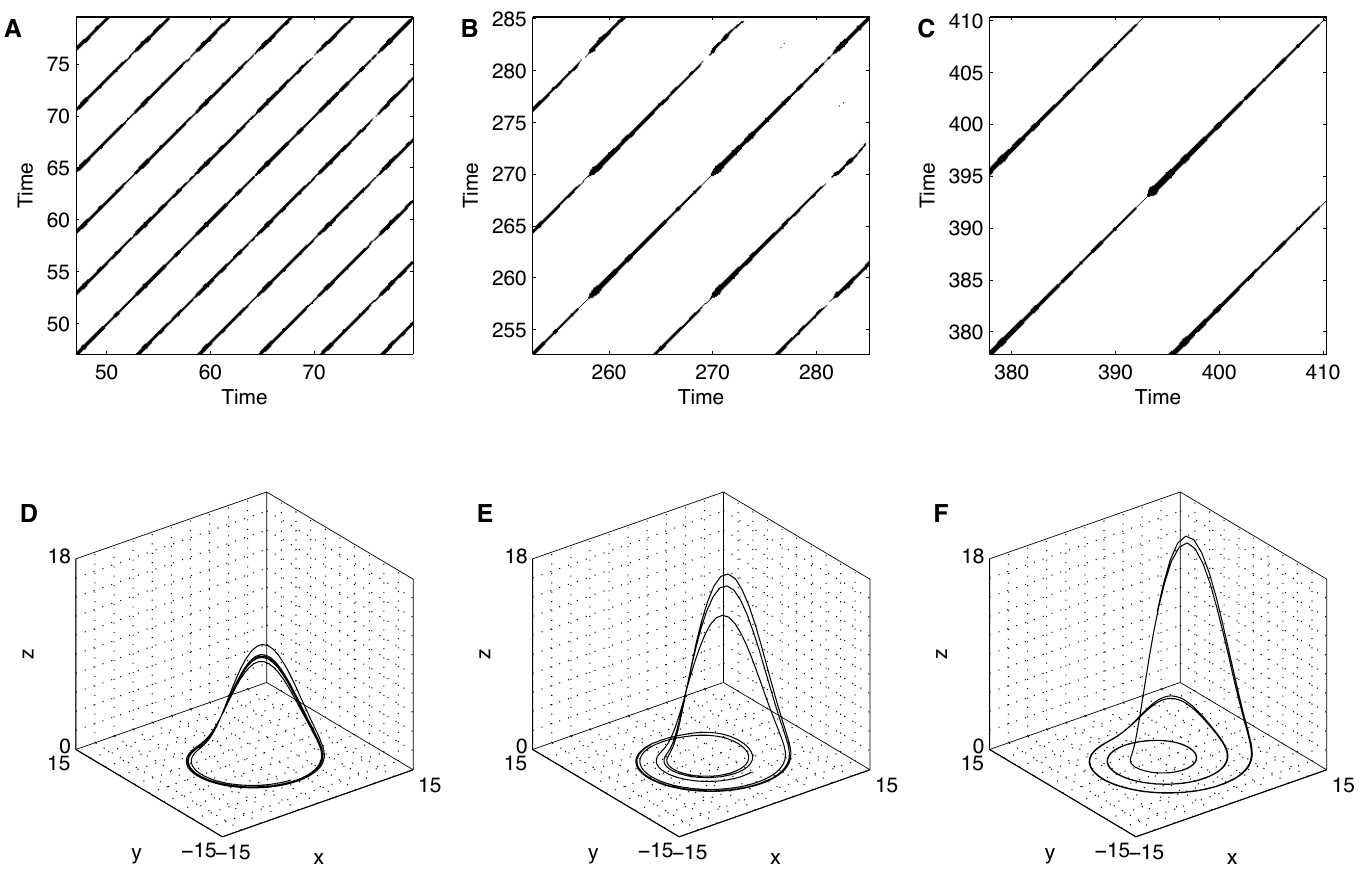}
\caption{(A--C) Magnifications of different periodic windows of the 
RP presented in Fig.~\ref{fig_rp_roessler}, and (D--F) the corresponding 
segments of the trajectory. The trajectories reveal the UPOs
of period 1 (D), period 2 (E) and period 3 (F).}
\label{fig_upos_roessler}
\end{figure}

For example, we can consider the R\"ossler system, Eq.~(\ref{eq_roessler}), with 
$a=0.15$, $b=0.2$ and $c=10$.
In its RP, many of such ``periodic'' windows can be 
identified (Fig.~\ref{fig_rp_roessler}). Moreover, 
the distance between the diagonal lines can 
vary from window to window, indicating a different UPO with a 
different period.  The period can be estimated by the vertical
distances between the recurrence points (e.\,g.~by
$\langle T^{(2)} \rangle$ or the mean length of the white vertical
lines) in the periodic window multiplied by the sampling time of the 
integrated trajectory. The UPO, which the trajectory of the 
R\"ossler system is close to,
can easily be identified in the magnifications of three different
windows of the RP as well as from the corresponding 
segments of the trajectory (Fig.~\ref{fig_upos_roessler}). 

The problem of using this method to localise UPOs is that a 
finite-length trajectory cannot visit all UPOs embedded 
in the attractor. Hence, it is still an open problem how to localise 
all of them using RPs.  However, first results indicate that there might 
be ways to overcome this difficulty. Furthermore, the authors in 
\cite{bradley2002} claim that RPs are not appropriate 
to find UPOs of high periods or UPOs which are very 
unstable. Nevertheless, the other existing methods to 
detect UPOs based on time series have the same problem \cite{so1997}.

%% file: meth_noise.tex
%
%
%
%

In this section, an estimation of the errors due to
observational noise on the quantification of RPs is derived. 
Based on this estimation it is possible to give a criterion to choose an
optimal threshold $\varepsilon$ to minimise these errors
\cite{thiel2002}.

Assume that we have a given scalar time series $x_i$ corrupted by
observational independent Gaussian noise $\xi_i$, i.\,e.~we have measured
$y_i=x_i+\xi_i$. Then, in the simplest case of non-embedding, 
the recurrence matrix, Eq.~(\ref{eq_rp}), becomes
\begin{equation} 
\mathbf{\tilde{R}}_{i,j} =
    \Theta\left(\varepsilon-\left|x_i-x_j+\xi_i-\xi_j \right|\right).
\end{equation} 
In order to estimate how the observational noise changes the structures 
in $\mathbf{\tilde{R}}_{i,j}$ with respect to $\mathbf R_{i,j}$, 
we consider the probability $\mathbf{P}_{i,j}$ to find a recurrence point
at the coordinates $(i,j)$.

\begin{figure}
\centering\includegraphics[width=0.55\textwidth]{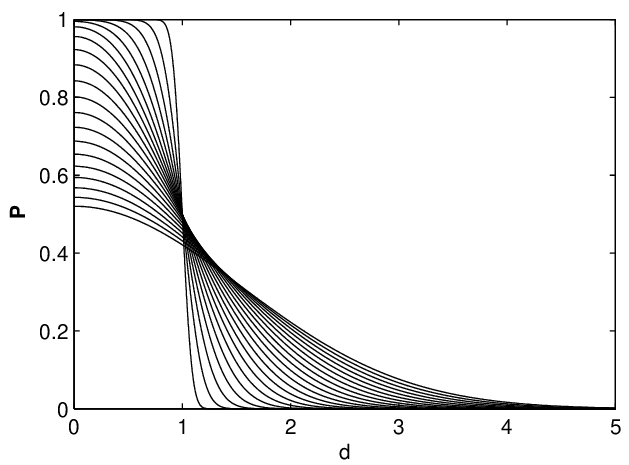}
\caption{Dependence of $\mathbf{P}$ on $d=\mathbf{D}/ \sigma$ for
different $\sigma \in[0.05\varepsilon \ \varepsilon]$ in steps of
$0.05\varepsilon]$ \cite{thiel2002}.}\label{fig_prob_rp_IJ}
\end{figure}

For an ensemble of realisations of the noise 
$\mathcal{W}(0,\sigma^2)$, the observation $y_i$ is Gaussian 
distributed with mean $x_i$ and standard deviation $\sigma$. 
Furthermore, we assume without loss of generality that $x_i=0$ 
and $x_j=-\mathbf{D}_{i,j}$ (cf.~Eq.(\ref{eq_DP})). Then the
probability to find a recurrence point at the coordinates $i,j$
is given by  
\begin{eqnarray}\label{eq_prob_rp_IJ} 
\mathbf{P}_{i,j} & = &
      \left( \frac{1}{\sqrt{2\pi}\sigma} \right)^2
      \int\limits_{-\infty}^{\infty}e^{-\frac{\xi_i^2}{2\sigma^2}}
      \int\limits_{\xi_i-\varepsilon}^{\xi_i+\varepsilon}e^{-\frac{(\xi_j-\mathbf{D}_{i,j})^2}{2\sigma^2}}d\xi_j\;d\xi_i
\nonumber \\
   & = &
      \frac{1}{8}\Bigg\{\mathrm{erfc}^2\left(\frac{\mathbf{D}_{i,j} - \varepsilon}{2\sigma}\right) - 
                       \mathrm{erfc}^2\left(-\frac{\mathbf{D}_{i,j}-\varepsilon}{2\sigma}\right) + 
                       \mathrm{erfc}^2\left(-\frac{\mathbf{D}_{i,j}+\varepsilon}{2\sigma}\right) -
\nonumber \\
   &  & \quad
                       \mathrm{erfc}^2\left(\frac{\mathbf{D}_{i,j}+\varepsilon}{2\sigma}\right)
                 \Bigg\} 
\end{eqnarray}

where $\text{erfc}(\cdot)=1-\text{erf}(\cdot)$.
Eq.~(\ref{eq_prob_rp_IJ}) maps the distance matrix $\mathbf{D}_{i,j}$ to the
probability matrix $\mathbf{P}_{i,j}$ to find a recurrence point in the RP
of $y_i=x_i+\xi_i$ at the coordinates $(i,j)$ for a fixed
underlying process and an ensemble of realisations of the
observational noise.
The smaller the values of $\sigma$ the closer $\mathbf{P}$ comes to a
Heaviside function (Fig.~\ref{fig_prob_rp_IJ}).
In the presence of noise, recurrence points
of the underlying system $x_i$ are recognised as such with
probability less than one, and analogously for non-recurrence
points.

%
%

Already a small amount of noise reduces the reliability of the
quantification of RPs, e.\,g.~for a noise level of $4\%$ the
probability that $\left |y_i-y_j\right| < \varepsilon$ for $\left
|x_i-x_j\right| \approx \varepsilon$ is reduced from one to less
than $0.6$, i.\,e.~more than $40\%$ of the recurrence points are
not recognised. 

If the density $\varrho$ of the differences $x_i-x_j$ is
given, the percentage of recurrence points
$p_b(\varepsilon,\sigma)$ that are  properly recognised in the
presence of observational noise can be calculated
\begin{equation}\label{Fehler1}
p_{b}(\varepsilon,\sigma) =
     \frac{\int\limits_{-\varepsilon}^{\varepsilon}\mathbf{P}(\mathbf{D},\varepsilon,\sigma)\varrho(\mathbf{D})dD}
          {\int\limits_{-\varepsilon}^{\varepsilon}\varrho(\mathbf{D})dD}.
\end{equation}  

Analogously, it is possible to compute the
percentage of properly recognised non-recurrence points
\begin{equation}\label{Fehler3}
p_{w}(\varepsilon,\sigma) =
      \frac{\int\limits_{-\infty}^{-\varepsilon}  \left(1-\mathbf{P}(\mathbf{D},\varepsilon,\sigma)\right)  \varrho(\mathbf{D})dD +
                  \int\limits_{\varepsilon}^{\infty}  \left(1-\mathbf{P}(\mathbf{D},\varepsilon,\sigma)\right)   \varrho(\mathbf{D})dD}
           {1-\int\limits_{-\varepsilon}^{\varepsilon}\varrho(\mathbf{D})dD}
\end{equation} 
Hence, as usually, two types of errors can be distinguished: 
\begin{enumerate}
\item false negative: a recurrence point is not recognised as such with
probability $1-p_b$ and 
\item false positive: a non-recurrence point is recognised
as a recurrence point with probability $1-p_w$.
\end{enumerate}

In order to minimise these errors, the threshold $\varepsilon$
should be chosen in such a way that $p_b$ and $p_w$ are maximised simultaneously. 
Even though the results depend on the distribution of the time
series, numerical simulations show that the choice of the threshold
should be at least five times the standard deviation of the 
observational noise $\varepsilon \approx 5\sigma$. This minimal 
choice is appropriate for a vast class of processes (e.\,g.~maps)
\cite{thiel2002}; higher-dimensional systems may require 
even a higher threshold $\varepsilon > 5\sigma$.
If $\varepsilon$ is smaller, effects
of the observational noise will have a dominant influence on the
detection of recurrence points. On the other hand, if
$\varepsilon$ is near the standard deviation of the underlying
process, the density of recurrence points will be too high to
detect detailed structures of the underlying process.

\begin{figure}[htbp]
\centering\includegraphics[width=\textwidth]{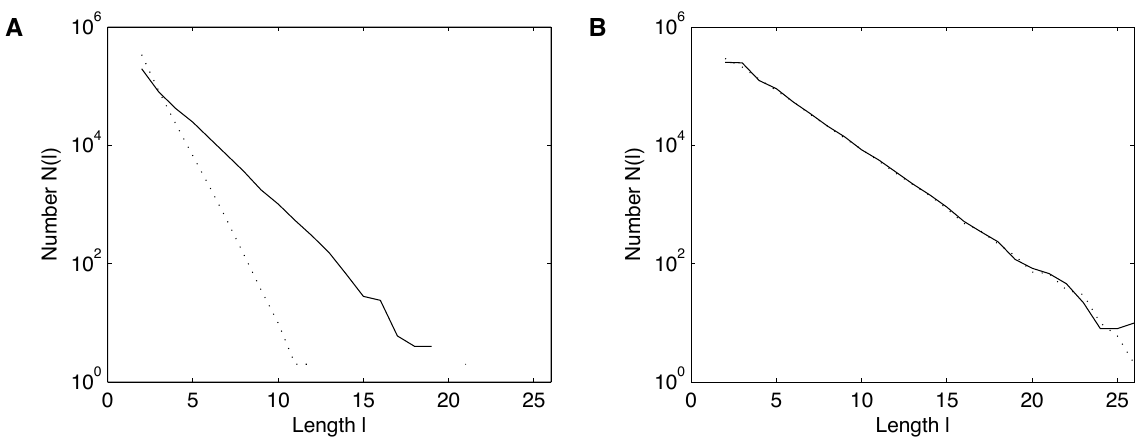}
\caption{
Cumulative distribution of diagonal lines for the logistic map, 
Eq.~(\ref{eq_logistic_map}), for $a=4$ and 3,000 data points, using
(A) the usual choice of $\varepsilon \approx \sigma$ and
(B) optimal choice of $\varepsilon \approx 5\sigma$
(solid line: without noise, dashed line: with $10\%$ noise) \cite{thiel2002}.} \label{fig_prob_diaglines_logistic}
\end{figure}

The impact of the proper choice of the threshold can be demonstrated
by studying the cumulative distribution of diagonal lines in the RP, e.\,g.~for
the logistic map, Eq.~(\ref{eq_logistic_map}), contaminated with
Gaussian white noise. For the choice of the threshold as the
standard deviation of the observational noise, $\varepsilon=\sigma$,
the distribution of the diagonal lines of the underlying process
is biased when observational noise is present (Fig.~\ref{fig_prob_diaglines_logistic}A). 
In contrast, with the proper choice $\varepsilon=5 \sigma$ the
distribution of the diagonal lines in the presence of noise coincides
with the distribution of the diagonal lines in the absence of noise 
(Fig.~\ref{fig_prob_diaglines_logistic}B).

Furthermore, noise has an influence on the estimation
of the dynamical invariants. Using the knowledge about
the relationship between the dynamical invariants 
and the noise as well as the threshold $\varepsilon$, 
the noise level can be estimated. Such a relationship has been
developed for the correlation entropy \cite{urbanowicz2003},
where a fit of the analytical function $K_2(\varepsilon)$ 
to the obtained entropy computed by RPs is used to estimate the noise-level $\sigma$.

%% file: appl_general.tex
%
%
%
%

The search for applications of RP based methods in the World Wide Web  
reveals numerous works (at present,
the Scirus search engine finds over 1,000 different works). 
RPs and the RQA are most popular
in physiology. However, various successful applications in other fields of
life science, as neuroscience and genomics, but also in
ecology, physics, chemistry, earth science and astrophysics, 
engineering and economy have also been published. In the following, 
some overview about the potentials of RPs for applications is presented by
means of a few selected examples.

One of the first applications of RPs was the analysis of heart beat
intervals \citep{zbilut91}. This study has revealed typical features in RPs 
for cardiac transplant patients and cardiomyopathy patients who 
underwent volume loading. Applying the RPs, it was inferred 
that the dimensionality and entropy of the heart beat variations
decrease during a significant cardiac event like myocardial infarction or
ventricular tachycardia. The investigation of the cardiac system
is one of the classical application fields of RPs and RQA. Numerous 
studies used, e.\,g., the RQA in order to monitor disease \cite{naschitz2004b}
or to detect predecessors of cardiac arrhythmia \cite{marwan2002herz}.

In further life science research RPs, OPRPs as well as RQA have been 
applied to, e.\,g., electromyography data \citep{webber95}, 
measurements based on eye movements \citep{shelhamer97},
data of postural fluctuations \citep{riley99}, 
EEG data \citep{babloyantz91,thomasson2001,marwan2004,acharya2005,groth2005} or
other neuronal signals \citep{faure2001}, 
in order to study the interacting physiological processes.
Characteristic patterns and rather fine 
frequency modulations in voice streams can be visualised by
the means of RPs \cite{facchini2005}. 

An RQA was applied to a DNA sequence of the genome  
{\it Caenorhabditis elegans} \citep{frontali99}, 
which is a small (approximately 
1 mm long) soil nematode found in temperate regions. This 
analysis has revealed long-range correlations in the
introns and intergenic regions, which are caused by
the frequent recurrence of oligonucleotides (a short sequence of some
hundreds of nucleotides) in these regions.
The recurrence of the oligonucleotides has been discovered by computing
the recurrence rate for overlapping windows which cover the 
DNA sequence. Other studies confirm such long-range correlations
in DNA sequences \cite{wu2004}.

An analysis based on RPs has been used to study monopole giant resonance in atomic 
nuclei \citep{vretenar99}. Due to the fact that a nucleus consists
of protons and neutrons, the oscillations can be divided into
two modes: (1) the densities of protons and neutrons oscillates in phase 
(isoscalar mode) and (2) the two densities have opposite phases (isovector
mode). Both of these modes exhibit significantly different RPs. Where the
oscillation of the isoscalar mode has an RP typical for 
regular oscillations, the RP for the isovector mode uncovers non-stationary
and chaotic dynamics. Other applications in physics were
performed in order to estimate the signal to noise ratio in
laser systems \cite{thiel2002}, or to estimate dynamical invariants,
as the dimensionality, of fluid flow systems \cite{thiel2004a}.

In chemistry
the RQA was applied to data from the Belousov-Zhabotinsky reaction 
especially to study
the transitions during its chemical evolution in an unstirred batch
reactor \cite{rustici99}. 
Using the RQA measures, the transitions between periodic, 
quasi-periodic and chaotic states could be observed. 
By means of JRPs, the synchronisation between
electrochemical oscillators were studied \cite{romano2005}.
Other applications
in chemistry/molecular biology concern the dynamics of chemical processes, for
example in molecular dynamics simulations of poly-peptides 
\citep{giuliani96,manetti2001} or to detect chaotic transitions of 
Nicotinamide adenine dinucleotide \cite{castellini2004}. 
Applying the RQA to glycoproteins of a virus
has uncovered the interaction between specific 
glycoprotein partners \citep{giuliani2002}.

Applications in earth science are yet rare. CRPs have been used
in order to compare similar dynamical evolutions in palaeo-climate
and modern rainfall data in NW Argentina \cite{marwan2003climdyn}.
The ability of CRPs to align time scales of geophysical profiles
were demonstrated on rock- and palaeo-magnetic data from
marine sediments \cite{marwan2002npg}. The estimation of $K_2$
by means of RPs was used to compare a general circulation model (GCM)
with re-analysis data \cite{vonbloh2005}.
A recent study
applied RPs on geomagnetic activity data represented by
several measurements (proxies for eastward and westward 
flowing polar currents) and derived the mutual information from
these RPs \cite{march2005}.

An astrophysical application of RPs has used radiocarbon 
data of the last 7,000 years \citep{kurths94}. The atmospheric
radiocarbon is influenced by the variation of solar activity
and exhibits century-scale variations of chaotic nature.
The main findings based on the RP analysis and a surrogate test reveal 
that these variations are indeed different from linear processes and that
there are different types of large events affecting their tendency to recur 
(e.\,g.~the Maunder minimum seems to be unique, whereas the Oort and Dalton 
minima as well as the Medieval maximum tend to recur). Moreover, the
authors have found that the present-day data are similar to 
the Medieval maximum. 
Besides, RPs were also used in order to investigate the synchronisation and 
phase difference in annual sunspot areas \cite{ponyavin2005,zolotova2006}.

The stability of the orbits of
terrestrial planets in the habitable 
zone of five extra-solar systems were investigated by computing
dynamical invariants derived from RPs \cite{asghari2004}.
By using the $K_2$ entropy, one 
full stable system, three systems with planets whose orbits are
stable for a long time and one unstable system were detected.

In engineering, applications of RPs or RQA are yet also rather rare. One of the
first has applied RPs to time series generated by models of
the Twin-T, Wien-bridge and other chaos generating electronic oscillator
circuits \cite{elwakil99}. Through visual inspection of the RPs, the chaotic behaviour 
of the model results has been confirmed. 
RPs have been used to estimate optimal embedding parameters and vicinity
threshold which are used for a noise reduction scheme in human speech
signals \citep{matassini2002b}.
A recent work proposed the method of JRPs as a very effective diagnosis 
tool in order to detect damage-induced changes in materials \cite{nichols2006}.

RPs have been used for research in economics.
For example, RPs have been inspected visually in order to 
find chaos in economics time series \citep{gilmore93,gilmore2001}. 
Whereas these visual inspections could not find chaos in the
considered economic time series (e.\,g.~unemployment rate, private
domestic investment, foreign exchange rate), a combined ``close returns'' and
surrogate test seems to reveal nonlinear dependencies among data of exchange 
rates. Other studies of foreign exchange data have used the RQA and have also 
found significant correlations between various currencies, which were
not obvious in the raw exchange data \citep{strozzi2002}.
In contrast to the results in \cite{gilmore2001}, the research of others 
who used the RQA has revealed chaos in exchange data \citep{holyst2001,belaire2002}.
Other applications on stock indices tried to predict ``bubble bursts''
on the stock market \cite{fabretti2005}. Moreover, the 
attempts to predict lottery numbers by the means of recurrences
should also be mentioned here, even this work seems to be not really 
serious (numerous sites in the WWW).

In the following subsections, we show in detail some exemplary 
applications of RPs, CRPs, JRPs and their quantification 
(RQA and dynamical invariants) to different 
kinds of experimental data.

%% file: appl_rqa.tex
%
%
%
%

First, we illustrate the capabilities of the RQA to detect
transitions in measured physiological data.

The activity of rather large ensembles of 
neurons (which act like nonlinear devices) is macroscopically 
measurable in the electroencephalogram 
(EEG) of the human scalp, which results from a spatial integration of 
post-synaptic potentials. Applying nonlinear 
techniques like the estimation of the correlation dimension to EEG 
has a long tradition (e.\,g.~\cite{babloyantz85, 
rapp86, gallez91, lutzenberger92, pritchard92}). 
However, most techniques are only well-defined for stationary 
time series generated by a low dimensional dynamical system,
and, hence, they fail in investigating event-related 
brain potentials (ERPs) because they are non-stationary by definition
\cite{sutton65}. Event-related potentials are characteristic
changes in the EEG of a subject during and shortly after a certain (e.\,g.~visual or acoustic) 
stimulus (surprising moment). The notation for an ERP is, e.\,g., N100
or P300, where N means a negative and P a positive potential and
the number corresponds to the time of ERP onset after the stimulus
in milliseconds.

\begin{figure}[b!]
\centering \includegraphics[width=.7\columnwidth]{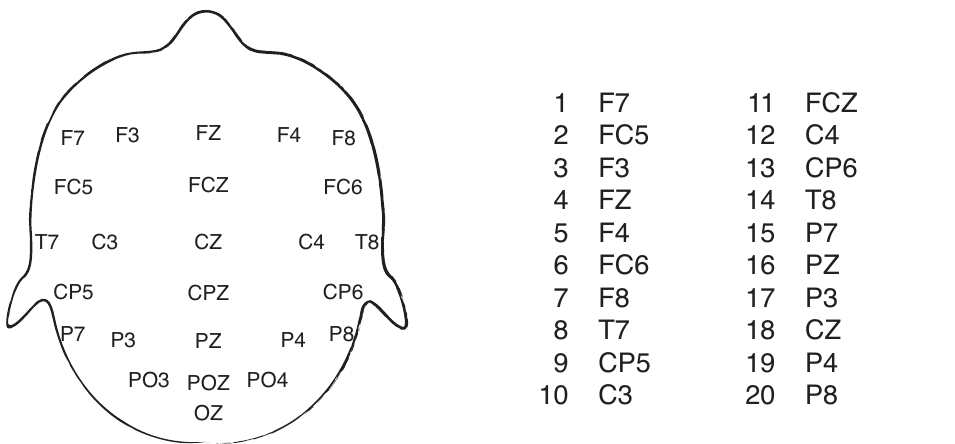} \\
\caption{Localisation of the electrodes on the head.}\label{fig_electrodes}
\end{figure}

%

Traditionally, ERP waveforms are determined by computing an ensemble 
average of a large collection of EEG trials that are stimulus time locked. 
By averaging the data points, which are time locked to the stimulus 
presentation, it is possible to filter out some signal (ERP) 
of the noise (spontaneous activity). This way,
the P300 component of the ERP (i.\,e.~brain activity 300~ms after
the onset of the stimulus) was the first potential discovery to vary in     
dependence on subject internal factors, like attention and expectation, instead   
on physical characteristics \citep{sutton65}. The amplitude of the P300    
component is highly sensitive to the novelty of an event and its relevance 
(surprising moment). So this 
component is assumed to reflect the updating of the environmental model of the  
information processing system (context updating) \cite{donchin81,donchin88}.
There are two main disadvantages of the averaging method. On the one hand, 
the number of trials needed to reduce the signal-to-noise-ratio is rather high.
This disadvantage is crucial for example in clinical studies, studies with children 
or studies in which repeating a task would influence the performance.
On the other hand, several high frequency structures reflecting other
important brain activities are filtered out by using the averaging method.
It is, therefore, desirable to find new approaches for analysing event related 
activity on a single trial basis. Applying the concepts of the RQA
to electro-physiological data could be one way of dealing with 
this problem.

\begin{figure}[b!]
\centering \includegraphics[width=\columnwidth]{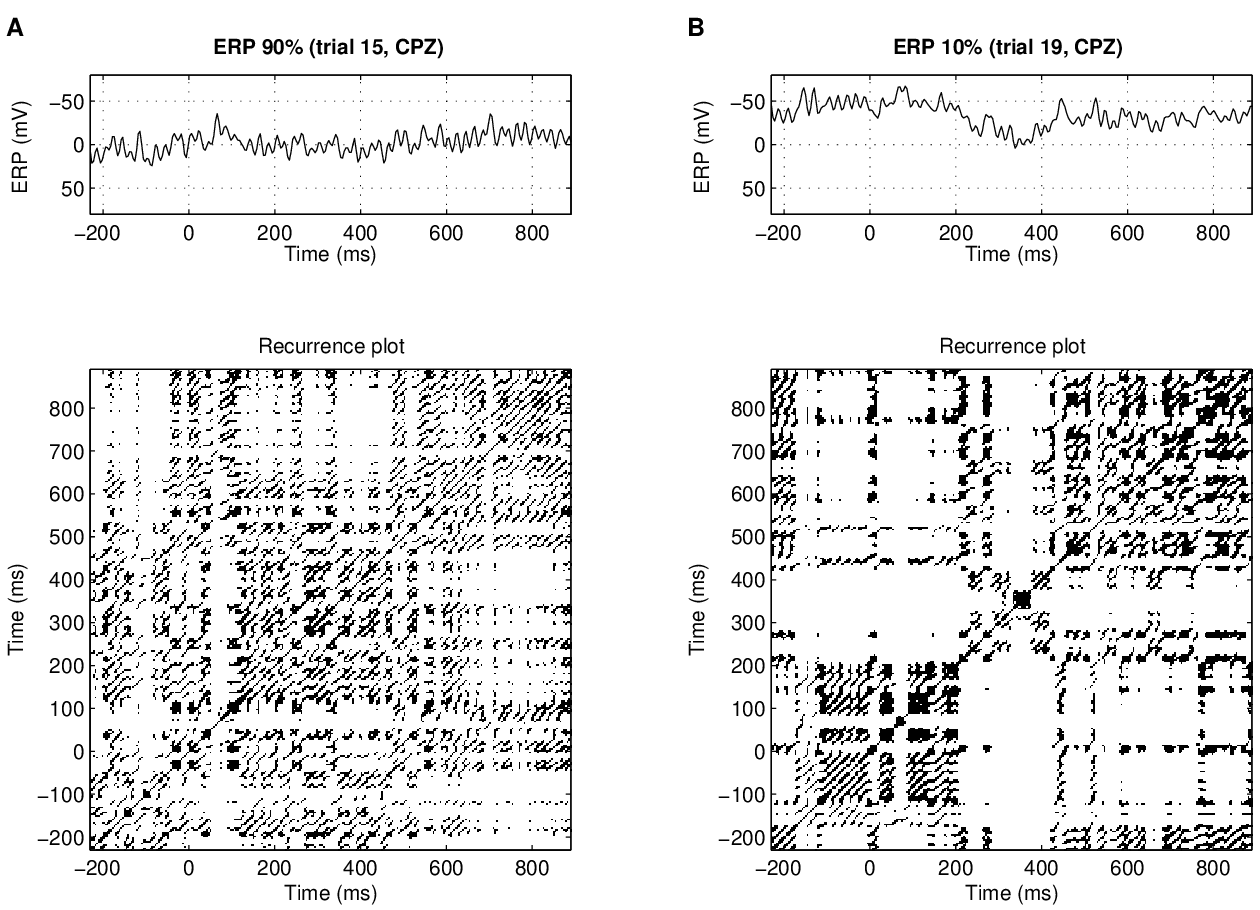} \\
\caption{ERP data for event frequencies of 90\% (A) and 
10\% (B), and their corresponding recurrence plots. 
For the lower event frequency (B) more clusters of
recurrence points occur at 100~ms and 300~ms.  Furthermore, a white 
band marks a transition in the process \cite{marwan2004}.
}\label{fig_rp_eeg}
\end{figure}

\begin{figure}[b]
\centering \includegraphics[width=\columnwidth]{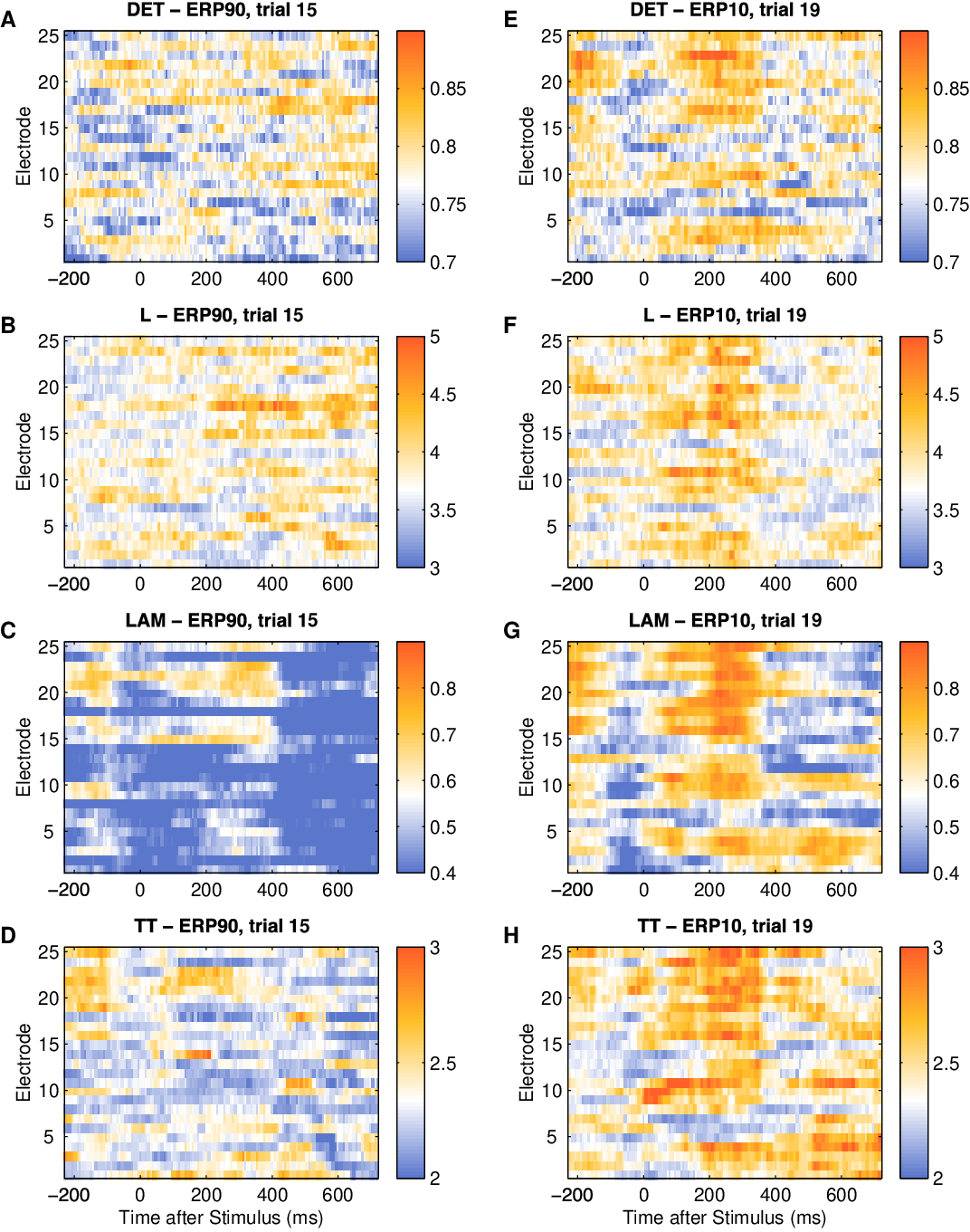} \\
\caption{RQA measures for selected trials for event frequencies 
of 90\% (A--D) and 10\% (E--H). The P300 component reflects the
surprising moment and can be detected in single trials by the measures 
$LAM$ (G) and $TT$ (H), which base on the vertical structures in the RP. 
The measures based on diagonal structures, $DET$ and $L$, are less apparent
\cite{marwan2004}.}
\label{fig_rqa_eeg}
\end{figure}

For this study, we analyse measurements of an Oddball experiment \cite{marwan2004}.
It studies brain potentials during a stimulus presentation; here acoustic
stimuli are used. In the analysis of a set of 40 trials of ERP data for an
event frequency of 90\% (ERP90) and a second set of 31 trials
for an event frequency of 10\% (ERP10), the RQA measures $DET$ and $L$,
and the vertical structures based measures $LAM$ and $TT$ are 
computed. 
The ERPs is measured at 25 electrodes (\reffig{fig_electrodes})
with a sampling rate of 4~ms.
Our aim is to study single trials in order
to find transitions in the brain processes as a consequence of
unexpected stimulation. 
Due to the N100 and the P300 components in the data,
the RPs show varying structures changing in time (\reffig{fig_rp_eeg}).
Diagonal structures and clusters of black points occur. The
non-stationarity of the data around the N100 and P300 causes
extended white bands along these times in the RPs. However,
the clustered black points around 300~ms occur in almost all
RPs of the ERP10 data set.
The application of the measures of complexity to these ERP data 
discriminates the single trials with a distinct
P300 component resulting from a low surprise
moment (high frequent events, ERP90 data) in favour of such trials with a
high surprise moment (less frequent events, ERP10 data; Fig.~\ref{fig_rqa_eeg}). 
The $LAM$ is the most distinct parameter
in this analysis. In the ERP data the $LAM$ reveals transitions 
from less to more laminar states after the occurrence
of the event and a transition from more laminar states
to less laminar ones after approximately 400~ms. These transitions occur
inside bounded brain areas (parietal to frontal along the central axis).
The comparable measures $DET$ and $LAM$ as well as $L$ and $TT$ are quite different in their
amplitudes. There are also differences in time and  
brain location of the found transitions.

These results show that the measures based on vertical RP structures
uncovers transitions, which are not found by the RQA measures based 
only on the diagonal RP lines. The RQA indicate transitions in the brain processes
into laminar states due to the surprising moment of observed events.

%% file: appl_exchangerates.tex
%
%
%
%

The study of economic systems by time series analysis is a 
challenging effort. Economic cycles are usually characterised 
by irregular variations in amplitude and period length,
rising the discussion whether the economical dynamics is, e.\,g., 
stochastic or chaotic, and whether it is stationary. Several
approaches from nonlinear data analysis have been applied in order to
discern the type of dynamics in economic time series, like the BSD test
\cite{barnett1997} or the Kaplan's test \cite{kaplan1994}. In the
last years, RPs and their quantitative analysis have
also become popular for testing for nonlinearity of economic dynamics
\citep{gilmore2001,holyst2001}. In the following
example the RQA is applied to financial exchange rates in order to
assess a nonlinearity in them \cite{belaire2002}.

Many open-economic theoretical models require the fulfilment of the
{\it purchasing power parity (PPP)}. This is an equilibrium
assumption in the market for tradeable goods: a good in a country
should sell for the same price as in any other country. One consequence
is that the real exchange rate must be stationary. 
Using $P$ as the domestic price, $P^{*}$ as the foreign currency
price and $S$ as the exchange rate, a formal description of this assumption is 
\begin{equation}
P = S P^{*}.
\end{equation}
The real exchange rate, i.\,e.~the ratio of the price of the domestic to foreign goods,
is 
\begin{equation}
E = \frac{S P^{*}}{P},
\end{equation}
and measures the price competitiveness. If the PPP assumption holds, this real exchange rate $E$
must be stationary. Therefore, it is
important to test for stationarity of $E$. Several
applied tests for stationarity are based on the assumption that the process
driving the dynamics of the exchange rates is inherently linear.
Hence, if the underlying process is nonlinear, 
these tests might erroneously reject the null-hypothesis that $E$ is a stationary
process, even though it is stationary indeed. In order to assess nonlinearity 
in exchange rates, RQA in connection with a surrogate test can be used.

The analysed data are quarterly US Dollar based exchange rates for 16 foreign
countries (Austria, Belgium, Canada, Denmark, Finland, France, Germany,
Greece, Ireland, Italy, Japan, The Netherlands, Norway, Spain, 
Switzerland and United Kingdom) spanning the time from January 
1957 to April 1998 (168 observations). First, the data are
transformed with a stationary-in-mean series and then filtered in order
to remove linear stochastic dependencies by fitting an AR model, Eq.~(\ref{eq_ar1}).
For details of the data aquisition and preprocessing, as well
as a comprehensive explanation of all tests applied to these
data, see \cite{belaire2002}.

\begin{table}[b]
\caption{Used threshold $\varepsilon$ and RQA measures of exchange rates.
RQA measures exceeding the 95$^{\text{th}}$ percentiles of
the empirical distribution based on surrogates (in parenthesis) are
shown in bold font. $RR$ and $DET$ values are given in \%.}\label{tab_rqa_exchangerates}
\scriptsize
\centering \begin{tabular}{lrrrrrrrrr}
\hline
Country		&\multicolumn{1}{c}{$\varepsilon$}	&\multicolumn{2}{c}{$RR$}	&\multicolumn{2}{c}{$DET$}	&\multicolumn{2}{c}{$ENT$}	&\multicolumn{2}{c}{$L_{\max}$}\\
\hline
\hline
Austria 	&0.086 			&\bf 5.02	&(2.77) 	&\bf 53.5	&(44.7)		&\bf 3.39	&(2.01) 	&\bf 35		&(26)\\
Belgium 	&0.048 			&\bf 4.62	&(1.59) 	&\bf 60.6 	&(30.8) 	&\bf 2.98 	&(1.38) 	&\bf 33 	&(18)\\
Canada 		&0.066 			&\bf 5.10 	&(4.08) 	&\bf 72.9 	&(65.7) 	&\bf 3.03 	&(2.99) 	&\bf 32 	&(31)\\
Denmark 	&0.127 			&4.98 		&(5.91) 	&58.3 		&(84.1) 	&2.10 		&(3.59) 	&25 		&(35)\\
Finland 	&0.167 			&5.01 		&(8.33) 	&19.5 		&(79.2) 	&2.59 		&(4.25) 	&20 		&(38)\\
France 		&0.155 			&\bf 4.93 	&(3.59) 	&\bf 81.3 	&(68.3) 	&1.82 		&(2.63) 	&23 		&(25)\\
Germany 	&0.089 			&\bf 4.98 	&(3.22) 	&\bf 85.4 	&(60.6) 	&\bf 3.28 	&(2.50) 	&\bf 31 	&(26)\\
Greece 		&0.062 			&\bf 5.02 	&(4.86) 	&60.4 		&(67.6) 	&\bf 3.25 	&(3.22) 	&\bf 37 	&(36)\\
Ireland 	&0.177 			&5.02 		&(8.15) 	&46.4 		&(83.5) 	&1.14 		&(4.02) 	&20 		&(40)\\
Italy 		&4.350 			&4.86 		&(8.19) 	&\bf 46.0 	&(36.2) 	&2.65 		&(3.02) 	&\bf 35 	&(32)\\
Japan 		&0.079 			&\bf 5.08 	&(2.62) 	&\bf 56.7 	&(54.7) 	&\bf 3.52 	&(2.52) 	&\bf 35 	&(26)\\
The Netherlands &0.110 		&\bf 5.08 	&(3.21) 	&\bf 74.0 	&(60.8) 	&\bf 3.35 	&(2.55) 	&\bf 32 	&(28)\\
Norway 		&0.075 			&\bf 5.02 	&(2.74) 	&\bf 63.6 	&(50.6) 	&\bf 3.38 	&(2.32) 	&\bf 37 	&(25)\\
Spain 		&0.182 			&5.05 		&(5.60) 	&20.3 		&(72.7) 	&1.92 		&(3.46) 	&22 		&(31)\\
Switzerland &0.055 			&\bf 5.70 	&(1.33) 	&\bf 86.6 	&(24.3) 	&\bf 4.30 	&(0.05) 	&\bf 40 	&(17)\\
United Kingdom &0.166 		&5.11 		&(7.88) 	&42.4 		&(84.2) 	&0.62 		&(4.01) 	&17 		&(39)\\
\hline
\end{tabular}
\end{table}

The RQA measures $RR$, $DET$, $ENT$ and
$L_{\max}$ (see Sec.~\ref{sec:RQA}) are calculated for each time series. The embedding
parameters used are $m=20$ and $\tau=1$; the recurrence
threshold $\varepsilon$ is determined such that
it is the smallest value at which all RQA parameters are
non-zero (Tab.~\ref{tab_rqa_exchangerates}). Then, 500 surrogate series for each currency are produced
by simply shuffling the exchange data. This shuffling destroys 
dependencies in time but preserves the distribution of the original series. 
The same RQA measures are then calculated from
these surrogates using the same embedding and RP parameters
as for the corresponding original data series. This leads to 
an empirical distribution of the RQA measures under the 
null-hypothesis of independence in time and an identical distribution. 
The authors claim that if all RQA measures $RR$, $DET$, $ENT$ and $L_{\max}$ of the
original data series exceed the 95$^{\text{th}}$ percentile of
the empirical distribution, the assumption of linearity must
be rejected. This is indeed the case for the currencies of Austria,
Belgium, Canada, Germany, Greece, Japan, The Netherlands, Norway 
and Switzerland (Tab.~\ref{tab_rqa_exchangerates}). The assumption
of a linear stochastic process cannot be rejected for the currencies
of Denmark, Finland, Ireland, Spain and United Kingdom. These
findings are supported by an application of the Kaplan's test
\cite{belaire2002}.

In conclusion, the authors of \cite{belaire2002} claim that the real exchange rates are 
probably driven
by a nonlinear mechanism. This can help to sustain the PPP assumption
even using data which pretend to be non-stationary, but analysed
under linear assumptions.

%% file: appl_damagedetection.tex
%
%
%
%

Early damage detection of mechanical systems is a crucial task for preventing catastrophic
failures or minimise maintenance costs. Vibration based structural health
monitoring (SHM) is one approach for damage detection. The main idea is
to look for early signatures in the dynamical response (like frequencies, 
phase ratios, mode shapes) of excitations (applied load) before serious problems arise. 
In a recent work, RQA has been employed for SHM simulation and proved to perform
better than standard frequency based measures \cite{nichols2006}.

\begin{figure}[htb]
\centering 
\includegraphics[width=0.6\columnwidth]{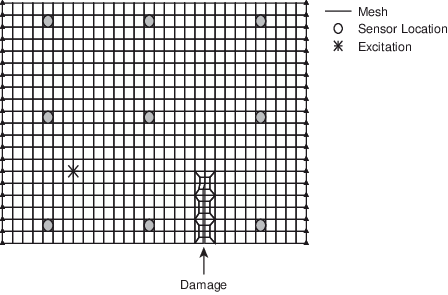}
\caption{Schema of the plate model with sensor, load and damage locations \cite{nichols2006}.
}\label{fig_nichols_1}
\end{figure}

A thin plate of hot-rolled steel is used in a finite element model in order
to simulate dynamic response under different damage conditions. The model
contains 624 elements, representing a plate of size $0.660 \times 0.408 \times 3.175\ 10^{-3}$~m$^3$,
and with clamp-free boundary conditions (Fig.~\ref{fig_nichols_1}). The simulated damage is imposed as a cut
in the plate with increasing length extending to 25\% in increments of 5\%. 
The strain is measured by using nine sensors, and the forcing input is applied at one
node near the left corner (Fig.~\ref{fig_nichols_1}), which is excited 
by a chaotic vibration, obtained from the Lorenz oscillation, Eq.~(\ref{eq_lorenz}).
Detailed explanations about the basic assumptions, the model and parameters
can be found in \cite{nichols2006}. 

The phase space vector is constructed by using all nine sensor measurements
as components, i.\,e.~the phase space is 9-dimensional. Next, CRPs between
each damage stage and the undamaged case are computed. As the simulated damage increases,
the phase space trajectory differs more and more from the trajectory of the undamaged case, what is reflected by 
a diminishment of the recurrence points and diagonal lines in the CRP. Therefore,
the RQA measures $RR$, $DET$ and $ENTR$ are computed within sliding windows
of length 10,000 (shift of 1,400 points) from the CRPs. 

\begin{figure}[htb]
\centering 
\includegraphics[width=0.6\columnwidth]{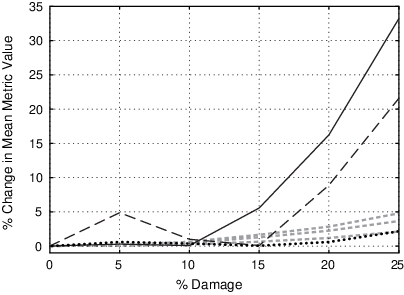}
\caption{Change of the RQA measures with damage along with the first three modal
frequencies: solid -- $RR$, dashed -- $DET$, dotted -- $ENTR$ and dashed-gray -- $f_1$, $f_2$, $f_3$ \cite{nichols2006}.
}\label{fig_nichols_2}
\end{figure}

The measures $RR$ and $DET$ clearly identify and quantify the damage scenarios
for damages larger than 10\% (Fig.~\ref{fig_nichols_2}). The $ENTR$ measure is not able to
detect the damage clearly (only for 25\%). The sensitivity of the $RR$'s and $DET$'s changes 
with damage can be compared with the plate's modal frequencies, which are often used
in SHM as benchmarks for comparison to the proposed technique. Both $RR$ and $DET$ 
exhibit larger changes due to the damage than the first three modal frequencies $f_1$,
$f_2$ and $f_3$ (Fig.~\ref{fig_nichols_2}). The highest changes are 33\% for $RR$ and
22\% for $DET$, where the highest changes for $f_1$, $f_2$ and $f_3$ are less than 
10\%. 

In conclusion, the RQA based measures appear to be a more effective tool for monitoring subtle changes
in mechanical structures. Their advantage lies in a high sensitivity and simple computation.
Moreover, the probabilistic nature of this method does not require assumptions
about the underlying dynamics, like stationarity or linearity.

%% file: appl_rescaling.tex
%
%
%
%

The problem of adjustment of data series with various time scales occurs in
many occasions, e.\,g., data preparation of
tree rings or geophysical profiles.
Often a large set of geophysical data series is gained at
various locations (e.\,g.~sediment cores).
Therefore, these data series have different lengths and time scales.
The first step in the analysis of these time series is
the synchronisation of both time scales.
Usually, this is done visually by comparing and
correlating each maximum and minimum in both data sets by hand
(``wiggle matching''), which includes the human factor of 
subjectiveness and is a lengthy process. The use of CRPs
can make this process more objective and automatic.

CRPs contain information about the time transformation
which is needed to align the time scales of two data series.
This is revealed by the distorted main diagonal, the LOS
(Subsecs.~\ref{sec:StructuresinRecurrencePlots} and 
\ref{sec:CrossRecurrencePlots}). 
A nonparametric rescaling function is provided by isolating 
this LOS from the CRP, which can be used for the re-alignment
of the time scales of the considered time series. 

\subsubsection{Time scale alignment of geological profiles}

\begin{figure}[htbp]
\vspace*{2mm}
\centering \includegraphics[width=\columnwidth]{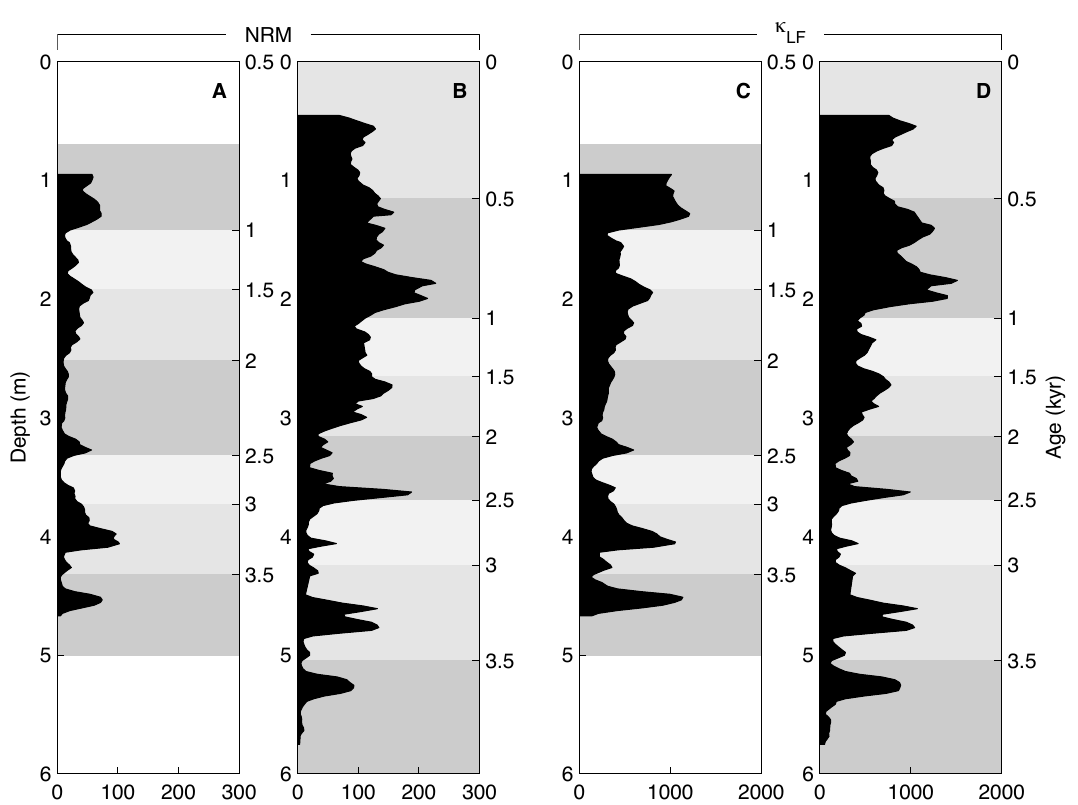}
\caption{Rock-magnetic measurements of lake sediments with different
time scales. Corresponding sections are marked with different grey values
\cite{marwan2005}.}
\label{fig:sample_raw}
\end{figure}

Here we apply this technique to re-adjust two geological
profiles (sediment cores) from the Italian lake {\it Lago di Mezzano} \citep{brandt99,marwan2005}.
The profiles cover approximately the same geological processes but have 
different time scales due to variations in the sedimentation rates
at the different sites.
The first profile (LMZC) has a length of about 5~m and the second
one (LMZG) of about 3.5~m (\reffig{fig:sample_raw}). From both profiles
a huge number of geophysical and chemical parameters were measured. 
Here we focus on the rock-magnetic measurements of the normalised
remanent magnetisation intensity (NRM) and the susceptibility $\kappa$.

\begin{figure}[htbp]
\vspace*{2mm}
\centering \includegraphics[width=10cm]{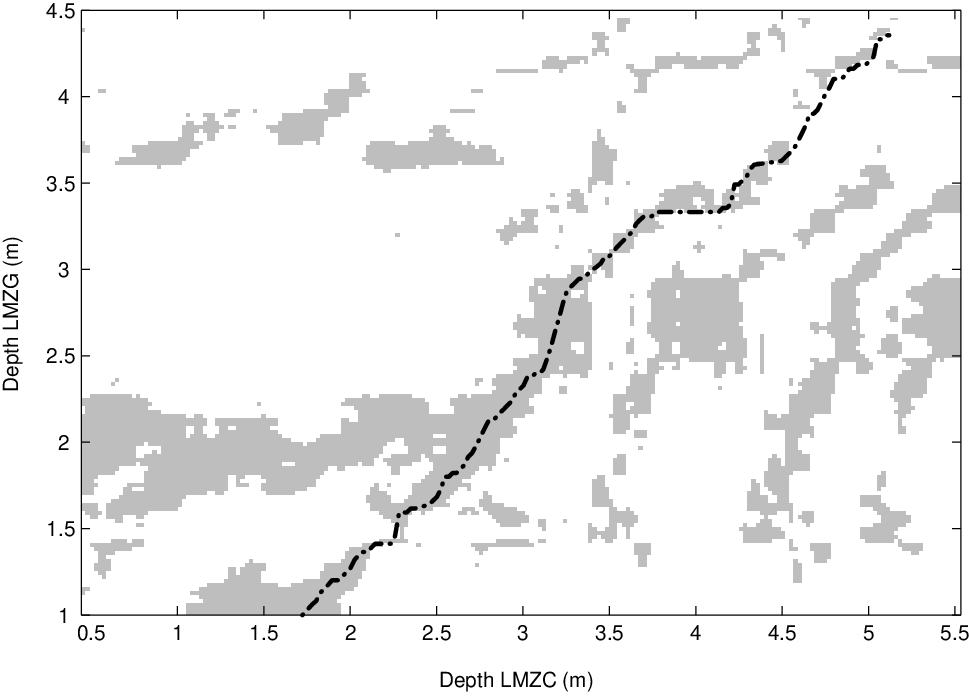}
\caption{Cross recurrence plot between rock-magnetic data shown in Fig.~\ref{fig:sample_raw}.
The dash-dotted line is the resolved LOS which can be used for re-adjustment of the time scales
of both data sets \cite{marwan2005}.}\label{fig:sample_rp}
\end{figure}

\begin{figure}[htbp]
\vspace*{2mm}
\centering \includegraphics[width=\columnwidth]{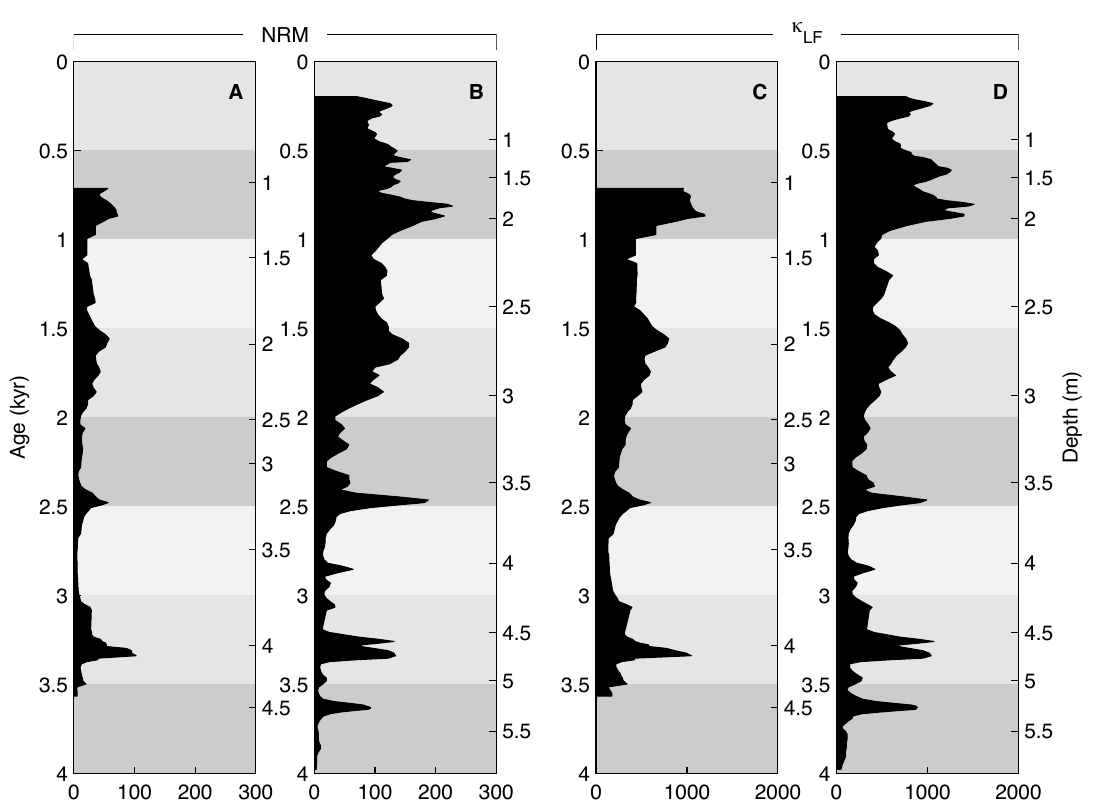}
\caption{Geological profiles after re-adjustment using the LOS which was found with
the CRP shown in Fig.~\ref{fig:sample_rp}. Corresponding sections are marked with 
different grey values \cite{marwan2005}.}\label{fig:sample_adjust}
\end{figure}

We use the time series NRM and $\kappa$ as components for the
phase-space vector, resulting in a two-dimensional system.
However, we apply an additional embedding using the time-delay
method \citep{theiler92}. 
A rather small embedding decreases the amount
of line structures representing the progress with negative time \citep{marwan2003diss}.
Using the embedding parameters dimension $m=3$ and delay $\tau=5$ (empirically found for these
time series), the final dimension of the reconstructed system is six.
The corresponding CRP reveals a partly disrupted, swollen and bowed LOS (\reffig{fig:sample_rp}).
This LOS can be automatically resolved, e.\,g.~by using the LOS-tracking algorithm
as described in the Appendix~\ref{apdx:LOS}. The application of this LOS
as the time-transfer function to the profile LMZG re-adjusts its time series
to the same time scale as LMZC (\reffig{fig:sample_adjust}).
This method offers a helpful tool for an automatic adjustment
of different geological profiles, in contrast to the rather subjective
method of ``wiggle matching'' (adjustment by harmonising maxima and minima by eye) used so far.

\subsubsection{Dating of a geological profile (magneto-stratigraphy)}\label{sec_frank}

From a sediment profile (Olguita profile, Patagonia, Argentina; 
\cite{warkus2002})
a measurement of the palaeo-polarity 
of the Earth's magnetic field (along with other measurements) is available. 
The starting point for any geological investigation of such a
profile is determining the time at which these sediments were deposited.
By applying the magneto-stratigraphic approach and a geomagnetic polarity
reference with known time scale, the polarity measurements can be used 
to determine a possible time scale for the profile. In \cite{cande95},
such a geomagnetic polarity reference is provided, which covers the last 
83~Myr. The Olguita profile contains seven reversals. The polarity
data consist of values one, for the polarity direction as today,
and zero, for the inverse polarity. Unfortunately, this
data series is too short (only 16 measurements) for a reliable
analysis. Nevertheless, for our purpose of demonstration 
we enlarge this data by interpolation. The Olguita profile
is transformed to an equidistant scale of 300 data points and
the reference data is transformed to an equidistant scale 
of 1,200 data points.

\begin{figure}[htbp]
\centering \includegraphics[width=\columnwidth]{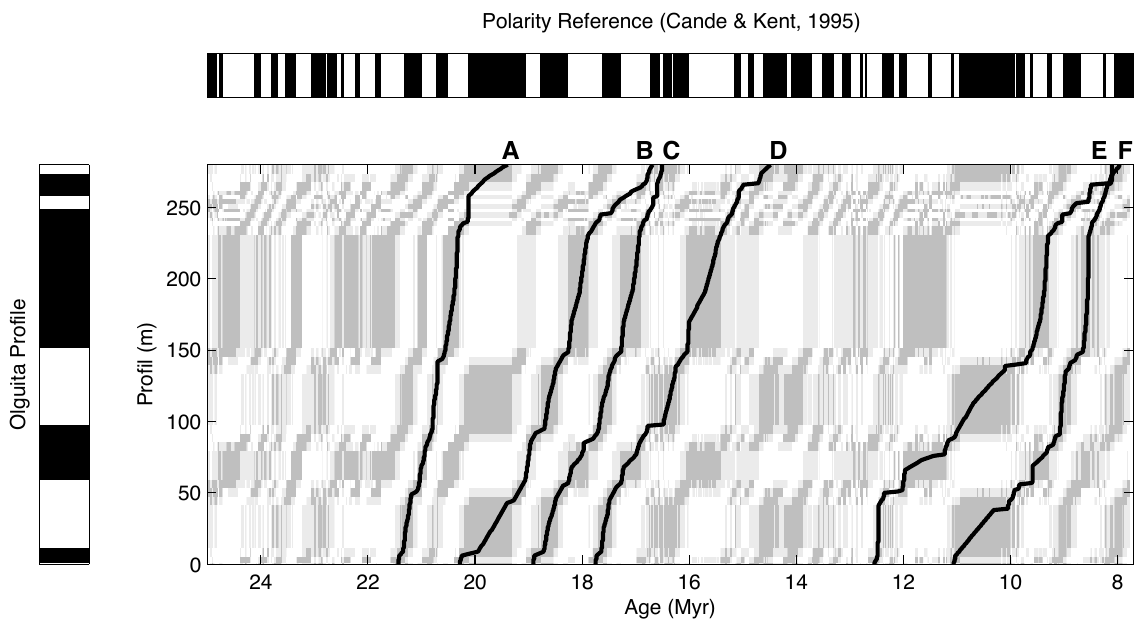} 
\caption{CRP between the polarity data of the Olguita profile and
the reference data according to \cite{cande95}. The used neighbourhood
criterion is FAN with 30\% RR (grey RP) and 40\% RR (bright grey RP). 
In the polarity data
the white colour marks a polarity of the Earth's magnetic field
in the present, whereas the black colour marks a reversal. Six potential
LOS are found (A--F, corresponding to the
potential LOS given in \reffig{fig_los_frank}) \cite{marwan2003diss}.}\label{fig_crp_frank}
\vspace{1cm}
\end{figure}

\begin{figure}[htbp]
\centering \includegraphics[width=\columnwidth]{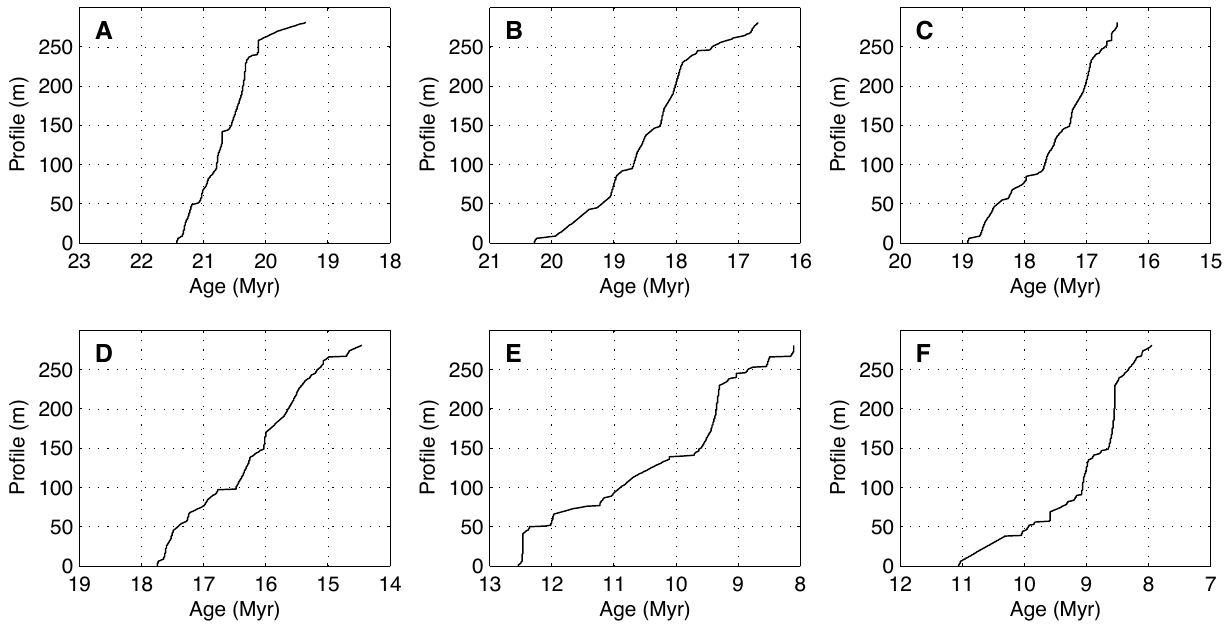} 
\caption{Potential LOS of the CRP presented in \reffig{fig_crp_frank}.
They correspond to the potential sedimentation rates of the
Olguita profile and mark sequences in the polarity reference, which match with the
Olguita profile \cite{marwan2003diss}. Due to this matching, the Olguita profile can be dated.}\label{fig_los_frank}
\end{figure}

A CRP is computed from these two data series by using an 
embedding dimension $m=4$, a delay of $\tau=6$ and 
a neighbourhood criterion of FAN
(30\% recurrence rate). Varying degrees of continuous 
lines between 21 and 16~Myr~BP and between 12 and 
8~Myr~BP occur in the CRP, which can be interpreted as
the desired LOS (\reffig{fig_crp_frank}).  We 
will analyse six of these possibilities for the LOS.
The search for the potential LOS is conducted using the algorithm
described in the Appendix \ref{apdx:LOS}. Moreover, we can 
evaluate the quality of these potential LOS by introducing
a quality factor that takes into consideration 
the amount of gaps $N_{\circ}$ and black dots $N_{\bullet}$ on this line
\begin{equation}
Q=\frac{N_{\bullet}}{N_{\bullet}+N_{\circ}}\ 100\%.
\end{equation}
The larger $Q$ is, the better the LOS is; $Q=100$\% stands for an
absolute continuous line. Moreover, the obtained LOS can be
interpreted as the sedimentation rate (\reffig{fig_los_frank}). Abrupt changes
in the sedimentation rate are not expected, thus, 
the potential LOS should not change abruptly.
As a criterion we can use hence the averaged
second derivative with respect to the time $\langle \partial_t^2\rangle$.

\begin{table}
\caption{Possible ages of the Olguita profile, which are based on the found potential 
LOS (\reffig{fig_los_frank}) and characteristics of these potential LOS \cite{marwan2003diss}.}\label{tab_q_frank}
\vspace{.5cm}
\centering \begin{tabular}{crcccr}
\hline
Plot	&Age (Myr)	&$N_{\bullet}$	&$N_{\circ}$	&$Q$ (\%)	&$\langle \partial_t^2\rangle$\\
\hline
\hline
A	&19.4--21.4	&345		&23		&93.8		&5.5\\
B	&16.7--20.3	&407		&43		&90.4		&12.5\\
C	&16.5--18.9	&351		&15		&95.9		&5.0\\
D	&14.4--17.8	&392		&16		&96.1		&20\\
E	&8.1--12.6	&482		&16		&96.8		&23\\
F	&7.9--11.1	&399		&13		&96.8		&18\\
\hline
\end{tabular}
\end{table}

The potential LOS differ slightly in the $Q$ factor, but 
strongly in the occurrence of abrupt changes in their slope 
(\reffig{fig_los_frank} and Tab.~\ref{tab_q_frank}).
The LOS in \reffig{fig_los_frank}C has the smallest 
$\langle \partial_t^2\rangle$ and could be, therefore,
a good LOS for the dating
of the Olguita profile. Regarding this result, the Olguita profile
would have an age between 16.5 and 18.9~Myr and an age-depth relation
as it is represented by the LOS in \reffig{fig_los_frank}C.
Warkus' investigation reveals the same result \citep{warkus2002}, 
although he also mentioned that
the dating based on the polarity data is ambiguous.
The alternative profile A has a similar good $Q$ value,
but reveals an abrupt change in its slope around 20~Myr.

The stated results are only
potential sequences and do not lay claim to absolute correctness.
The example was presented just to illustrate the potentials of CRPs.
In general, for such geological tasks as presented in the two previous 
applications, the distance matrix, Eq.~(\ref{eq_DP}), might be more 
appropriate.

%% file: appl_crp_synchro.tex
%
%
%
%

Now we apply CRPs to geology and palaeo-climatology, 
where, as we have already seen, data are characterised by 
short length and non-stationarity.
Such kind of data is rather typical in earth science,
because of the unique character of, e.\,g., outcrops or drilling cores,
which does not usually allow to repeat or refine a measurement.

A higher variability in rainfall and river discharge was discussed
to be a reason for a sudden increase of the amount of landslides in 
NW Argentina 30,000~$^{14}$C years ago. A potential cause of 
the higher variability in rainfall is the El Ni\~no/Southern 
Oscillation, represented by the Southern 
Oscillation Index (SOI). In order to support this hypothesis, annual layered
lake sediments from the Santa Maria Basin (Province
Salta, NW Argentina) with an age of 30,000~$^{14}$C years
were compared with the El Ni\~no dynamics from today.
The colour variation of the lake sediments comes from
reworked older sediments which are eroded and deposited only during
extreme rainfall events, and, therefore, provides an 
archive of the precipitation variability \citep{trauth99,trauth2000}. 

We compare the present-day SOI data with the
palaeo-rainfall data in two steps. At first, the
similarity between the SOI and modern rainfall is 
studied by using CRPs. Then, the similarity structure derived
from the CRP of the SOI and the palaeo-rainfall is
computed and compared with the similarity structure 
between the present-day data \cite{marwan2003climdyn}.

\begin{figure}[tb]
\centering \includegraphics[width=.6\columnwidth]{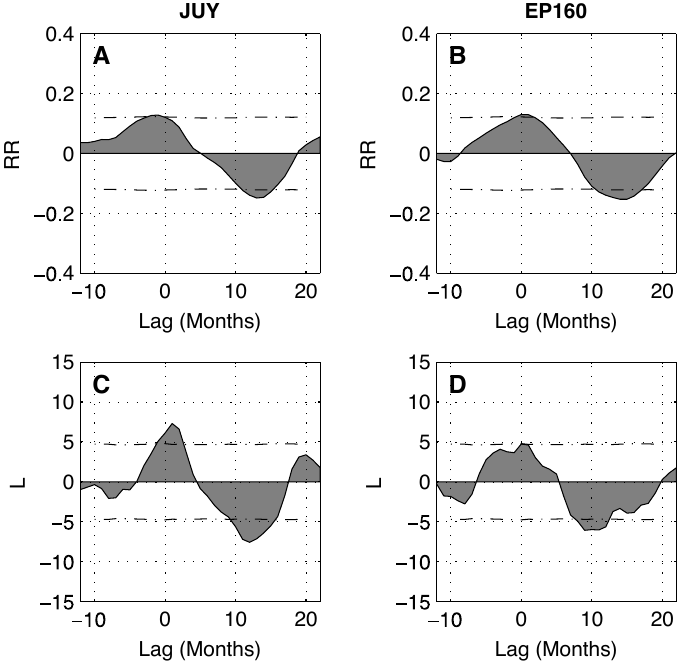} 
\caption{$RR^c_{\tau}$ and $L^c_{\tau}$ measures of the CRPs between
SOI and precipitation in Jujuy (A, C) and palaeo-precipitation (B, D). 
Extreme values reveal high
similarity between the dynamics of the rainfall and the ENSO
\cite{marwan2003climdyn}.}\label{figs_crqa_varves}
\end{figure}

For the assessment of the modern El Ni\~no/Southern 
Oscillation influence on local rainfall in NW
Argentina, the monthly precipitation data from the station San
Salvador de Jujuy (JUY)
is used. The CRP analysis of JUY and SOI reveals clear positive values
for the measures $RR^c_{\tau}$ and $L^c_{\tau}$
around a lag of zero and negative values after  8--12~months, which
suggests a significant link between Jujuy rainfall and the El Ni\~no/Southern 
Oscillation (~\reffig{figs_crqa_varves}\,A, C).  
The comparison between SOI and the 30,000~$^{14}$C year old precipitation 
data yields one maximum and two minimum values
for $RR^c_{\tau}$ and $L^c_{\tau}$ for delays of about zero and ten months,
similar to those found for JUY  (\reffig{figs_crqa_varves}B, D).

The similarity between the time series of the modern rainfall data
and the palaeo-precipitation record from the lake sediments suggests
that an El Ni\~no-like oscillation was active around 30,000~$^{14}$C years
ago (roughly corresponding to 34,000~cal.~years BP), which corresponds
with the results of the investigation of Coccolithophores production
\citep{beaufort2001}. In the semiarid basins of the NW Argentine
Andes, the El Ni\~no-like  variation could have caused significant
fluctuations in local rainfall around 30,000~$^{14}$C years ago
similar to modern conditions, and, hence, could help to explain more frequent
landsliding approximately 34,000~years ago in the semiarid basins of
the Central Andes.

%% file: appl_planets.tex
%
%
%
%

The stability of extra-solar planetary systems is
a central question of astrobiology. This kind of studies are important for future
space missions  dedicated to find habitable
terrestrial planets in other stellar systems.
The extra-solar planetary systems
Gl~777~A, HD~72659, Gl~614, 47~Uma and HD~4208 are
examined using extensive numerical experiments, 
concerning the question of whether they could host
terrestrial-like planets in their habitable zones
(HZ) \cite{asghari2004}. 

Besides the study of the mean motion resonance
between fictitious terrestrial planets and the
existing gas giants in these five extra-solar systems, 
the stability of their orbits are investigated.  A
fine grid of initial conditions for a potential 
terrestrial planet within the HZ is chosen for each
system, from which the stability of orbits is then
assessed by direct integration over a time interval
of one million years. For each of the five systems the
two-dimensional grid of initial conditions contains 80
eccentricity points for the Jovian planet and up to
160 semimajor axis points for the fictitious planet.
The equations of motion are integrated using a
Lie-series integration method with an adaptive step
size control. 

The stability of orbits is examined by means of two different methods: $K_2$
estimated from RPs (Subsec.~\ref{sec:k2_d2}) and 
the maximum eccentricity achieved by the planet over the one million
year integration. The eccentricity is an indication of
the habitability of a terrestrial planet in the HZ;
any value of $e>0.2$ produces a significant
temperature difference on a planet's surface between
apoapse and periapse.
Here we summarise the results obtained for the
extra-solar planetary system Gl~777~A, which is
a wide binary with a very large separation (3000~AU).
Hence, there is no need to take the
perturbations of the very far companion into account.

\begin{figure}[htbp] 
\centering
\includegraphics[width=.7\textwidth]{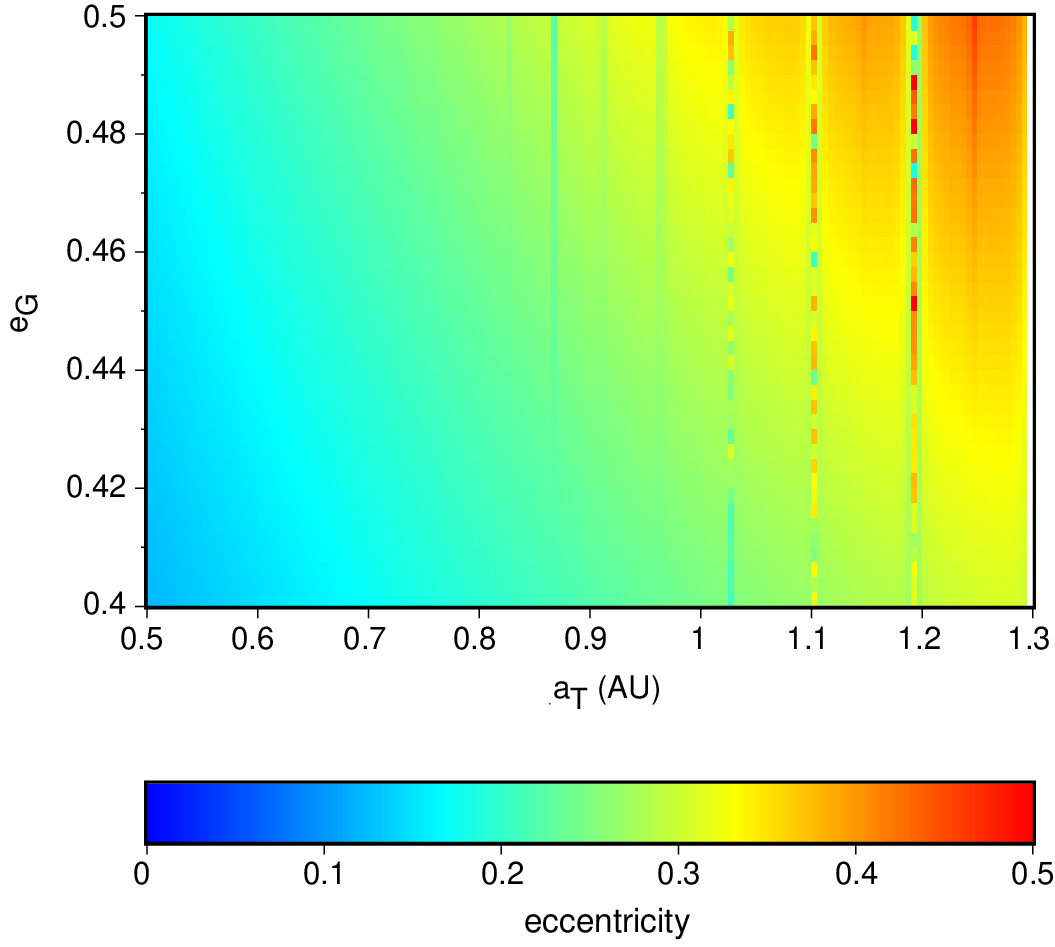}
\caption{Initial condition diagram for fictitious
planets in the system Gl~777~A: initial semimajor axes
of the planet versus the eccentricity of the Jovian
planet. The maximum eccentricity of an orbit during
its dynamical evolution is colour-coded.
} \label{fig_planets_max_ecc}
\end{figure}

\begin{figure}[htbp]
\centering
\includegraphics[width=.7\textwidth]{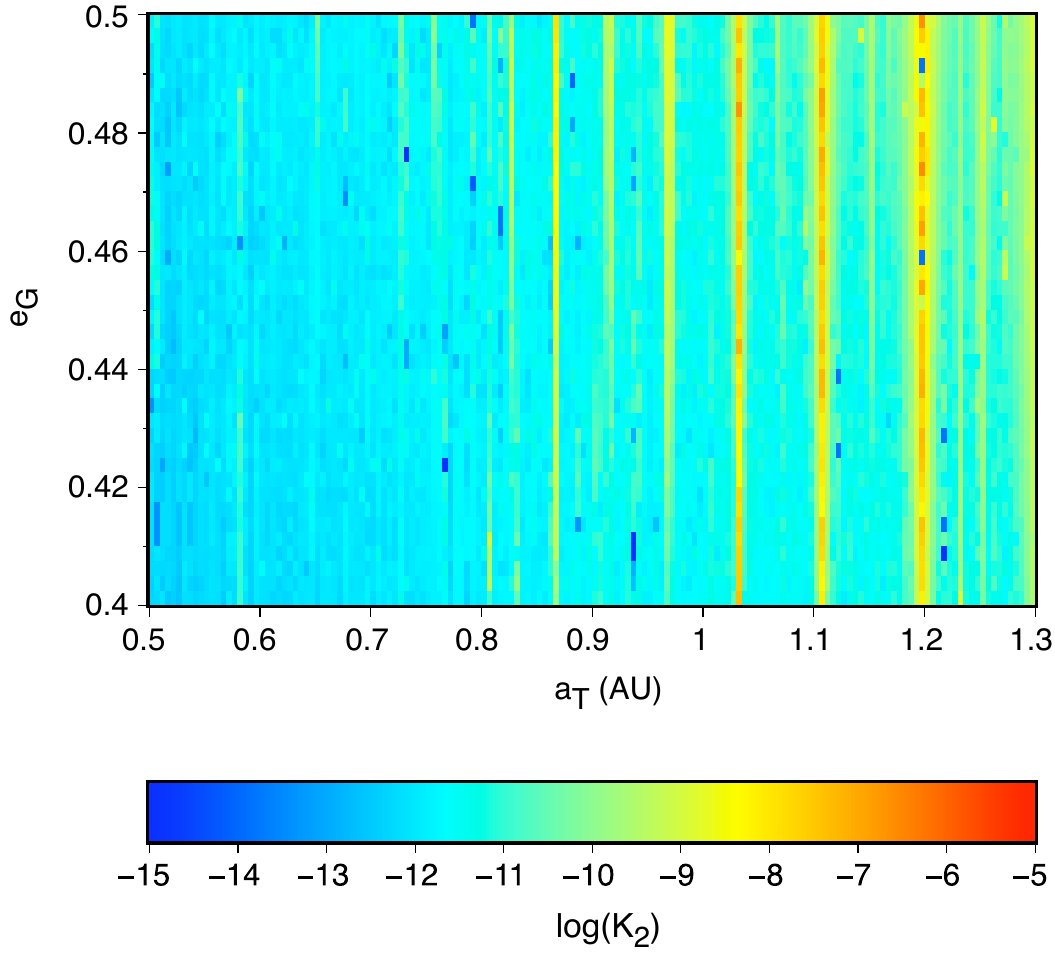}
\caption{Initial condition diagram for fictitious
planets in the system Gl~777~A: initial  semimajor
axes of the planet versus the eccentricity of the
Jovian planet. The value of the $K_2$ of an orbit is
colour-coded.} \label{fig_planets_k2} 
\end{figure}

Considering the maximum eccentricity, high order resonance
can be found (Fig.~\ref{fig_planets_max_ecc}). Furthermore, unstable orbits
due to high eccentricity and high semimajor axes values can
be determined (red or yellow colours in Fig.~\ref{fig_planets_max_ecc}). 
The latter feature is due to the larger perturbations near to the existing planet.   
The maximum eccentricity indicates the variable
distance to the central star and consequently it is an
indirect measure of the differential energy flux
(insolation)  on the planet. Therefore, it is possible
to determine where the variation of this distance does
not exceed 50\%, corresponding to an eccentricity of $e=0.2$. 

On the other hand, the correlation entropy $K_2$  is a more sensitive measure
for the predictability of the orbit (Fig.~\ref{fig_planets_k2}). In particular,
high order resonance is clearly indicated using this method, even in the case
that the resonance is acting when the eccentricity of the planet is as low as
$e=0.4$ (bottom of Fig.~\ref{fig_planets_k2}). $K_2$ was estimated using the
automated algorithm described in the Appendix~\ref{apdx:automated_k2}.

Hence, we can conclude that both methods for the quantification of the
stability of the orbits complement each other.
From this analysis it can be inferred, that 
the planets in the system Gl~777~A will last
long enough in the HZ to acquire the necessary
conditions for life in the region with $a < 1$ AU \cite{asghari2004}.

%% file: appl_synchro.tex
%
%
%
%

As mentioned in Subsec.~\ref{sec:Synchro}, many natural and laboratory
systems are ill-phase defined or non-phase-coherent, i.\,e.~they posses
multiple time scales. In such cases, the RP based method for the
synchronisation analysis is appropriate. To illustrate the application
of this method to experimental data, we first show the results of the
synchronisation analysis of two coupled electrochemical oscillators
which exhibit non-phase coherent dynamics \cite{romano2005,kiss2005,kurths2006}. In
this experiment the coupling strength between the electrochemical
oscillators can be systematically varied. Hence, it is an active
experiment.

Second, we apply the recurrence based synchronisation analysis to a
passive experiment. We consider the synchronisation between the left and
right fixational eye movements. The coupling strength
between both eyes cannot be changed systematically, and hence, a
hypothesis test has to be performed to get statistical significant
results.

\subsubsection{Synchronisation of electrochemical oscillators} 

\begin{figure}[b]
\centering
\includegraphics[width=.7\textwidth]{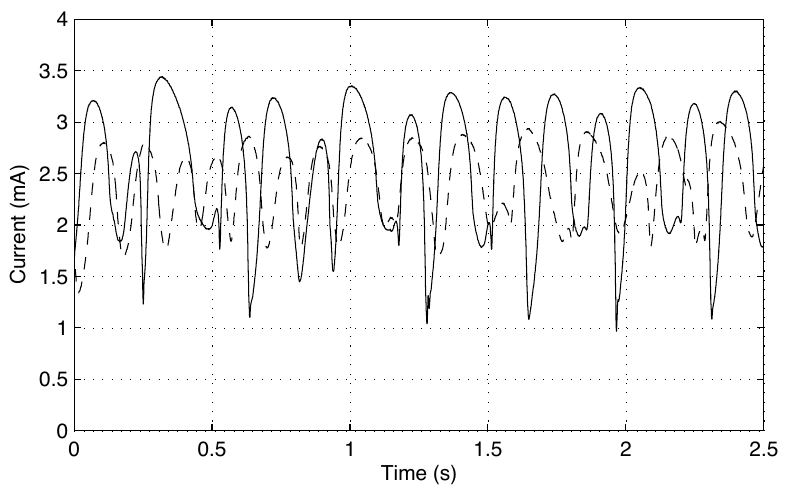}
\caption{Current of two non-phase-coherent electrochemical 
oscillators over time for coupling strength 
$\mu=0$ (solid: first, dashed: second electrode). 
}\label{fig_chem}
\end{figure}

We consider a laboratory experiment in which the synchronisation between
electrochemical oscillators can be studied \cite{kiss2002}. In this experiment,
the electro-dissolution of iron in sulphuric acid cause non-phase coherent,
chaotic current oscillations (Fig.~\ref{fig_chem}).  
A standard three-compartment electrochemical cell consisting of two iron 
working electrodes, a Hg/Hg$_2$ 
SO$_4$/K$_2$SO$_4$ reference electrode, and a Pt mesh counter-electrode is used
(the detailed experimental setup can be found in \cite{kiss2000,kiss2002b}). 
The experiment is carried out in H$_2$SO$_4$.
The applied potential of both electrodes is held at the same 
value using a potentiostat. Zero restiance ammeters are used 
to measure the currents of the electrodes. The
coupling strength $\mu$ between the electrochemical oscillators can be
varied by changing the connected resistors between the electrodes
and the potentiostat \cite{kiss2005}. We compute the two indices $CPR$ and $JPR$ for PS
respectively GS for six different values of the coupling strength
$\mu=\{0.0,0.2,0.4,0.6,0.8,1.0\}$.

%
%

%
%

\begin{figure}[t!]
\centering
\includegraphics[width=.6\textwidth]{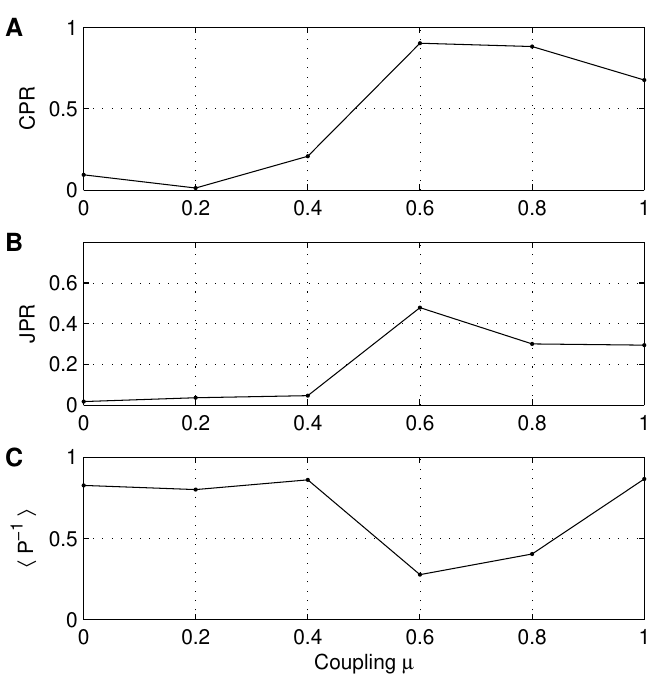}
\caption{
(A) $CPR$ index for PS,
(B) $JPR$ index for GS and
(C) $\langle P^{-1} \rangle$ index for GS (method of mutual 
false nearest neighbors)
in dependence on the coupling strength $\mu$ 
of two non-phase-coherent electrochemical 
oscillators.
}\label{results_chem}
\end{figure}

The transition to PS and GS is detected simultaneously at the coupling
strength $\mu=0.6$ (Fig.~\ref{results_chem}A, B). This is in accordance with
theoretical results, which confirm that the transition to PS and GS occurs
almost simultaneously for non-phase coherent oscillators \cite{osipov2003}.

%
%
%

In contrast, the method of mutual false nearest neighbors 
(MFNN) \cite{rulkov1995,arnhold1999,boccaletti2000} does not 
yield plausible results in this case. The index $\langle P^{-1} \rangle$ 
is the inverse of the mean value of the MFNN parameter. This 
index is zero for systems that are not in GS, and approximately 
one for systems in GS \cite{rulkov1995,boccaletti2000,arnhold1999}. 
The index $\bar P^{-1}$ indicates GS for no coupling,$\mu=0$ (Fig.~\ref{results_chem}C),
but for coupling strength $\mu=0.6$, where the other two 
indices $CPR$ and $JPR$ indicate synchronisation 
(Fig.~\ref{results_chem}A, B), $\langle P^{-1} \rangle$ does 
not. Hence, in the case of non-coherent oscillators the method of
MFNN does not yield reliable results.
(We used a total of 25,000 data 
points, $T_r=T_d=0.035~s$, i.\,e.~a sampling rate 
of 2~kHz), $d_r=5$ and $d_d=12$. The measure
$P(n,d_r,d_d)$ was computed at 10,000 different locations 
on the attractor and was used to evaluate the 
average values $\langle \bar P(d_r,d_d) \rangle$.)

\subsubsection{Synchronisation analysis of cognitive data}
Now we present the application of the RP based synchronisation analysis 
to a passive experiment, namely, the synchronisation between 
right and left eye movements during the fixation on one point. 
In this cognitive experiment, the coupling strength between both 
eyes cannot be varied systematically. Hence, a hypothesis test based 
on twin surrogates (Subsec.~\ref{sec:Surrogates}) is performed to 
get the statistical significance of the obtained results.

During fixation of a stationary target, our eyes perform small
involuntary and allegedly erratic movements to counteract
retinal adaptation. If these eye movements are experimentally
suppressed, retinal adaptation to the constant input induces
very rapid perceptual fading \cite{riggs53}. 
The fixational movements of the
left and right eye are correlated very poorly at best
\cite{ciuffreda95}. Therefore, it is highly desirable to
examine these processes from a perspective of PS. 

\begin{figure}[t]
\centering
\includegraphics[width=\textwidth]{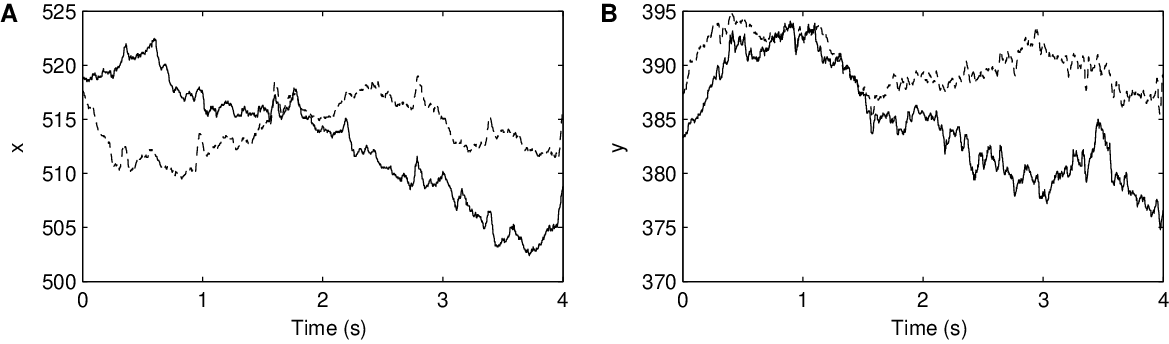}
\caption{
Simultaneous recording of left (solid) and right
(dashed) fixational eye movements: (A) horizontal component (B)
vertical component.}\label{fig_eyes_data}
\end{figure}

The analysis of several trials and subjects has been  presented in
\cite{romano2006wiley}; here we concentrate  on the results for one
subject and one trial. In each trial the subject fixates a small
stimulus (black square on a white background, $3 \times 3$ pixels on a
computer display) with a spatial extent of $0.12^\circ$.
Eye movements are achieved using an EyeLink-II system (SR Research,
Osgoode, Ontario, Canada) with a sampling rate of  500~Hz and an
instrumental spatial resolution less than $0.005^\circ$.  The horizontal
and vertical component of  the eye movements are recorded
(Fig.~\ref{fig_eyes_data}).

\begin{figure}[th]
\centering \includegraphics[width=1.0\textwidth]{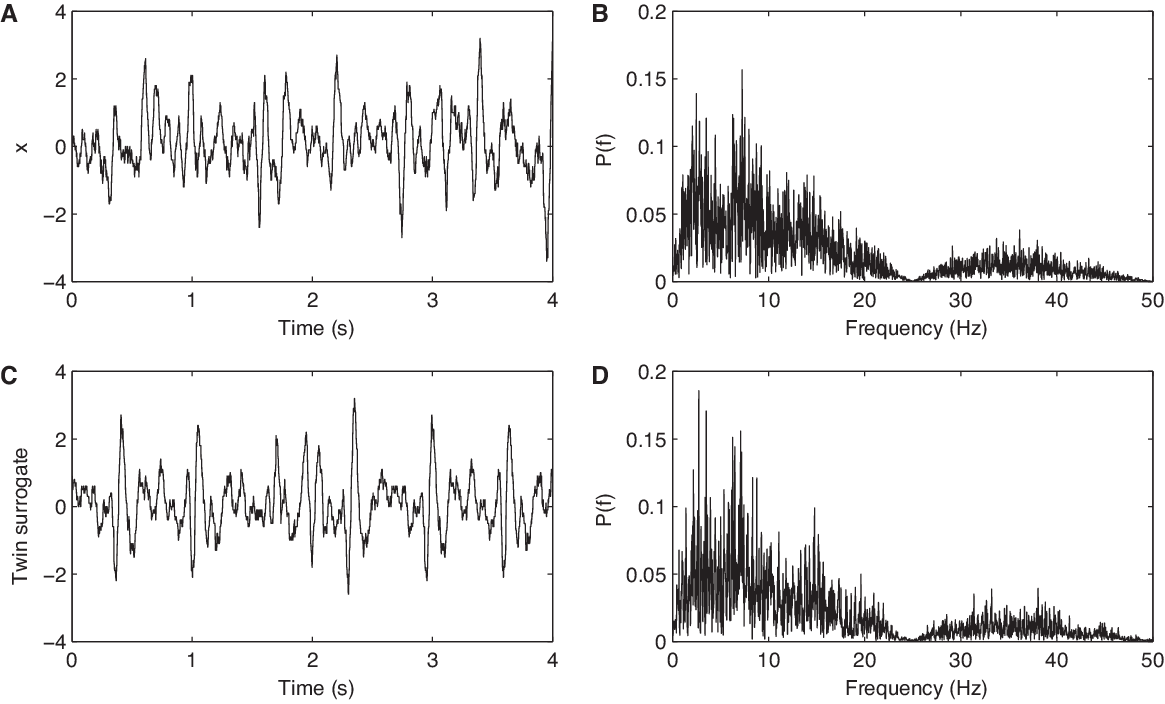}
\caption{
Filtered horizontal component of the left eye
of one participant (A) and its corresponding periodogram (B).
(C) Horizontal component of one surrogate of the left
eye and (D) its corresponding periodogram. }
\label{fig_eyes_surrogate}
\end{figure}

The data are first high-pass filtered applying a difference
filter $\tilde x(t)=x(t)-x(t-\tau)$ with  $\tau=40$~ms in order
to eliminate the slow drift of the data. After this filtering,
the trajectory is oscillating with maximum spectral
power in the frequency range between 3 and 8~Hz
(Fig.~\ref{fig_eyes_surrogate}A, B). However, the trajectories of
the eye movements are rather noisy and non-phase coherent. Therefore, it
is cumbersome to estimate the phase of these data and, hence, the
application of the recurrence based measure $CPR$ is appropriate.

For one trial of one participant, we obtain $CPR=0.911$. Even 
though this value is high (the maximal value that $CPR$ can take 
is 1), a hypothesis test should be performed in order to get statistically significant results.
The hypothesis test is carried out by computing
200 twin surrogates of the left eye's trajectory 
(cf.~Subsec.~\ref{sec:Surrogates}). At a first glance, the
characteristics of the original time series are well reproduced
by the twin surrogate (Fig.~\ref{fig_eyes_surrogate}C).
The structure of the corresponding periodogram 
is also qualitatively reproduced (Fig.~\ref{fig_eyes_surrogate}D).
Note that the periodogram of the twin surrogate is not
identical with the one of the original time series. 
This is because the twin surrogates correspond 
to another realisation of
the same underlying process (respectively another trajectory
starting at different initial conditions of the same underlying
dynamical system) and the periodiograms of different realisation of a 
process differ slightly.

\begin{figure}[b]
\centering \includegraphics[width=.6\textwidth]{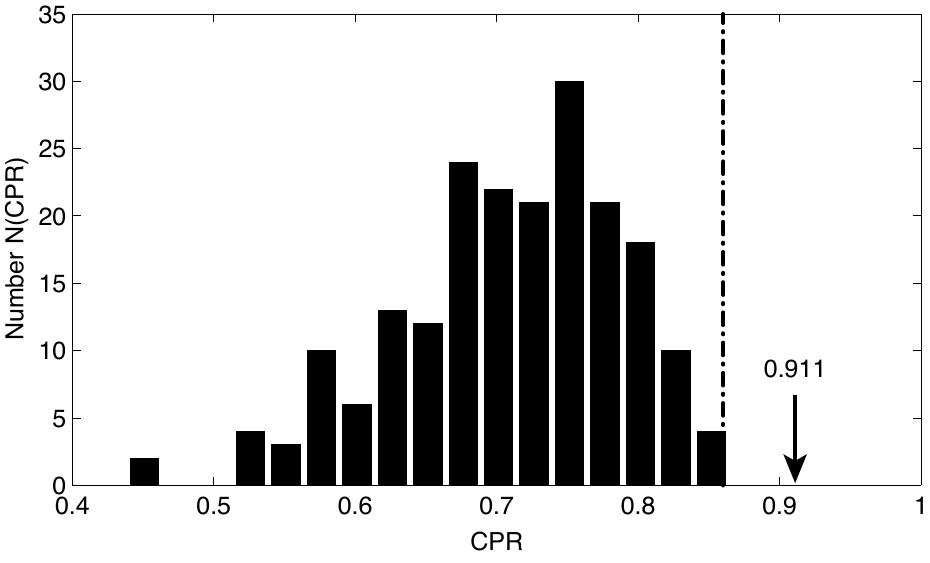}
\caption{Histogram of the PS index $CPR$ of 200 twin surrogates
(for one trial of one participant). The $CPR$ 
index for the original data (arrow) is significantly 
different from the one of the surrogates (dashed: 99\% 
significance border of the rank order statistics).}\label{fig_eyes_test}
\end{figure}

Then, the synchronisation index
$CPR$ between each twin surrogate of the left eye and
the measured right eye's trajectory is computed.
The PS index of the original data is
significantly different ($p<0.01$) from those of the surrogates
(Fig.~\ref{fig_eyes_test}), which
strongly indicates that the concept of PS can be successfully
applied to study the interaction between the trajectories of
the left and right eye during fixation. This result also
suggests that the physiological mechanism in the brain that
produces the fixational eye movements controls both eyes
simultaneously, i.\,e.~there might be only one centre in the
brain that produces the fixational movements in both eyes or a
close link between two centres. This finding of PS between left
and right eyes is in good agreement with current knowledge of
the physiology of the oculomotor circuitry.


%% file: apdx_models.tex
Here we list the mathematical models used in the examples of this paper:

\begin{itemize}
\item 
Auto-regressive process first order
\begin{equation}\label{eq_ar1}
x_{i} = a\,x_{i-1} + \xi_i,
\end{equation}
where $\xi_i$ is white noise.

\item 
2D auto-regressive process second order
\begin{equation}\label{eq_2D-ar2}
x_{i, j} = \sum_{k,l=1}^2 a_{k,l}\,x_{i-k,j-l} + \xi,
\end{equation}
where $\xi$ is white noise.

\item 
The logistic map
\begin{equation}\label{eq_logistic_map}
x_{i}=a\,x_{i-1}\left( 1-x_{i-1}\right).
\end{equation}

\item
The Bernoulli map
\begin{equation}\label{eq_bernoulli}
x_{i}=2\,x_{i-1}\mod(1).
\end{equation}

\item
The R\"ossler system \cite{roessler1976}
\begin{eqnarray}\label{eq_roessler}
\dot x &=& - y - z, \nonumber \\ 
\dot y &=& x + a \, y, \\ 
\dot z &=& b + z \, (x - c).\nonumber
\end{eqnarray} 

\item
The Lorenz system \cite{lorenz63}
\begin{eqnarray}\label{eq_lorenz}
\dot x &=& -\sigma(x-y), \nonumber\\
\dot y &=& -xz+rx-y,\\
\dot z &=& xy-bz.\nonumber
\end{eqnarray}

\item
Two R\"ossler systems mutually coupled by the first component
\begin{eqnarray} \label{eq_2roessler1}
\dot x_1 &=& - (1+\nu) x_2 - x_3 + \mu(y_1-x_1), \nonumber \\ 
\dot x_2 &=& (1+\nu) x_1 + a \, x_2,\\ 
\dot x_3 &=& b + x_3 \, (x_1 - c), \nonumber 
\end{eqnarray} 
\begin{eqnarray}\label{eq_2roessler2}
\dot y_1 &=& - (1-\nu) y_2 - y_3 + \mu(x_1-y_1), \nonumber \\ 
\dot y_2 &=& (1-\nu) y_1 + a \, y_2, \\ 
\dot y_3 &=& b + y_3 \, (y_1 - c).\nonumber
\end{eqnarray} 

\item
Two R\"ossler systems mutually coupled by the second component
\begin{eqnarray} \label{eq_2roessler2_1}
\dot x_1 &=& - (1+\nu) x_2 - x_3, \nonumber \\ 
\dot x_2 &=& (1+\nu) x_1 + a \, x_2+\mu(y_{2}-x_{2}),\\ 
\dot x_3 &=& b + x_3 \, (x_1 - c), \nonumber 
\end{eqnarray} 
\begin{eqnarray}\label{eq_2roessler2_2}
\dot y_1 &=& - (1-\nu) y_2 - y_3 , \nonumber \\ 
\dot y_2 &=& (1-\nu) y_1 + a \, y_2+\mu(x_{2}-y_{2}), \\ 
\dot y_3 &=& b + y_3 \, (y_1 - c).\nonumber
\end{eqnarray}

\item
The Lorenz system driven by a R\"ossler system
\begin{eqnarray}\label{roess_driver}
\dot x_1&=&b+x_1(x_2-c),\nonumber\\
\dot x_2&=&-x_1-x_3,\\
\dot x_3&=&x_2+a x_3,\nonumber
\end{eqnarray}

\begin{eqnarray}\label{lorenz_driven}
\dot y_1&=&-\sigma(y_1-y_2),\nonumber\\
\dot y_2&=&r\,u-y_2-u\,y_3,\\
\dot y_3&=&u\,y_2-b\,y_3,\nonumber
\end{eqnarray}
where $u=x_1+x_2+x_3$.
\end{itemize}

%% file: apdx_algorithms.tex
\subsection{Algorithm to fit the LOS}\label{apdx:LOS}

The extraction of the LOS from the CRP in 
Subsecs.~\ref{sec:CrossRecurrencePlots} and \ref{sec:appl_rescaling}
was performed by using the following simple two-step algorithm. 
The set of indices of recurrence points 
$\mathbf{R}_{i_{\alpha}, j_{\beta}} = 1$ is denoted by
$\{(i_{\alpha}, j_{\beta})\}_{\alpha, \beta \in \mathds{N}}$.
Those recurrence points $\mathbf{R}_{i_{\tilde \alpha}, j_{\tilde \beta}}$
belonging to the LOS are denoted by 
$\{(i_{\tilde \alpha}, j_{\tilde \beta})\}_{\tilde \alpha, \tilde \beta \in \mathds{N}}$.

\begin{enumerate}
\item Find the recurrence point $(i_1,j_1)$ next to the axes origin. This is
      the first point of the LOS.
\item Find the next recurrence point at $(i_{\alpha},j_{\beta})$ 
      after a previous determined LOS point $(i_{\tilde \alpha},j_{\tilde \beta})$
      by looking for recurrence points in a squared window of size $w=2$, 
      located at $(i_{\tilde \alpha},j_{\tilde \beta})$.
      If the edge of the window meets a recurrence point $(i_{\alpha},j_{\beta})$,
      we follow we step (\ref{crp_algorithm_w_end}), else we iteratively increase the size of the window.
      \label{crp_algorithm_w_start}
\item If there are subsequent recurrence points in $y$-direction 
($x$-direction), the size $w$ of the window is iteratively increased
in $y$-direction ($x$-direction) until a predefined size 
$(w+\Delta w) \times (w+\Delta w)$
($\Delta w < dx, \Delta w < dy$) or until no new recurrence points are
met. Using $\Delta w$ we compute 
the next LOS point $(i_{\tilde \alpha+1},j_{\tilde \beta+1})$ by 
determination of the centre of mass of the cluster of recurrence points
with $i_{\tilde \alpha+1}=i_{\tilde \alpha}+(w + \Delta w)/2$ and 
$j_{\tilde \beta+1}=j_{\tilde \beta}+(w + \Delta w)/2$.
This avoids that the algorithm places the LOS around 
widespread areas of recurrence points, but locates the LOS 
within these areas. This can be controlled by
the two additional parameter $dx$ and $dy$.
The next step is to set the LOS point 
$(i_{\tilde \alpha+1},j_{\tilde \beta+1})$ to the new 
starting point and to return to the step (\ref{crp_algorithm_w_start}).
These steps are repeated until the end of the RP is reached.
      \label{crp_algorithm_w_end}
\end{enumerate}

This algorithm is merely one of many
possibilities. Its application should be carefully performed.
The following criteria should be met in order
to obtain a good LOS. The amount of targeted recurrence points by
the LOS $N_{\bullet}$ should converge to the maximum and the amount of gaps in
the LOS $N_{\circ}$ should converge to the minimum. An analysis with various
estimated LOS confirms this requirement. The correlation between 
two LOS-synchronised data series arises with $N_{\bullet}$
and with $1/N_{\circ}$ (the correlation coefficient correlates 
best with the ratio $N_{\bullet}/N_{\circ}$).

This algorithm for the reconstruction of the LOS is implemented
in the CRP toolbox for Matlab$^{\tiny{\textregistered}}$
(provided by TOCSY: http://tocsy.agnld.uni-potsdam.de).

\subsection{Algorithm for the reconstruction of a time series from its RP}\label{apdx:alg_recons}

The reconstruction algorithm (cf.~Subsec.~\ref{sec:Reconstruction}) consists of three main phases 
and a total of nine steps: 
\renewcommand{\theenumi}{\Roman{enumi}}
\renewcommand{\labelenumi}{\theenumi.}
\renewcommand{\theenumii}{\arabic{enumii}}
\renewcommand{\labelenumii}{(\theenumii)}
\renewcommand{\theenumiii}{\Alph{enumiii}}
\renewcommand{\labelenumiii}{\theenumiii: }
\begin{enumerate} 
\item Determine the set difference $\mathcal{R}_{j} \setminus \mathcal{R}_{i}$
\begin{enumerate} 
\item If $n$ columns of the matrix $\mathbf R_{i,j}$ are identical, 
store their indices and remove $n-1$ of them, so that every column is
unique. \label{algorithm_iteration_doublecols}
\item Determine for each recurrence point at $(i,j)$ the number of
neighbours of $x_j$ which are not neighbours of $x_i$, i.\,e.~the
number of elements in the set difference $\mathcal{R}_{j} \setminus \mathcal{R}_{i}$. 
This number is denoted as $\mathbf{N}_{i,j}$ and can be calculated by
$\sum_k (1-\mathbf{R}_{i,k})\,\mathbf{R}_{j,k}$.
\item There exist exactly two points at $j_1$ and $j_2$, such  
that $\mathbf{N}_{i,j_{1/2}}=0 \quad \forall i$. 
These two points correspond to the maximum and the minimum of the time
series. Choose one of these two indices as starting point $k$
and assign the first element of the rank order series $r_1^{\text{ord}} = k$.
\end{enumerate} 

\item Iteration
\begin{enumerate} 
\setcounter{enumii}{3}
\item Look for the position of the minimum in the set 
$\{\mathbf{N}_{i,k}\}_i$. \label{algorithm_iteration_start}
\begin{enumerate} 
    \item If the minimum is unique, i.\,e.~there is one $j$ such that
    $\mathbf{N}_{j,k}<\mathbf{N}_{i,k}\quad \forall i\ne j$, 
    the next point is then $k=j$.
    \item If the minimum is not unique, i.\,e.~there is a set of 
    $m$ indices $\{j_1,\ldots,j_m\}$, so that 
    $\mathbf{N}_{j_1,k}=\ldots=\mathbf{N}_{j_m,k}\le \mathbf{N}_{i,k} \quad \forall i$,
    look for the position of the minimum in the set $\{\mathbf{N}_{k,j_i}\}_i$. This
    position will be the next point $k$.
\end{enumerate}
\item Add $k$ to the rank order series $r_i^{\text{ord}}$.
\item Go to step (\ref{algorithm_iteration_start}) until all indices are ranked. 
\end{enumerate} 

\item Final Reconstruction 
\begin{enumerate} 
\setcounter{enumii}{6}
\item Generate random numbers so that for each entry in the 
ordered series $r_i^{\text{ord}}$ is one random number, and rank order these numbers
to the series $y_i$. 
\item Generate the time series $x_i$ by assigning 
$x_{r_i^{\text{ord}}} = y_i$.
\item Reintroduce at the position of the ``identical columns''
obtained in step (\ref{algorithm_iteration_doublecols}) 
the values of the points at the corresponding
indices which remained in the RP. 
\end{enumerate} 
\end{enumerate} 

\renewcommand{\theenumi}{\arabic{enumi}}
\renewcommand{\labelenumi}{(\theenumi)}
\renewcommand{\theenumii}{\alph{enumii}}
\renewcommand{\labelenumii}{(\theenumii)}
\renewcommand{\theenumiii}{\roman{enumiii}}
\renewcommand{\labelenumiii}{(\theenumiii)}

\subsection{Twin Surrogates algorithm}\label{apdx:alg_twins}
A twin surrogate (cf.~Subsec.~\ref{sec:Surrogates}) trajectory 
$\vec {x}^s_i$ of $\{\vec {x}_i \}_{i=1}^N$ is 
generated in the following way: 

\renewcommand{\theenumii}{\Alph{enumii}}
\renewcommand{\labelenumii}{\theenumii: }

\begin{enumerate}
\item Identify all pairs of twins. 
\item Choose an arbitrary starting point, 
      e.\,g.~$\vec{x}^s_1=\vec{x}_j$. Set index $i=2$.
\item \label{algorithm_surrogate_start}
\begin{enumerate}
    \item If $\vec{x}_j$ has no twin, the next point of the surrogate
          trajectory is $\vec{x}^s_i=\vec{x}_{j+1}$. 
    \item If $\vec{x}_j$ has a twin at $\vec{x}_k$, either proceed with
          $\vec{x}^s_i=\vec{x}_{j+1}$ or $\vec{x}^s_i=\vec{x}_{k+1}$ 
          with equal probability (if triplets occur proceed analogously).
\end{enumerate}
\item Increase $i = i + 1$ and go back to step (\ref{algorithm_surrogate_start}) 
      until the surrogate time
      series has the same length as the original one.
\end{enumerate}

\renewcommand{\theenumii}{\alph{enumii}}
\renewcommand{\labelenumii}{(\theenumii)}

\subsection{Automatisation of the $K_2$ estimation by RPs}\label{apdx:automated_k2}
For many applications, e.\,g., if spatio-temporal data
has to be analysed, it is desirable to automate the
algorithm to estimate $K_2$ based on RPs
(Subsec.~\ref{sec:k2_d2}). Such an automated algorithm
is also more objective, as otherwise the choice of the
proper scaling regions of $p_c(\varepsilon,l)$ (Eq.~\ref{eq_estimator_k2}) depends
to some extent on the choice of the data analyst.

For the practical application, at first, 
the cumulative distribution of diagonals
$p_c(\varepsilon,l)$ (the probability to find a diagonal of at 
least length $l$ in the RP) has to be calculated for different thresholds
$\varepsilon$. The question arises, which values of
$\varepsilon$ should be considered. As each system has
its proper amplitude, which may differ from one system
to another one, the choice will be different for each
case and it is subjected to some arbitrariness. To
overcome this problem, the value of the
recurrence rate $RR$, Eq.~(\ref{eq_rr}), should be fixed, because it is normalised, and
then calculate the corresponding $\varepsilon$ 
(cf.~Subsec.~\ref{sec:Selectionofthethreshold}).  
Using this choice of $\varepsilon$, the arbitrariness of choosing
appropriate values for $\varepsilon$ is avoided
and the same procedure of the estimation of 
$p_c(\varepsilon,l)$ can be applied for all systems. 

The next step is crucial for the automatisation: the
scaling region of $\ln p_c(RR,l)$ vs.~$l$ and the
plateau in $K_2(RR)$ vs.~$RR$ must be estimated
automatically. In both cases it is possible to apply a
cluster dissection algorithm \cite{spaeth92}, which
divides the set of points into distinct clusters. In
each cluster a linear regression is performed. The
algorithm minimises the sum of all square residuals in
order to determine the scaling region and the plateau.
To find both regions automatically, the following
settings have been found to be appropriate
\cite{asghari2004}: 

\begin{itemize} 
\item Only
diagonal lines up to a fixed length $l_{\max}$ are
considered. Longer lines are excluded because of
finite size effects. Reasonable values of $l_{\max}$
are at about 10\% of the length of the time series.
\item Only values of $p_c(RR,l)$ with $N^2 p_c(RR,l)> 500$
are regarded to obtain a reliable statistic. 

\item
About 100 different values for $\varepsilon$ might be
considered, corresponding to 100 equally spaced
recurrence rates $RR$ between 1\% and 99\%, to have a
good defined plateau in $K_2(RR)$ vs.~$RR$. 
\item
Furthermore,  the number of clusters have to be
specified when applying the cluster dissection
algorithm: for the detection of the scaling region in
$\ln p_c(RR,l)$ vs.~$l$, two different clusters seem to
be a rather good choice. Then, the slope of the
largest cluster should be used. For the detection of the plateau in
$K_2(RR)$ vs.~$RR$, three clusters should be chosen and the
value of the cluster with the minimum absolute slope is then
used.
\end{itemize}